\documentclass[longauth,traditabstract]{aa}

\usepackage[nonamebreak]{natbib}
\usepackage[stable]{footmisc}
\usepackage{amsmath}
\usepackage{amssymb}
\usepackage{txfonts}
\usepackage{placeins}
\usepackage{natbib}
\bibpunct{(}{)}{;}{a}{}{,} %
\usepackage{graphicx}
\usepackage{epstopdf}
\usepackage{ifthen}
\usepackage[table,usenames,dvipsnames]{xcolor}
\definecolor{linkcolor}{rgb}{0.6,0,0}
\definecolor{citecolor}{rgb}{0,0,0.75}
\definecolor{urlcolor}{rgb}{0.12,0.46,0.7}
\usepackage[breaklinks, colorlinks, urlcolor=urlcolor, linkcolor=linkcolor,citecolor=citecolor,pdfencoding=auto]{hyperref}
\hypersetup{linktocpage}
\usepackage{overpic}
\usepackage{fixltx2e}
\usepackage{rotating}
\usepackage{booktabs}
\usepackage{enumitem}
\usepackage{etoolbox}

\usepackage[flushleft]{threeparttable}
\usepackage{multirow}

\def\setsymbol#1#2{\expandafter\def\csname #1\endcsname{#2}}
\def\getsymbol#1{\csname #1\endcsname}

\def\Planck{\textit{Planck}}

\newbox\tablebox    \newdimen\tablewidth
\def\leaderfil{\leaders\hbox to 5pt{\hss.\hss}\hfil}
\def\endPlancktable{\tablewidth=\columnwidth 
    $$\hss\copy\tablebox\hss$$
    \vskip-\lastskip\vskip -2pt}

\def\tablenote#1 #2\par{\begingroup \parindent=0.8em
    \abovedisplayshortskip=0pt\belowdisplayshortskip=0pt
    \noindent
    $$\hss\vbox{\hsize\tablewidth \hangindent=\parindent \hangafter=1 \noindent
    \hbox to \parindent{$^#1$\hss}\strut#2\strut\par}\hss$$
    \endgroup}
\def\doubleline{\vskip 3pt\hrule \vskip 1.5pt \hrule \vskip 5pt}

\def\L2{\ifmmode L_2\else $L_2$\fi}

\def\DeltaT{\ifmmode \Delta T\else $\Delta T$\fi}
\def\deltat{\ifmmode \Delta t\else $\Delta t$\fi}
\def\fknee{\ifmmode f_{\rm knee}\else $f_{\rm knee}$\fi}
\def\Fmax{\ifmmode F_{\rm max}\else $F_{\rm max}$\fi}
\def\solar{\ifmmode{\rm M}_{\mathord\odot}\else${\rm M}_{\mathord\odot}$\fi}
\def\Msolar{\ifmmode{\rm M}_{\mathord\odot}\else${\rm M}_{\mathord\odot}$\fi}
\def\Lsolar{\ifmmode{\rm L}_{\mathord\odot}\else${\rm L}_{\mathord\odot}$\fi}
\def\inv{\ifmmode^{-1}\else$^{-1}$\fi}
\def\mo{\ifmmode^{-1}\else$^{-1}$\fi}
\def\sup#1{\ifmmode ^{\rm #1}\else $^{\rm #1}$\fi}
\def\expo#1{\ifmmode \times 10^{#1}\else $\times 10^{#1}$\fi}
\def\,{\thinspace}
\def\lsim{\mathrel{\raise .4ex\hbox{\rlap{$<$}\lower 1.2ex\hbox{$\sim$}}}}
\def\gsim{\mathrel{\raise .4ex\hbox{\rlap{$>$}\lower 1.2ex\hbox{$\sim$}}}}

\def\simprop{\mathrel{\raise .4ex\hbox{\rlap{$\propto$}\lower 1.2ex\hbox{$\sim$}}}}
\def\deg{\ifmmode^\circ\else$^\circ$\fi}
\def\pdeg{\ifmmode $\setbox0=\hbox{$^{\circ}$}\rlap{\hskip.11\wd0 .}$^{\circ}
          \else \setbox0=\hbox{$^{\circ}$}\rlap{\hskip.11\wd0 .}$^{\circ}$\fi}
\def\arcs{\ifmmode {^{\scriptstyle\prime\prime}}
          \else $^{\scriptstyle\prime\prime}$\fi}
\def\arcm{\ifmmode {^{\scriptstyle\prime}}
          \else $^{\scriptstyle\prime}$\fi}
\newdimen\sa  \newdimen\sb
\def\parcs{\sa=.07em \sb=.03em
     \ifmmode \hbox{\rlap{.}}^{\scriptstyle\prime\kern -\sb\prime}\hbox{\kern -\sa}
     \else \rlap{.}$^{\scriptstyle\prime\kern -\sb\prime}$\kern -\sa\fi}
\def\parcm{\sa=.08em \sb=.03em
     \ifmmode \hbox{\rlap{.}\kern\sa}^{\scriptstyle\prime}\hbox{\kern-\sb}
     \else \rlap{.}\kern\sa$^{\scriptstyle\prime}$\kern-\sb\fi}
\def\ra[#1 #2 #3.#4]{#1\sup{h}#2\sup{m}#3\sup{s}\llap.#4}
\def\dec[#1 #2 #3.#4]{#1\deg#2\arcm#3\arcs\llap.#4}
\def\deco[#1 #2 #3]{#1\deg#2\arcm#3\arcs}
\def\rra[#1 #2]{#1\sup{h}#2\sup{m}}

\def\dots{\relax\ifmmode \ldots\else $\ldots$\fi}
\def\WHzsr{\ifmmode $W\,Hz\mo\,sr\mo$\else W\,Hz\mo\,sr\mo\fi}
\def\mHz{\ifmmode $\,mHz$\else \,mHz\fi}
\def\GHz{\ifmmode $\,GHz$\else \,GHz\fi}
\def\mKs{\ifmmode $\,mK\,s$^{1/2}\else \,mK\,s$^{1/2}$\fi}
\def\muKs{\ifmmode \,\mu$K\,s$^{1/2}\else \,$\mu$K\,s$^{1/2}$\fi}
\def\muKRJs{\ifmmode \,\mu$K$_{\rm RJ}$\,s$^{1/2}\else \,$\mu$K$_{\rm RJ}$\,s$^{1/2}$\fi}
\def\muKHz{\ifmmode \,\mu$K\,Hz$^{-1/2}\else \,$\mu$K\,Hz$^{-1/2}$\fi}
\def\MJysr{\ifmmode \,$MJy\,sr\mo$\else \,MJy\,sr\mo\fi}
\def\MJysrmK{\ifmmode \,$MJy\,sr\mo$\,mK$_{\rm CMB}\mo\else \,MJy\,sr\mo\,mK$_{\rm CMB}\mo$\fi}
\def\microns{\ifmmode \,\mu$m$\else \,$\mu$m\fi}

\def\muK{\ifmmode \,\mu$K$\else \,$\mu$\hbox{K}\fi}
\def\microK{\ifmmode \,\mu$K$\else \,$\mu$\hbox{K}\fi}
\def\muW{\ifmmode \,\mu$W$\else \,$\mu$\hbox{W}\fi}
\def\kms{\ifmmode $\,km\,s$^{-1}\else \,km\,s$^{-1}$\fi}
\def\kmsMpc{\ifmmode $\,\kms\,Mpc\mo$\else \,\kms\,Mpc\mo\fi}
\providecommand{\sorthelp}[1]{}

\newcommand{\healpix}{{\tt HEALPix}}

\newcommand{\Herschel}{\textit{Herschel}}

\newcommand{\lcv}{\left(\begin{array}{r}}
\newcommand{\rcv}{\end{array}\right)}

\usepackage{bm}

\def\reff@jnl#1{{\rm#1\/}}
\def\apj{\reff@jnl{ApJ}}       
\def\apjs{\reff@jnl{ApJS}}     
\def\aaps{\reff@jnl{A\&AS}}    
\def\mnras{\reff@jnl{MNRAS}}   
\def\prd{\reff@jnl{Phys.\ Rev.\ D}}    

\newcommand{\beq}{\begin{equation}}
\newcommand{\eeq}{\end{equation}}

\newcommand{\be}{\begin{equation}}
\newcommand{\ee}{\end{equation}}

{\begin{enumerate}\setlength{\itemsep}{0mm}}%
{\end{enumerate}}

{\begin{enumerate}\setlength{\itemsep}{0mm}}%
{\end{enumerate}}

\def\tfrac#1#2{{\textstyle\frac{#1}{#2}}}

\begin{document}

\title{\textit{Planck} intermediate results. LV. Reliability and thermal properties of high-frequency sources in the Second Planck Catalogue of Compact Sources}

\author{\small
Planck Collaboration: Y.~Akrami\inst{12, 43, 45}
\and
M.~Ashdown\inst{51, 4}
\and
J.~Aumont\inst{74}
\and
C.~Baccigalupi\inst{62}
\and
M.~Ballardini\inst{17, 32}
\and
A.~J.~Banday\inst{74, 8}
\and
R.~B.~Barreiro\inst{47}
\and
N.~Bartolo\inst{20, 48}
\and
S.~Basak\inst{67}
\and
K.~Benabed\inst{42, 69}
% ...
}
\institute{\small
AIM, CEA, CNRS, Universit\'{e} Paris-Saclay, Universit\'{e} Paris-Diderot, Sorbonne Paris Cit\'{e}, F-91191 Gif-sur-Yvette, France\goodbreak
\and
APC, AstroParticule et Cosmologie, Universit\'{e} Paris Diderot, CNRS/IN2P3, CEA/lrfu, Observatoire de Paris, Sorbonne Paris Cit\'{e}, 10, rue Alice Domon et L\'{e}onie Duquet, 75205 Paris Cedex 13, France\goodbreak
\and
African Institute for Mathematical Sciences, 6-8 Melrose Road, Muizenberg, Cape Town, South Africa\goodbreak
\and
Astrophysics Group, Cavendish Laboratory, University of Cambridge, J J Thomson Avenue, Cambridge CB3 0HE, U.K.\goodbreak
\and
Astrophysics \& Cosmology Research Unit, School of Mathematics, Statistics \& Computer Science, University of KwaZulu-Natal, Westville Campus, Private Bag X54001, Durban 4000, South Africa\goodbreak
\and
CITA, University of Toronto, 60 St. George St., Toronto, ON M5S 3H8, Canada\goodbreak
\and
CMDLABS, Cambridge Machines Deep Learning and Bayesian Systems Ltd, 22 Wycombe End, Beaconsfield, Buckinghamshire HP9 1NB, United Kingdom\goodbreak
\and
CNRS, IRAP, 9 Av. colonel Roche, BP 44346, F-31028 Toulouse cedex 4, France\goodbreak
\and
California Institute of Technology, Pasadena, California, U.S.A.\goodbreak
\and
Computational Cosmology Center, Lawrence Berkeley National Laboratory, Berkeley, California, U.S.A.\goodbreak
\and
D\'{e}partement de Physique Th\'{e}orique, Universit\'{e} de Gen\`{e}ve, 24, Quai E. Ansermet,1211 Gen\`{e}ve 4, Switzerland\goodbreak
\and
D\'{e}partement de Physique, \'{E}cole normale sup\'{e}rieure, PSL Research University, CNRS, 24 rue Lhomond, 75005 Paris, France\goodbreak
\and
Departamento de Astrof\'{i}sica, Universidad de La Laguna (ULL), E-38206 La Laguna, Tenerife, Spain\goodbreak
\and
Departamento de F\'{\i}sica, Universidad de Oviedo, C/ Federico Garc\'{\i}a Lorca, 18 , Oviedo, Spain\goodbreak
\and
Department of Astrophysics/IMAPP, Radboud University, P.O. Box 9010, 6500 GL Nijmegen, The Netherlands\goodbreak
\and
Department of Physics \& Astronomy, University of British Columbia, 6224 Agricultural Road, Vancouver, British Columbia, Canada\goodbreak
\and
Department of Physics \& Astronomy, University of the Western Cape, Cape Town 7535, South Africa\goodbreak
\and
Department of Physics, Gustaf H\"{a}llstr\"{o}min katu 2a, University of Helsinki, Helsinki, Finland\goodbreak
\and
Department of Physics, Princeton University, Princeton, New Jersey, U.S.A.\goodbreak
\and
Dipartimento di Fisica e Astronomia G. Galilei, Universit\`{a} degli Studi di Padova, via Marzolo 8, 35131 Padova, Italy\goodbreak
\and
Dipartimento di Fisica e Scienze della Terra, Universit\`{a} di Ferrara, Via Saragat 1, 44122 Ferrara, Italy\goodbreak
\and
Dipartimento di Fisica, Universit\`{a} La Sapienza, P. le A. Moro 2, Roma, Italy\goodbreak
\and
Dipartimento di Fisica, Universit\`{a} degli Studi di Milano, Via Celoria, 16, Milano, Italy\goodbreak
\and
Dipartimento di Fisica, Universit\`{a} degli Studi di Trieste, via A. Valerio 2, Trieste, Italy\goodbreak
\and
Dipartimento di Fisica, Universit\`{a} di Roma Tor Vergata, Via della Ricerca Scientifica, 1, Roma, Italy\goodbreak
\and
European Space Agency, ESAC, Planck Science Office, Camino bajo del Castillo, s/n, Urbanizaci\'{o}n Villafranca del Castillo, Villanueva de la Ca\~{n}ada, Madrid, Spain\goodbreak
\and
European Space Agency, ESTEC, Keplerlaan 1, 2201 AZ Noordwijk, The Netherlands\goodbreak
\and
Gran Sasso Science Institute, INFN, viale F. Crispi 7, 67100 L'Aquila, Italy\goodbreak
\and
HEP Division, Argonne National Laboratory, Lemont, IL 60439, USA\goodbreak
\and
Haverford College Astronomy Department, 370 Lancaster Avenue, Haverford, Pennsylvania, U.S.A.\goodbreak
\and
Helsinki Institute of Physics, Gustaf H\"{a}llstr\"{o}min katu 2, University of Helsinki, Helsinki, Finland\goodbreak
\and
INAF - OAS Bologna, Istituto Nazionale di Astrofisica - Osservatorio di Astrofisica e Scienza dello Spazio di Bologna, Area della Ricerca del CNR, Via Gobetti 101, 40129, Bologna, Italy\goodbreak
\and
INAF - Osservatorio Astronomico di Padova, Vicolo dell'Osservatorio 5, Padova, Italy\goodbreak
\and
INAF - Osservatorio Astronomico di Trieste, Via G.B. Tiepolo 11, Trieste, Italy\goodbreak
\and
INAF, Istituto di Radioastronomia, Via Piero Gobetti 101, I-40129 Bologna, Italy\goodbreak
\and
INAF/IASF Milano, Via E. Bassini 15, Milano, Italy\goodbreak
\and
INFN - CNAF, viale Berti Pichat 6/2, 40127 Bologna, Italy\goodbreak
\and
INFN, Sezione di Bologna, viale Berti Pichat 6/2, 40127 Bologna, Italy\goodbreak
\and
INFN, Sezione di Ferrara, Via Saragat 1, 44122 Ferrara, Italy\goodbreak
\and
INFN, Sezione di Roma 2, Universit\`{a} di Roma Tor Vergata, Via della Ricerca Scientifica, 1, Roma, Italy\goodbreak
\and
Institut d'Astrophysique Spatiale, CNRS, Univ. Paris-Sud, Universit\'{e} Paris-Saclay, B\^{a}t. 121, 91405 Orsay cedex, France\goodbreak
\and
Institut d'Astrophysique de Paris, CNRS (UMR7095), 98 bis Boulevard Arago, F-75014, Paris, France\goodbreak
\and
Institute Lorentz, Leiden University, PO Box 9506, Leiden 2300 RA, The Netherlands\goodbreak
\and
Institute of Astronomy, University of Cambridge, Madingley Road, Cambridge CB3 0HA, U.K.\goodbreak
\and
Institute of Theoretical Astrophysics, University of Oslo, Blindern, Oslo, Norway\goodbreak
\and
Instituto de Astrof\'{\i}sica de Canarias, C/V\'{\i}a L\'{a}ctea s/n, La Laguna, Tenerife, Spain\goodbreak
\and
Instituto de F\'{\i}sica de Cantabria (CSIC-Universidad de Cantabria), Avda. de los Castros s/n, Santander, Spain\goodbreak
\and
Istituto Nazionale di Fisica Nucleare, Sezione di Padova, via Marzolo 8, I-35131 Padova, Italy\goodbreak
\and
Jet Propulsion Laboratory, California Institute of Technology, 4800 Oak Grove Drive, Pasadena, California, U.S.A.\goodbreak
\and
Jodrell Bank Centre for Astrophysics, Alan Turing Building, School of Physics and Astronomy, The University of Manchester, Oxford Road, Manchester, M13 9PL, U.K.\goodbreak
\and
Kavli Institute for Cosmology Cambridge, Madingley Road, Cambridge, CB3 0HA, U.K.\goodbreak
\and
Laboratoire d'Oc{\'e}anographie Physique et Spatiale (LOPS), Univ. Brest, CNRS, Ifremer, IRD, Brest, France\goodbreak
\and
Laboratoire de Physique Subatomique et Cosmologie, Universit\'{e} Grenoble-Alpes, CNRS/IN2P3, 53, rue des Martyrs, 38026 Grenoble Cedex, France\goodbreak
\and
Laboratoire de Physique Th\'{e}orique, Universit\'{e} Paris-Sud 11 \& CNRS, B\^{a}timent 210, 91405 Orsay, France\goodbreak
\and
Lawrence Berkeley National Laboratory, Berkeley, California, U.S.A.\goodbreak
\and
Low Temperature Laboratory, Department of Applied Physics, Aalto University, Espoo, FI-00076 AALTO, Finland\goodbreak
\and
Max-Planck-Institut f\"{u}r Astrophysik, Karl-Schwarzschild-Str. 1, 85741 Garching, Germany\goodbreak
\and
Mullard Space Science Laboratory, University College London, Surrey RH5 6NT, U.K.\goodbreak
\and
NAOC-UKZN Computational Astrophysics Centre (NUCAC), University of KwaZulu-Natal, Durban 4000, South Africa\goodbreak
\and
National Centre for Nuclear Research, ul. L. Pasteura 7, 02-093 Warsaw, Poland\goodbreak
\and
Purple Mountain Observatory, No. 8 Yuan Hua Road, 210034 Nanjing, China\goodbreak
\and
SISSA, Astrophysics Sector, via Bonomea 265, 34136, Trieste, Italy\goodbreak
\and
San Diego Supercomputer Center, University of California, San Diego, 9500 Gilman Drive, La Jolla, CA 92093, USA\goodbreak
\and
School of Chemistry and Physics, University of KwaZulu-Natal, Westville Campus, Private Bag X54001, Durban, 4000, South Africa\goodbreak
\and
School of Physics and Astronomy, Cardiff University, Queens Buildings, The Parade, Cardiff, CF24 3AA, U.K.\goodbreak
\and
School of Physics and Astronomy, Sun Yat-sen University, 2 Daxue Rd, Tangjia, Zhuhai, China\goodbreak
\and
School of Physics, Indian Institute of Science Education and Research Thiruvananthapuram, Maruthamala PO, Vithura, Thiruvananthapuram 695551, Kerala, India\goodbreak
\and
Simon Fraser University, Department of Physics, 8888 University Drive, Burnaby BC, Canada\goodbreak
\and
Sorbonne Universit\'{e}, CNRS, UMR 7095, Institut d'Astrophysique de Paris, 98 bis bd Arago, 75014 Paris, France\goodbreak
\and
Sorbonne Universit\'{e}, Observatoire de Paris, Universit\'{e} PSL, \'{E}cole normale sup\'{e}rieure, CNRS, LERMA, F-75005, Paris, France\goodbreak
\and
Space Science Data Center - Agenzia Spaziale Italiana, Via del Politecnico snc, 00133, Roma, Italy\goodbreak
\and
Space Sciences Laboratory, University of California, Berkeley, California, U.S.A.\goodbreak
\and
The Oskar Klein Centre for Cosmoparticle Physics, Department of Physics, Stockholm University, AlbaNova, SE-106 91 Stockholm, Sweden\goodbreak
\and
Universit\'{e} de Toulouse, UPS-OMP, IRAP, F-31028 Toulouse cedex 4, France\goodbreak
% ...
}

\abstract{
We describe an extension of the most recent version of the Planck Catalogue of Compact Sources (PCCS2), produced using a new multi-band Bayesian Extraction and Estimation Package ({\tt BeeP}). {\tt BeeP} assumes that the compact sources present in PCCS2 at 857\,GHz have a dust-like spectral energy distribution (SED), which leads to emission at both lower and higher frequencies, and adjusts the parameters of the source and its SED to fit the emission observed in \Planck's three highest frequency channels at 353, 545, and 857\,GHz, as well as the IRIS map at 3000\,GHz. In order to reduce confusion regarding diffuse cirrus emission, {\tt BeeP}'s data model includes a description of the background emission surrounding each source, and it adjusts the confidence in the source parameter extraction based on the statistical properties of the spatial distribution of the background emission. {\tt BeeP} produces the following three new sets of parameters for each source: (a) fits to a modified blackbody (MBB) thermal emission model of the source; (b) SED-independent source flux densities at each frequency considered; and (c) fits to an MBB model of the background in which the source is embedded.  {\tt BeeP} also calculates, for each source, a reliability parameter, which takes into account confusion due to the surrounding cirrus. This parameter can be used to extract sub-samples of high-frequency sources with statistically well-understood properties.  We define a high-reliability subset ({\tt BeeP/base}), containing 26\,083 sources (54.1\,\% of the total PCCS2 catalogue), the majority of which have no information on reliability in the PCCS2.  We describe the characteristics of this specific high-quality subset of PCCS2 and its validation against other data sets, specifically for: the sub-sample of PCCS2 located in low-cirrus areas; the Planck Catalogue of Galactic Cold Clumps (GCC); the \Herschel\ GAMA15-field catalogue; and the temperature- and spectral-index-reconstructed dust maps obtained with \Planck's Generalized Needlet Internal Linear Combination ({\tt GNILC}) method. The results of the {\tt BeeP} extension of PCCS2, which are made publicly available via the Planck Legacy Archive, will enable the study of the thermal properties of well-defined samples of compact Galactic and extragalactic dusty sources.
}

\keywords{catalogues -- cosmology: observations -- ISM: clouds --
submillimeter: general}

\authorrunning{Planck Collaboration}
\titlerunning{PCCS2 Reliability and Thermal Properties}

\maketitle

\section{Introduction}
\label{sec:introduction}

The \Planck\footnote{\Planck\ (\url{http://www.esa.int/Planck}) is a project of the European Space Agency (ESA) with instruments provided by two scientific consortia funded by ESA member states and led by Principal Investigators from France and Italy, telescope reflectors provided through a collaboration between ESA and a scientific consortium led and funded by Denmark, and 
additional contributions from NASA (USA).} 
satellite \citep{planck2014-a01} was designed to image the temperature anisotropies of the cosmic microwave background (CMB) with a precision limited only by astrophysical foregrounds. To achieve its objectives, \Planck\ observed the entire sky in nine broadband channels between 30 and 857\,GHz. The \Planck\ all-sky maps contain not only the CMB, but also a variety of diffuse sources of ``foreground'' emission -- especially the Milky Way from radio to far-infrared wavelengths, as well as extragalactic backgrounds such as the cosmic infrared background (CIB) and Sunyaev-Zeldovich emission from clusters of galaxies. In addition to diffuse emission, the \Planck\ maps contain emission from compact Galactic objects (cold dense clumps, supernova remnants, etc.) and a wide variety of unresolved external galaxies.

The Planck Catalogue of Compact Sources \citep[PCCS;][]{planck2013-p05} contains compact sources extracted from the \Planck\ maps using the first 15 months of data. The source-detection algorithm was independent at each frequency and consequently the PCCS comprises nine independent lists.  The second version of the catalogue (PCCS2; \citealt{planck2014-a35}) was produced using the full-mission data, obtained between 13 August 2009 and 3 August 2013. 

At the frequencies observed by the High Frequency Instrument (HFI; 100--857\,GHz), the diffuse sky background consists mainly of cirrus, i.e., dust emission from our own Galaxy, which covers a large part of the sky, is bright, and spatially fluctuates in a complex way \citep{Low1984}. The presence of this cirrus significantly complicates the detection and validation of compact sources, particularly because the statistical properties of this background are poorly understood, and since this cirrus contains localized structures that can be easily confused with genuinely compact sources. In addition, most of the compact sources expected in the frequency range 217--857\,GHz, both Galactic and extra-galactic, have a dust-dominated spectrum similar to that of the cirrus. 
 
\begin{figure*}[htbp!]
	\begin{center}
		\leavevmode
		\includegraphics[width=0.49\textwidth]{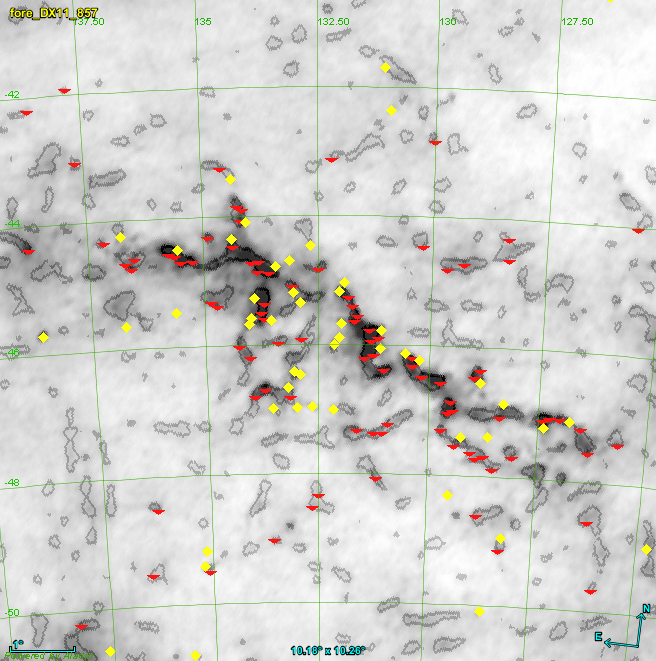}
		\includegraphics[width=0.49\textwidth]{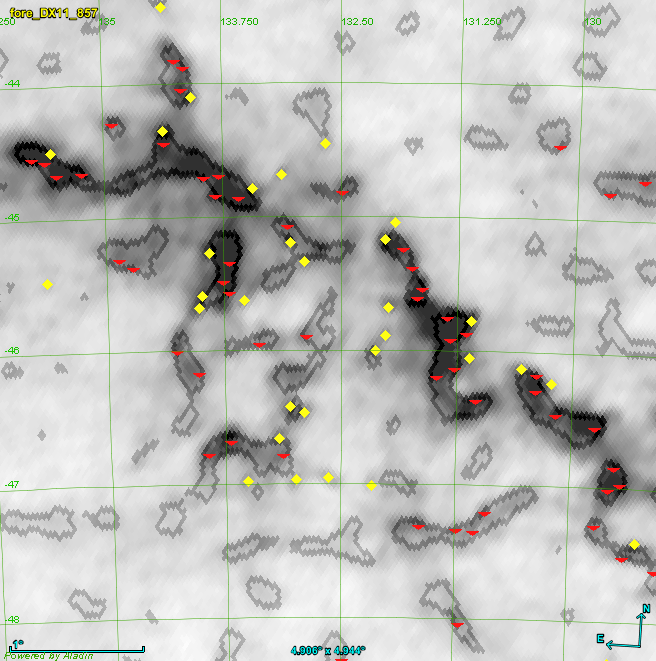}
		\caption{{\it Left:} A $10\deg \times 10\deg$ mid-Galactic-latitude ($b^{\rm II}\approx45\deg$) region of the \Planck\ 857-GHz map superimposed on the PCCS2+2E filament mask (grey contours).  PCCS2 sources are yellow diamonds and PCCS2E sources are red triangles.  The selected region contains complex backgrounds with localized features such as filaments and cirrus, causing the mask to break up into numerous islands.  Many PCCS2+2E sources trace these structures, suggesting that some of the sources are parts of filamentary structures broken up by the source-finder and not genuine compact sources.
{\it Right:} The central $4.9\deg \times 4.9\deg$ of the picture on the left, showing more clearly the spatial distribution of the PCCS2 and PCCS2E sources relative to the mask. 
		}
		\label{fig:IntroCirrusRegion}
	\end{center}
\end{figure*}

The approach of PCCS2 to this problem was the cautious but simple one of defining a set of masks within which the cirrus emission was bright or complex, and labelling all compact sources found within these masks as ``suspicious.'' The masks were derived at each frequency from: (a) brightness-thresholded total emission maps; and (b) maps of filamentary emission derived from a difference-of-Gaussians technique.  All the compact sources detected in the union of these two masks were put into separate lists, referred to as PCCS2E (E for ``Excluded''), and their reliability was not determined. The Exclusion masks include the Galactic plane and the low-Galactic-latitude regions, and cover from 15\,\% of the sky at 100\,GHz to 66\,\% of the sky at 857\,GHz.  PCCS2E contains 2487 (43290) sources at 100\,GHz (857\,GHz), to be compared to 1742 (4891) sources in the PCCS2 ``proper.''\footnote{In the rest of this paper we shall refer to PCCS2 as the list of sources {\it not\/} included in PCCS2E, and we shall call the union of both ``PCCS2+2E.''}
The vast majority of compact sources detected in the HFI maps therefore reside in the PCCS2E. While it is likely that many of the sources within PCCS2E are not genuine compact sources, but rather bumps or filaments in the cirrus background, inspection by eye of the maps clearly reveals that many of the sources are very probably genuine.  Figure~\ref{fig:IntroCirrusRegion} shows a $10\deg \times 10\deg$ patch of sky on which the locations of both PCCS2 and PCCS2E sources are displayed.  The lack of information on the reliability of the PCCS2E sources diminishes the overall utility of the PCCS2+2E.  This new study addresses that problem.
 
We do this by making use of two kinds of information available in the \Planck\ maps but not used by PCCS2.  First, we use data from multiple frequencies simultaneously.  The vast majority of high-frequency compact sources in PCCS2+2E, both Galactic and extragalactic, radiate thermal dust emission, which can be adequately modelled with a modified blackbody (MBB) spectral energy distribution (SED) characterized by a temperature and a spectral index ($T$, $\beta$). This smooth spectral behaviour can be used to improve the detectability and reliability of individual sources at high frequencies, while at the same time determining the parameters of the corresponding SEDs.  This technique has been used to construct several previous \Planck\ catalogues, including: the Catalogue of Galactic Cold Clumps \citep{planck2014-a37}; the Catalogue of Sunyaev-Zeldovich Sources \citep{planck2014-a36}; the List of High-Redshift Source Candidates \citep{planck2015-XXXIX}; the band-merged version of the Early Release Catalogue of Compact Sources \citep{ChenBandMerge}; and the Multi-frequency Catalogue of Non-thermal Sources \citep{planck2018-LIV}.

The second piece of information is that the brightness distribution of the diffuse cirrus emission varies relatively slowly and smoothly across the sky.  This implies that its spatial-statistical properties are likely to be homogeneous within relatively large patches. In addition, since the cirrus itself has an SED of the MBB type, its spatial distribution is correlated across frequency channels.  The statistical properties of the background can therefore be determined locally with good precision, and this information can be used to help separate sources from backgrounds.

We have carried out a re-analysis of all the sources contained in PCCS2+2E at 857\,GHz,\footnote{We have not attempted to re-detect and extract sources from the map, but instead use as starting point of our analysis the locations of all sources already existing in PCCS2+2E. However, all the photometric data used in our analyses are re-extracted from the \Planck\ maps.} which assumes that a single compact source is responsible for the emission observed across a range of frequencies, both below and above 857\,GHz.
We further assume that each source can be distinguished from the diffuse background in which it is embedded, either by being an outlier (in the sense that its spatial distribution does not match the statistical properties of the background) or by exhibiting a significantly different SED.  We combine multi-channel information re-extracted from \Planck\ and IRAS maps to: (a) assess the reliability of detection of each source, taking into account potential confusion with the background; (b) re-determine the flux density of each source at frequencies from 353 to 857\,GHz; (c) evaluate the spatial parameters (location and extension) of the compact source; and (d) estimate the parameters of an MBB fit to the emission across all the frequencies considered.

The results of this re-analysis are included in the \Planck\ Legacy Archive\footnote{\url{https://www.cosmos.esa.int/web/planck/pla}} (PLA) as an extension of the PCCS2 and PCCS2E 857-GHz catalogues, appending the values of the new parameters to the original files. This extension of PCCS2 enables extraction of sub-samples that have well understood statistical properties, which in turn enables the study of the thermal properties of compact Galactic and extragalactic sources. 

The outline for this paper is as follows.  In Sect.~\ref{sec:DataMain}, we present the data that we use as input to the analysis.
In Sect.~\ref{sec:Methodology}, we detail the model that we use to describe the sources and associated backgrounds, and we outline the Bayesian algorithm that we use to analyse each source and the main parameters that it outputs (details are given in Appendix~\ref{sec:beep}). 
In Sect.~\ref{sec:Simulations}, we describe the simulations that we have built and used to tune and validate the algorithm and some of the main results.
In Sect.~\ref{sec:CatalogueProduction}, we describe how we produce and filter the new information added to the PCCS2+2E catalogue.
In Sect.~\ref{sec:CatalogueCharacteristics}, we carry out a global characterization of the results of this analysis.
In Sect.~\ref{sssec:XternalCatalogues}, we validate the results of this analysis against PCCS2 and other catalogues, and (for diffuse emission parameters) against dust maps derived from \Planck\ data. 
In Sect.~\ref{sec:Conclusions}, we summarize our results, and provide recommendations for users of the new source information.

We have also included several appendices as follows. In Appendix~\ref{sec:beep} we detail the statistical machinery that we use. In Appendix~\ref{sssec:Simulations} we describe how we have used our simulations to characterize and test the results. In Appendix~\ref{sec:ContaminationBayes}  we comment on our Bayesian approach to contamination analysis, as opposed to a more classical frequentist approach. Finally, in Appendix~\ref{sssec:Examples} we include for reference the resulting SEDs that we obtain for a small number of well-known sources. 

Parts of this paper describe details of our methods, and are necessarily long and technical.  For readers whose main interests are the use of our results, we recommend to focus on Sects.~\ref{sec:MethSourceModel} and \ref{subsec:backgmodel}, which describe our source and background models, and Sects.~\ref{sec:CatalogueProduction} and \ref{sec:CatalogueCharacteristics}, which describe how we generate catalogue information, and how we then select a ``base'' catalogue of reliable sources. Section~\ref{sssec:XternalCatalogues} compares our results to other catalogues, and can be skimmed unless such comparisons are important to the reader. Our main results are summarized in the final section, and Appendix~\ref{sssec:Examples} provides some specific examples of well-studied or interesting sources extracted from our catalogue.

\section{Data}
\label{sec:DataMain}

We use the 857-GHz source list of the Second \Planck\ Catalogue of Compact Sources \cite{planck2014-a35} to provide the initial source locations for our multifrequency Bayesian analysis. The angular resolution of \Planck\ was highest at 857\,GHz (corresponding to 4\parcm7), and this list contains the largest number of sources of any individual frequency in PCCS2.  The 857-GHz source list contains flux densities for each source detected at 857\,GHz, as well as estimates of flux densities at 545 and 353\,GHz at the same locations. We note that the 857-GHz list does not contain any indication of the reliability of individual sources; the highest frequency at which such an indication is given is 353\,GHz. 

Our analysis then uses the \Planck\ all-sky temperature maps at 353, 545, and 857\,GHz from the \Planck\ 2015 release \citep{planck2014-a01} to derive the characteristics of sources and their surrounding background.  These maps are provided in the Planck Legacy Archive in \healpix\ \citep{HEALPix} format with $N_{\rm side}\,{=}\,2048$. The description of these maps can be found in \cite{planck2014-a08}. In addition, we use the 3000-GHz IRIS map, a reprocessed IRAS map described in \cite{IRISMap}, with the same pixelization as the \Planck\ maps\footnote{For clarity, we do not use the DIRBE-inpainted maps which filled in the IRAS gaps.}.

Since the start of this work, a new generation of \Planck\ maps has been released, which is referred to as the 2018 or Legacy release \citep{planck2016-l01}. However, a new catalogue of compact sources has not been extracted from the Legacy maps. Therefore, we continue using the \Planck\ 2015 maps that are the source of PCCS2.

\section{Methodology}
\label{sec:Methodology}

There is a long history of astronomers constructing catalogues, and many different approaches have been implemented, depending on the source and background properties.  When the sources are unresolved and the background has no correlations, then the optimal approach is simply to use a point-spread-function filter
\citep[e.g.,][]{Stetson1987} or thresholding methods appropriate for
isolated sources, perhaps with varying noise levels, using software
such as {\tt SExtractor} \citep{BertinArnouts1996}.
When the statistical properties of the
background are known, one can instead use a matched-filter approach
\citep[e.g.,][]{Tegmark1998,Barreiro2003}.  If the background is more complex,
if the sources themselves are partially resolved, or if the observed fields
are crowded, the task of making a reliable catalogue becomes much more
difficult.  Several methods have been used to extract compact sources
from confused Galactic regions,
for example, using second derivatives and multi-Gaussian fitting
as in {\tt CuTEx} \citep{Molinari2011}, using higher-resolution data and
multi-scale extraction as in {\tt getdist} \citep{Menshchikov2013} applied
to {\it Herschel\/} data, a similar multi-scale approach with
{\tt Gaussclumps} applied to LABOCA data \citep{Csengeri2014}, or
associating contiguous bright regions as a single
source in {\tt Clumpfind} \cite{Williams1995} or 
{\tt FellWalker} \citep[e.g.,][]{Nettke2017} for SCUBA-2 data.
A completely different strategy focuses on estimating the background properties
simulaltaneously with the source properties, and that is the approach we
follow here.

We carry out an independent Bayesian likelihood analysis \citep[see e.g.,][]{hobson_rocha_savage_2009} for each source contained in the 857-GHz catalogue of PCCS2+2E, and for the background surrounding it.  The likelihood analysis takes as input four maps (353, 545, and 857\,GHz from \Planck\ 2015, and 3000\,GHz from IRIS).  We implement this analysis in software called the Bayesian Estimation and Extraction Package, and refer to it as {\tt BeeP}.  The analysis of each source assumes a model of the signal due to the source, and another due to the background.

\subsection{Source model}
\label{sec:MethSourceModel}

We model the signal $\vec{s}_j$ due to the $j$th source as 
\begin{equation}
\label{eq:SourcesModel1}
\vec{s}_j(\vec{x};\vec{\Theta}_j) = A_j \vec{f}(\vec{\phi}_j)
\vec{\tau}(\vec{x}-\vec{X}_j;\vec{a}_j),
\end{equation}
where $A_j$ is an overall amplitude for the source at some chosen reference frequency, which we take to be 857\,GHz,\footnote{The reference frequency does not need to be the centre of one of the data channels.} $\vec{f}$ contains the emission coefficients at each frequency, which depend on the emission-law parameter vector $\vec{\phi}_j$ of the source (see below), and $\tau(\vec{x}-\vec{X}_j;\vec{a}_j)$ is the convolved spatial template at each frequency of a source centred at the position $\vec{X}_j \equiv \{X_j,Y_j\}$ and characterized by the shape parameter vector $\vec{a}_j$.  Thus, the parameters to be determined for the $j$th source are its overall amplitude, position, shape, and emission law, which we denote collectively by $\vec{\Theta}_j = \{A_j,\vec{X}_j,\vec{a}_j,\vec{\phi}_j\}$.

If we make explicit the dependence of the source signal with the frequency channel ($i$), we have
\begin{equation}
\label{eq:SourcesModel1}
\vec{s}_{ji}(\vec{x};\vec{\Theta}_j) = A_j f_i(\vec{\phi}_j)
\left[\vec{\widehat{\tau}}(\vec{x}-\vec{X}_j;\vec{a}_j) * \vec{B}_i(x)\right],
\end{equation}
where $\vec{B}_i(x)$ is the beam point-spread function of channel $i$.
In this study we are mostly targeting completely unresolved objects, i.e., beam-shaped ``point sources''; however, since PCCS2+2E also includes extended objects, we model the intrinsic shape of a source as a symmetrical two-dimensional Gaussian,
\begin{equation}
\label{eq:SourcesModel2}
\vec{\widehat{\tau}}(\vec{x};\vec{a}\equiv r) \equiv \frac{1}{2 \pi\, r^2} \exp\left(-\frac{x^2+y^2}{2\, r^2}\right),
\end{equation}
where $\vec{a} \equiv r$ is the source radius.

The intrinsic spatial profile of the source $\vec{\widehat{\tau}}(\vec{x};\vec{a}_j)$ (before any instrumental distortion) is assumed to remain unchanged across frequencies.\footnote{The source shape is also convolved with the pixel window function at each frequency and this is taken into account in our analysis. In this particular case the pixel window function does not change across maps.}
To allow the intrinsic source size to vary with frequency would require more parameters and increased uncertainties to account for a situation that corresponds to a minority of sources. We have therefore chosen to impose a single, constant size parameter for a given source.

As mentioned in Sect.~\ref{sec:introduction}, the frequency spectra of most of the compact objects found in the \Planck-HFI maps can be well-represented by an MBB spectrum \citep{planck2014-a35}; however, the SEDs of a minority of sources, for instance blazars, are not well-described by a modified blackbody.  Therefore, we fit all sources with {\it both\/} MBB and ``Free'' models.  In the latter, the emission coefficient $f_{\nu_i}$ at each channel is a free parameter.  The MBB spectrum is written as
\begin{equation}
\label{eq:MBB_SED}
\ln f_{\nu} = \beta ~ \ln \left(\frac{\nu}{\nu_0} \right)
+ \ln \left[\frac{B_\nu(T)}{B_{\nu_0}(T)}\right],
\end{equation}
where the spectral parameters $\vec{\phi}=\{\beta,T\}$ are the dust emissivity spectral index and temperature, respectively, $B_\nu(T)$ is the Planck law of blackbody radiation, and $\nu_0$ is once again the reference frequency.  We normalize $\vec{f}$ so that $f_\nu=1$ at $\nu=\nu_0$.

The Free model is written as
\begin{equation}
\vec{f} = [f_{\nu_1}, \cdots ,f_{\nu_n}]^{\sf T}\,,
\label{eq:Free_MBB}
\end{equation}
where the emission coefficients $f_{\nu_i}$ are free parameters.  In effect, this model is a way to estimate source flux densities in each channel without imposing an SED, but still assuming that there is a single source at all frequencies.  This extra flexibility comes at the cost of a larger model complexity, since it requires more free parameters. The flux-density estimates for the Free model are those that can most closely be compared to the ones already present in PCCS2+2E.

\begin{figure}[htbp!]
	\begin{center}
		\leavevmode
		\includegraphics[width=0.49\textwidth]{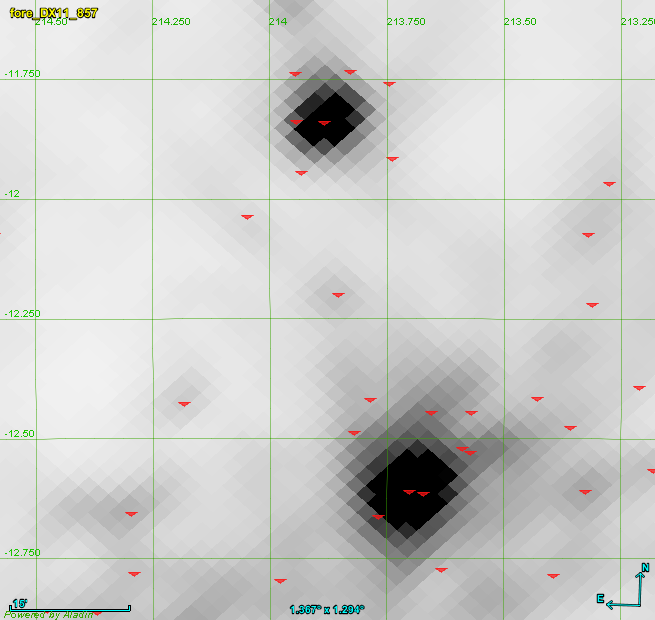}
		\caption{Small patch ($1\pdeg3 \times 1\pdeg3$) taken from the \Planck\ 857-GHz map. The red triangles are PCCS2+2E catalogue positions. One can see that around the two bright ``blobs'' there are many associated PCCS2+2E sources, but their locations do not reflect the actual brightness peaks. 
		}
		\label{fig:IntroMovement}
	\end{center}
\end{figure}

The location of the centre of the source is represented in Eq.~\eqref{eq:SourcesModel1} by $\vec{X}_j$. Our analysis initially assumes that the source is centred at the location defined in the 857-GHz list of PCCS2+2E. However, the source centre may be expected to vary slightly from channel to channel, and for this reason we allow our method to deviate from the initial values in an attempt to find the best overall location. Furthermore, during this investigation we realized that many of the source locations listed in PCCS2+2E are not well determined: in many cases we see that the centres of one or more sources are located around the edge of a well-defined blob of emission (e.g., Fig.~\ref{fig:IntroMovement}). This problem affects about 10\,\% of all sources in PCCS2+2E for the higher-frequency channels, and is inherent to the Mexican-hat wavelet 2 (MHW2) algorithm used to perform the detection. This wavelet, when used as a filter, is known to maximize the S/N of the objects, but it is also known to produce artefacts at a fixed distance from the centre of the source, Such artefacts related to the shape of the filter can be identified and removed particularly well in the cleaner regions of the sky. This additional cleaning step was performed for the lower-frequency channels of PCCS2, where the beamwidths are larger and these ringing effects are more prominent, but it was not performed for the higher-frequency channels because it was not considered necessary. Moreover, the MHW2 algorithm is well suited for the detection of point-like objects; however, when dealing with slightly extended structures such as those found at 857\,GHz, the artefacts introduced by this filter are more evident, and a two-step cleaning procedure is definitely needed.

In our analysis we allow a new location to be determined from all the frequencies considered.  As a result, in a number of cases several PCCS2+2E sources will be associated with the same physical source location.\footnote{There are $8269$ ($17.2\,\%$) PCCS2+2E locations that are associated with a different source. Of those $162$ ($3.3\,\%$ of the PCCS2) are in the PCCS2 region.} However, there are also many genuinely independent sources that are relatively close to each other, and there is a risk that the algorithm would ``merge'' them. We have therefore compromised by allowing our algorithm to move the location by at most 3~pixels (4\parcm5) away from its starting point.  If this extreme is reached without an optimal solution being found, a flag indicating this is set in the final parameters.

\subsection{Background model}
\label{subsec:backgmodel}

We now need to account for the astronomical background $\vec{b}(x)$ and the instrumental noise $\vec{n}(x)$.  A strong assumption of our framework is that the joint background in which the sources are immersed ($\vec{b}(x) + \vec{n}(x)$) is a two-dimensional, statistically isotropic
Gaussian random field. Such a field is fully defined by its covariance matrix, which we use in our method as the mathematical representation of the background.  The full-sky maps observed by \Planck\, however, are neither statistically isotropic nor Gaussian.  At high Galactic latitudes, the diffuse emission from Galactic dust is faint, and the (mostly extra-galactic) brighter compact sources stand out easily against it. However, the situation changes rapidly at low Galactic latitudes, as the diffuse emission competes in brightness with even the brightest compact sources. In this situation, confusion between ``genuine'' sources and the diffuse emission leads to difficulties in estimating the statistical properties of the background alone.

To improve our estimation of the properties of the background, we first reduce the size of the sky patch analysed around the source such that we can assume that statistical isotropy applies locally.\footnote{The ``field'' size we select is $3\pdeg69 \times3\pdeg69$. The motivation for this choice can be found in Appendix~\ref{psssec:ArtificialSimulations}.}  Second, we use the covariance matrix of the {\it cross}-power spectra across frequency channels. This improves the situation, since the instrumental noise $\vec{n}(x)$ is mostly uncorrelated across channels, and the astronomical background $\vec{b}(x)$ is better-determined by the larger data volume.  The determination of an accurate cross-spectrum covariance matrix turns out to be a key element in our method.  To improve the estimation of the off-diagonal components of this matrix, we filter out the noise component using the theory of random covariance matrices \citep[][chapter~9]{RiskAnalys}.  We have found that we also need to weight the off-diagonal elements (which represent the correlated part of the background) with respect to the diagonal elements (which represent the ``noise'') in order to accommodate the very large dynamic range of sources. The weighting factor that we use is tuned on simulations to reduce bias in the recovery of source parameters. More details on these analysis choices are described in Appendix~\ref{sec:beep}.

In practice, PCCS2+2E provides a list of {\it potentially\/} genuine sources that are embedded in the background whose properties we are estimating.  For each of these sources, we create ``background'' maps (see Sect.~\ref{sec:AlgoImpl}) by masking all surrounding PCCS2+2E sources\footnote{For this purpose we merge all three source lists between 353 and 857\,GHz.} and inpainting the masked areas (see Sect. 5.3 of \citealt{InPaint}). We use a 7\arcm\ masking and inpainting radius to provide a good balance between effective source-brightness removal and preservation of the statistical properties of the background (see Figs.~\ref{fig:Masks} and \ref{fig:ForeBackmaps}, and the discussion in Sect.~\ref{subsubsec:CovarianceMatrixEstimation}), especially at low Galactic latitudes where the density of sources is very high.  Close to the Galactic plane, a large fraction of the background patch (up to 74\,\% near the Galactic centre) is masked and inpainted, which might be expected to have a significant effect on the estimation of the detection significance.\footnote{The fraction of inpainted pixels in the patch is reported in one of the columns of the catalogue and can be used to filter the selection.}  More generally, we expect that inpainting may bias the estimation of the background properties, but it cannot be avoided because the effect of unremoved bright sources or of corresponding holes would certainly be much higher.  The impact of inpainting cannot be modelled analytically, and the only way to assess it is through simulations.  Simulations with different degrees of inpainting are discussed in Sect.~\ref{sec:Simulations}, and show that the effect on source parameters is indeed small (as discussed further in
Sect.~\ref{psssec:InjectSimulations}).

\subsection{Combined model and its analysis} 
\label{sec:SingleSrcModel}

In this section we present the principles of our Bayesian analysis methodology.  Appendix~\ref{sec:beep} gives technical details of the approach and its practicalities.

We first combine our models for sources and background into a model of the observed maps.  A realistic model would have to include the entirety of sources and the full sky together. However, as described in Sect.~\ref{subsec:characterizationMethod}, under the assumption that the sources do not blend together, it is possible to simplify the problem and model each source independently:
\begin{equation}
\label{eq:dataModel}
\vec{d}_j(\vec{x}) = 
\vec{s}_j(\vec{x};\vec{\Theta}_j) + \vec{b}_j(\vec{x}) +
\vec{n_j}(\vec{x}),
\end{equation}
where $\vec{d}_j$ is the data vector (pixel values), and $\vec{b}_j$ and $\vec{n}_j$ represent astrophysical and noise backgrounds in the neighbourhood of the source ($j$).

We can now build the likelihood of a single compact object as
\begin{equation}
\label{eq:LikelihoodCompleteMain}
\mathcal{L}(\vec{\Theta})=
\frac{\exp\left\{-\frac{1}{2}\left[\vec{d} - \widehat{\vec{b}} - \vec{s}(\vec{\vec{\Theta}}) \right]^{\sf T} \tens{N}^{-1} \left[\vec{d}- \widehat{\vec{b}} - \vec{s}(\vec{\vec{\Theta}}) \right] \right\}}
{\left(2\pi\right)^{N_{\rm pix}/2} \left|\tens{N}\right|^{1/2}},
\end{equation}
where $\widehat{\vec{b}}$ is the generalized background ($\vec{b} + \vec{n}$), $\tens{N}$ is the generalized background covariance matrix, and all individual source parameters have been concatenated into $\vec{\Theta}$ for convenience. For clarity we have dropped the source index here.

The above expression allows us to consider the likelihood of a ``no-source'' model $\mathcal{L}_0$, when $A$, the source amplitude is $0$.  $\mathcal{L}_0$ is a constant, since it does not contain any parameters.  The expression that we seek to maximize is the $\log$ of the $\mathcal{L}(\vec{\Theta})/\mathcal{L}_0$ ratio, which represents the likelihood that there is a source in addition to the background.

\begin{figure}[htbp!]
	\begin{center}
		\leavevmode
		\includegraphics[width=0.40\textwidth]{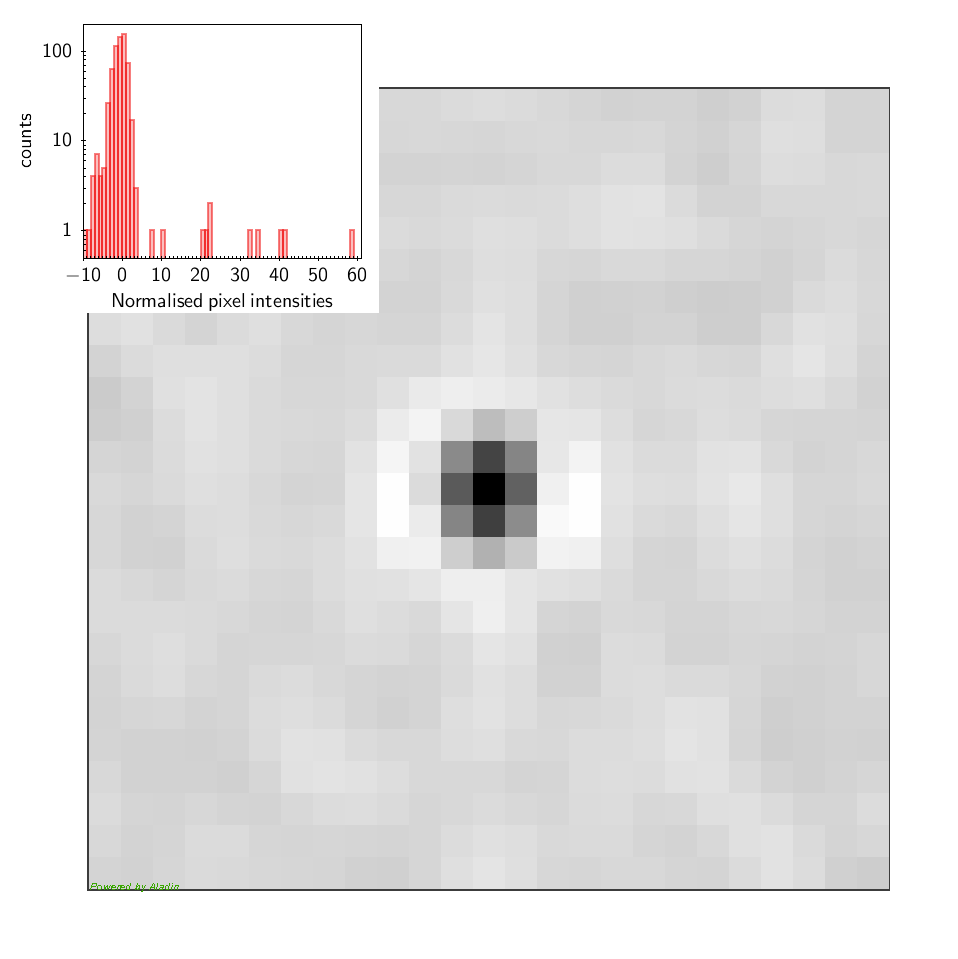}\\
		\includegraphics[width=0.395\textwidth]{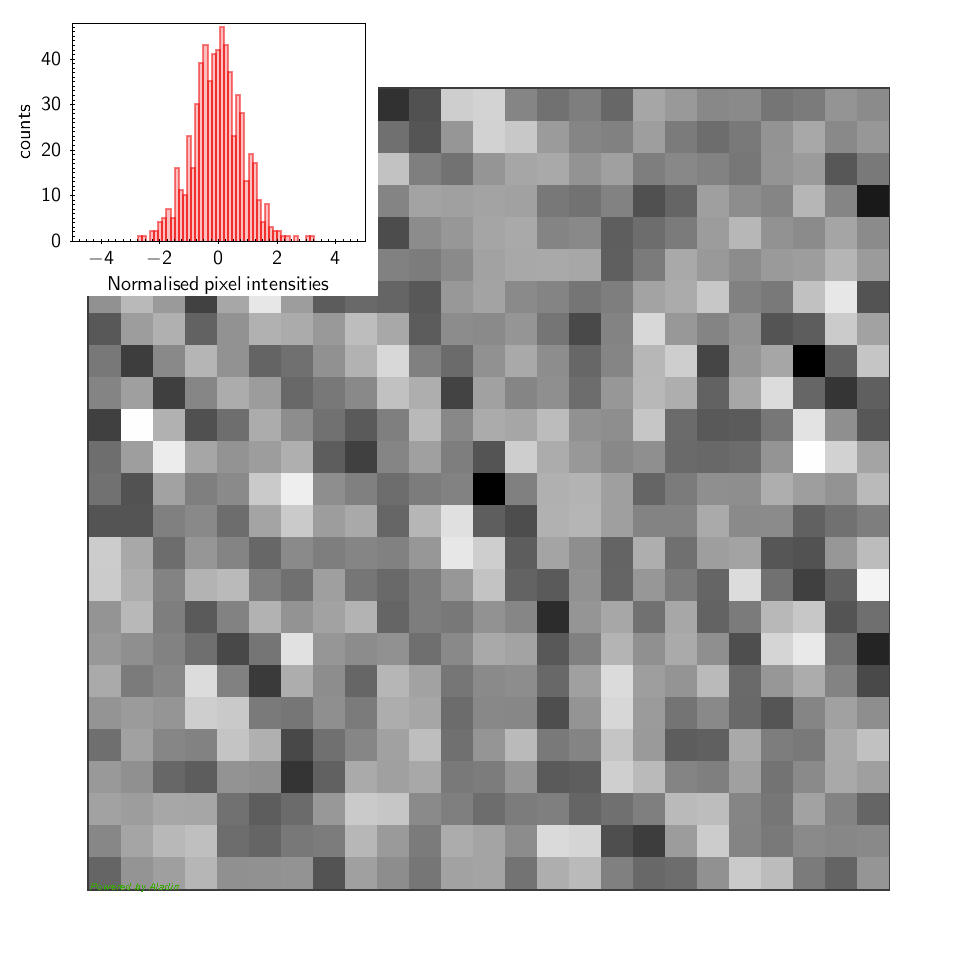}
		\caption{ Examples of potential analysis fields. The upper panel depicts
				a high significance source (PCCS2 857 G172.20+32.04). The histogram (shown in the inset; Y-scale is log)
				is a mixture of a Gaussian component from the background
				pixels, plus a strong upper tail generated by the source in the
				centre. The lower panel is a field with no detected sources in it (PCCS2 857 G172.20+32.04; Y-scale is linear).
				This time only the Gaussian component is present. The tails of
				the distribution are compatible with ``just background''.
				Each field is $25 \times 25$ pixels (1 pixel ~ $1\parcm717$). The pixel intensities (Eq. \ref{eq:FilteredField}) are unitless.}
		\label{fig:Histogram}
	\end{center}
\end{figure}

If $\widehat{\vec{\Theta}}$ is the parameter set that maximizes the likelihood ratio (Eq.~\ref{eq:LikelihoodCompleteMain}), then we define the quantity ${\cal R}$, corresponding to \textsf{NPSNR} in the catalogue\footnote{Identifiers in sans-serif capital letters correspond to column labels in {\tt BeeP} output catalogues.} (the Neyman-Pearson signal-to-noise ratio), by
\begin{equation}
\label{eq:NPSNR_defMain}
\ln\left(\frac{\mathcal{L}(\widehat{\vec{\Theta}})}{\mathcal{L}_0}\right) =
\tfrac{1}{2}{\cal R}^{2}.
\end{equation}
${\cal R}$ is the detection significance level that expresses the number of sigmas of the detection, and is given in the \textsf{NPSNR} column of the {\tt BeeP} catalogues.  In the case that all of our assumptions hold, and all source parameters are known except amplitude, $A$, then ${\cal R}$ would in fact be the inverse of the fractional error on the amplitude, $A/\Delta A$.  However, in practice, as we shall see, typical values of ${\cal R}$ are considerably higher than $A/\Delta A$. This is the result of either broken assumptions or uncertainties on the other estimated parameters that propagate into the source amplitude.  In particular, the presence of cirrus produces strong positive-tail events in the likelihood, and this might be interpreted (erroneously) as generated by the source of interest (see Fig.~\ref{fig:Histogram} for examples).

To account for this effect, we build an estimate of the non-Gaussianity of the background that is independent of the likelihood, which we refer to as $\textsf{RELTH}$.  Essentially we look in the background patch for outliers to a white-noise, unitary ($\sigma = 1$) Gaussian random field in pixel space ($\vec{X}$), which is what we would expect if all our assumptions hold, in other words, under the null hypothesis of our model.  We assume that the positive outlier pixels created by the source itself are no more than a small fraction of the total number of pixels in a small patch around the source.  Using the definition of quantiles, one would expect that
\begin{equation}
\label{eq:QuantileDefMain}
\int_{-\infty}^{\textsf{RELTH}} \,\, \frac{\exp\left[-\frac{1}{2}\left(\frac{x}{\sigma}\right)^2\right]}{\sqrt{2 \pi} \sigma} \textrm{d}x = 1-\alpha,
\end{equation}
where $\textsf{RELTH}$ is the $1-\alpha$ distribution quantile, and $\sigma$ is the width of the Gaussian.  Using simulations, we have verified that the fraction of outlier pixels created by the source is less than 5\,\%\ of the total, so we use $\alpha$ = 5\,\%.


$\textsf{RELTH}$ can be read directly from the histogram of the actual field, and then Eq.~\eqref{eq:QuantileDefMain} solved for $\sigma$.  If the background pixels ($[1-\alpha]$\,\% of the patch pixels) comply with the assumptions of the background model, then they will follow a unitary Gaussian distribution and the solution of Eq.~\eqref{eq:QuantileDefMain} is $\sigma = 1$.  However, as a result of the intrinsic non-Gaussianity of the background, the tails of the background histogram are expected to be larger than those of the unitary Gaussian distribution.  This distribution of background pixel brightness with extended tails can then be approximated by a Gaussian, but with $\sigma > 1$ to account for the larger tails.  Solving Eq.~\eqref{eq:QuantileDefMain},
\begin{equation}
\label{eq:RelthSigma}
\sigma = k \,\, \textsf{RELTH},
\end{equation}
where $k$ is a pure numerical constant given by 
\begin{equation}
\label{eq:RelthKdefMain}
k = \frac{1}{\sqrt{2} \, \textrm{erfc}^{-1}(2\alpha)},
\end{equation}
and $\textrm{erfc}^{-1}$ is the inverse complementary error function.

We can now correct our ``naive'' significance $\textsf{NPSNR}$ and define a new source significance variable as
\begin{equation}
\label{eq:MRelthDef}
\textsf{SRCSIG} = \frac{1}{k} \,\, \frac{\textsf{NPSNR}}{\textsf{RELTH}},
\end{equation}
where $k$ is a constant given by Eq.~\eqref{eq:RelthKdefMain}, which is the same for all sources.
SRCSIG expresses the likelihood that there is a source in the patch being analysed.  If the histogram of the background patch is Gaussian, then $\sqrt{2} \, \textrm{erfc}^{-1}(2\alpha) = \textsf{RELTH}$ by definition and $\textsf{SRCSIG} = \textsf{NPSNR}$.  If our initial assumptions hold, as predicted, then $\textsf{NPSNR}$ is the detection significance.  However when there is non-Gaussianity in the background, either from diffuse components or localized features, then $\textsf{RELTH}$ increases and a penalty is applied to the Gaussian criterion.  The penalty is reduced towards high Galactic latitudes away from cirrus, where the isotropy and Gaussian assumptions hold well.  In the neighbourhood of the Galactic plane, or inside cirrus structures, the criterion becomes more stringent in order to avoid false positives induced by the non-Gaussianity of the background.\footnote{See Sect.~\ref{subsec:characterizationMethod}.}

Finally, we note that RELTH depends on the detailed statistics of the field brightness.  Therefore its ability to provide an estimate of the relative level of non-Gaussianity in the background is not uniform across the sky. However, tests using simulations show that it is effective both at low and high Galactic latitudes, and it can safely be used to correct \textsf{NPSNR}. On the other hand, it should probably not be used to directly compare levels of non-Gaussianity in regions that differ significantly in complexity.

\section{Simulations}
\label{sec:Simulations}

We have tested our method extensively using simulations. These tests have allowed us to tune parameters intrinsic to the method, and to assess the quality of the extracted source descriptors.  There are four types of simulations, as follows.
\begin{enumerate}
\item {\it Synthetic\/} simulations (Appendix~\ref{psssec:ArtificialSimulations}) comprise data that mimic a basic assumption of the method as closely as possible, namely that the background is a homogeneous Gaussian random process.  To make these simulations, we combine CMB map realizations based on the \Planck\ 2015 best-fit cosmological model, with noise consistent with that of the \Planck\ detectors as described in \citet{planck2014-a14}.  To these we add Gaussian sources whose thermal emission characteristics are taken from a preliminary {\tt BeeP} extraction. We use these simulations to test the algorithm, and fix some of its basic parameters, such as the optimal size of the patch analysed around each source, and to check the impact of some systematics such as projection distortions. 

\item{\it Injection\/} simulations (Appendix~\ref{psssec:InjectSimulations}) attempt to reproduce the properties of the diffuse backgrounds that are seen by \Planck.  The basic principle is to use the 2015 \Planck\ maps and add to them a known set of sources.  We have produced three distinct types of these simulations: (a) we remove from the observed maps the sources present in PCCS2+2E, inpaint the holes, and inject at the same locations point-like sources whose thermal emission parameters are those of the original source (as extracted by {\tt BeeP} in a preliminary run); (b) as in (a), but the fake sources are injected in the vicinity of the original ones rather than at the PCCS2+2E location; and (c) the locations of the fake sources are randomly drawn from a uniform distribution over the high-latitude sky, and their thermal properties are drawn from the distribution present in PCCS2.  In this case the original PCCS2+2E sources are not removed from the maps.  In addition, we have also produced realizations of the above three types that include known source extensions.  As detailed further in Appendix~\ref{psssec:InjectSimulations}, these simulations allow us to:
\begin{itemize}
\item assess the effect of inpainting on the results;
\item determine an optimal level for the covariance matrix cross-correlation factor;
\item assess biases in the recovered source parameters, e.g., temperature and spectral index;
\item assess the accuracy of the estimated source locations, and on this basis establish a correction to the estimated location uncertainties; and 
\item assess biases and establish corrections to both the estimated flux densities and their uncertainties (see Sect.~\ref{ssec:FluxDensityAccuracy}).
\end{itemize}

\item {\it FFP8\/} simulations (Appendix~\ref{psssec:FFP8Simulations}) are the most realistic realizations of the all-sky maps as observed by \Planck\ and processed through the PR2 pipelines,\footnote{The PCCS2+2E source catalogues were produced from the PR2 maps. The newer PR3 maps released by \Planck\ in 2018 are based on significantly different pipelines, and have not been used to generate source catalogues; for this reason we cannot use the newest FFP10 simulations associated with PR3.} and are fully independent of the observed maps.  In particular they reproduce the variation across the sky and in frequency of the \Planck\ beams, which is something that we do not include in our injection simulations. However, an important drawback is that a corresponding simulation of the IRIS sky is not available and therefore we cannot extract thermal-emission parameters in order to compare them directly to {\tt BeeP}'s results on \Planck\ maps. Nonetheless, we are able to use these simulations to assess the impact of the beam variation on the recovery of flux densities and on the positional error, and on this basis we establish a correction to the flux-density estimates.

\item {\it No-source} simulations (see Sect.~\ref{subsec:Reliability}) use a list of locations that are {\it not\/} present in PCCS2+2E, and on which we run {\tt BeeP}. Under the assumption that such locations contain only background emission,\footnote{This assumes that for the level of sensitivity we are aiming at, the PCCS2+2E catalogues are almost complete \citep{planck2014-a35}.} these simulations allow us to estimate the number of spurious sources generated by {\tt BeeP}, i.e., the background-related contamination fraction of the resulting catalogue.  The empty locations are selected in the neighbourhood of the catalogue positions in order to preserve the distribution of sources on the sky.  We have placed the sources at a random location within an annulus of radii $12'$ and $14'$, enforcing that each injection location is at least $12'$ from any other. We then mask and inpaint the original source.
\end{enumerate}

All of the above tests and their results are described in detail in the Sects.~\ref{subsec:Reliability} and \ref{ssec:EstimateValuesQualityCriterion}, as well as Appendix~\ref{sssec:Simulations}.

\section{Catalogue production}
\label{sec:CatalogueProduction}

The basic principles of the production methodology for the catalogue are described in Sect.~\ref{sec:Methodology} and implementation details in Appendix~\ref{sec:beep}. The {\tt BeeP} software takes as input a catalogue of sources and associated maps, and processes all sources. The output is an extension of the input catalogue, in effect adding to each source a number of new parameter fields.

As described in Sect.~\ref{sec:DataMain}, the input catalogue is the union of the 857-GHz PCCS2 and PCCS2E (PCCS2+2E) source lists, which contains $48\,181$ entries. The input data are the 2015 \Planck\ full-mission frequency maps between 353 and 857\,GHz, and the IRIS map.  The IRIS map does not cover the full sky, and therefore a small subset of sources ($650$) has been processed with \Planck\ channels only.  This restriction seriously impairs the constraining capabilities of the likelihood, and hence a downgraded quality status has been assigned to these sources.
As a consequence, the output catalogue contains $47\,531$ complete entries.  Of those, $42\,869$ \, (about  $90\,\%)$ are in the PCCS2E, and only $4\,662 \, (10\,\%)$ in the PCCS2.

\subsection{Reliability assessment}
\label{subsec:Reliability}

Once we have processed the entire input catalogue through {\tt BeeP}, we can apply filters to select subsets of sources. The first and most critical filter is reliability. For this purpose, we interpret our detection significance statistic \textsf{SRCSIG} in terms of reliability. 

In a classical frequentist framework, we would draw the test receiver operational characteristic curve \citep[ROC,][chapter~2]{VTrees}.
The ROC curve shows the balance between ``completeness,'' or true positive rate, and the false positive or ``spurious'' rate, when varying the threshold of the detection significance statistic.  However, since we are not adding any new entries to the PCCS2+2E catalogue, we will always be limited by the initial catalogue's completeness.  Our focus will therefore be on the spurious error rate or ``contamination.''  The spurious error rate is the probability of classifying a source as real when only background is present for a given \textsf{SRCSIG},\footnote{In Appendix~\ref{sec:ContaminationBayes} we present the procedure using the ``dialect'' of the orthodox hypothesis testing framework.}
\begin{equation}
\label{eq:ReliabilityFormula}
\rm{Contamination} \equiv \Pr\left({\rm real} \, | \, {\rm only\, background} ; \textsf{SRCSIG}\right).
\end{equation}
Owing to the complexity of the data, the most practical way of estimating contamination is through simulations.  For this purpose, we use the no-source simulations described in Sect.~\ref{sec:Simulations}.  We run the {\tt BeeP} algorithm on the no-source catalogues, and compute the \textsf{SRCSIG} statistic.  We then compute the percentage of locations where there is not a source for which the \textsf{SRCSIG} statistic is larger than a certain threshold.  This gives an estimate of the contamination (Eq.~\ref{eq:ReliabilityFormula}),\footnote{The uncertainty in the contamination estimate is $\sqrt{p(1-p)/(n+3)}$, where $p$ is the contamination and $n$ is the number of ``false sources'' ($n \approx 20\,000$).  Even for large $n$, as in our case, some care must be used when selecting very low contaminations. An estimated contamination of $0.005$ already carries an uncertainty of about $10\,\%$.} under the assumption that there are no sources.

\begin{figure*}[htbp!]
	\begin{center}
		\leavevmode
		\includegraphics[width=0.49\textwidth]{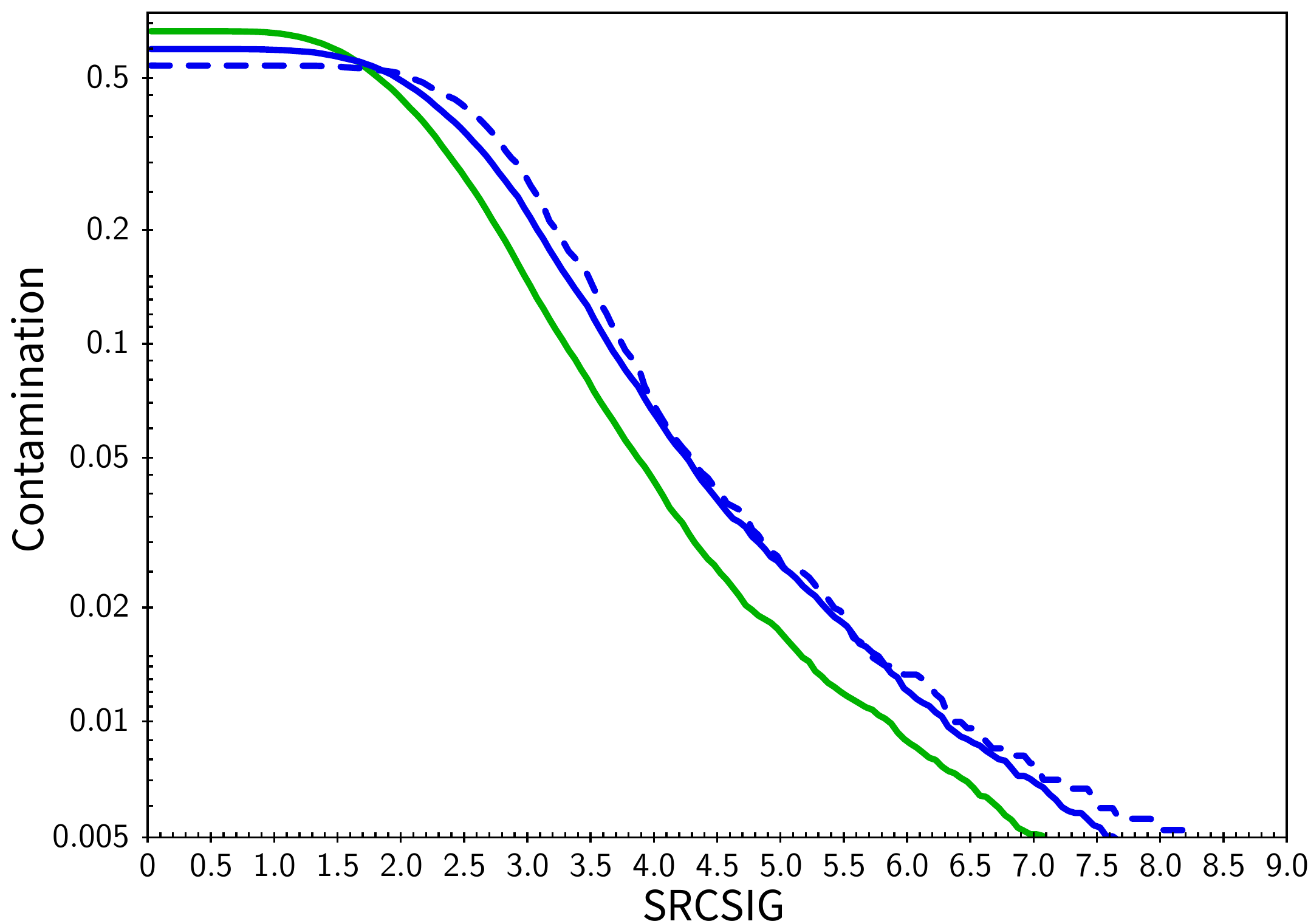}
		\includegraphics[width=0.49\textwidth]{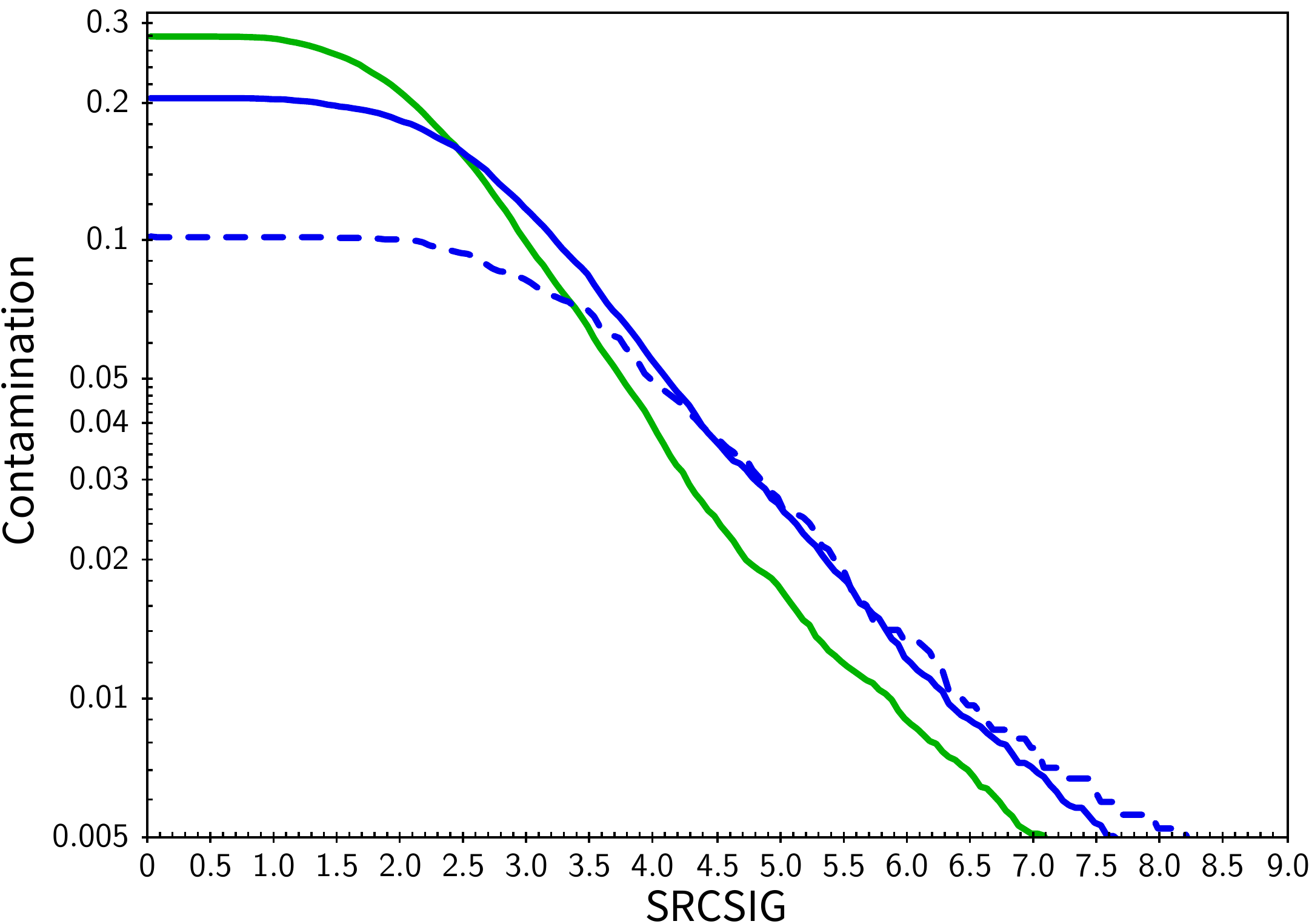}
		\caption{Contamination (Eq.~\ref{eq:ReliabilityFormula} and Appendix~\ref{sec:ContaminationBayes}) when \textsf{SRCSIG}$\,{\geq}\,x$, in the cases of $\textsf{NPSNR}\,{>}\,3$ (left) and $\textsf{NPSNR}\,{>}\,5$ (right). The blue curves, solid for full sky (PCCS2+2E) and dashed for high Galactic latitudes (PCCS2), display the contamination for simulations when the ``no sources'' are located in the neighbourhood of actual catalogue positions. The original sources are masked and inpainted.  The solid green line shows the estimated contamination of a simulation exactly like that of the solid blue curve (full sky), but this time the original sources are not removed and inpainted. 
		In this figure and several others in this paper, we label the axis with the name of the corresponding field in the output of {\tt BeeP} analysis (in capital roman letters), e.g., here ``\textsf{SRCSIG}.''}
		\label{fig:Reliability}
	\end{center}
\end{figure*}

Figure~\ref{fig:Reliability} shows how this estimate of the contamination varies with the \textsf{SRCSIG} threshold, for two different thresholds of \textsf{NPSNR}.  Solid blue lines are full-sky results, and dashed lines correspond to a catalogue restricted to PCCS2 sources.  The solid (full-sky) green line is obtained similarly, but the original source is not removed.  This test is carried out to show that the presence of the original source in the background significantly and systematically modifies the non-Gaussianity of the background in the area being analysed, reducing in a systematic way the \textsf{SRCSIG} distribution.  As can be seen in Fig.~\ref{fig:Reliability}, this effect would artificially (and incorrectly) reduce the contamination for a given \textsf{SRCSIG} threshold.

Figure~\ref{fig:Reliability} shows that if the catalogue is restricted to the more reliable sources,
there is very little difference in the contamination levels of the PCCS2+2E full catalogue (solid line) and the PCCS2 subset (dashed line); this indicates that {\tt BeeP} accounts adequately for the non-Gaussianity of the background. We select $\textsf{SRCSIG} > 3.7$ as an interesting threshold, which leads to a contamination level between 5\,\%  and 10\,\% (Fig.~\ref{fig:Reliability}).

Our simulation-based estimate of contamination relies on the prior assumption that there are no sources at the locations analysed, which is probably not correct for PCCS2+2E where crowding becomes significant.  This makes the estimate of Eq.~\eqref{eq:ReliabilityFormula} a conservative one. The curves in Fig.~\ref{fig:Reliability} should then be read as the maximum contamination for a given \textsf{SRCSIG} threshold. To make it more realistic, the estimate should be reduced taking into account the catalogue completeness, as described in Appendix~\ref{sec:ContaminationBayes}. However, for high values of \textsf{NPSNR}, the correction is very small;\footnote{Indeed, by comparing the curves with the two \textsf{NPSNR} thresholds shown in the left and right panels of Fig.~\ref{fig:Reliability}, it can be deduced that the correction must already be very small at \textsf{NPSNR}$\,{>}\,3$, a very low value for \textsf{NPSNR}.} in this case one can safely use Fig.~\ref{fig:Reliability} as a reasonable estimate of the catalogue contamination.
Comparison of the solid and dashed lines in Fig.~\ref{fig:Reliability} also shows the effect of crowding on contamination, which is at most 10\,\% for low SRCSIG. 

With the above considerations, a catalogue can be selected to have a given reliability level by adopting thresholds in \textsf{SRCSIG} and \textsf{NPSNR}. For example, if we define the condition
\begin{equation}
\textsf{SRCSIG} > 3.7 \wedge \textsf{NPSNR} > 5.0,
\label{eq:reliabilityCriterion}
\end{equation}
where the symbol ``$\wedge$'' means ``logical and,'' the resulting catalogue has a maximum contamination between 5 and 10\,\%.\footnote{If we did not impose \textsf{NPSNR}$\,{>}\,5$, then \textsf{SRCSIG} ${>}\,3.7$ alone would set contamination to approximately 10\,\%.} The reliability condition in Eq.~\eqref{eq:reliabilityCriterion} is one of the important components for building the ``{\tt BeeP/base}'' catalogue (see Sect.~\ref{subsubsec:BeePbasecatalogue})

\subsection{Rejection of outliers}
\label{ssec:EstimateValuesQualityCriterion}

As a result of the large range of source flux densities and the background conditions, it is reasonable to expect that under extreme conditions the simplified data model, and the likelihood, become a sub-optimal description of the statistical properties of the data, and that significant outliers will arise.  As one of our goals is to have a well-defined set of statistical descriptors for the catalogue estimates, these extreme outliers need to be identified and removed to avoid biasing or distorting the characterization.

The extensive set of simulations described in Sect.~\ref{sec:Simulations} was used to identify such cases (see Appendix~\ref{psssec:InjectSimulations} for more details).  We find that any sources whose estimates do \textit{not} meet the following ``outlier-rejection criterion'' must be considered unreliable:
\begin{align}
&\textsf{EXT} > 1.46 \wedge \textsf{TEMP} < 60 \wedge \textsf{BETA} < 5\nonumber\\
&\quad \wedge 
(\textsf{TH2SB} - \textsf{TL2SB}) >0.8\nonumber\\
&\quad \wedge (\textsf{BETAH2SB}-\textsf{BETAL2SB})> 0.25,
\label{eq:QualityCriterion}
\end{align}
where \textsf{EXT}, \textsf{TEMP}, and \textsf{BETA} are the estimated source extension, temperature, and spectral index, respectively.  The differences (\textsf{TH2SB} $-$ \textsf{TL2SB}) and (\textsf{BETAH2SB} $-$ \textsf{BETAL2SB}) are the estimated uncertainties of the temperature and spectral index\footnote{We reject sources where the recovered parameter uncertainties are extremely low, indicating that the likelihood sampler has not been able to explore the parameter space adequately. There may be some exceptional cases where the uncertainties are very low because the model fits the data extremely well, and these will also be rejected. One such example can be seen in Fig. \ref{fig:SecondExample}. It is possible, by examining the results of {\tt BeeP}, especially the $\chi^2$ of the free model fit, to decide that the case should not be rejected.}.  The value of \textsf{EXT} that we use to create the filter is the ``uncorrected'' source size parameter (see Sect.~\ref{ssec:SourceSize}, Appendix~\ref{ssec:AlgoImplSrcDetec}, and Appendix~\ref{ssec:AlgoImplPostMultiModal}).

The criterion of Eq.~\eqref{eq:QualityCriterion} selects a very small fraction of the catalogue sources (2462, or about 5\,\%).  Of those, 1463 would also have been rejected by the reliability criterion (Eq.~\ref{eq:reliabilityCriterion}).  Thus only 999 or 2\,\% of the sources that pass the reliability criterion are rejected by the outlier-rejection criterion (Eq.~\ref{eq:QualityCriterion}).

\subsection{Convergence filter}
\label{ssec:Convergence}

\begin{figure}[htbp!]
	\begin{center}
		\leavevmode
		\includegraphics[width=0.495\textwidth]{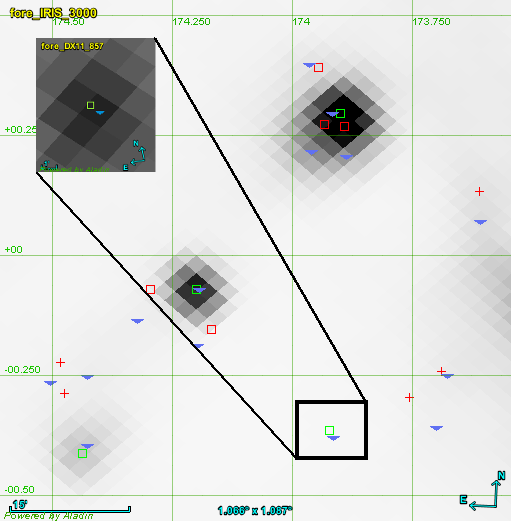}
		\caption{Patch of $1\pdeg0\times 1\pdeg0$ area centred on $l=173\pdeg91, b= +00\pdeg25$, from the IRIS 3000\,GHz map. 
		}
		\label{fig:MaxFound}
	\end{center}
\end{figure}

Our logical framework assumes a binary classification scheme, such that each region of interest is either diffuse background or a compact source. However, a binary classification model, regardless of the significant advantage of its simplicity, is not complete enough to explain the full complexity of the data set.  In fact, as described in Sect.~\ref{subsec:Reliability} (see also Appendix~\ref{sec:ContaminationBayes}), we compute the probability of a set of pixels \textit{not} being part of the diffuse background (rejection of the null hypothesis), and the \textsf{SRCSIG} statistic acts as the discriminating variable.  This mathematical machinery requires us to find a likelihood maximum in the proximity of the source position.  However, in some cases, e.g., at low Galactic latitudes or along very extended sources, that condition may not be met.  For example, in Fig.~\ref{fig:MaxFound} there are some PCCS2E positions (blue triangles) that are well separated from the actual centre of the compact object, which coincides with the likelihood maximum.  Since we have limited the likelihood ``travel'' distance to three pixels from the original PCCS2E+2E location (see Sects.~\ref{sec:MethSourceModel} and \ref{sec:AlgoImpl}), in some of these cases {\tt BeeP} fails to find a maximum.  The code then assumes that the original PCCS2+2E position is correct, and samples the likelihood field around it.  For extended sources where {\tt BeeP} could not find a likelihood maximum, such as those shown in Fig.~\ref{fig:MaxFound}, \textsf{SRCSIG} can still attain a high value because the location does not have background-like properties.  For this reason we have introduced a new catalogue field \textsf{MAXFOUND}, that flags when a likelihood maximum was found.  Considering that being above a given \textsf{SRCSIG} threshold means, it is likely that this is not part of the background.  \textsf{MAXFOUND} then allows one to discriminate between a compact object (value 1, Fig.~\ref{fig:MaxFound}, green squares) or something else (value 0, Fig.~\ref{fig:MaxFound}, red squares). 

\begin{figure}[htbp!]
	\begin{center}
		\leavevmode
		\includegraphics[width=0.495\textwidth]{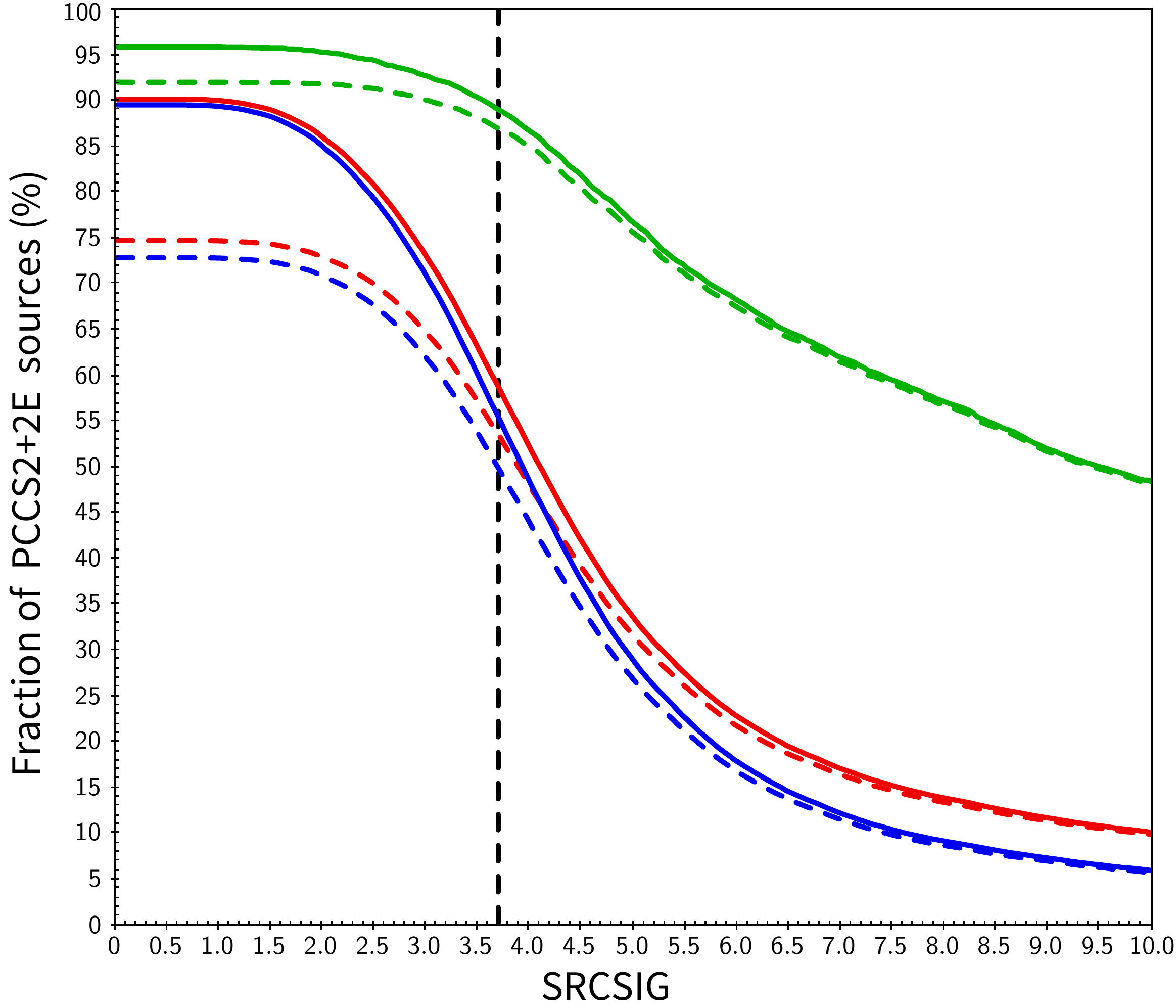}
		\caption{Fraction of PCCS2+2E sources with \textsf{NPSNR} $ > 5$ and above a given \textsf{SRCSIG} threshold.  Green curves show PCCS2 sources, blue curves show PCCS2E sources, and red curves show the full PCCS2+2E.  Dashed lines are the result of imposing \textsf{MAXFOUND}$ = 1$.  The dashed black vertical line (\textsf{SRCSIG} $= 3.7$) is the reliability criterion threshold that we have selected for the {\tt BeeP/base} catalogue.
}
		\label{fig:CatSrcSigHist}
	\end{center}
\end{figure}

In Fig.~\ref{fig:CatSrcSigHist} we show the total fraction of PCCS2+2E sources with \textsf{NPSNR}$\,{>}\,5$ and above a given \textsf{SRCSIG}.  The dashed curves in Fig.~\ref{fig:CatSrcSigHist} show the impact of adding the condition of \textsf{MAXFOUND}$\,{=}\,1$. The intersection of the curves with the \textsf{SRCSIG}$\,{=}\,0$ axis shows the fraction of sources with \textsf{NPSNR}$\,{>}\,5$.

\subsection{Quality filter}
\label{ssec:QualityFilter}

We summarize the quality of the source parameter estimates in a new field, \textsf{EST\_QUALITY}, which assigns five points to each source and subtracts penalties from this maximum value if certain quality criteria are not met.  \textsf{EST\_QUALITY} = 5 means that the estimates of source parameters are highly reliable.  Penalties subtracted if specific quality criteria are not met are listed in Table~\ref{table:QualityPenalties}.  When \textsf{MAXFOUND} $\neq 1$ (no likelihood maximum), it is not possible to guarantee an optimal extraction of source parameter estimates.  However, sources that fail only the \textsf{MAXFOUND} condition may still be used in many cases where a rigorous statistical characterization is not required.  For this reason the associated penalty was set to half of the other criteria.  Source estimates not meeting the ``outliers criterion,'' or that were examined in only the \Planck\ channels (because they are located in the IRAS gaps), should be used with great caution.

\begin{table}[htbp!]
\begingroup
\newdimen\tblskip \tblskip=5pt
\caption{Penalties applied to sources whose parameter estimates do \textit{not} meet the quality criteria (note that the maximum quality level is 5).
}
\label{table:QualityPenalties}
\nointerlineskip
\vskip -3mm
\footnotesize
\setbox\tablebox=\vbox{
   \newdimen\digitwidth
   \setbox0=\hbox{\rm 0}
   \digitwidth=\wd0
   \catcode`*=\active
   \def*{\kern\digitwidth}
   \newdimen\signwidth
   \setbox0=\hbox{+}
   \signwidth=\wd0
   \catcode`!=\active
   \def!{\kern\signwidth}
\halign{\tabskip 0pt #\hfil\tabskip 1em&
         \hfil#\hfil\tabskip 0pt\cr
\noalign{\doubleline}
Sources \textit{not} meeting:& Penalty\cr
\noalign{\vskip 3pt\hrule\vskip 5pt}
\textsf{EXT} $> 1.46 \,\wedge$ \textsf{TEMP} $< 60 \, \wedge$ \textsf{BETA} $< 5 \, \wedge$& $-2$\cr
 \quad(\textsf{TH2SB} - \textsf{TL2SB}) $> 0.8 \, \wedge$& \cr
 \quad(\textsf{BETAH2SB} - \textsf{BETAL2SB}) $> 0.25$ (outliers, Eq.~\ref{eq:QualityCriterion})& \cr
\noalign{\vskip 3pt}
\textsf{MAXFOUND} $= 1$ (no likelihood maximum)& $-1$\cr
\noalign{\vskip 3pt}
All four channels used (IRIS data missing)& $-2$\cr
\noalign{\vskip 5pt\hrule\vskip 4pt}}}
\endPlancktable                    
\endgroup
\end{table}

\subsection{{\tt BeeP/base} catalogue}
\label{subsubsec:BeePbasecatalogue}

Let us now examine the sub-catalogue defined by the conditions given in Eq.~\eqref{eq:reliabilityCriterion}.  If we require \textsf{EST\_QUALITY}$\,{\geq}\,4$, this sub-catalogue contains 24\,511 of the 43\,290 objects in the PCCS2E (56.6\,\%).  If we require \textsf{EST\_QUALITY}$\,{=}\,5$, however, we still find 21\,997 sources (50.8\,\% of the PCCS2E objects).  We therefore add this condition and define a ``reliable and accurate'' sub-catalogue based on the three following conditions:

\begin{equation}
\label{eq:catalogGoodFilter}
\textsf{NPSNR} > 5 \, \wedge \, \textsf{SRCSIG}> 3.7 \, \wedge \, \textsf{EST\_QUALITY} = 5.
\end{equation}
This sub-catalogue, which we shall refer to as {\tt BeeP/base}, contains 26\,083 (54.1\,\% of the full PCCS2+2E) objects. Unless otherwise stated, all figures in the rest of this paper are based on it.  If we require a more stringent contamination level, say below 1\,\%, (\textsf{SRCSIG}$\,{>}\,7.0$ and \textsf{EST\_QUALITY}$\,{=}\,5$), there remain 5\,077 (11.7\,\%) compact objects in the PCCS2E.	

Although in the PCCS2+2E there is no indication of the source-detection significance, for comparison we computed one by dividing the MHW2 estimates of the source flux density and its uncertainty, $\textsf{DETFLUX} / \textsf{DETFLUX\_ERR}$.  The median value of the PCCS2+2E-estimated S/N (8.96) is considerably lower than the equivalent value of \textsf{NPSNR} in the {\tt BeeP} catalogue (12.82).  However, one must remember that {\tt BeeP} is a multi-channel method, and jointly analysing more than one frequency strengthens the background-rejection criterion.

\subsection{Beyond {\tt BeeP/base} }
\label{subsec:BeyondBase}

In Sect.~\ref{subsubsec:BeePbasecatalogue} we have described how we have extracted a subset of the sources in PCCS2+2E ({\tt BeeP/base}) that we consider to be ``reliable and accurate.'' Based on our analysis, this means that:
\begin{itemize}
\item the uncertainties on the extracted model parameters are realistic;
\item the number of false detections is low.
\end{itemize}

We caution the user of {\tt BeeP/base} that the parameter uncertainties for many sources in this catalogue are relatively large. For example, Fig.~\ref{fig:SREFfractional} (supported by simulations in Appendix~\ref{sssec:Simulations}, see e.g., Fig.~\ref{fig:ConsistencySystematics}) shows that sources with the lowest \textsf{NPSNR}s  have flux-density extraction uncertainties larger than about 40\,\%. At first glance this does not seem consistent with a naive interpretation of \textsf{NPSNR} as an ``SNR-like" quantity, but we remind the reader that \textsf{NPSNR} reflects the uncertainties of {\it all\/} model parameters, not only flux-density determination. Figure~\ref{fig:CornerPlot} shows in particular that the flux-density determination is correlated with other parameters (in particular the size, temperature, and spectral index), and this certainly contributes significantly to increasing the uncertainties.

We have selected {\tt BeeP/base} as a good approach for studying the broad characteristics of the results of our analysis. However, we expect that each user of these results will select a specific subset of sources based on their own needs. For example, if low flux-extraction uncertainties are required, then the threshold on \textsf{NPSNR} should be correspondingly increased, and we suggest using Fig.~\ref{fig:SREFfractional} as a guideline. Similarly, Fig.~\ref{fig:Reliability} can be used to set a threshold related to contamination by false detections. Each user of our results should determine the specific criteria that need to be applied to meet their objectives.

\section{Base catalogue characteristics}
\label{sec:CatalogueCharacteristics}

We now describe and characterize the {\tt BeeP/base} catalogue.
As mentioned previously, all the results of this analysis (i.e., for all PCCS2+2E sources, not only those in {\tt BeeP/base}) are available online via the Planck Legacy Archive. The Explanatory Supplement \citep{planck2016-ES}, which accompanies the results, includes an annotated list of all the parameters provided for each source. In this paper, we provide a summary of the key parameters in Table~\ref{tab:CatPars}. Some of these are described in more detail in this section.

\begin{table}[htbp!]
\begingroup
\newdimen\tblskip \tblskip=5pt
\caption{Summary of the key parameters generated by {\tt BeeP} for each source in PCCS2+2E and available online via the Planck Legacy Archive. All physical parameters include corresponding uncertainties.}
\label{tab:CatPars}
\nointerlineskip
\vskip -3mm
\footnotesize
\setbox\tablebox=\vbox{
   \newdimen\digitwidth
   \setbox0=\hbox{\rm 0}
   \digitwidth=\wd0
   \catcode`*=\active
   \def*{\kern\digitwidth}
   \newdimen\signwidth
   \setbox0=\hbox{+}
   \signwidth=\wd0
   \catcode`!=\active
   \def!{\kern\signwidth}
\halign{\tabskip 0pt #\hfil\tabskip 1em&
         \hfil#\hfil\tabskip 0pt\cr
\noalign{\doubleline}
Component & Extracted Parameters\cr
\noalign{\vskip 3pt\hrule\vskip 5pt}
Source& New location \cr
(thermal model) & Extension \cr
  & Thermal SED properties \cr
   & [Temperature, Spectral index, Ref. Flux Density]\cr
    & Flux density in \Planck\ and IRAS channels\cr
   & Extraction quality parameters\cr
    & [\textsf{NPSNR, RELTH, SRCSIG, EST\_QUALITY}]\cr
\noalign{\vskip 3pt}
Source & New location \cr
(free model)    & Flux density in \Planck\ and IRAS channels\cr
  & Thermal SED properties $^{\rm a}$\cr
   & [Temperature, Spectral index, Ref. Flux Density]\cr
    & \cr
\noalign{\vskip 3pt}
Background& Surface brightness in \Planck\ and IRAS channels\cr
(32x32 pixel patch)     & Signal to noise ratios (source/background)\cr
  & Thermal SED properties \cr
   & [Temperature, Spectral index, Ref. Flux Density]\cr
\noalign{\vskip 5pt\hrule\vskip 4pt}}}
\endPlancktable                    
\tablenote {{\rm a}} Fitted to the flux densities {\it after\/} extraction.\par
\endgroup
\end{table}

\subsection{Reliability and quality parameters}

The set of reliability and quality parameters includes: 
\begin{itemize}

\item \textsf{NPSNR}, which measures the S/N of the combined detection (Eq.~\ref{eq:NPSNR_defMain});

\item \textsf{SRCSIG}, which measures the likelihood that the source is a real compact object distinct from the background (Eq.~\ref{eq:MRelthDef});

\item \textsf{EST\_QUALITY}, which measures the trustworthiness of the source descriptor estimates extracted by {\tt BeeP} (see Sect.~\ref{ssec:QualityFilter}).

\end{itemize} 

It is important not to confuse the roles of \textsf{SRCSIG} and \textsf{EST\_QUALITY}.  \textsf{SRCSIG} indicates the likelihood of a source being real, whereas \textsf{EST\_QUALITY} provides an assessment of the quality of the estimated source parameters, given that the source is real.  For instance, a bright nearby object may have a very large \textsf{SRCSIG} because we are sure it is a real object. Nonetheless it might still fail the \textsf{EST\_QUALITY} criteria if, for example, {\tt BeeP} cannot find the likelihood peak. In that case there is no guarantee that the recovered parameter estimates are optimal.

\subsection{Source properties}
\label{subsec:SourceProperties}

This set of parameters gives the position and properties of the sources and their uncertainties.

\subsubsection{Thermal properties}
\label{ssec:ThermalProperties}

\begin{figure*}[htbp!]
	\begin{center}
		\leavevmode
		\includegraphics[width=0.80\textwidth]{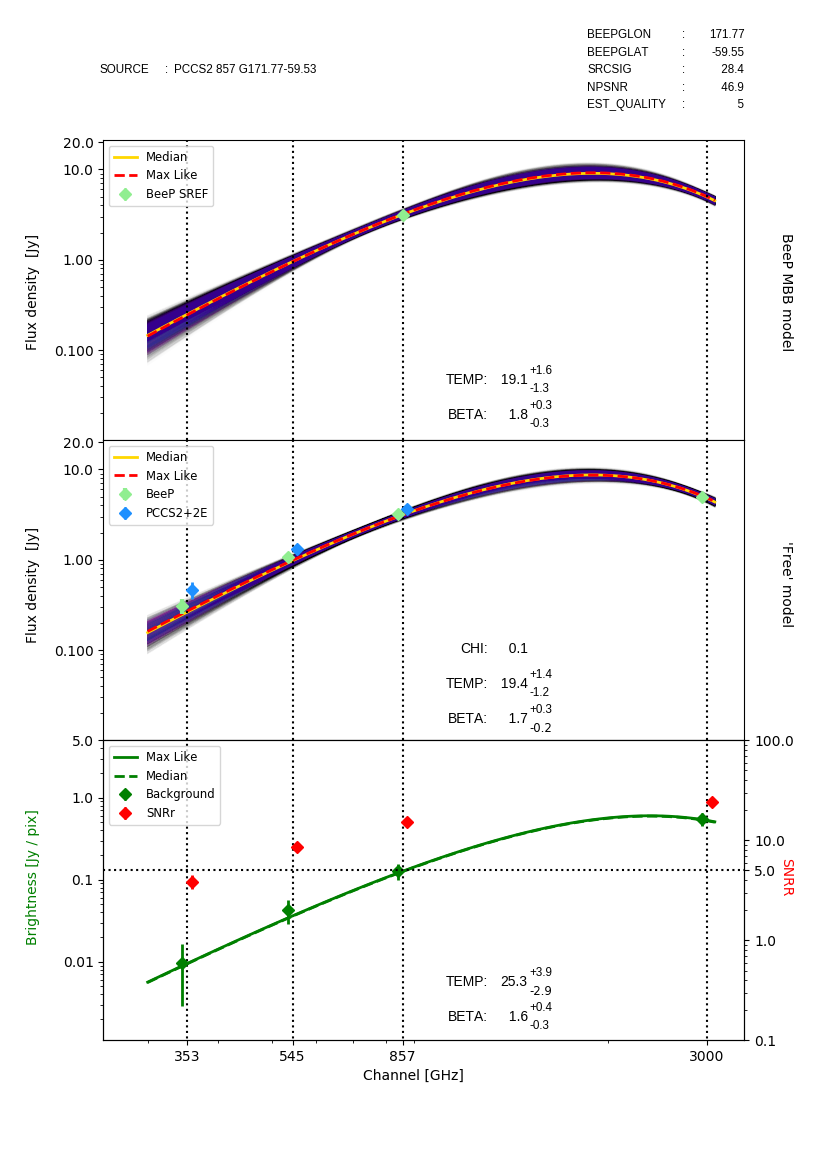}
		\caption{Example of fitting the MBB (upper panel) and Free (middle panel) SED models to the data for one source  (NGC 895).  The background is given in the bottom panel.  The yellow and red dashed curves are the median and maximum-likelihood fits, respectively. The purple and black bands are the $\pm1\,\sigma$ and $\pm2\,\sigma$ regions, respectively, of the posterior density.  Blue diamonds are the PCCS2+2E flux-density estimates (\textsf{APERFLUX}).  The green diamonds are: in the upper panel {\tt BeeP}'s estimate of the flux density at 857\,GHz, and in the middle panel  {\tt BeeP}'s Free estimates of the flux density at each frequency.  In the lower panel, dark green diamonds are the background brightness estimates at each frequency, and the green curves are the maximum likelihood (dashed) and the median (solid) models.  Red diamonds are the average source brightness divided by the background rms brightness in that patch, i.e., raw S/N.  The data points are slightly displaced from their nominal frequencies to avoid overlaps. A similar plot is provided in the Planck Legacy Archive for each source in the {\tt BeeP} catalogue; see the Planck Explanatory Supplement for further information (\url{http://www.cosmos.esa.int/web/planck/pla/}).  We note that this figure is reproduced exactly as it will be delivered to the user from the online archive. In Appendix~\ref{sssec:Examples} we provide some representative examples of spectra for different kinds of sources, to show some of the results obtained by {\tt BeeP}.}  
		\label{fig:SEDPlot}
	\end{center}
\end{figure*}

We fit the multifrequency data for a given source with two SED models (see Fig.~\ref{fig:SEDPlot}), each of which requires an independent run of the likelihood.

\begin{itemize}

\item \textbf{Modified Blackbody (MBB) model}. The source brightness levels are colour-corrected to account for the detector bandpasses.  The following parameters are optimized by the likelihood:
	\begin{itemize}
		\item $X$ and $Y$ position coordinates, with origin at the PCCS2+2E position;
		\item ${\textsf{EXT}}$, source extension;
		\item ${\textsf{SREF}}$, source reference flux density;
		\item ${\textsf{TEMP}}$, source temperature;
		\item ${\textsf{BETA}}$, source spectral index.
	\end{itemize}
All source parameters, geometrical and physical, are sampled jointly. The reference flux density is given at 857\,GHz.  The reference flux density at 857\,GHz is not the flux density measured in the 857-GHz channel; it is rather a scaling factor for the model that could be specified at any frequency.  We have chosen 857\,GHz for convenience (see Eq.~\ref{eq:MBB_SED}).  For this model we also provide the flux densities in the individual channels, computed from the fitted model.

\vskip 3mm
	
\item \textbf{Free model}.  The \textsf{FREE} columns are developed in two steps. First, samples are drawn from the geometrical parameters and flux densities at each channel.  The flux densities at individual channels are optimized by the likelihood.  All source parameters, geometrical and physical, are sampled jointly.  From the flux-density samples at each frequency we compute a best-fit value and an uncertainty.  The following parameters are optimized by the likelihood:
	\begin{itemize}
		\item $X$ and $Y$ position coordinates, with the origin at the PCCS2+2E position;
		\item ${\textsf{EXT}}$, source extension;
		\item ${\textsf{FREES3000}}$, flux density at 3000\,GHz;
		\item ${\textsf{FREES857}}$, flux density at 857\,GHz;
		\item ${\textsf{FREES545}}$, flux density at 545\,GHz;
		\item ${\textsf{FREES353}}$, flux density at 353\,GHz.
	\end{itemize}
We then fit an MBB model to the four data pairs ($S_\nu,\,\sigma_{S_\nu}$), using a Gaussian likelihood with colour-correction, resulting in a source reference flux density given at 857\,GHz.
\end{itemize}

\begin{figure*}[htbp!]
	\begin{center}
		\leavevmode
		\includegraphics[width=0.99\textwidth]{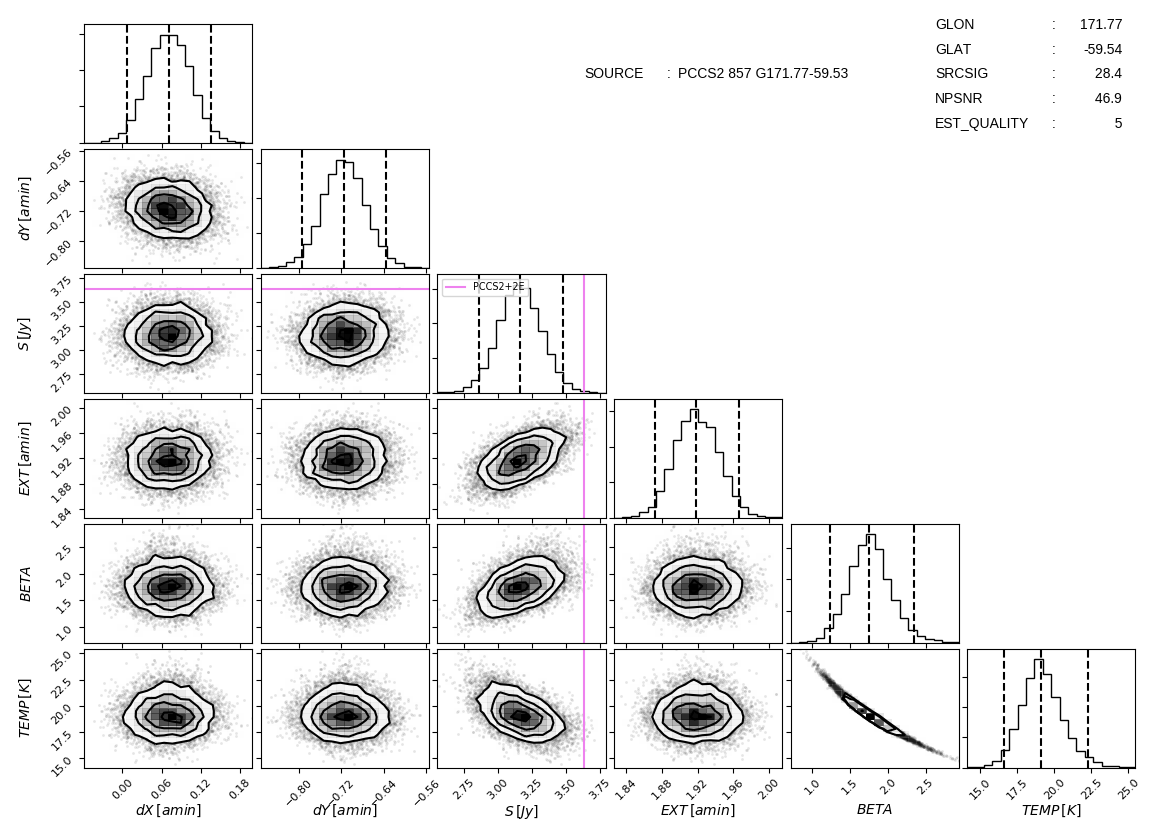}
		\caption{Corner plot \citep{cornerPlot} of parameter posterior distributions for one source (NGC 895).  Off-diagonal positions show marginalized bi-dimensional posterior distributions of the parameter samples defining the row and the column.  Diagonal positions contain posterior marginalized distributions.  The magenta lines mark the PCCS2+2E catalogue flux density in the 857-GHz channel.  There is one such plot for each source in {\tt BeeP}'s catalogue.  The source extension (\textsf{EXT}) samples shown have not been corrected for the narrower beams employed in the likelihood.  See the Planck Explanatory Supplement for further information (\url{http://www.cosmos.esa.int/web/planck/pla/}). This figure is reproduced exactly as it will be delivered to the user from the online archive.}
		\label{fig:CornerPlot}
	\end{center}
\end{figure*}

{\tt BeeP} also provides, as an output, plots of the source-parameter posterior distributions for the MBB model (see an example in Fig.~\ref{fig:CornerPlot}).

\subsubsection{Size}
\label{ssec:SourceSize}

The spatial extent of source-related emission peaks in the maps results from the convolution of the source size and the beam. These are degenerate variables over the relatively narrow range of variation of beam size in the \Planck\ maps.
 {\tt BeeP} uses a source-extension parameter \textsf{EXT} which represents the intrinsic radius of the source in Eq.~\eqref{eq:SourcesModel2}. However, in Appendix~\ref{ssec:AlgoImplSrcDetec}, we explain that  {\tt BeeP} artificially narrows the beams to allow for emission bumps in the maps that are narrower than the beam size. Therefore \textsf{EXT} does not correspond to the actual intrinsic source size; however, \textsf{EXT} is easily corrected to a new parameter \textsf{R}, which is the intrinsic source radius corresponding to the real beam sizes. Both parameters are provided in the  {\tt BeeP} results. Furthermore, we remind the reader that we have simplified the source model by assuming that it is a symmetrical 2D Gaussian. The parameter \textsf{R} thus gives a useful indication of whether the source is extended, but it does not reflect any potential source elongation and should therefore be used with appropriate caution.

\begin{figure}[htbp!]
	\begin{center}
		\leavevmode
		\includegraphics[width=0.49\textwidth]{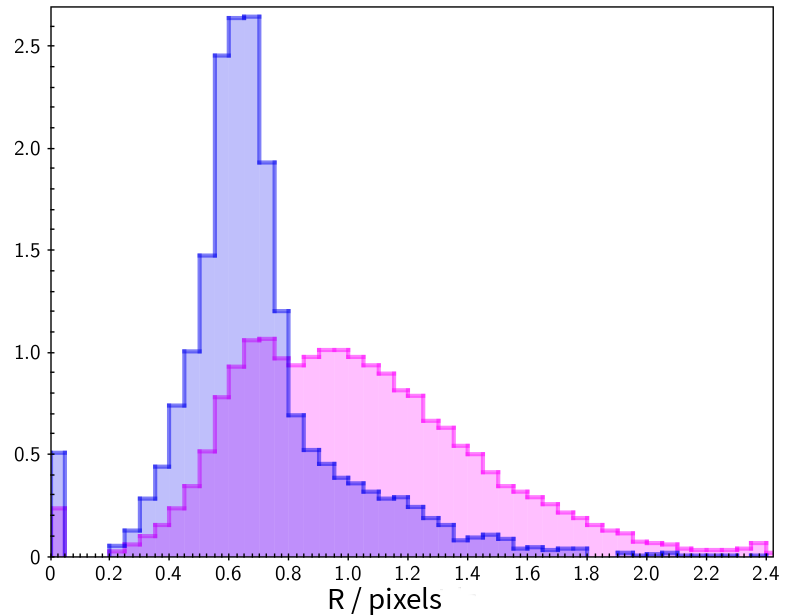}
		\caption{Normalized histograms of the recovered source size \textsf{R}
		PCCS2 sources are shown in blue and the full catalogue in purple.  \textsf{R} has been corrected for the excess resulting from using narrower beams in the likelihood.  Beam-sized objects appear in the figure at $\textsf{R} \sim 0$.  One pixel here corresponds to 1\parcm72.
		}
		\label{fig:realR_histogram}
	\end{center}
\end{figure}

The distribution of source radii (\textsf{R}) found by {\tt BeeP} is shown in Fig.~\ref{fig:realR_histogram}.  The PCCS2 subset (shown in blue), is compatible with a population overwhelmingly dominated by unresolved sources (the size distribution peaks at 1\parcm2).  Instead, the full PCCS2+2E (purple) set peaks at 1\parcm7. This is expected, since a large fraction of the PCCS2E objects are nearby and Galactic, and many of them show more extended shapes.

\subsubsection{Position}
\label{ssec:CatPositionalAccuracy}

\begin{figure}[htbp!]
	\begin{center}
		\leavevmode
		\includegraphics[width=0.49\textwidth]{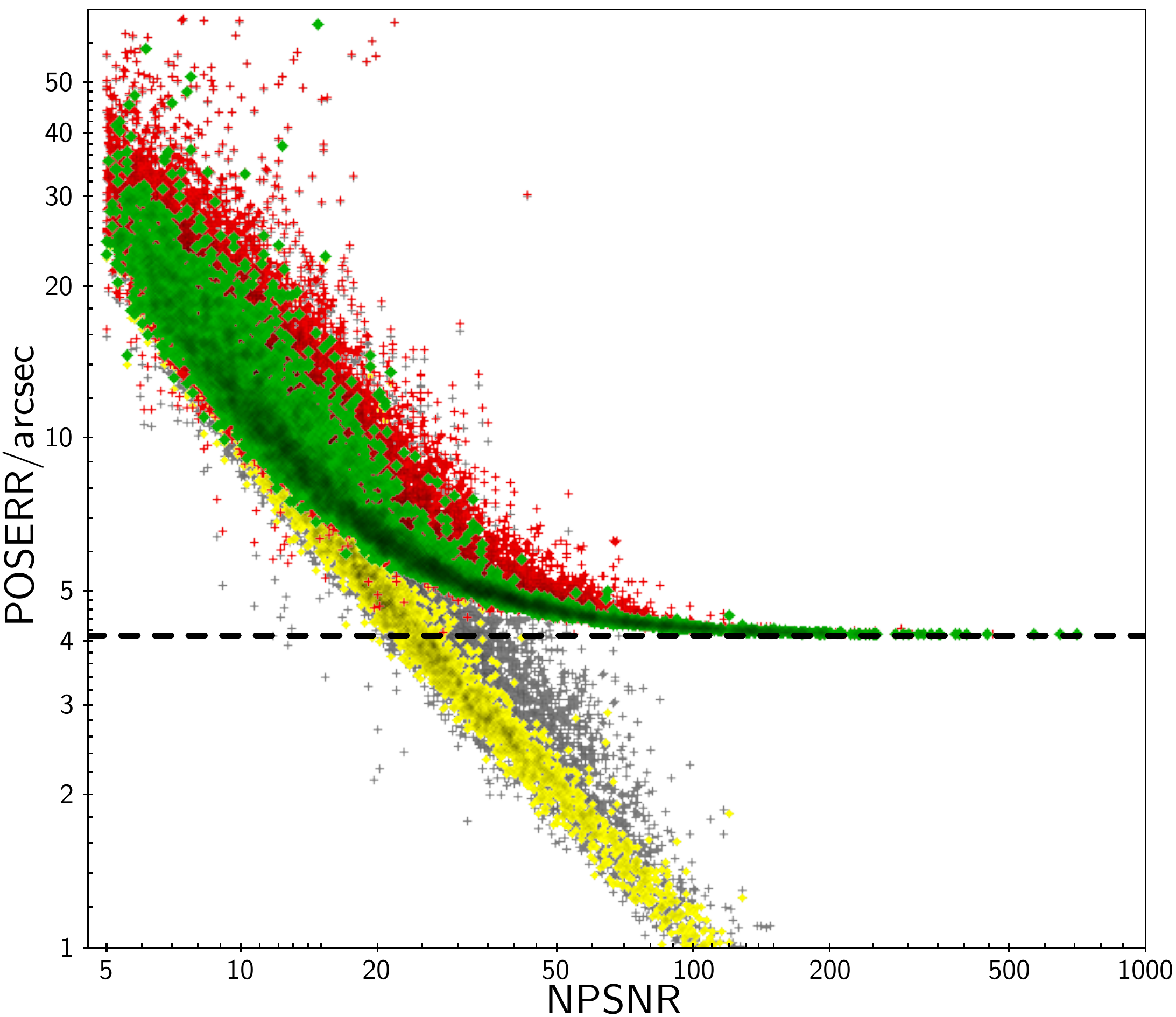}
		\caption{Radial position error \textsf{POSERR} versus \textsf{NPSNR}.  Grey and yellow points mark sources in the PCCS2+2E and PCCS2, respectively, before correction.  Red and green points mark sources in the PCCS2+2E and PCCS2, respectively, after correction.  The horizontal dashed line is the saturation constant added to correct the position uncertainty, 4\parcs11.}
		\label{fig:CatPositionalAccuracy}
	\end{center}
\end{figure}

One of the important characteristics of {\tt BeeP} is its ability to determine an effective sub-pixel source position.  Since the position is determined from a multifrequency analysis, it does not in general correspond to any of the positions found in PCCS2+2E.  \textsf{POSERR} is the uncertainty radius around the position.  Its probability density function is a Rayleigh distribution with a scaling parameter equal to \textsf{POSERR}. If $Z = \sqrt{X^2 + Y^2}$ and \{X,Y\} are independent and both normally distributed with a standard deviation $\sigma$, then $Z$ follows a Rayleigh distribution with a scaling parameter equal to $\sigma$.  {\tt BeeP}'s sub-pixel accuracy significantly reduces the large negative kurtosis usually imposed by the pixelization on the error distributions, as can be seen in Fig.~\ref{fig:CornerPlot}. \textsf{POSERR} is computed as the 95th percentile of the samples' radial offset distribution divided by 2.45, to give $\sigma$, the Rayleigh scale factor.  The probability that the true source position is inside a radius of (1$\times$, 2$\times$, 3$\times$) \textsf{POSERR} is (39.3\,\%, 86.5\,\%, 98.9\,\%).  Figure~\ref{fig:CatPositionalAccuracy} shows the dependence of \textsf{POSERR} on \textsf{NPSNR}.

Simulations show that \textsf{POSERR} is significantly underestimated in a subset of cases, predominantly those with high values of \textsf{NPSNR}.  A detailed description of this issue is given in Sect.~\ref{psssec:InjectSimulations} and shown in Fig.~\ref{fig:PosAccuracySNR}.  To address this problem, we correct the position errors using the procedure developed in Sects.~\ref{psssec:InjectSimulations} and \ref{psssec:FFP8Simulations}, which follows closely that used for PCCS2 \citep[see equation~7 and table~8 of][]{planck2014-a35}.  The correction consists of adding a term in quadrature to \textsf{POSERR}, which causes small values to saturate at a minimum level of $\sigma_0 = 4\parcs4$ (see Fig.~\ref{fig:CatPositionalAccuracy}).  This level was determined through simulations, as described in Sects.~\ref{psssec:InjectSimulations} and \ref{psssec:FFP8Simulations}.

To verify that the correction determined through simulations applies to the {\tt BeeP/base} catalogue, we examined the PCCS2 subset.  The correlation seen in Fig.~\ref{fig:CatPositionalAccuracy} (yellow dots) is very high ($-0.98$), and its slope $a = -1.09$ is very close to what is seen in the simulations. This high degree of consistency between the simulated data and the real data justifies application of the  correction to the data.

The median positional error of the full corrected catalogue is 11\parcs5 (1/9 of a \Planck\ pixel).  For the PCCS2 subset it is 7\parcs9, or less than $1/12$ of a pixel.

\subsubsection{Flux density}
\label{ssec:FluxDensityAccuracy}

To obtain an unbiased estimate of a flux density, one must know the shape of the instrumental beam and the morphology of the source.  By using a constant Gaussian shape to model the beam, equal to the average \Planck\ Gaussian effective beam \citep{FeBeCop}, we introduce a systematic bias in estimates of the flux density \citep[see, e.g.,][section~2 and table~2]{planck2014-a35}.  Furthermore, in any multi-channel analysis such as {\tt BeeP}, the beam shape is not as clearly defined as in the case of a single-channel catalogue.  The effective beam is in fact a combination of the individual channel beams, and it changes with the beam spatial Fourier mode (via the covariance) and source SED parameters.  A simple correction such as the one suggested in \cite{planck2014-a35} is insufficient in this case.  Instead, our approach is to ``calibrate'' the bias in the output of {\tt BeeP} using simulations. This is explained in detail in Sect.~\ref{psssec:FFP8Simulations} (see also Sect.~\ref{ssec:AlgoImplSrcDetec}).  The simulations that we use are the \Planck\ FFP8 simulations, which are the most complete and realistic for \Planck\ 2015 data, and which contain accurate sky and instrument models.  Using the FFP8 simulations (Sect.~\ref{psssec:FFP8Simulations}), and comparing recovered values to input values, we estimate that {\tt BeeP}'s reference flux-density estimator is biased high by about 11.0\,\%, which reflects the lack of realism of our model regarding source extension.
An $11\,\%$ reduction in the reference flux densities produced by {\tt BeeP} is therefore applied to both SED models (MBB and Free).  Specifically, flux densities in all four channels are reduced by this same factor for the Free model.

\begin{figure*}[htbp!]
	\begin{center}
		\leavevmode
		\includegraphics[width=0.33\textwidth]{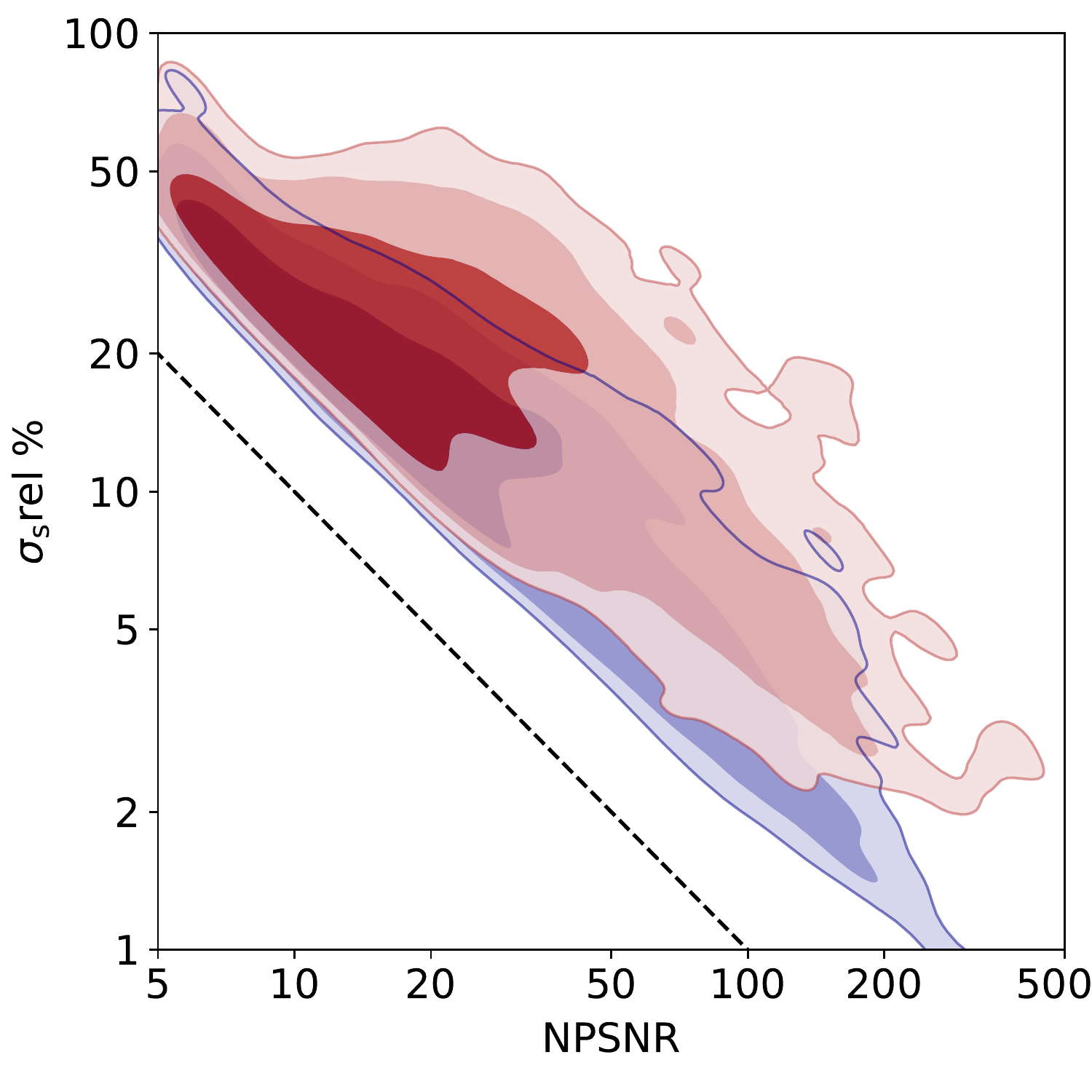}
		\includegraphics[width=0.33\textwidth]{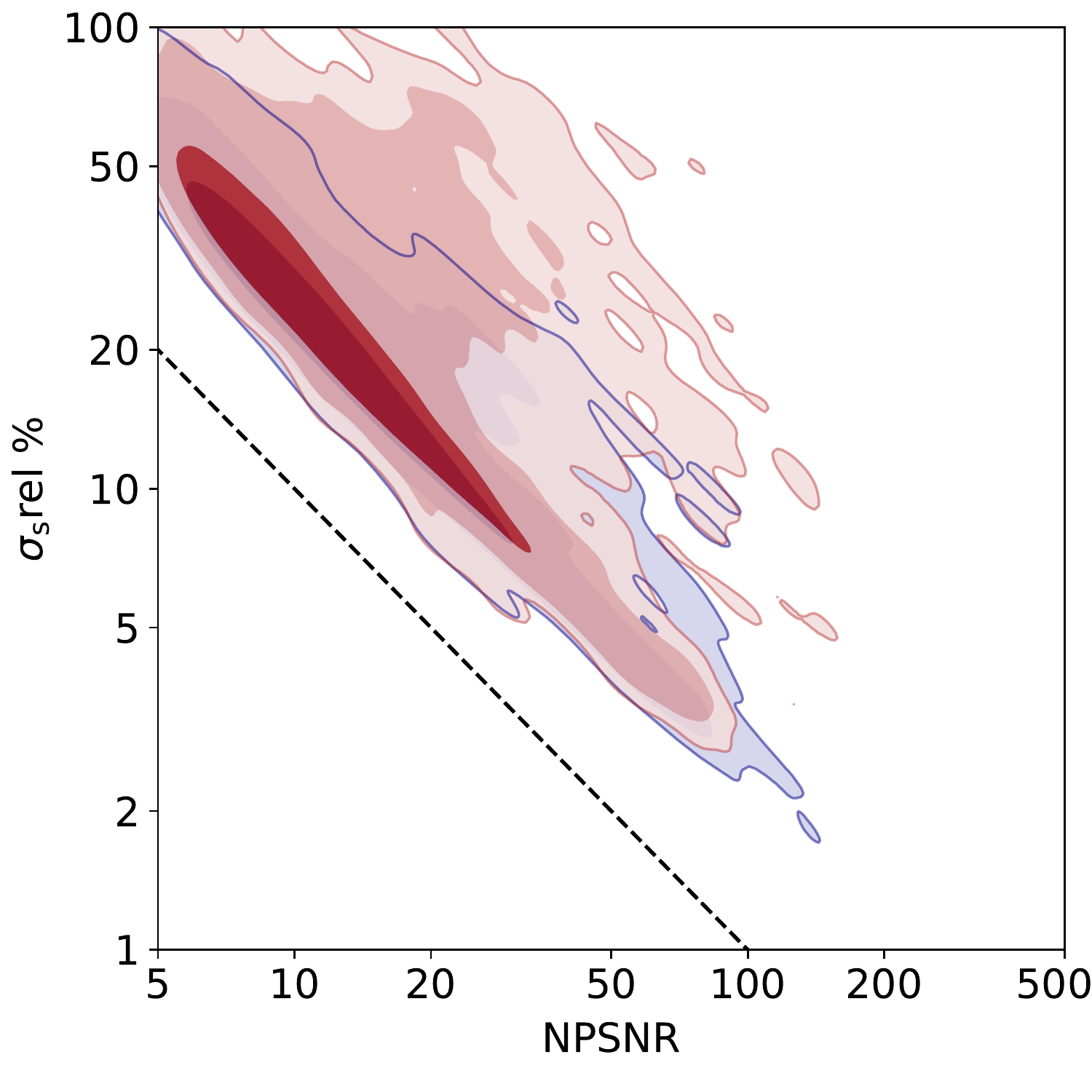}
		\includegraphics[width=0.33\textwidth]{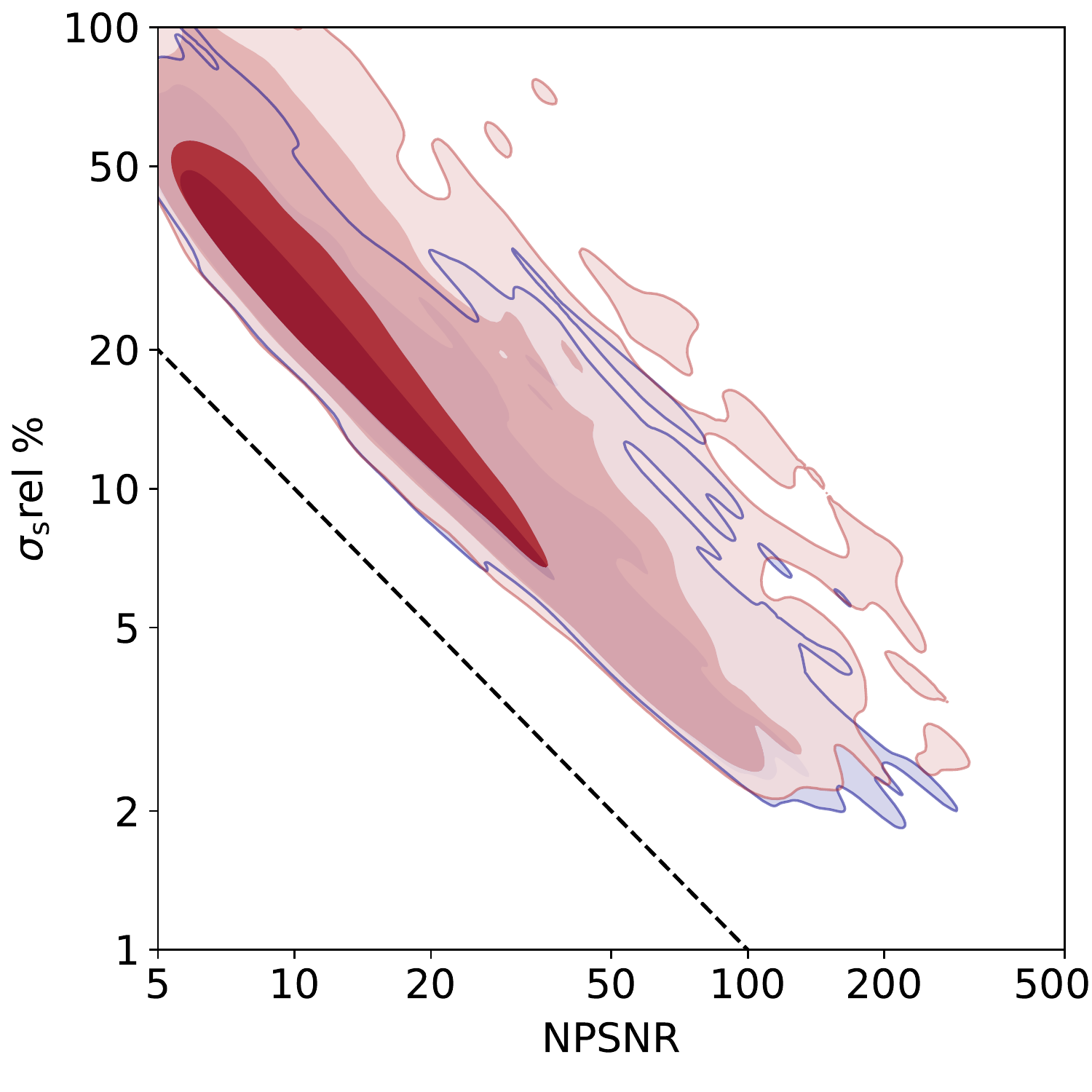}
		\caption{Flux density uncertainties  ($\sigma^{\rm rel}_S$, see Eq.~\ref{eq:S-SNR}, as a percentage) versus \textsf{NPSNR}.  Blue contours ([$68,95,99$]\%) show the distribution of uncorrected values (as presented in the catalogue), while red contours show the distribution of values after the corrections suggested in Sect.~\ref{psssec:InjectSimulations}.  The left panel depicts sources in the PCCS2, the middle panel shows sources in the PCCS2E with Galactic latitude greater than 10\deg, and the right panel displays the sources close to the Galactic plane, with $b \leq 10^\circ$.  For reference, we show $\texttt{NPSNR}^{-1}$ (black dashed line), the theoretical lower boundary for $\sigma^{\rm rel}_S$, which can only be achieved if flux density is the sole parameter in the model.}
		\label{fig:SREFfractional}
	\end{center}
\end{figure*}

The estimated flux-density accuracy is also subject to systematic effects caused by beam and source shapes. Figure~\ref{fig:SREFfractional} displays the variation of the relative flux-density error bar $\sigma^{\rm rel}_S$, defined as
\begin{equation}
\label{eq:S-SNR}
\sigma_S^{\rm rel} \equiv \frac{\Delta S}{S}\,, 
\end{equation}
where $S$ is the estimated flux density, $\Delta S$ is the estimated flux-density uncertainty, and $\sigma^{\rm rel}_S$ is the inverse of the measured S/N.  For reference, the black dashed line on the left lower corner is the $\textsf{NPSNR}^{-1}$ line. This is the theoretical lower boundary for $\sigma^{\rm rel}_S$ that would be expected if the only unknown parameter were the flux density.  Figure~\ref{fig:SREFfractional} shows that the catalogue's flux-density uncertainties are much higher ($\sigma^{\rm rel}_S \gg \textsf{NPSNR}^{-1}$) than the lower boundary, which should be expected from the fact that there are five more unknown parameters, whose individual uncertainties propagate into the flux-density estimate.
However, not all of the additional parameters contribute equally.  Inspecting the posteriors in Fig.~\ref{fig:CornerPlot}, it becomes clear that $\textsf{EXT}$ and the MBB parameters $\{T, \beta \}$ have a much larger contribution than the position parameters.  The correlation between the flux errors and the other parameter uncertainties explains the gap between the black dashed line and the green points in the figure.  However, with the help of simulations (see Sects.~\ref{subsec:IntroducingSourceExtension} and \ref{psssec:FFP8Simulations}), we find that the estimated flux-density errors are overly optimistic for a fraction of the high \textsf{NPSNR} population. The situation is similar to that for the positional accuracy estimates (see Sect.~\ref{ssec:CatPositionalAccuracy}).  For most purposes the (uncorrected) flux-density estimates and uncertainties found in the catalogue can be used without concern.  But if a more rigorous statistical characterization is required, we suggest correcting the flux-density uncertainty estimates using the procedure developed in Appendix~\ref{subsec:IntroducingSourceExtension}
%
There is a modest penalty in flux-density accuracy for applying this correction (Fig.~\ref{fig:SREFfractional}, red contours). 

\begin{figure}[htbp!]
	\begin{center}
		\leavevmode
		\includegraphics[width=0.49\textwidth]{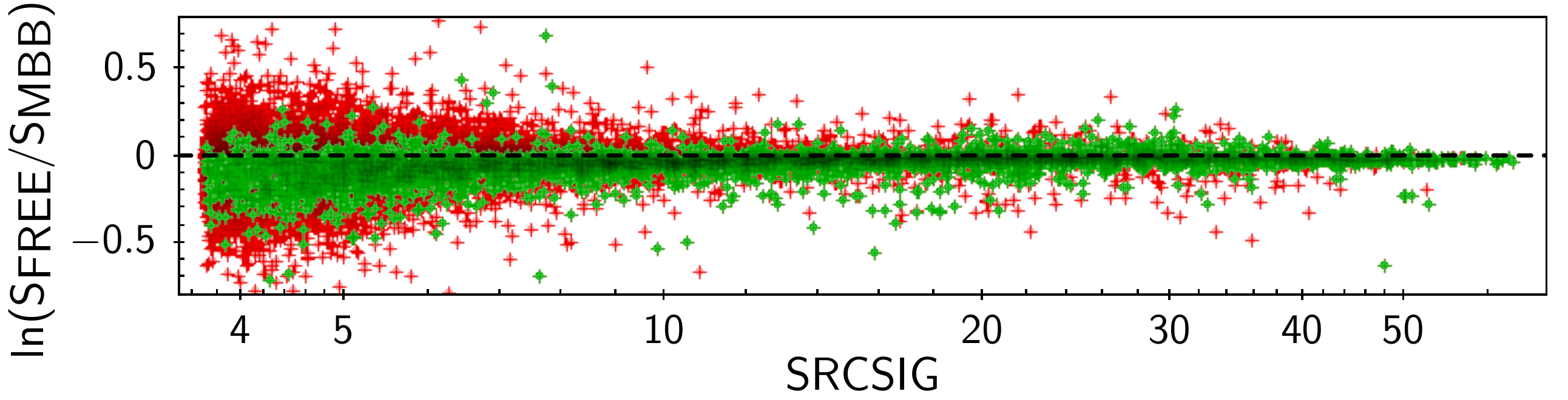}
		\caption{Ratio of Free flux density to MBB model flux density at 857\,GHz, as a function of \textsf{SRCSIG}.  PCCS2 sources are shown in green, while PCCS2+2E are in red.  Only sources whose Free and MBB positions are within 0\parcm8 (half a pixel) of each other, and whose MBB fit to the independent flux measurements has reduced $\chi^2 < 5$ are included.  The total number of sources included (25\,236) are $\approx 97\,\%$ of the {\tt BeeP/base} catalogue.
		}
		\label{fig:SrefMbbFree}
	\end{center}
\end{figure}

{\tt BeeP} produces two sets of flux-density estimates: the MBB \textsf{SREF} and the Free \textsf{FREESREF}.
In Fig.~\ref{fig:SrefMbbFree} we compare their values to test the consistency between the two models.  Instead of simply calculating percentage differences, we plot the logarithm of the output to input ratio.  If ${\rm out/in} \approx 1$, then $\ln({\rm out/in}) \sim ({\rm out} - {\rm in})/{\rm in}$, which corresponds closely to percentages.  But when ${\rm out/in}$ is far from 1, then $\ln({\rm out/in})$ keeps the symmetry between in and out, which would not be the case with the more common $({\rm out} - {\rm in})/{\rm in}$ formula.  We find this feature very convenient for visually identifying biases in the differences.

As expected, there is higher dispersion for sources drawn from the PCCS2E catalogue (shown in red), as a result of generally more complex backgrounds at low Galactic latitudes. Sources from the PCCS2 (shown in green) are less affected by this issue. We note the small (3.5\,\%) bias towards negative values of $\ln(S_{\rm Free}/S_{\rm MBB})$. This bias becomes more pronounced at lower values of \textsf{SRCSIG}.  A possible source of this bias is that inclusion of inter-frequency cross-correlations in the likelihood for the background model allows for better removal of background emission, on average raising $S_{\rm MBB}$.


\subsubsection{Spatial distribution of the source properties}
\label{ssec:CatPropsThermal}

\begin{figure*}[htbp!]
	\begin{center}
		\leavevmode
		\includegraphics[width=0.85\textwidth]{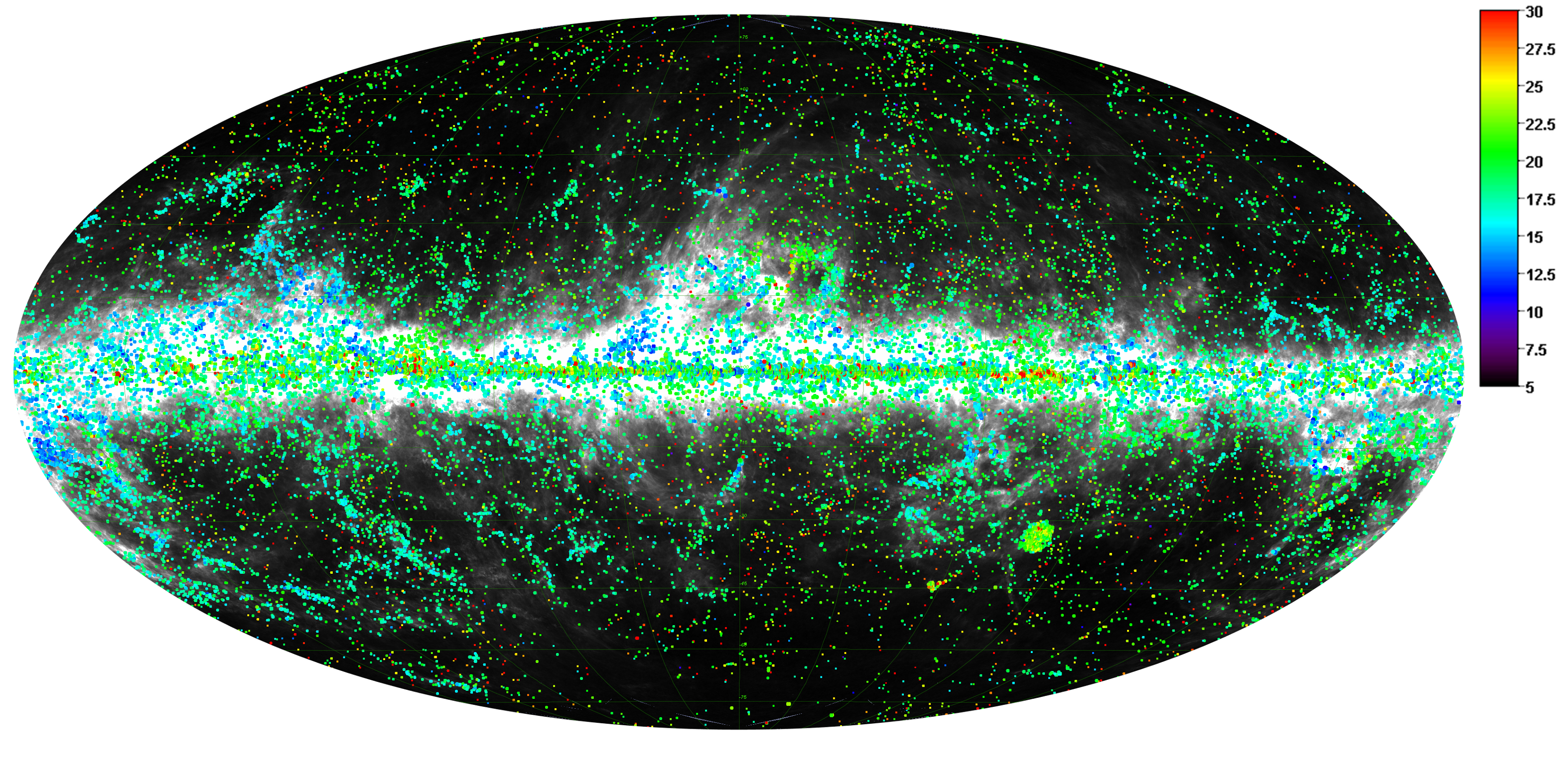}\\
		\includegraphics[width=0.85\textwidth]{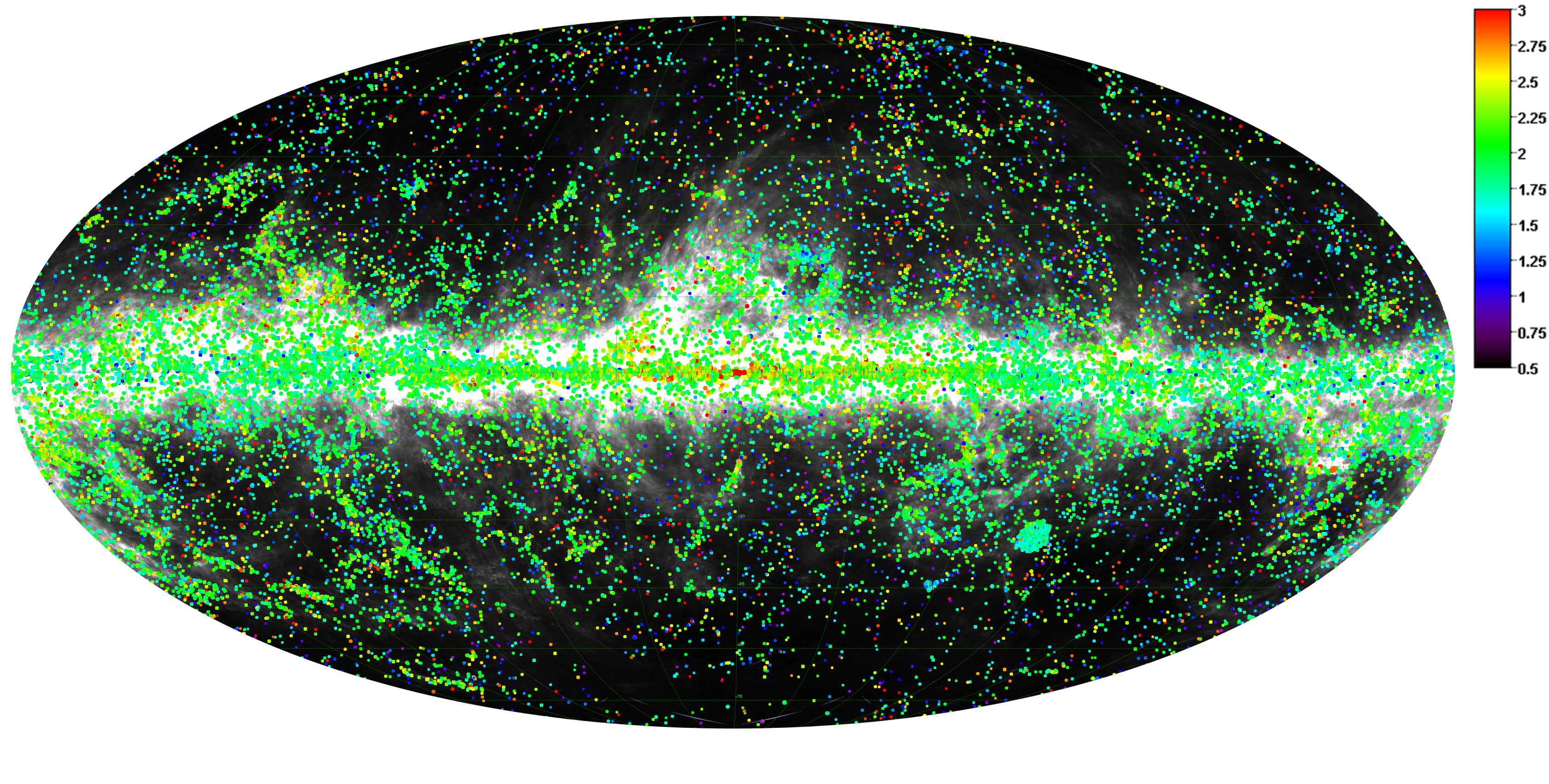}
		\caption{
			{\it Top:} Temperatures of sources in the catalogue (colour scale in thermodynamic kelvins).  {\it Bottom:} Spectral indices $\beta$ of the MBB SED model.  The catalogue was filtered using the condition of Eq.~\eqref{eq:catalogGoodFilter}.  The size of each circle representing an object is proportional to the logarithm of the source flux density in janskys. This figure also makes clear the extent that sources in PCCS2+2E trace cirrus; see also the smaller region in Fig.~\ref{fig:IntroCirrusRegion}.
}
		\label{fig:SourcesThermalProps}
	\end{center}
\end{figure*}

\begin{figure}[htbp!]
	\begin{center}
		\leavevmode
		\includegraphics[width=0.49\textwidth]{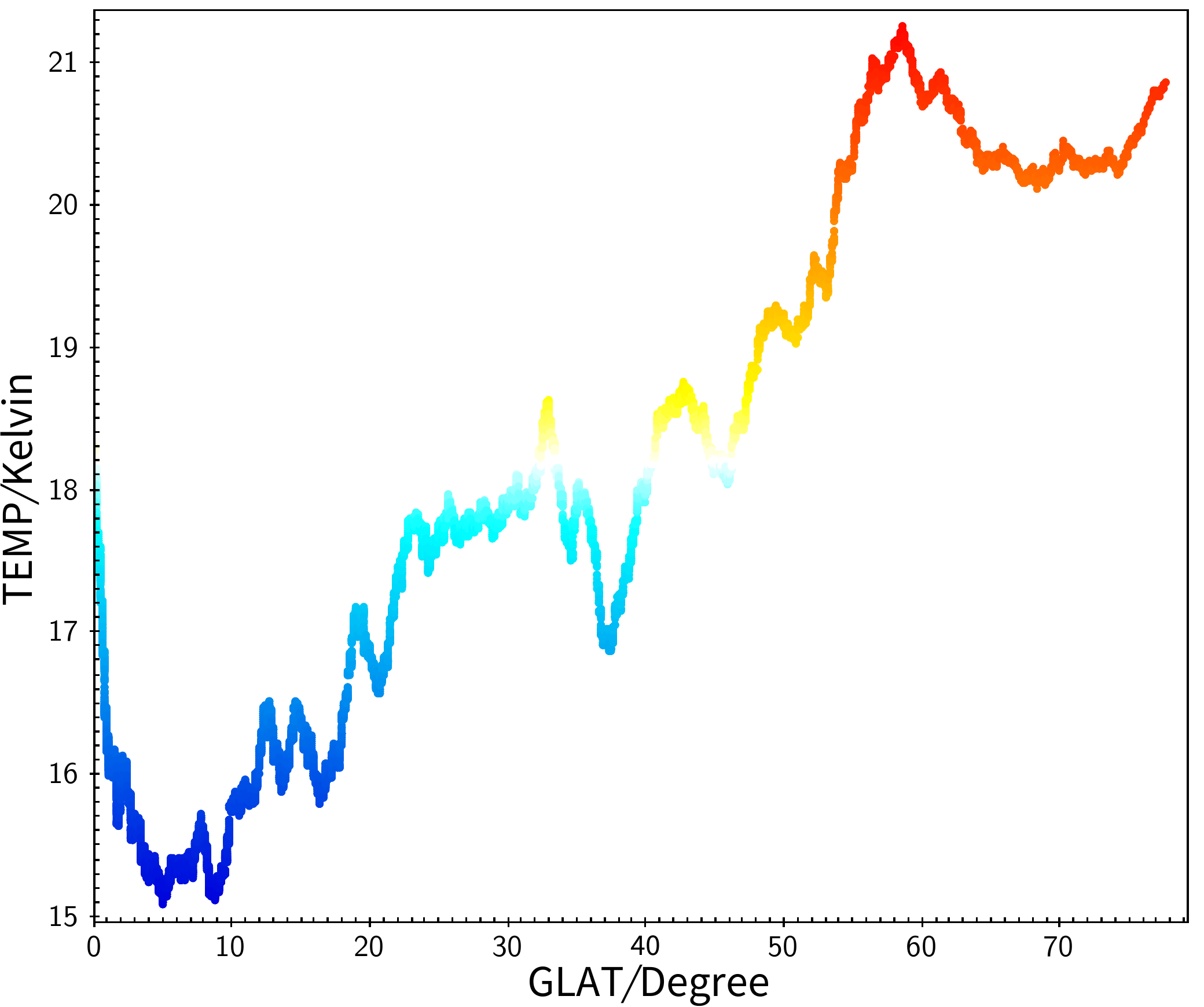}
		\caption{Boxcar average (window 500 samples) temperature of sources ordered by absolute Galactic latitude.  There is a clear trend.}
		\label{fig:SourcesGlatAverage}
	\end{center}
\end{figure}

Figure~\ref{fig:SourcesThermalProps} shows the spatial distribution on the sphere of MBB prameters $T$ and $\beta$ for the compact sources.  High Galactic latitudes show a larger percentage of warmer objects (see Fig.~\ref{fig:SourcesGlatAverage}) and very few cold sources (dark blue).  Cold sources are mostly Galactic in nature, and aligned with filaments of gas and dust.  They also match regions of intense star formation.  As a result of the strong correlation between $T$ and $\beta$
(see Sect.~\ref{ssec:CatSourcePopulations} and Fig.~\ref{fig:CornerPlot}), a higher density of sources with low $\beta$ is expected at higher Galactic latitudes.

\begin{figure*}[htbp!]
	\begin{center}
		\leavevmode
		\includegraphics[width=\textwidth]{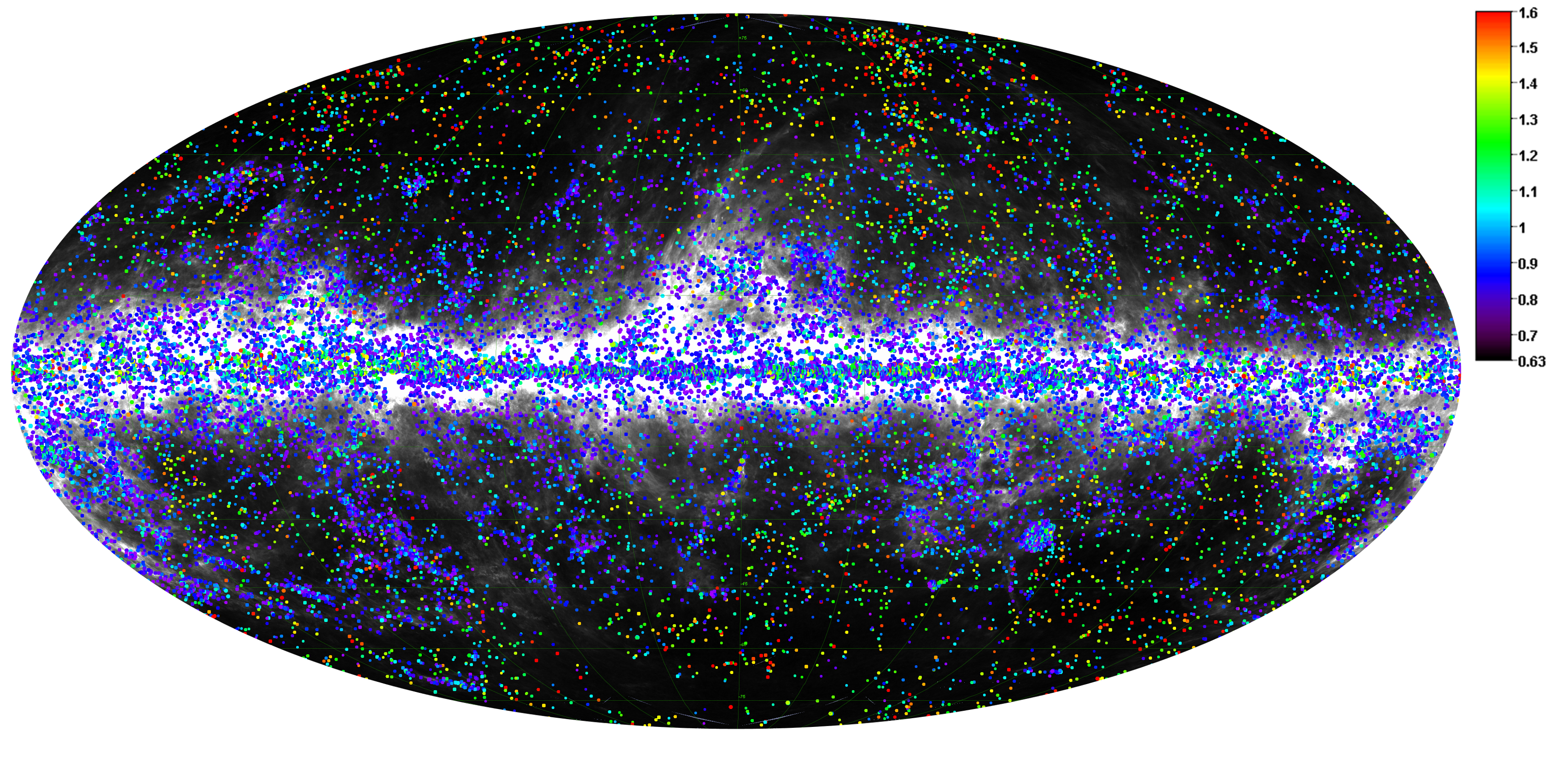}
		\caption{Spatial distribution of the significance statistic \textsf{SRCSIG}.  The colour bar represents \textsf{SRCSIG} on a logarithmic scale.
		}
		\label{fig:SourcesSrcsigProps}
	\end{center}
\end{figure*}

One problem with the extraction of this catalogue is the severe non-homogeneity of the background.  The brighter sources, represented with larger circles, are concentrated in the Galactic plane (see Fig.~\ref{fig:SourcesThermalProps}).  However, as one can see in Fig.~\ref{fig:SourcesSrcsigProps}, the regions with higher \textsf{SRCSIG} are preferentially located at high Galactic latitudes, roughly matching the PCCS2 domains.
This is the result of smoother backgrounds and less severe non-Gaussianity.

\subsubsection{Source populations}
\label{ssec:CatSourcePopulations}

\begin{figure*}[htbp!]
	\begin{center}
		\leavevmode
		\includegraphics[width=0.49\textwidth]{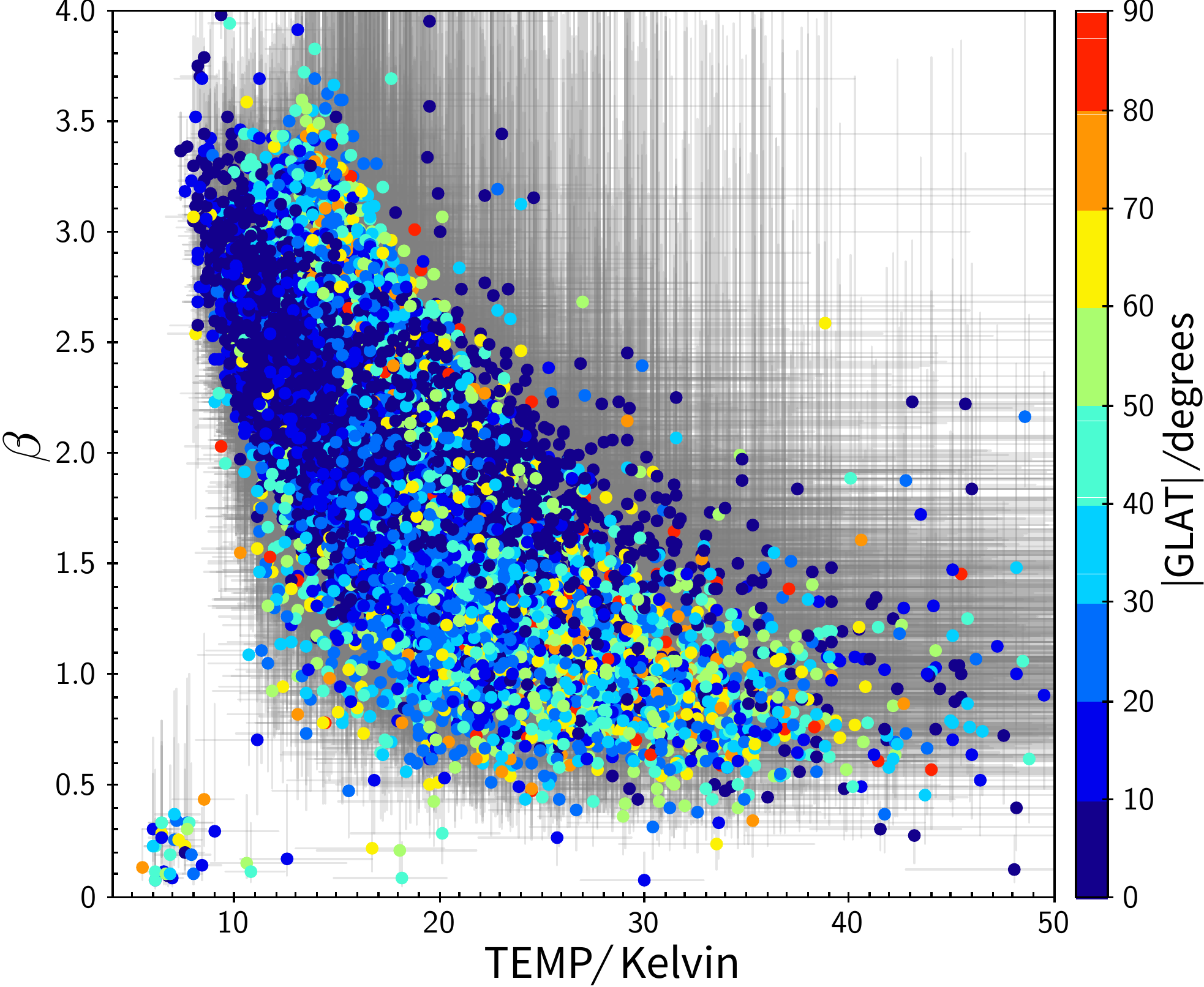}
		\includegraphics[width=0.49\textwidth]{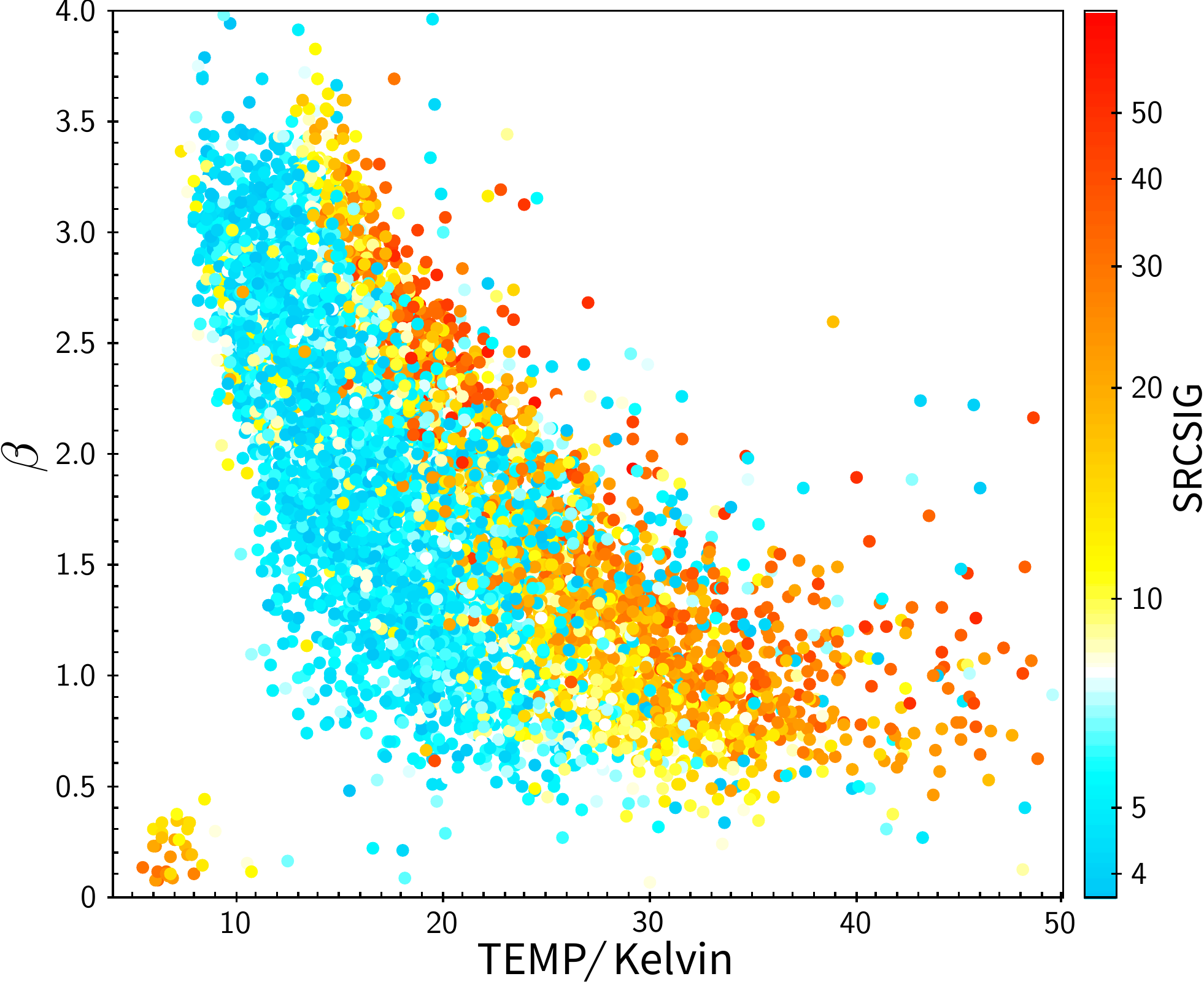}
		\caption{MBB parameters $\beta$ and $T$ for sources in the catalogue.  The sources are coloured by Galactic latitude in the left panel, with $1\,\sigma$ error bars in grey, and by the \textsf{SRCSIG} statistic in the right panel.  The small cluster of sources close to the lower left corner is the non-thermal population of flat-spectrum sources.
		}
		\label{fig:DistribTempBeta}
	\end{center}
\end{figure*}

Figure~\ref{fig:DistribTempBeta} (left panel) shows the catalogue MBB estimates on the $T$--$\beta$ plane, coloured by Galactic latitude.  The $T$--$\beta$ set forms a banana-shaped distribution with an excess of colder sources ($T < 18$\,K) at low Galactic latitudes.  This cold population was the main target of the Planck Catalogue of Galactic Cold Clumps \citep[GCC,][see Sect.~\ref{sssec:XternalCatalogues}]{planck2014-a37}.  {\tt BeeP}'s likelihood has a more inclusive selection criterion, since it is not limited to sources embedded in warmer backgrounds.  However, as may be seen in Sect.~\ref{sssec:XternalCatalogues}, the temperature contrast between source and background boosts the detection strength (\textsf{NPSNR}).  The $T$--$\beta$ uncertainty (in grey) can be important, particularly for warmer ($T \geq 18\,$K) and steeper sources ($\beta \geq 1.5$).  Most of the warmer sources are faint at 353 and 545\,GHz, such that they are just above or even below the background levels.  This severely reduces {\tt BeeP}'s constraining power, since then only the two higher frequency channels contribute significantly. 


Figure~\ref{fig:DistribTempBeta} (right panel) illustrates more clearly the influence of individual channels on the overall significance, which is affected by all the channels processed by {\tt BeeP}.  Colder sources are brighter, for the same reference flux density, at the three \Planck\ channels, which have much lower noise than IRIS.  Naively, one would thus expect these sources to have high reliability;
however, they are also much more likely to be found embedded in bright and complex background regions at low Galactic latitude.  This imposes upon them a penalty on source significance, not only because the background is stronger, but because the levels of non-Gaussianity are also much higher.  For this reason, colder sources have generally lower estimated \textsf{SRCSIG} than warmer ones.

There is a small group of synchrotron flat-spectrum sources characterized by their non-physical MBB parameter values (see Fig.~\ref{fig:DistribTempBeta}, bottom left corner of both panels). To identify this subset, we found all sources that satisfied $\textsf{BETA} < 0.5$, $\textsf{TEMP} < 15$, and $\textsf{EST\_QUALITY} \geq 4$.  We cross-matched the high-Galactic-latitude ($|b| > 20\deg$), flat-spectrum population (24 sources) with \Planck's PCCS2+2E 30-GHz catalogue.  The cross-match returned 23 common objects. Note that we are not removing any of these sources, but providing a simple way to identify them in the extended catalogue.  The remaining {\tt BeeP} object (PCCS2 857 G207.16-60.71) just misses the reliability criterion of {\tt BeeP/base}, with $\textsf{NPSNR} = 4.84 < 5$.

\subsection{Background properties}
\label{subsec:BackgroundProperties}

As a by-product of the {\tt BeeP} analysis, we obtain the MBB parameters of the background thermal emission around each source.  We compute the average brightness and standard deviation \{$I_{\nu}$, $\sigma_{I_{\nu}}$\} from the four background maps over a square patch $33\times33$ pixels ($56\parcm7\times56\parcm7$) across, centred on the PCCS2+2E source position. Reduced resolution was also employed in \citep{planck2013-p06b} to stabilize the evaluation of the $T$--$\beta$ pairs.  The CIB monopole, added to the \Planck\ 2015 maps as reported in \cite{planck2014-a09}, was then subtracted.  Offsets do not affect estimates of properties of compact objects, as they are subtracted before the likelihood evaluation. However, they are important when estimating the background thermal properties.  Uncertainties resulting from map calibration and CIB monopole errors are also added directly to $\sigma_{I_{\nu}}$.  Then, following exactly the same procedure as in the case of the Free source model, an MBB background model curve, with colour correction, was fitted to these data pairs using a Gaussian likelihood.  These curves are also shown in the SED plots, e.g., Fig.~\ref{fig:SEDPlot}.

\begin{figure*}[htbp!]
	\begin{center}
		\leavevmode
		\includegraphics[width=0.33\textwidth]{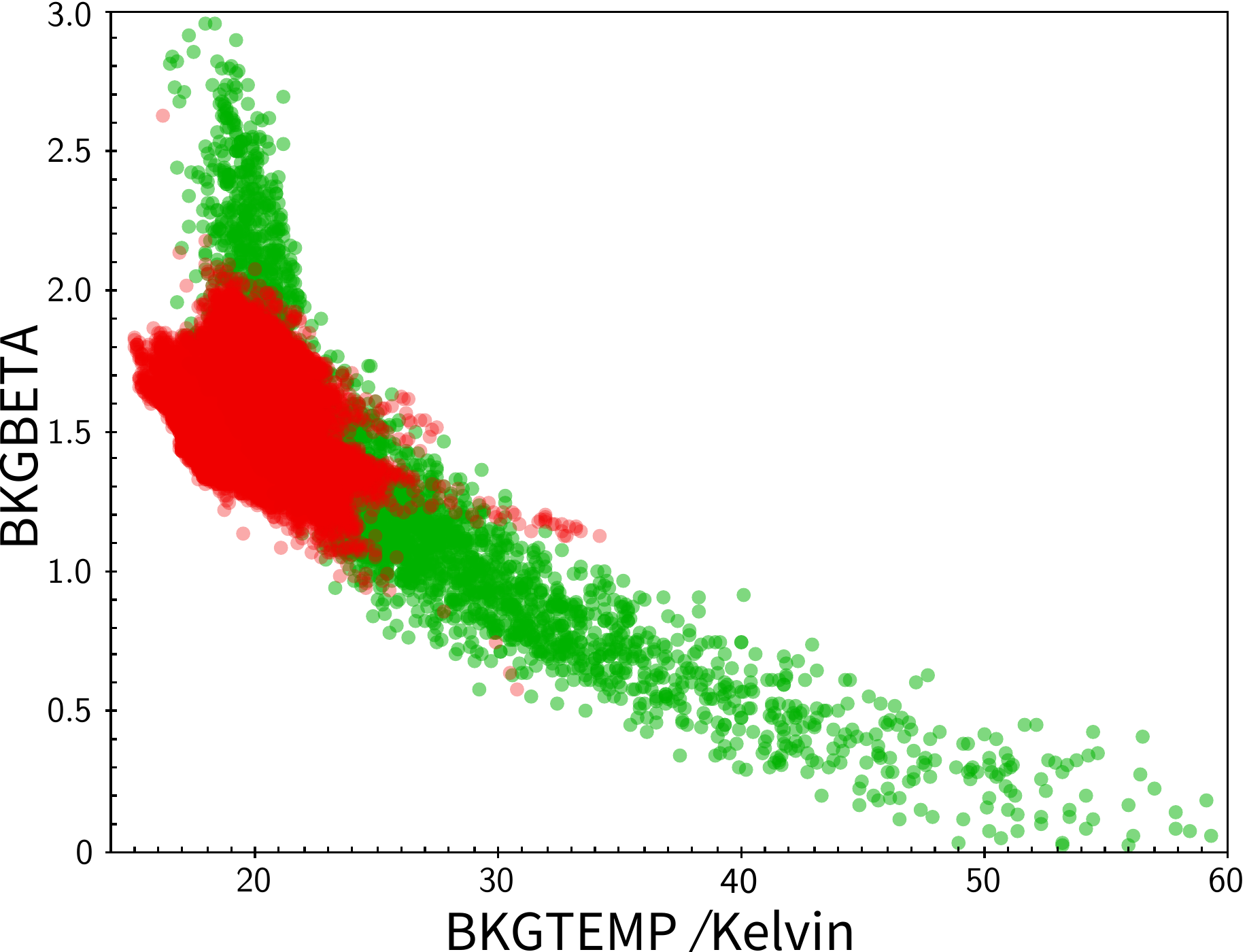}
		\includegraphics[width=0.33\textwidth]{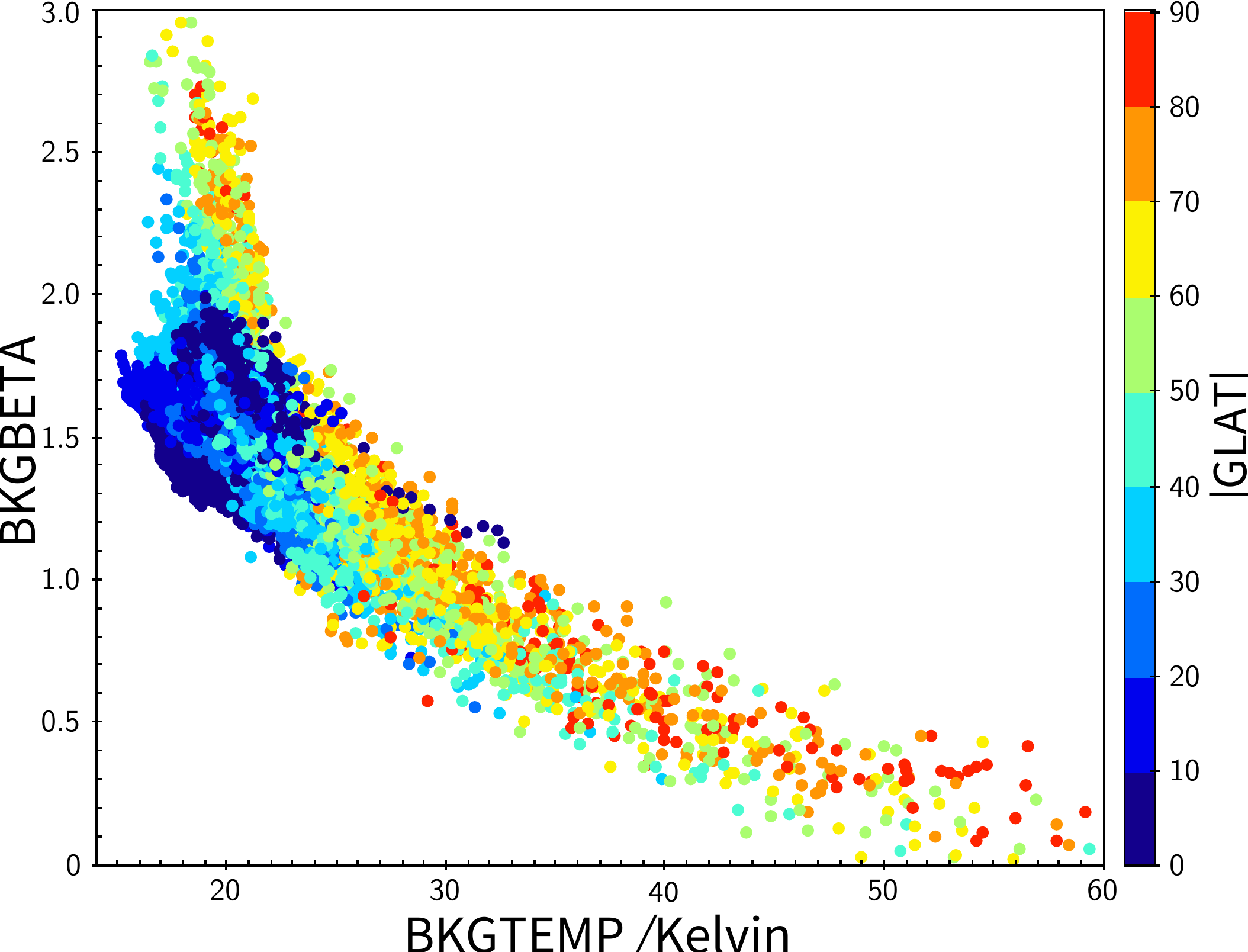}
		\includegraphics[width=0.33\textwidth]{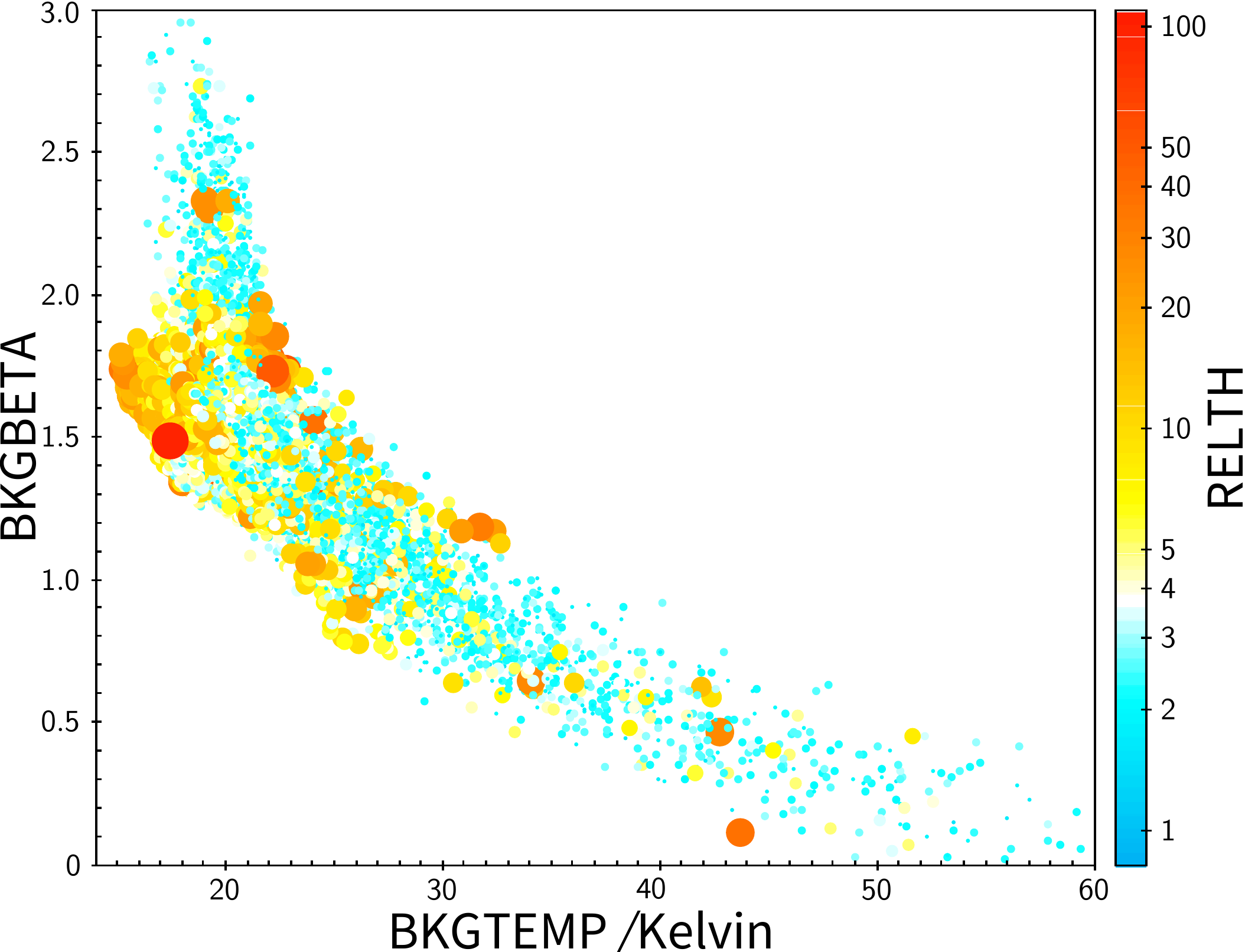}
		\caption{Three views of the relationship between $T$ and $\beta$ for the background, from the MBB model.  {\it Left:} PCCS2 background regions, which are at high Galactic latitude, are in green.  PCCS2E Galactic mask regions, which have strong, dominant dust emission, are in red.  {\it Middle:} Colours show Galactic latitude.  {\it Right:} Colours show \textsf{RELTH}, a measure of the non-Gaussianity of the patch.  The symbol size here is proportional to \textsf{RELTH}.
		}
		\label{fig:ValBackTempBeta}
	\end{center}
\end{figure*}

At these frequencies the dominant background component is dust, particularly for low Galactic latitudes (Fig.~\ref{fig:ValBackTempBeta} left and centre).  However the picture becomes slightly more complicated for high Galactic latitudes where CIB anisotropies, instrumental noise, and CMB anisotropies (especially at 353\,GHz) also make significant contributions.  CIB anisotropies are important only at scales of 1\deg\ and smaller \citep{planck2016-XLVIII,planck2013-p06b}, while instrumental noise is important only at even smaller scales.  In contrast, the CMB appears predominantly at 1\deg\ and larger scales. The CMB signal is faint compared to dust at 545\,GHz and above, but not at 353\,GHz at high Galactic latitude.

\begin{figure*}[htbp!]
	\begin{center}
		\leavevmode
		\includegraphics[width=0.49\textwidth]{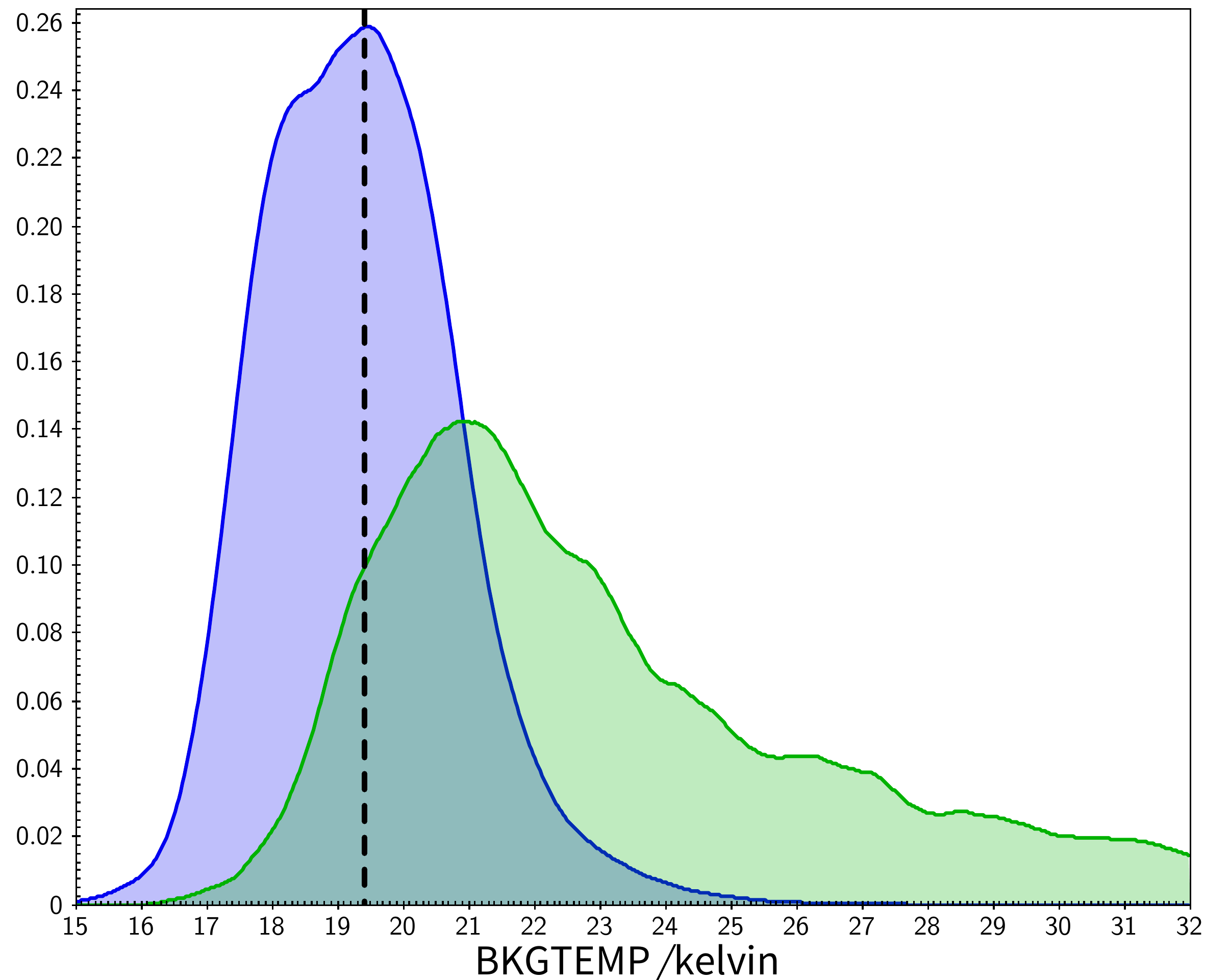}
		\includegraphics[width=0.49\textwidth]{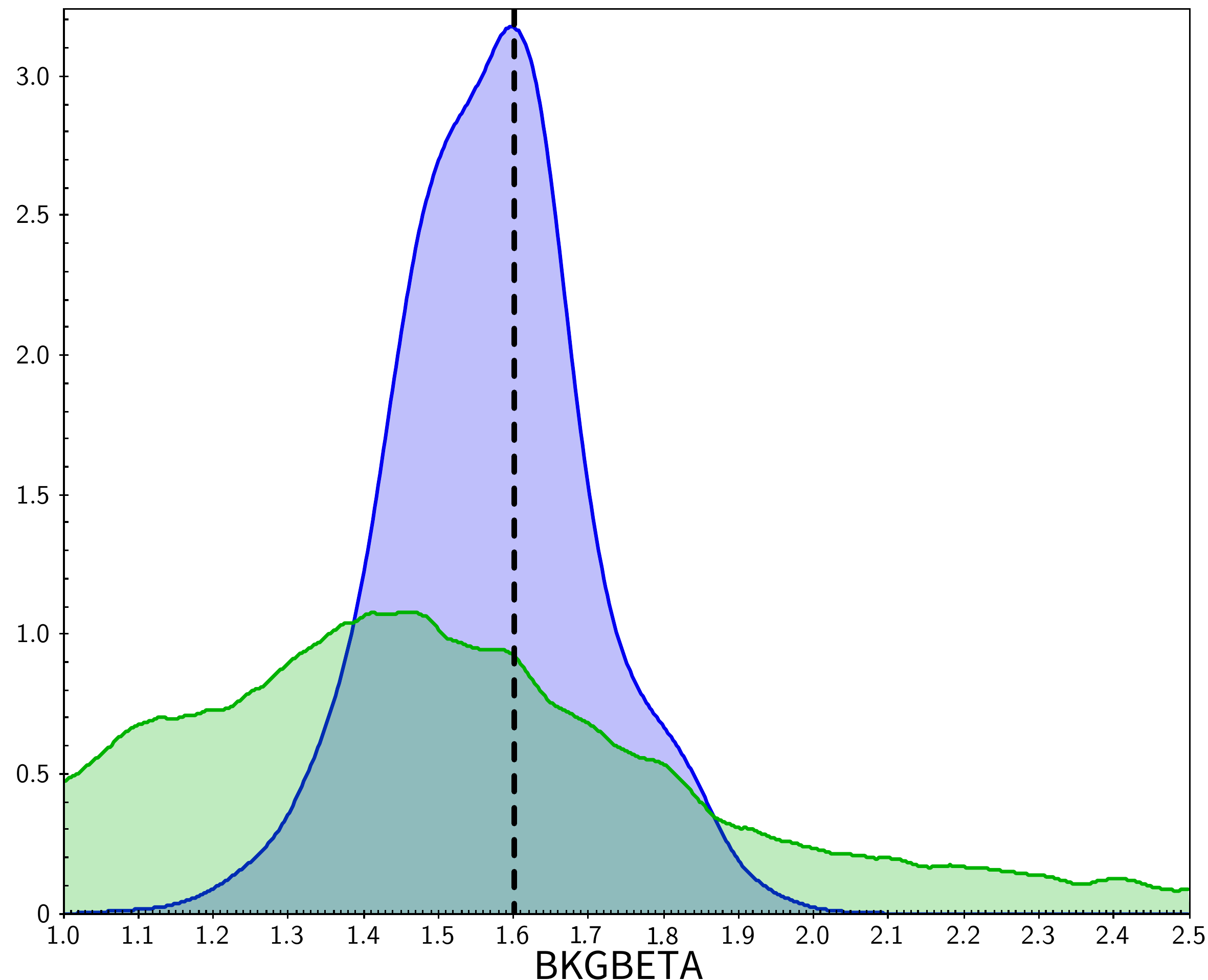}
		\caption{Distributions of MBB background temperature (left) and spectral index (right).  PCCS2 regions (high Galactic latitude) are green and zones where dust emission is dominant are blue, as defined by the PCCS2E cirrus and Galactic masks.  Dashed lines mark the best-fit values for dust temperature (19.4\,K) and $\beta$ (1.6), from \cite{planck2016-XLVIII}.}
		\label{fig:ValBackHist}
	\end{center}
\end{figure*}

Figure~\ref{fig:ValBackHist} shows the histogram of the MBB parameters $T$ and $\beta$ for dust-rich regions defined by the masks used in PCCS2E (blue), and for high Galactic latitudes (green).  The distributions are different as a result of their dissimilar composition (see also Fig.~\ref{fig:ValBackTempBeta} left). In regions where dust is dominant, the agreement with {\tt GNILC} \citep{planck2016-XLVIII} estimates of temperature and $\beta$ is excellent.  In regions of low column density, the agreement deteriorates significantly (see also Sect.~\ref{sssec:BackgroundEstimates} and Fig.~\ref{fig:gnilccomparison} for further explanation of these features).

\begin{figure*}[htbp!]
	\begin{center}
		\leavevmode
		\includegraphics[width=0.29\textwidth]{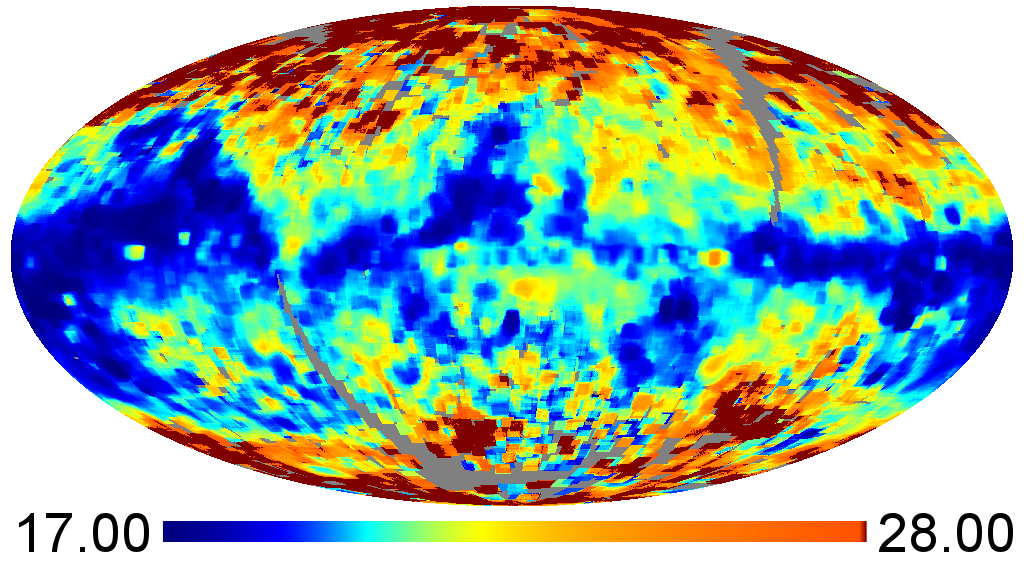}
		\includegraphics[width=0.29\textwidth]{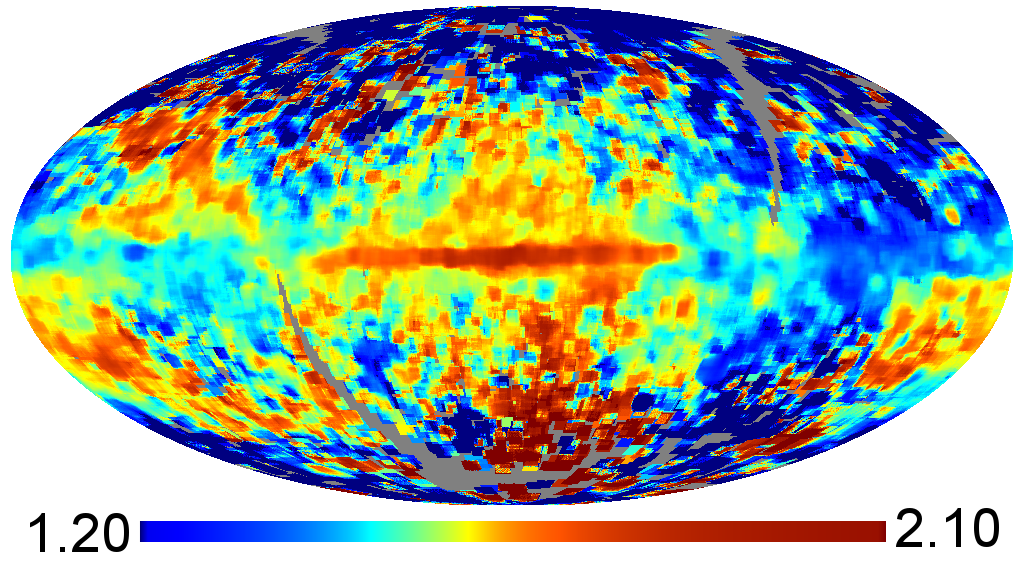}\\
		\includegraphics[width=0.29\textwidth]{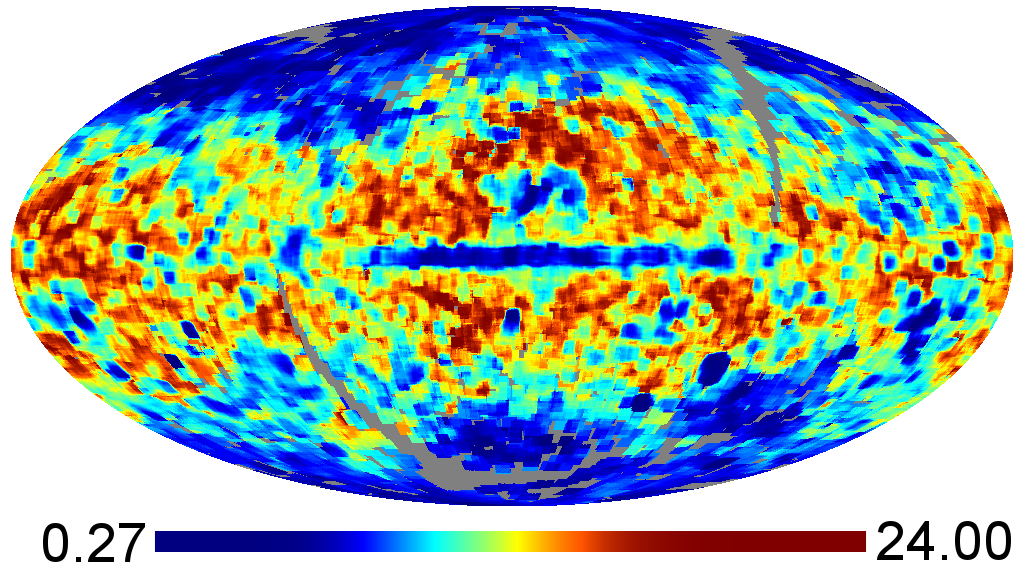}
		\includegraphics[width=0.29\textwidth]{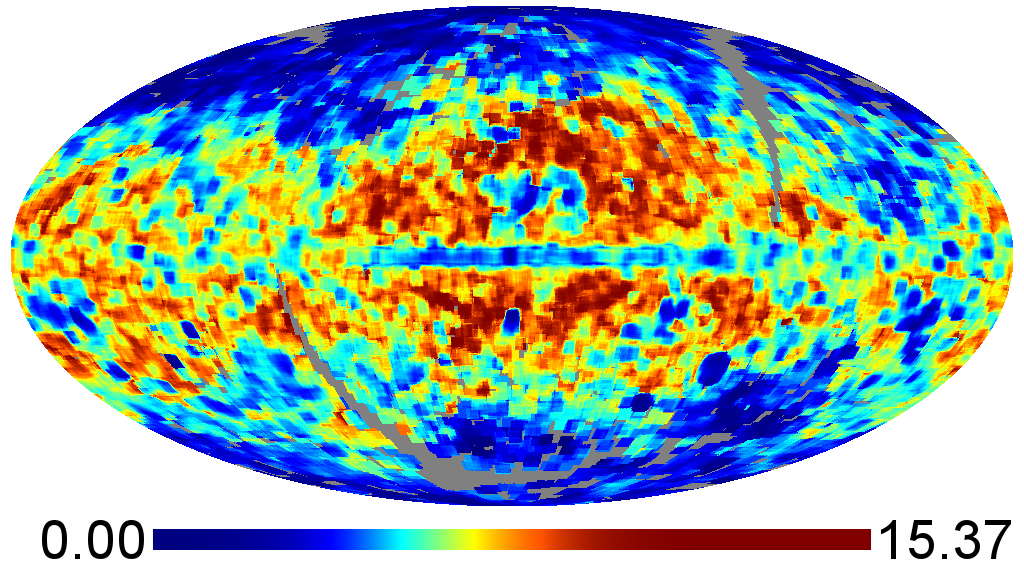}\\
		\includegraphics[width=0.29\textwidth]{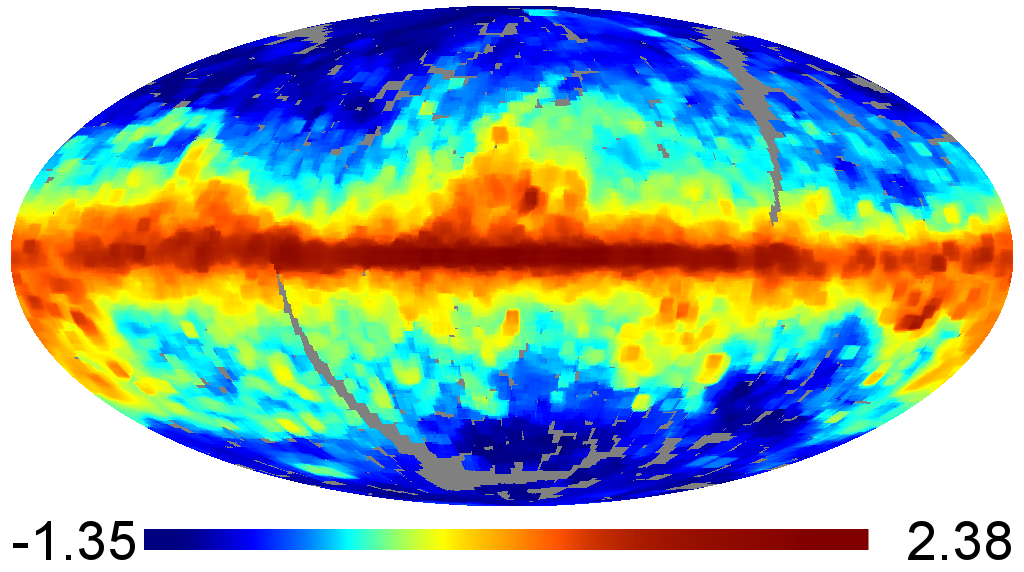}
		\includegraphics[width=0.29\textwidth]{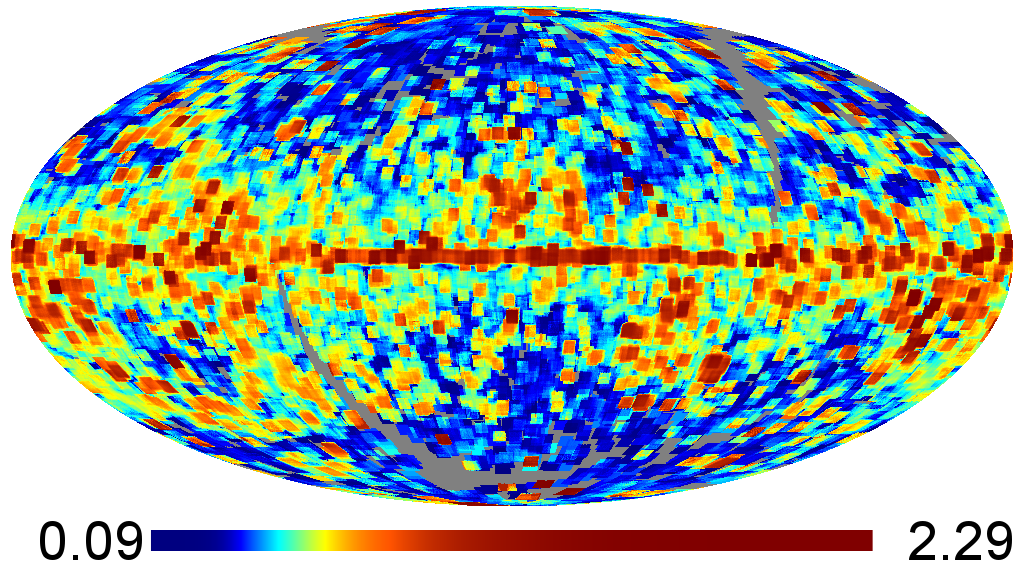}
		\caption{{\it Top:} MBB parameters fitted to the background, with $T$(K) on the left and $\beta$ on the right.
The high-temperature region at ($l=155^\circ; b=77^\circ$) is artificially created by strong artefacts in the IRIS data. {\it Middle:} Inverse relative uncertainty $T/\Delta T$ (left) and $\beta/\Delta\beta$ (right) of the MBB parameters. {\it Bottom left:} Reference background brightness, $\log({\rm Jy}\,{\rm pixel}^{-1}$, evaluated at 857\,GHz.  This is the value at 857\,GHz of a multifrequency fit, not the value directly measured at 857\,GHz.  {\it Bottom right:} $\log(\textsf{RELTH})$, computed with $\alpha = 5\,\%$ (see Eq.~\ref{eq:QuantileDefMain}).  On all panels there is a region at ($l = 208^\circ; b = -18^\circ$) of extreme values and uncertainties caused by artefacts present in the IRIS data.  The colour bars have been histogram-equalized.  regions are either inside the IRIS mask, or had insufficient data.}
		\label{fig:BackGroundTempBeta}
	\end{center}
\end{figure*}

Figure~\ref{fig:BackGroundTempBeta} shows the spatial distribution of the estimated background parameters.  High-Galactic-latitude zones show consistently higher temperature and lower $\beta$ than the dusty regions close to the Galactic plane.  This result is consistent with previous analyses \citep{planck2013-p06b,planck2011-7.0}.  Regions close to the Galactic centre have higher $\beta$ than those at larger longitudes (upper right panel).  The effect is less pronounced for $T$ (upper left panel).

One of the interesting background parameters estimated by {\tt BeeP} is \textsf{RELTH}, which measures the non-Gaussianity of the background. The spatial distribution of \textsf{RELTH} is shown in the bottom right panel of Fig.~\ref{fig:BackGroundTempBeta}.
As a consequence of the non-Gaussian nature of dust emission, it is expected that \textsf{RELTH} correlates with background emission, and this is indeed evident from the bottom panels. Nonetheless, \textsf{RELTH} depends on the detailed statistics of the field being analysed (Sect.~\ref{sec:SingleSrcModel}, Eq.~\ref{eq:QuantileDefMain}), therefore direct comparison of \textsf{RELTH} levels in regions of widely varying complexity is likely biased.  Figure~\ref{fig:ValBackTempBeta} (right panel) shows that although regions with high non-Gaussianity exist over the full range of thermal emission properties, the coldest background regions are {\it all\/} highly non-Gaussian.  This is at least partly due to the fact that they are located near to the Galactic plane, where there is the most confusion.

\section{Comparison with other catalogues}
\label{sssec:XternalCatalogues}

\subsection{\Planck\ PCCS2 catalogue.}\label{psssec:PCCS2E_catalogues}

The PCCS2 contains the most reliable sources in the full PCCS2+2E, because of their location in low-background regions.  The PCCS2+2E was built using a single channel MHW algorithm, which is of a very different nature than that of {\tt BeeP}. Therefore, comparison of PCCS2 and {\tt BeeP} source parameters provides an interesting cross-validation of the two methods. For reference, we also include PCCS2E in the comparisons.

\subsubsection{Flux density estimates}

To compare source flux-density estimates, we use the {\tt BeeP} Free values at all but 857\,GHz, since these were obtained with a data model more in line with the single-channel measurements of the PCCS2+2E.  At 857\,GHz, however, we compare the fitted MBB flux density, not the individual flux density (\textsf{FREES857}).  This allows for a broader validation because we are testing the full range of flux densities and the SED model all at once.  It is also a less noisy estimate.  At the same time, since so much more information goes into the {\tt BeeP} estimate than into the PCCS2 estimate, we should not a priori expect a very good match.  For PCCS2+2E, we use \textsf{APERFLUX} estimates, which were obtained using an aperture photometry algorithm.

\begin{figure*}[htbp!]
	\begin{flushleft}
		\leavevmode
		\includegraphics[width=0.49\textwidth]{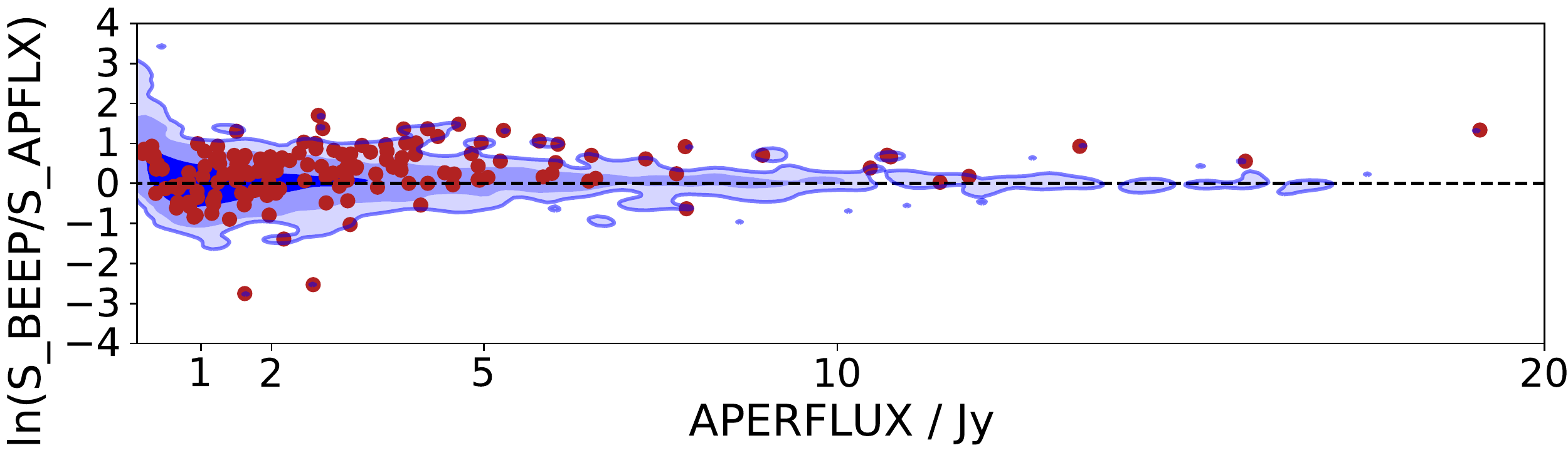}
		\includegraphics[width=0.49\textwidth]{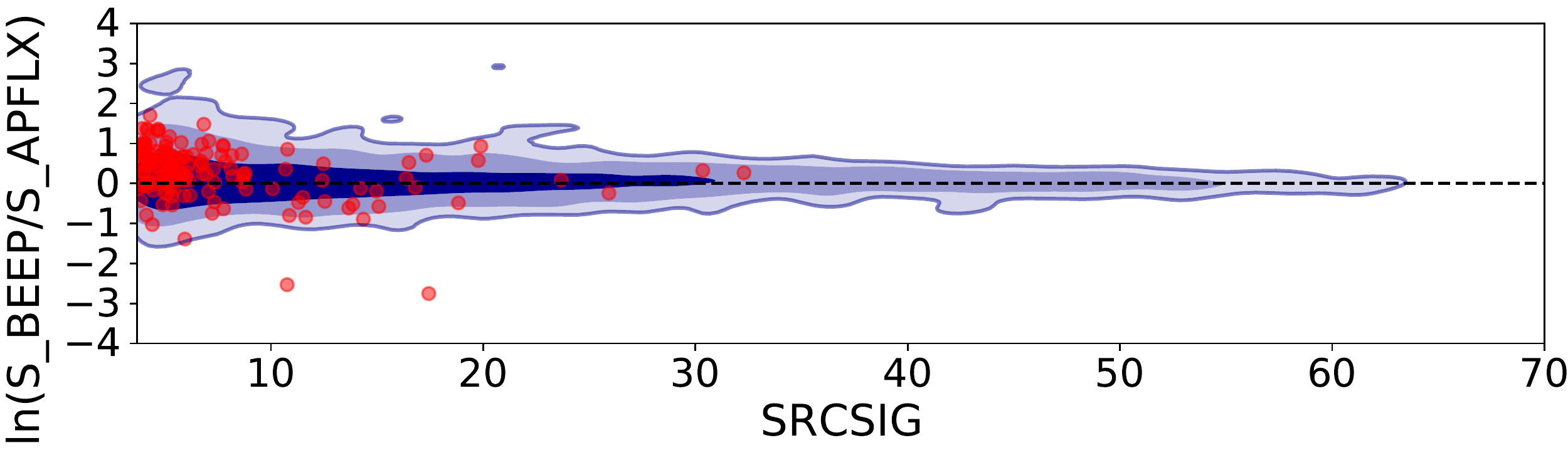}\\
		\includegraphics[width=0.49\textwidth]{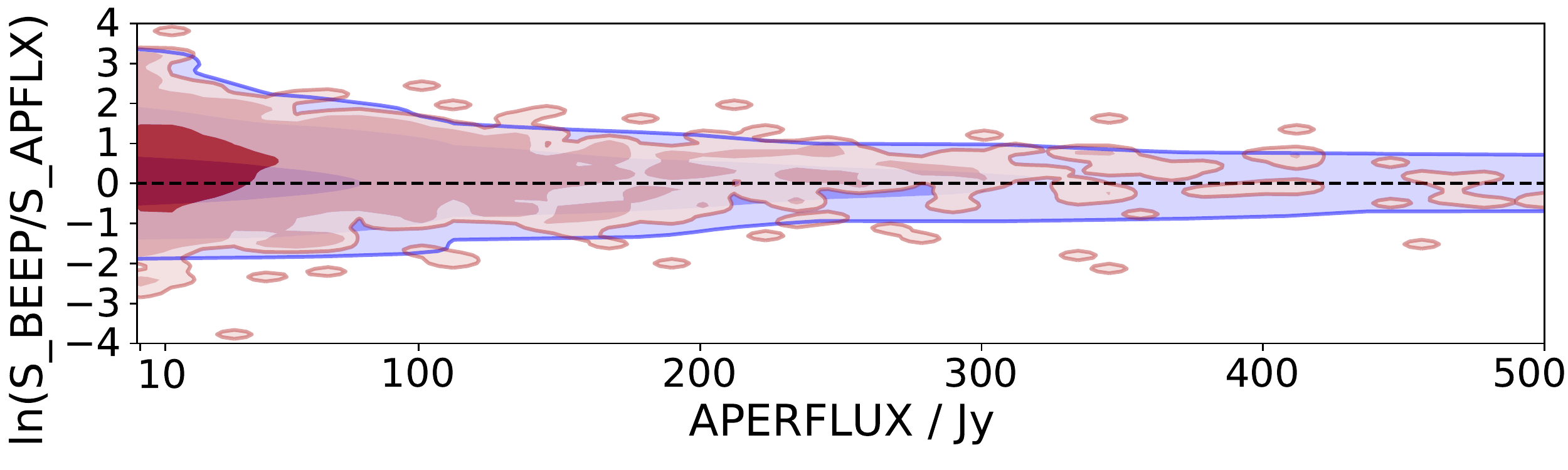}
		\includegraphics[width=0.49\textwidth]{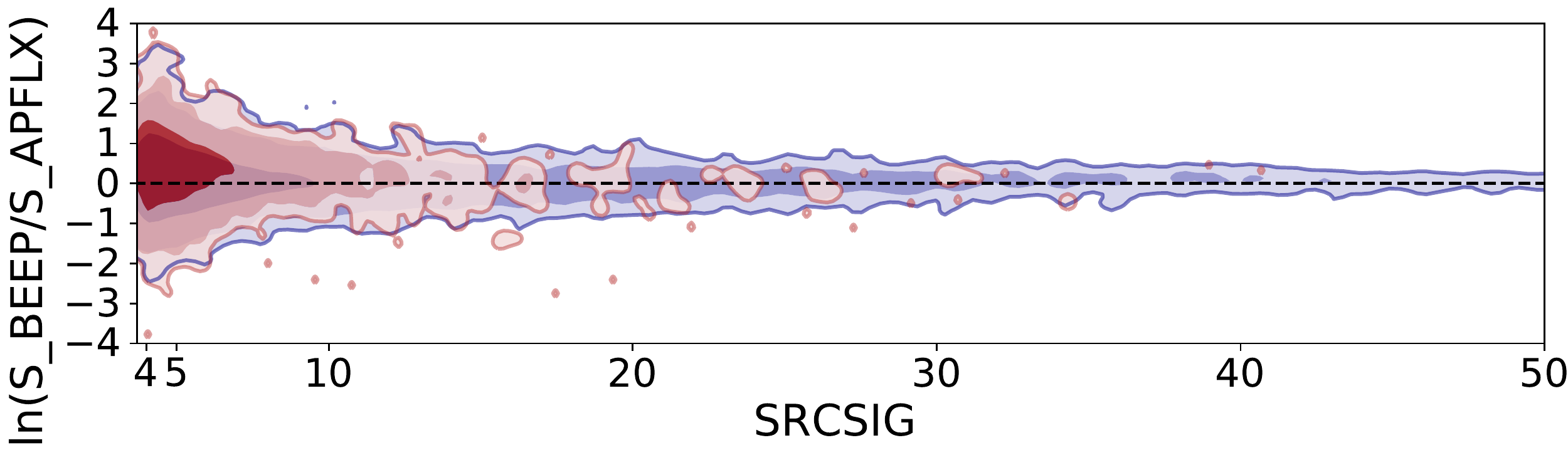}
		\caption{Comparison of {\tt BeeP}'s Free flux densities at 857\,GHz with aperture flux-density values (\textsf{APERFLUX}) from PCCS2+2E. We plot $\ln(S_{\textsf{BeeP}}/S_{\textsf{APERFLUX}})$ against \textsf{APERFLUX} (left) and \textsf{SRCSIG} (right).  The upper row shows the distribution of the PCCS2 values in blue (contours are [$68,95,99$]\%). The red dots are sources that moved by more than one pixel from the original PCCS2+2E position.
			The lower row includes the full PCCS2+2E. The red contours represent the distribution of sources that moved by more than one pixel, and the blue ones the remaining population. In the top row we show individual dots because there are too few of them to make a density plot.
		}
		\label{fig:ValPCCS2Comp}
	\end{flushleft}
\end{figure*}

Figure~\ref{fig:ValPCCS2Comp} shows the results of this comparison.  On the average, there is good consistency between {\tt BeeP}'s estimates and those in the PCCS2. There is, however, an overall bias with a median of about $+4.0$\,\% (mean $+5.8$\,\%) for PCCS2, which increases to $+5.0$\,\% (mean $+7.3$\,\%) for the full PCCS2+2E.

Although we use the PCCS2+2E source locations as a starting point, we allow {\tt BeeP} to search for a better effective location in the close neighbourhood if that increases the likelihood ratio.  Maximizing the likelihood ratio is equivalent to maximizing the flux density, because we assume that the background is homogeneous around the source.  One might expect that the population of sources that moves from its original position by a significant amount should, on average, show higher flux densities.  Figure~\ref{fig:ValPCCS2Comp} shows in red sources that moved by more than one pixel from their PCCS2+2E position.  As expected, the densities of this population are clearly biased high compared with those of PCCS2+2E.  Considering these effects, we remove from the comparison sources whose position changed by more than one pixel with regard to the PCCS2+2E estimate, and extended sources with $\textsf{R} \geq 1.64$ pixels.  Values of \textsf{R} in Fig.~\ref{fig:realR_histogram} were obtained from \textsf{EXT}, correcting for the excess that results from using narrower beams in the likelihood. Now the flux-density bias for the full PCCS2+2E becomes negligible: median $= -0.6\,\%$ (mean $= -0.8\,\%$).  Therefore from now on we only use this subset of sources to compare {\tt BeeP}'s flux-density estimates with those in the \textsf{PCCS2+2E} catalogue.

The second factor affecting the flux bias between {\tt BeeP} and PCCS2 is background removal.  For low \textsf{APERFLUX},  {\tt BeeP}'s flux densities seem to become increasingly biased high as we go down in flux.  
At 0.45\,Jy we are already at the sensitivity limit for single-channel aperture photometry.  At these very low flux densities, the effects of Eddington-type bias become important.  However, the multi-channel nature of {\tt BeeP} makes it more sensitive, with an efficient background removal even at these low signal regimes.  This effect is much more pronounced in the PCCS2 than in the PCCS2E, because the fraction of sources with flux densities below the 0.45-Jy threshold is larger.  A simple example to understand how this bias occurs is the following. Imagine a completely homogeneous but positively correlated background, containing valleys and crests. Now imagine a very faint source population, all of the same flux, embedded in it. Applying an aperture photometry method to recover the flux, the sources sitting in the valleys, would appear in the faint end group. A method that could reduce the background to zero would recover the true flux, which when compared with the faint end of the aperture photometry flux estimate would appear biased high.  To account for that, in addition to the previous filters, we further removed sources with \textsf{APERFLUX} $< 0.45$ Jy.
The comparison restricted to this PCCS2 subset now shows a rather small bias: median $= 1.1\,\%$ (mean $= -1.1\,\%$).

\begin{figure}[htbp!]
	\begin{flushleft}
		\leavevmode
		\includegraphics[width=0.49\textwidth]{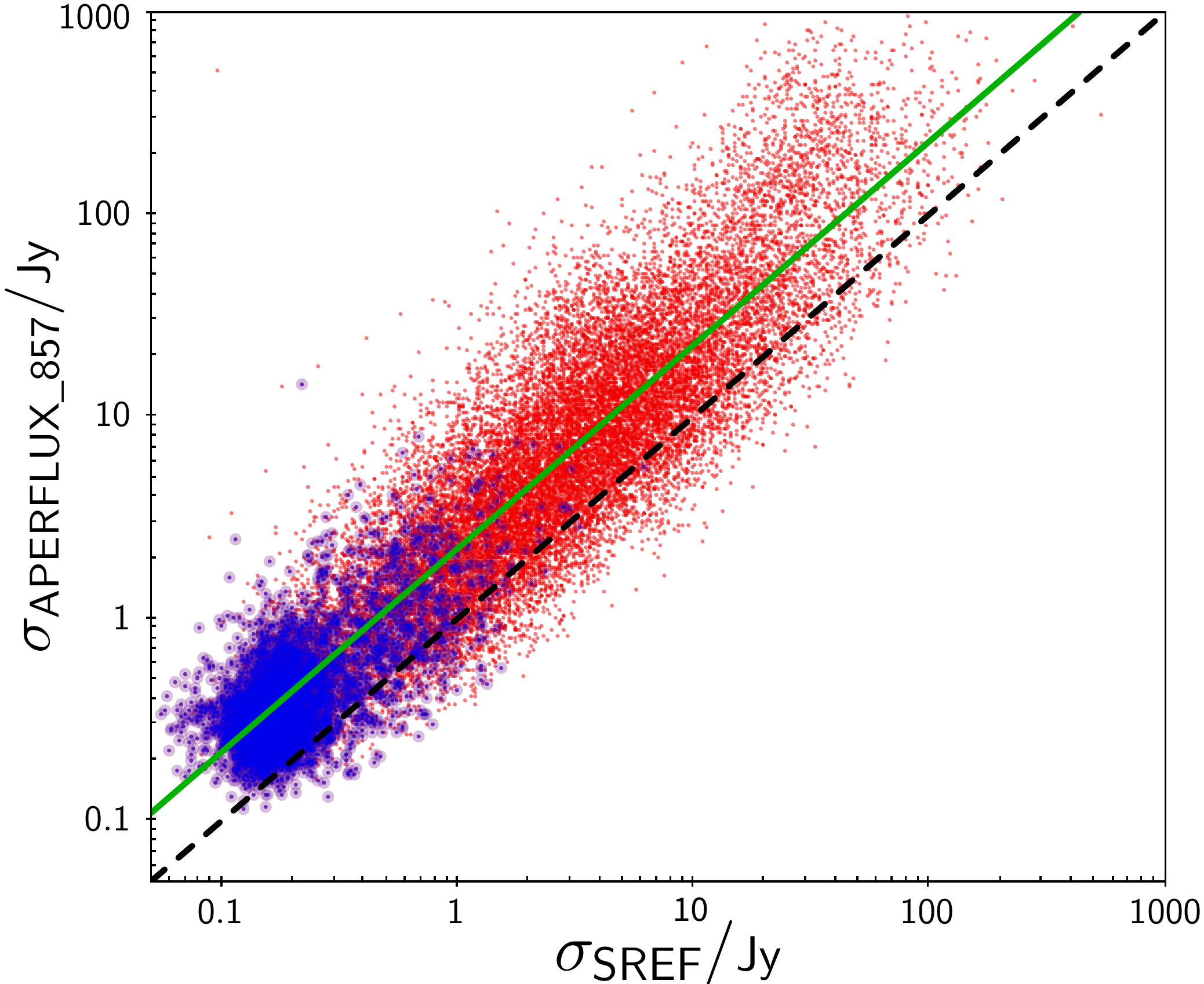}
		\caption{Comparison of the uncertainty determined in PCCS2+2E (red markers), and the PCCS2 (blue markers), on the aperture flux density at 857\,GHz (\textsf{APERFLUX\_857}) to the uncertainty on the reference flux density as obtained by {\tt BeeP} (\textsf{SREF}). The black dashed line represents equality, and the green solid line is the best fit, which has a slope very close to 2 (actually 2.2).}
		\label{fig:FluxComparison}
	\end{flushleft}
\end{figure}

On average, the uncertainties in {\tt BeeP} flux densities are a factor of about 2 smaller than those of the aperture flux estimates in PCCS2+2E (see Fig.~\ref{fig:FluxComparison}).
The combination of uncertainties obtained by {\tt BeeP} and those of PCCS2+2E explains the dispersion of Fig.~\ref{fig:ValPCCS2Comp} adequately.

\begin{figure}[htbp!]
	\begin{flushleft}
		\leavevmode
		\includegraphics[width=0.49\textwidth]{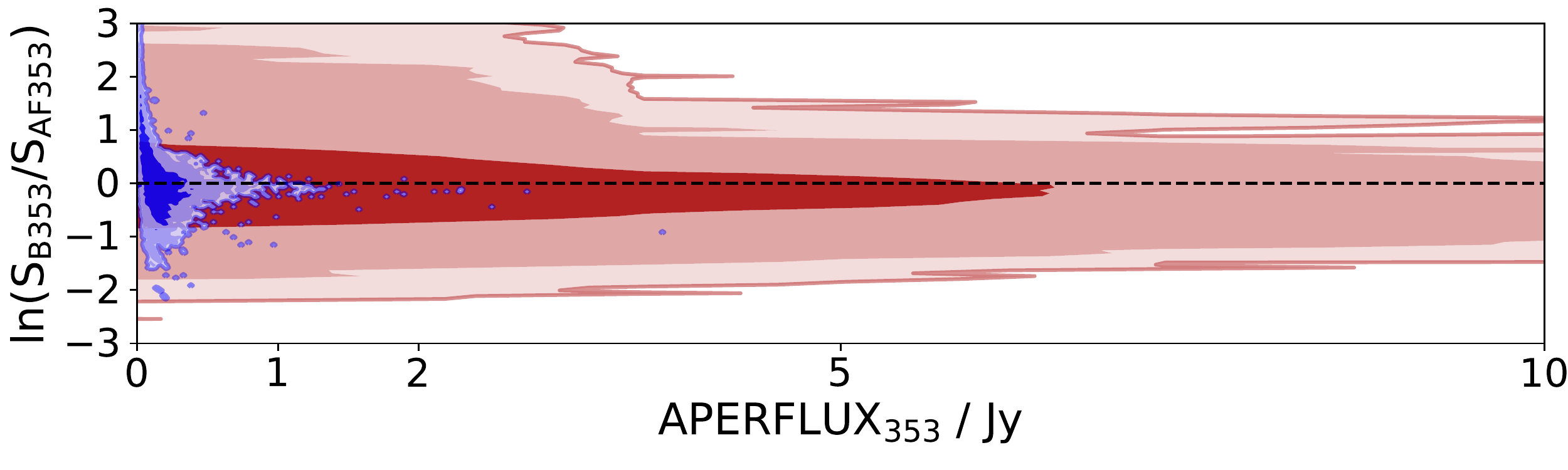}\\
		\includegraphics[width=0.49\textwidth]{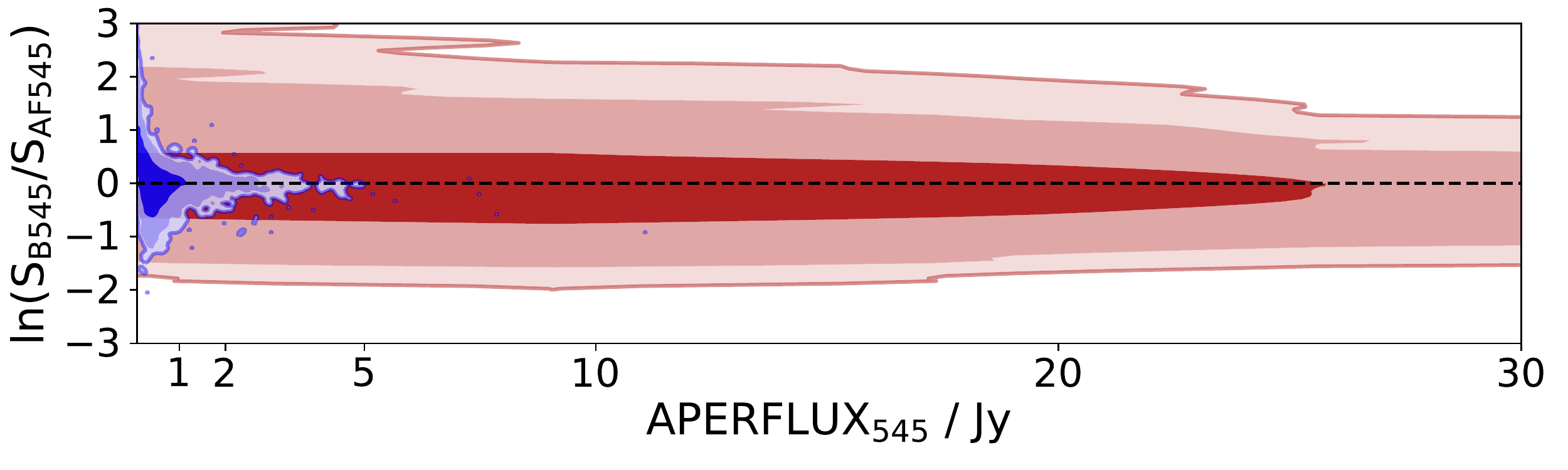}
		\caption{Comparison of {\tt BeeP}'s Free flux densities at 353\,GHz (upper row) and 545\,GHz (lower row) with aperture flux-density values (\textsf{APERFLUX}) from \textsf{PCCS2+2E}.
			The blue contours ([$68,95,99$]\%) show the distribution of the PCCS2 subset of sources and the red ones the remaining PCCS2E.
			 Unlike in Fig.~\ref{fig:ValPCCS2Comp}, {\tt BeeP} sources whose position shifted by more than one pixel from the original \textsf{PCCS2+2E} and those with $\textsf{EXT} \geq 1.64$ pixels (extended) are not included.}
		\label{fig:ValPCCS2353545}
	\end{flushleft}
\end{figure}

We further compared {\tt BeeP}'s Free flux-density estimates at 353 and 545\,GHz, \textsf{FREES353} and \textsf{FREES545}, with the \textsf{PCCS2+2E} equivalents, \textsf{APERFLUX\_353} and \textsf{APERFLUX\_545} (see Fig.~\ref{fig:ValPCCS2353545}).  The subset depicted was obtained by removing sources whose {\tt BeeP} position estimate changed by more than one pixel from the original \textsf{PCCS2+2E} and those that appear to be extended, with $\textsf{R} \geq 1.64$\,pixels (see Fig.~\ref{fig:realR_histogram}).  For the \textsf{PCCS2+2E} (in purple) the flux-density biases we find are $0.6\,\%$ (median) at 545\,GHz and $-2.6\,\%$ (median) at 353\,GHz.  The dispersion of the estimates is high, similar to that of the 857-GHz channel, but consistent with the combined uncertainties from {\tt BeeP} and \textsf{PCCS2+2E}.

\subsubsection{Source positions}

\begin{figure}[htbp!]
	\begin{center}
		\leavevmode
		\includegraphics[width=0.49\textwidth]{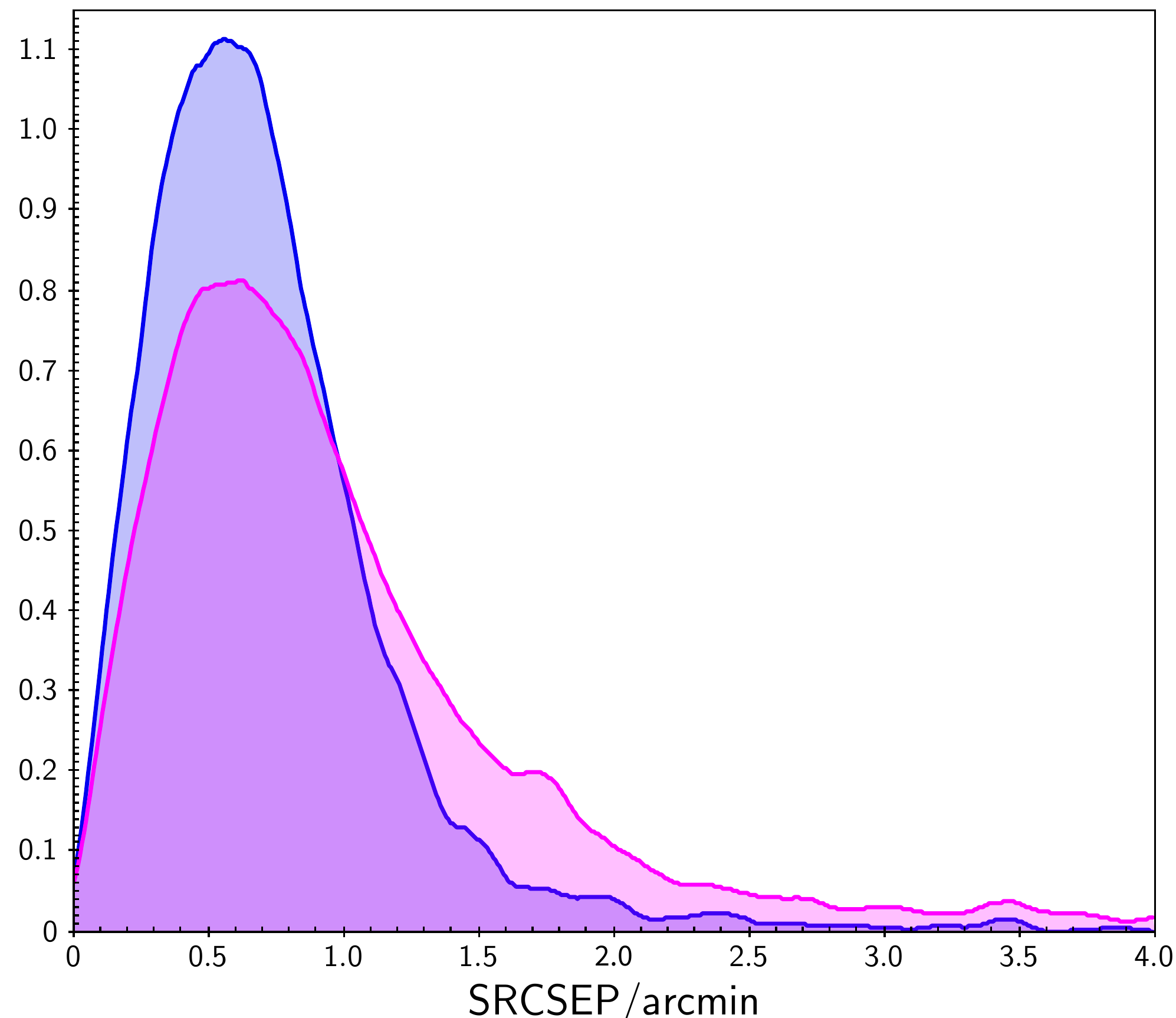}
		\caption{Normalized histograms of the differences between the PCCS2 (in blue) and the PCCS2E (in pink) position estimates and those from {\tt BeeP}.  The PCCS2E distribution shows a significantly more extended tail. The two small bumps in the distribution are at 1\parcm72 (1 pixel) and 3\parcm43 (2 pixels).  The PCCS2 histogram has a less extended tail compatible with smoother homogeneous backgrounds.  The peak of both distributions is at about 0\parcm65, the average position-error estimate for the 857-GHz channel in equation~7 and table~8 of \citet{planck2014-a35}.
		}
		\label{fig:PCCS2_PosComp}
	\end{center}
\end{figure}

Assuming that positional errors $\{X,Y\}$ are independent and Gaussian-distributed in both the PCCS2+2E and {\tt BeeP}, then the distance between both positions should follow a Rayleigh distribution (see Sect.~\ref{ssec:CatPositionalAccuracy}) with a scale factor $\sigma$ dependent on the positional accuracies of both catalogues.  Figure~\ref{fig:PCCS2_PosComp} shows the histogram of the distances between the {\tt BeeP} and PCCS2+2E positions.  The PCCS2 subset histogram (in blue) is a good match with the shape of a Rayleigh distribution.  As expected, the PCCS2E exhibits a wider tail.  The PCCS2E histogram also has bumps at 1\parcm72 (1 pixel)
and 3\parcm43 (2 pixels).  These small excesses are the natural result of the map pixel grid.  As may be seen in Fig.~\ref{fig:CornerPlot}, {\tt BeeP}'s positional uncertainty seems little affected by the map pixelization.  However, the presence of these small bumps at exact multiples of the pixel size, indicates a possible greater impact on the PCCS2+2E, which might add a small negative kurtosis in the PCCS2+2E $\{X,Y\}$ error distributions.  Given that {\tt BeeP}'s positional uncertainty is so small, if we take the {\tt BeeP} positions as the true values, then the histograms in Fig.~\ref{fig:PCCS2_PosComp} are consistent with the positional uncertainty characterization of the 857-GHz channel in the PCCS2+2E.  The distributions (PCCS2 and PCCS2E) peak at around 0\parcm65.  This value is a good match to the average 0\parcm65 position error estimate for the 857-GHz channel of the PCCS2 subset in equation~7 and table~8 of \citet{planck2014-a35}.


\subsubsection{Background complexity and reliability}

\begin{figure}[htbp!]
	\begin{center}
		\leavevmode
		\includegraphics[width=0.48\textwidth]{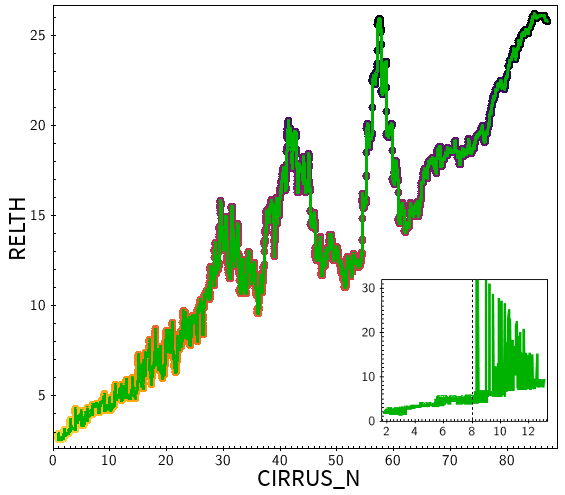}
		\caption{Correlation between the {\tt BeeP} \textsf{RELTH} parameter and the PCCS2+2E \textsf{CIRRUS\_N} parameter.  Sources in the {\tt BeeP} catalogue were sorted in ascending order of \textsf{RELTH}.  Then a boxcar average with window = 500 was calculated for both \textsf{RELTH} and the corresponding PCCS2+2E \textsf{CIRRUS\_N} values.  The relationship is particularly tight for low values of the parameters, as seen in the expanded detail window, obtained following the exact same procedure as the main picture but using PCCS2 data only.
		We also reduced the boxcar window to just 50 samples here.  The dashed vertical line is the PCCS2 \textsf{CIRRUS\_N} reliability threshold.}
		\label{fig:CirrusRelth}
	\end{center}
\end{figure}

The PCCS2+2E catalogue contains a field \textsf{CIRRUS\_N} that flags entries with a complex background, and therefore a higher probability of being spurious.  \textsf{CIRRUS\_N} is the number of sources detected at 857\,GHz within a circle centred on the source with a radius of $1^\circ$ \citep{planck2014-a35}.  {\tt BeeP}'s \textsf{RELTH} is a measurement of the local background non-Gaussianity, either intrinsic or as a result of localized structures (cirrus and filaments).  Its role is pivotal in defining {\tt BeeP}'s reliability criterion \textsf{SRCSIG} (see Eq.~\ref{eq:MRelthDef}): a higher value of \textsf{RELTH} implies a larger correction to \textsf{NPSNR}, or, similarly to \textsf{CIRRUS\_N}, a lower reliability of a putative source.  These two quantities, although different, should exhibit some degree of correlation if the background non-Gaussianity is indeed the main source of false positives.  Figure~\ref{fig:CirrusRelth} shows such a correlation between the variables.  The relationship is particularly tight for low values of both variables, as seen in the inset part of the figure, which was obtained by applying the same procedure as in the main picture to the PCCS2 subset.  The moving-average window was also reduced to 50 samples for greater resolution.  It is clear from this figure that below \textsf{CIRRUS\_N} $= 8$, there is a well defined correlation between the two quantities. The opposite happens above the threshold.  According to \cite{planck2014-a35}, \textsf{CIRRUS\_N} $= 8$ is the suggested source reliability threshold.

\subsubsection{Reliability}

\begin{figure}[htbp!]
	\begin{center}
		\leavevmode
		\includegraphics[width=0.48\textwidth]{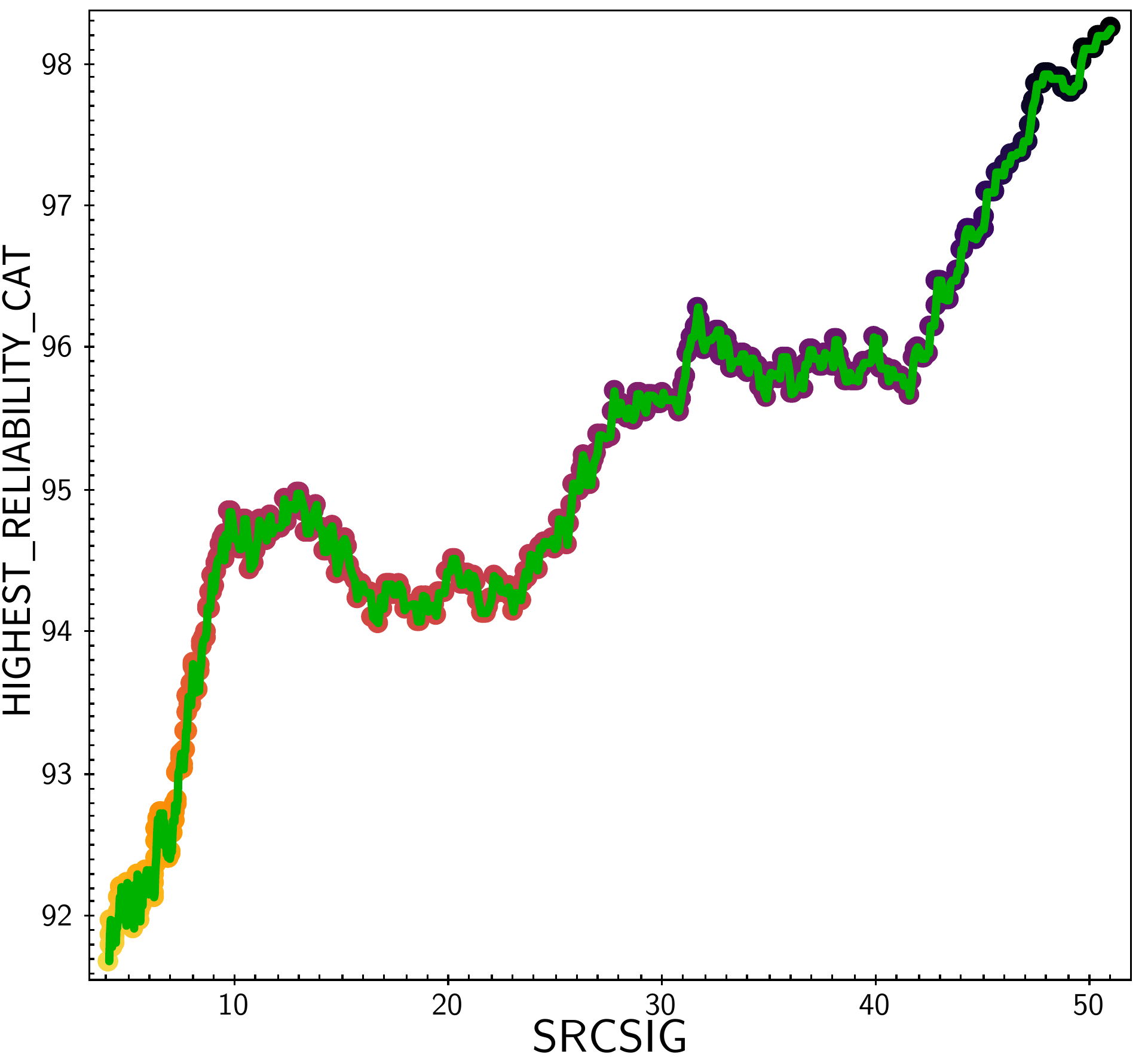}
		\caption{Correlation between the PCCS2 quantities \textsf{HIGHEST\_RELIABILITY\_CAT} and {\tt BeeP} \textsf{SRSIG}, for common sources, as described in the text.  We sorted the sources in \textsf{SRCSIG} ascending order, and computed a boxcar average over \textsf{SRCSIG} and \textsf{HIGHEST\_RELIABILITY\_CAT} with a 100-sample window.  There is a clear positive correlation between these two variables.}
		\label{fig:Reliability353}
	\end{center}
\end{figure}

PCCS2+2E contains no reliability information at 857\,GHz. The PCCS2 list at 353\,GHz is the closest in frequency to {\tt BeeP}'s reference channel (857\,GHz), which includes source reliability information.  PCCS2 sources at 353\,GHz are classified as having medium (80\,\%) to high (99\,\%) reliability.  We cross-matched (within a 5\arcm\ radius) PCCS2 353-GHz sources with the full {\tt BeeP} catalogue, and found 786 (58.5\,\% of the PCCS2 353\,GHz list)\footnote{Most of the unmatched 353-GHz sources do not have a counterpart in the PCCS2 at 857\,GHz, due to the important increase in the level and complexity of the background, which reduces the S/N and results in no detection.}.  
All but one of these 786 sources appear in the {\tt BeeP} catalogue with a contamination lower than $10\,\%$ ($\textsf{SRCSIG}\,{>}\,2$ and $\textsf{NPSNR}\,{>}\,5$).   Demanding an even lower contamination of 5\,\% in the {\tt BeeP} selection ($\textsf{SRCSIG}\,{>}\,4.0$, $\textsf{NPSNR}\,{>}\,5$), there remain 744 common sources (94.7\,\% of the common 786~sources).  
The highest reliability (99\,\%) subset of PCCS2 at 353\,GHz contains 427 sources, 416 (97.4\,\% of the PCCS2 353\,GHz sources) of which are in the {\tt BeeP} 5\,\% contamination subset.  Given that the majority of the PCCS2 353-GHz catalogue sources have positive spectral indexes \citep{planck2014-a35}, if they are already reliable at 353\,GHz they should be even more so at higher frequencies, where they are brighter.  However, one must also consider the effect of the embedding background. If the spectral index of the background is steeper than that of the source, the contrast between the source brightness and that of the background might actually decrease.  As a check, we cross-correlated PCCS2 353\,GHz \textsf{HIGHEST\_RELIABILITY\_CAT} with {\tt BeeP/base}'s \textsf{SRCSIG} (Fig.~\ref{fig:Reliability353}).  The well defined trend confirms the expected positive correlation.

\subsection{Planck Catalogue of Galactic Cold Clumps (GCC).}
\label{psssec:GCC_catalogue}

The Planck Catalogue of Galactic Cold Clumps (GCC) was constructed from the same input data used by {\tt BeeP}, namely, the \Planck\ 353-, 545-, and 857-GHz channels, and the 3000-GHz IRIS map \citep{planck2014-a37}.  Similarly to {\tt BeeP}, GCC is generated using a multi-channel algorithm (\texttt{CoCoCoDeT}) on the entire sky. However, this detection algorithm is very different than that of {\tt BeeP}, since it targets the temperature contrast between cold clumps (cold compact emission regions) and a warm background \citep{GCC_Algo}.  The difference in approach makes it interesting to compare the parameters estimated by {\tt BeeP} and GCC.

For this purpose, we cross-matched {\tt BeeP} and the GCC catalogues using a 5\arcm\ matching radius.  The common set contains 8690 entries (65.6\,\% of GCC).  Of these, only 47 are in the PCCS2 (0.54\,\%).  If we further require that the common sources are of good quality according to GCC estimation (\textsf{FLUX\_QUALITY} = 1), then the common set reduces to 5165 sources, with only 36 in the PCCS2. Of these, 73\,\% (3757 sources) are in {\tt BeeP/base}, and this is the set that we use for comparison.

\begin{figure*}[htbp!]
	\begin{center}
		\leavevmode
		\includegraphics[width=0.33\textwidth]{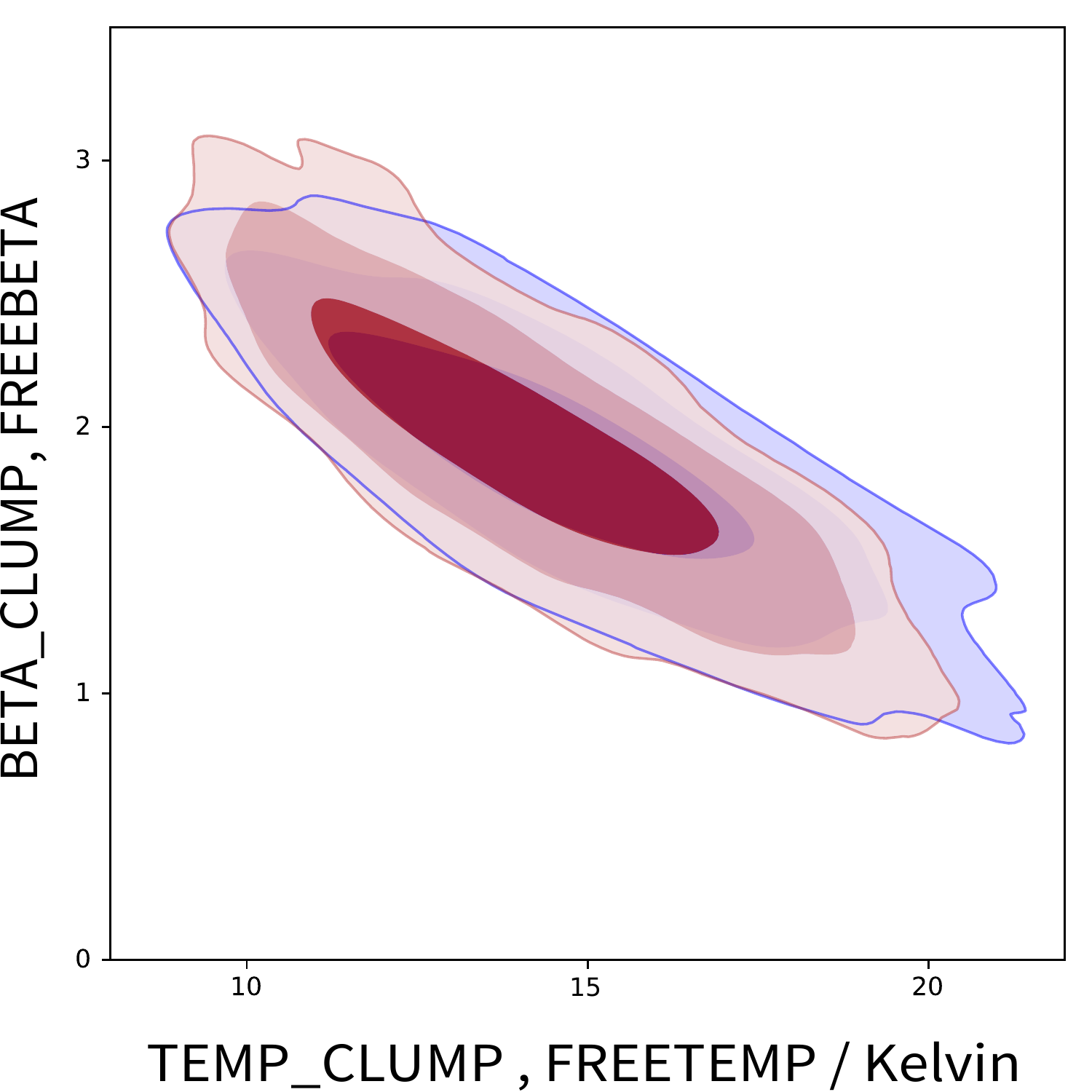}	
		\includegraphics[width=0.33\textwidth,height=0.325\textwidth]{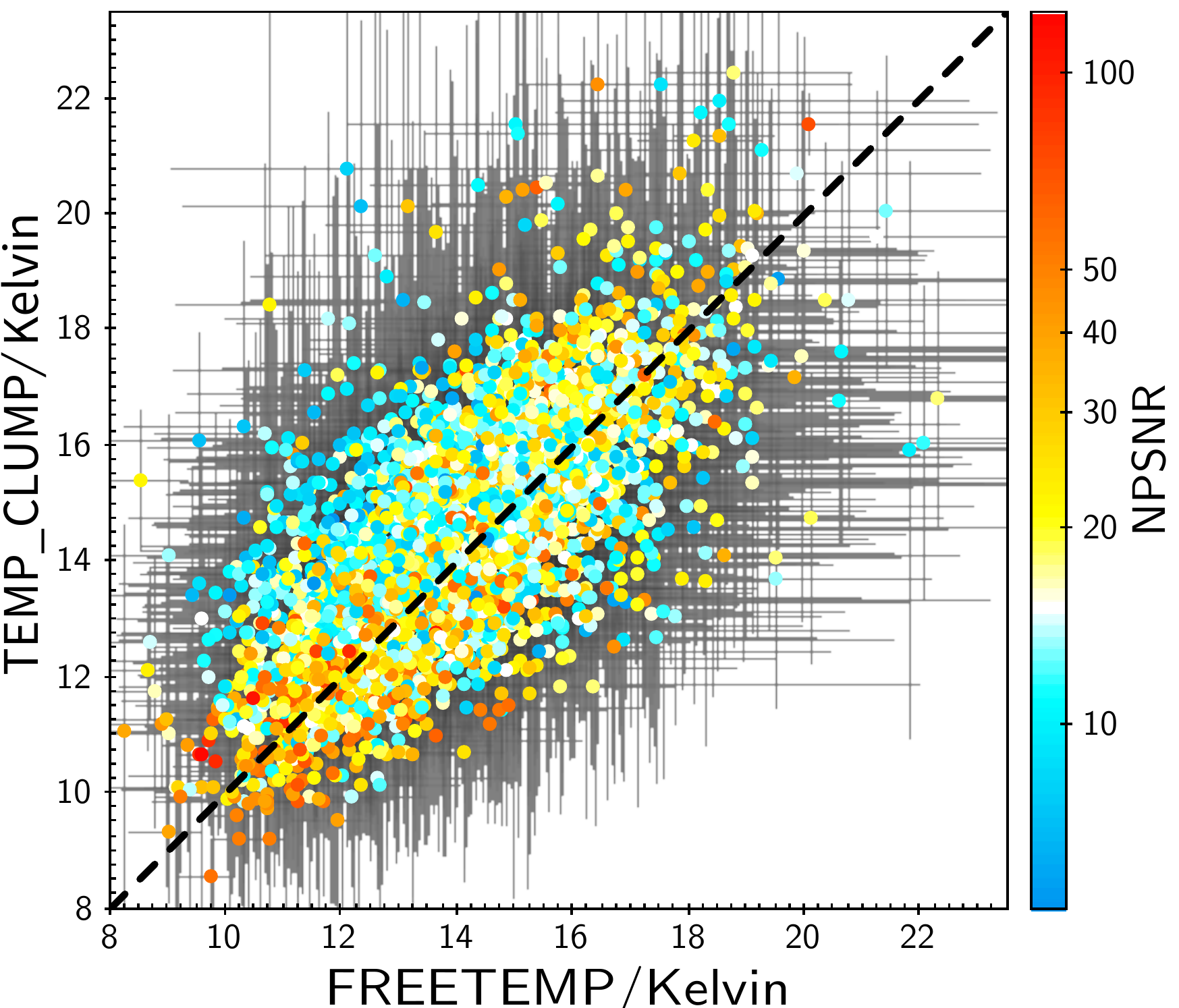}
		\includegraphics[width=0.33\textwidth,height=0.325\textwidth]{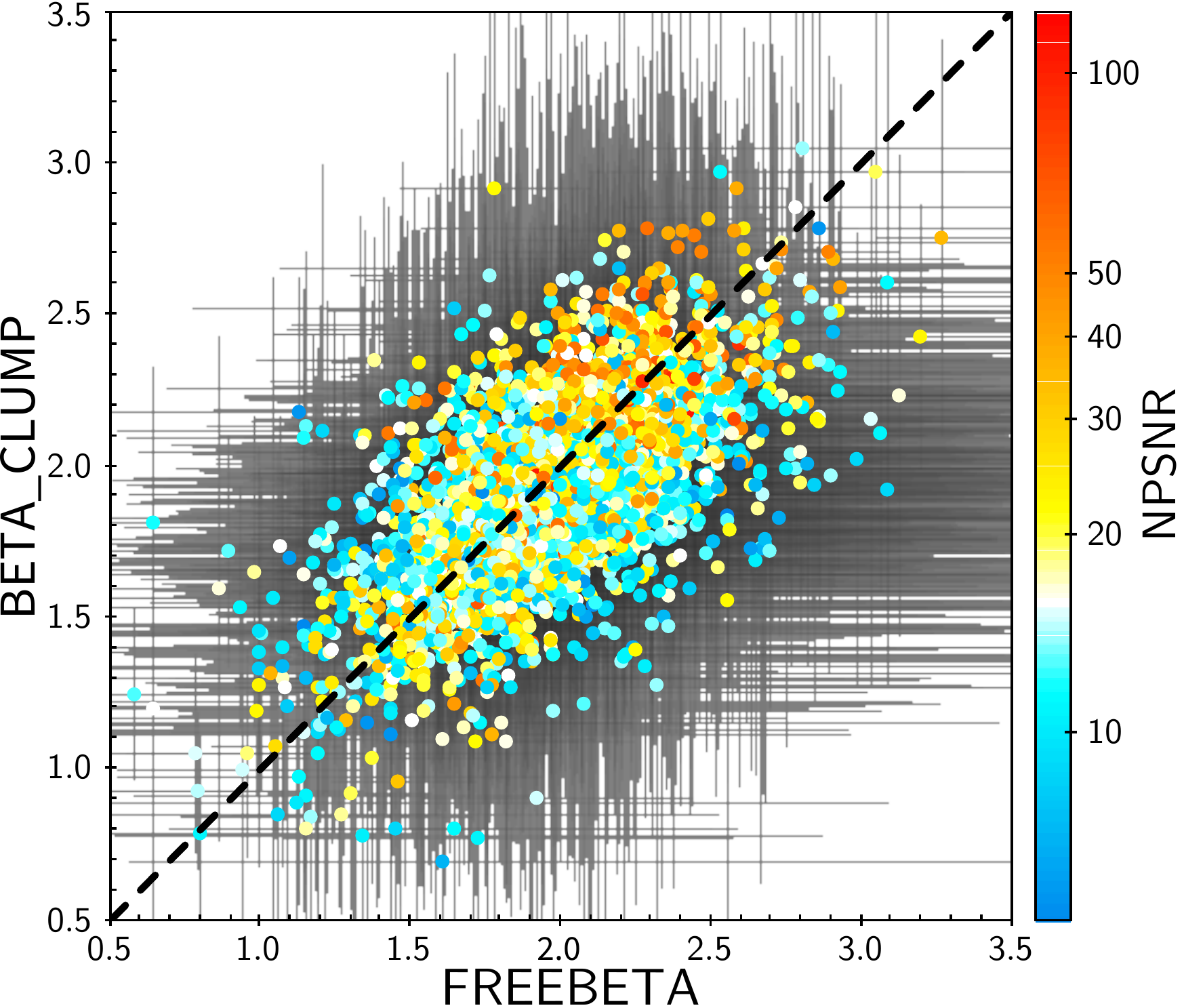}	
		\caption{Comparison of source properties as found by {\tt BeeP} and in the Galactic Cold Clumps  catalogues, for the cross-matched subset described in Sect.~\ref{psssec:GCC_catalogue}.  {\it Left:} $T$ versus $\beta$ for {\tt BeeP}, red contours ([$68,95,99$]\%), and GCC, blue contours. {\it Middle and right:} GCC \textsf{TEMP\_CLUMP} and \textsf{BETA\_CLUMP} versus {\tt BeeP} $T$ and $\beta$, with $1\,\sigma$ error bars. The black dashed line shows equality.}
		\label{fig:CompGCC_catalogue}
	\end{center}
\end{figure*}

Figure~\ref{fig:CompGCC_catalogue} shows a comparison between the {\tt BeeP} MBB parameter estimates and their equivalent in GCC.  There is good consistency between the two.
Both methods show large uncertainty in $T$ and $\beta$, which is not surprising, since we only have four frequencies and we are fitting a three-parameter model.  There is a small positive bias in the GCC temperatures with respect to {\tt BeeP} ($+2.8\,\%$ median, $+3.2\,\%$ mean).  A small negative bias is also seen in the GCC spectral indices with respect to {\tt BeeP} ($-2.2\,\%$ median and mean), which is also expected, considering the negative correlation between $T$ and $\beta$.

In order to further assess consistency, we examine the difference between the two estimates, normalized by the combined uncertainty:

\begin{equation}
\delta \gamma \equiv \frac{\gamma_{\rm BeeP} - \gamma_{\rm GCC}}{\sqrt{\sigma^2_{\gamma_{\rm BeeP}} + \sigma^2_{\gamma_{\rm GCC}}}}\,,
\label{eq:Norm_GCC_Beep}
\end{equation}
where $\gamma$ stands for either $T$ or $\beta$. The dispersion of this quantity should be of order unity. Instead, we find that
$\sigma_{\delta T}\approx 0.59$ and $\sigma_{\delta \beta}\approx 0.43$.  However, our simulations already indicated that {\tt BeeP} overestimates the error bars for both temperature and spectral index (see Table~\ref{table:TempBetaStats}). 
The extra deficit is probably the result of a positive correlation between the estimates of both methods, which arises from the fact that both use the same data.

We now turn to an assessment of the influence of the source-to-background temperature contrast on the estimation of significance, which is of interest because it is this contrast that drives the selection function of the GCC algorithm \texttt{CoCoCoDeT} \citep{planck2014-a37}.  We define
\begin{equation}
\Delta \xi \equiv \frac{\xi^{\rm source} - \xi^{\rm back}}{\sqrt{\sigma^2_{\textrm{source}} + \sigma^2_{\textrm{back}}}}
\label{eq:NormContrast}
\end{equation}
as the normalized contrast, where $\sigma_{\textrm{source}}$ and $\sigma_{\textrm{backg}}$ are the source and background $1\,\sigma$ errors (assumed to be uncorrelated), and $\xi$ can be either $T$ or $\beta$.

\begin{figure*}[htbp!]
	\begin{center}
		\leavevmode
		\includegraphics[width=0.33\textwidth]{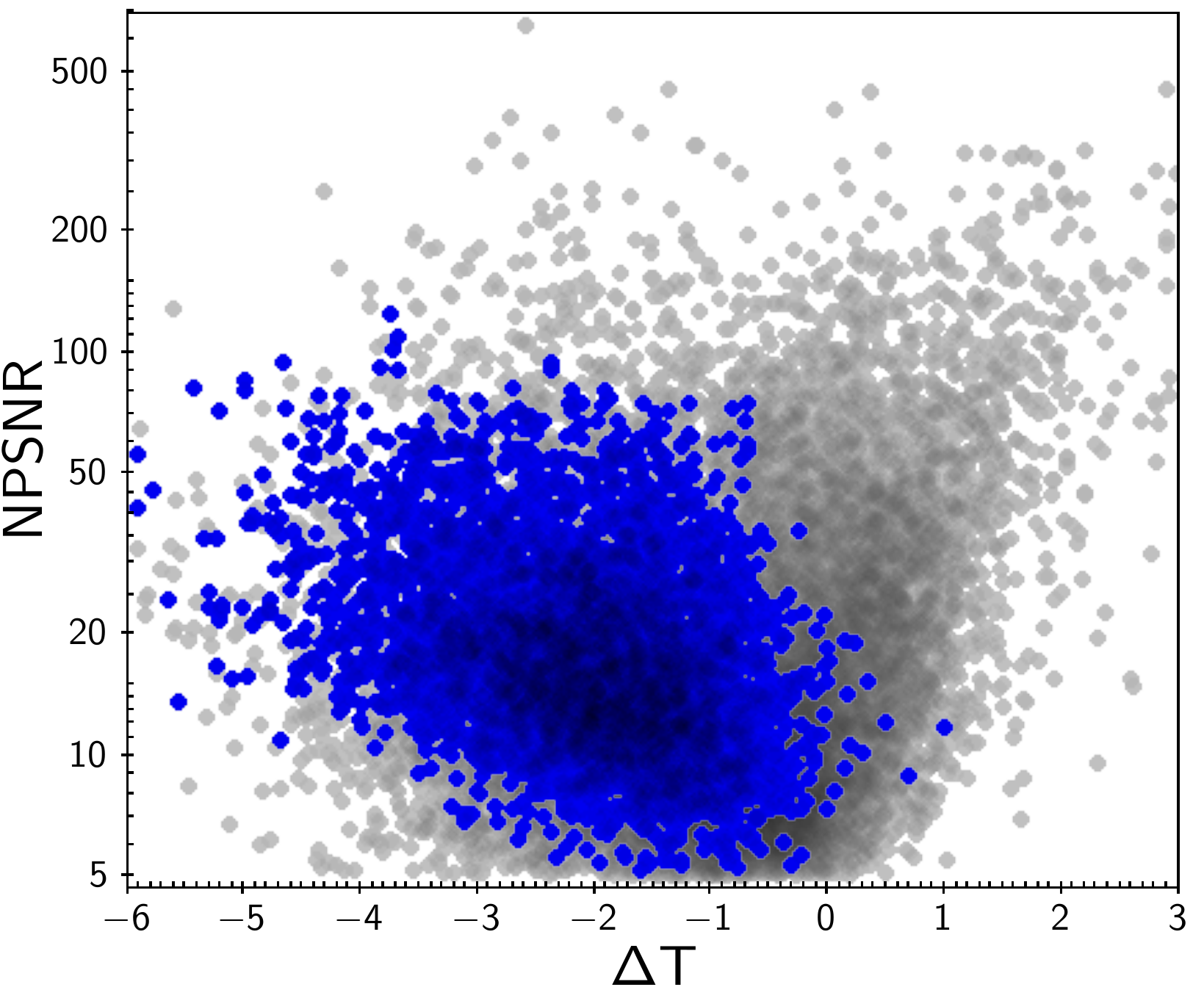}
		\includegraphics[width=0.33\textwidth]{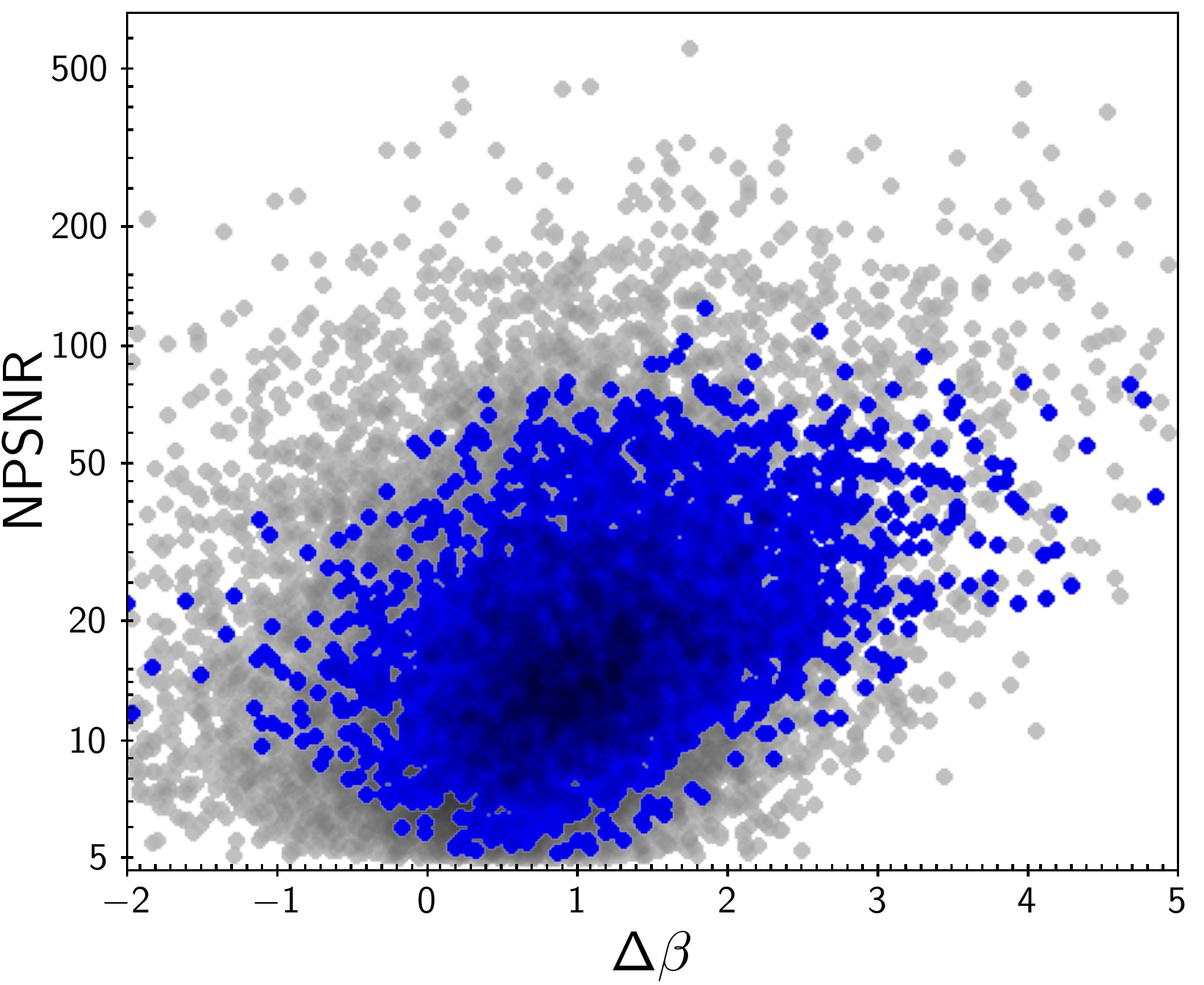}
		\includegraphics[width=0.33\textwidth]{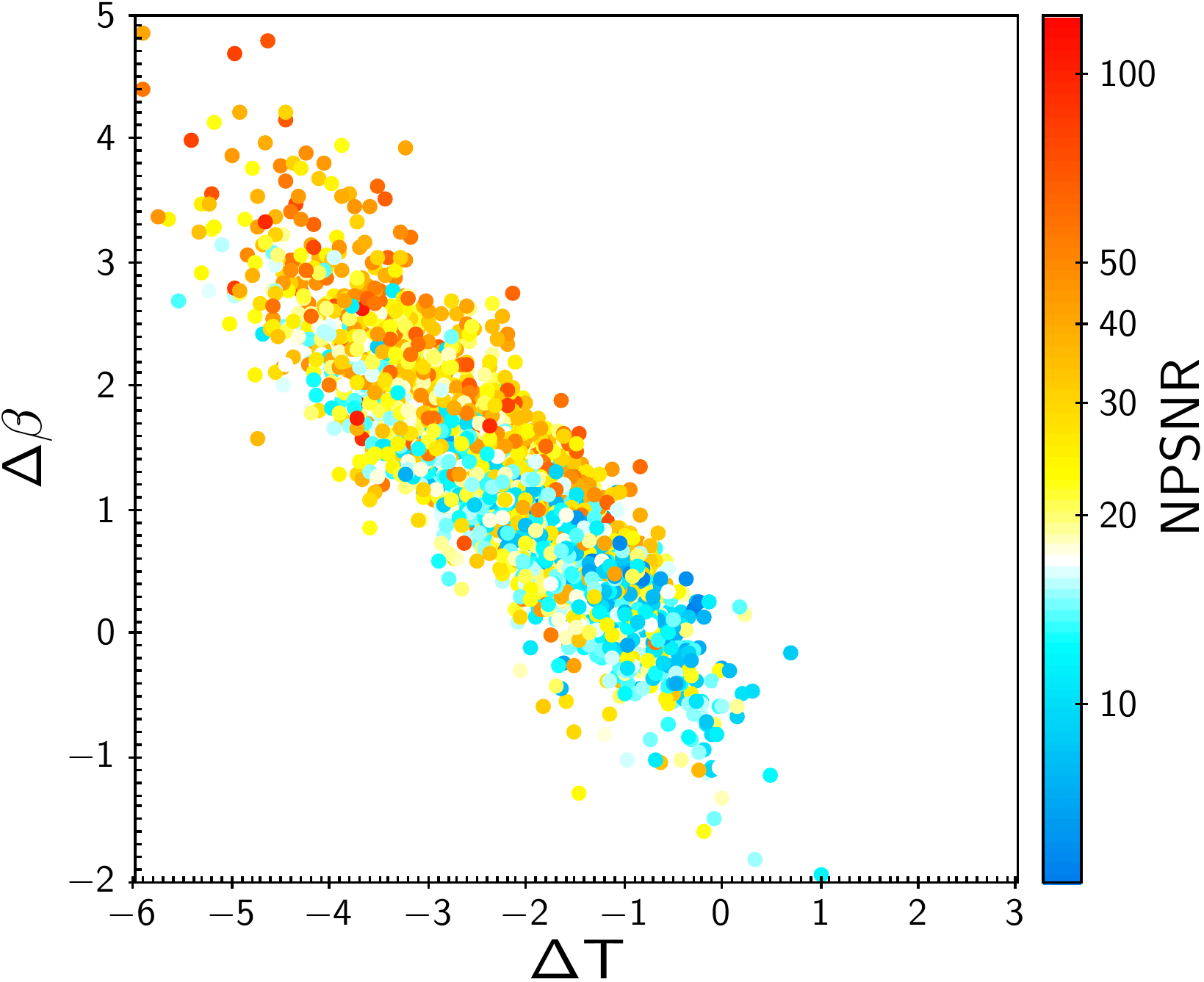}
		\caption{{\it Left and Middle:} Detection significance level \textsf{NPSNR} versus normalized contrast $\Delta T$ (left) and $\Delta\beta$ (middle), defined in Eq.~\eqref{eq:NormContrast}, for the set of sources in common between the {\tt BeeP} and GCC catalogues (blue points), and for the entire {\tt BeeP/base} catalogue (grey points).  Correlations are seen for the common subset (shown in blue), but not for the entire {\tt BeeP/base} catalogue (in grey).  {\it Right:} $\Delta T$ versus ($\Delta \beta$) for the GCC common sample, with colour showing \textsf{NPSNR}.  We see that $\Delta T$ and $\Delta \beta$ are highly correlated, and each is also correlated with \textsf{NPSNR}.}
		\label{fig:CompGCC_correlation}
	\end{center}
\end{figure*}

We note that a contrast in the thermal properties of source and background translates into varying brightness ratios in each frequency channel.  As a consequence, we expect that the MBB parameter source versus background contrast correlates with \textsf{NPSNR} via the {\tt BeeP} likelihood.  Indeed, when we limit the comparison to the GCC-matched subsample of {\tt BeeP/base} shown in blue in the middle and left panels of Fig.~\ref{fig:CompGCC_correlation}, a positive correlation appears between the contrast significance $\Delta T$ and $\Delta \beta$ (Eq.~\ref{eq:NormContrast}) and the {\tt BeeP} \textsf{NPSNR}.
In addition, the right panel shows that the significance of the parameter recovery is positively correlated with the magnitude of the contrast.  However, as may be seen from the grey points in the left and middle panels, when the entire {\tt BeeP/base} catalogue is included, the contrast as estimated by {\tt BeeP} shows only a mild correlation with \textsf{NPSNR}.

BeeP therefore recovers the source versus background contrast that the \texttt{CoCoCoDeT} algorithm uses to select GCC sources. However, the {\tt BeeP} selection function is not limited to cold compact objects immersed in warm backgrounds, and provides a larger population of cold objects, including many that are not found in GCC.  In fact, the {\tt BeeP} base catalogue contains 11\,145 cold objects ($T < 16$\,K) in the region defined by the Galactic mask, as compared to $5489$ with $\textsf{FLUX\_QUALITY = 1}$ in the GCC.

\subsection{Herschel H-Atlas catalogue (350\,$\mu$m)}
\label{psssec:Hatlas_catalogue}

\begin{figure}[htbp!]
	\begin{center}
		\leavevmode
		\includegraphics[width=0.49\textwidth]{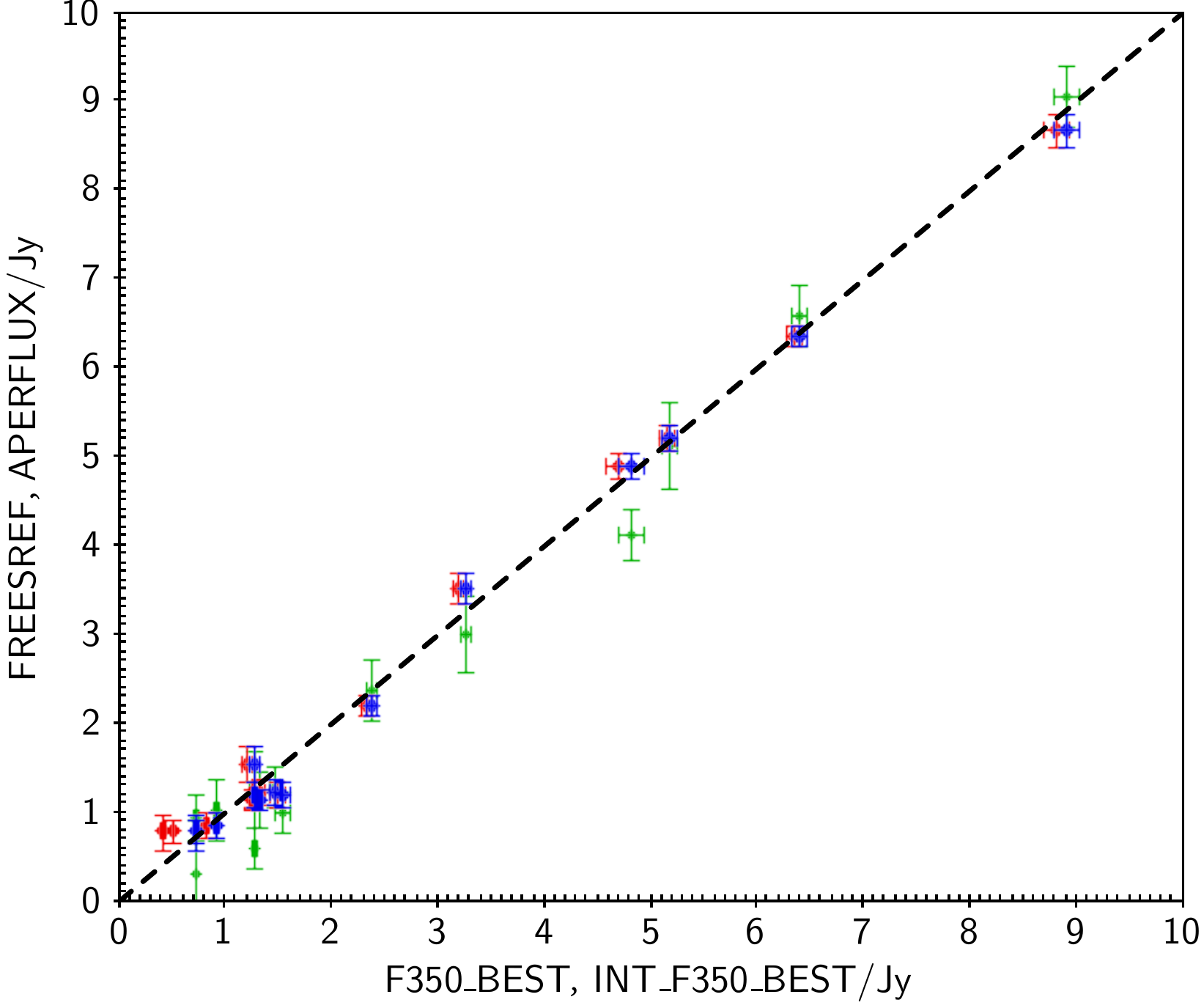}
		\caption{Flux densities (\textsf{FREESREF}) of {\tt BeeP} sources compared to those (\textsf{F350\_BEST}) sources in the \Herschel\ H-ATLAS GAMA15 field catalogue.  The brightest H-ATLAS sources within 5\arcm\ of the corresponding \Planck\ source are shown in red.  The beam-weighted sums of all sources within the corresponding \Planck\ 857-GHz beam are shown in blue.  Green symbols give a similar comparison between the \Herschel\ beam-weighted flux-density sum and the PCCS2 \textsf{APERFLUX} flux density.  The {\tt BeeP} errors are not corrected here (see Sect.~\ref{ssec:FluxDensityAccuracy})}
		\label{fig:HatlasFluxes}
	\end{center}
\end{figure}

We have compared {\tt BeeP} flux estimates with those in one field (GAMA15) of the \Herschel-ATLAS catalogue \citep{HATLAS1,HATLAS2}.  Because of the large disparity between the sensitivity and angular resolution of \Planck-HFI and \Herschel-SPIRE, we collated the catalogues by first selecting all H-ATLAS sources within a radius of $5'$ around each \Planck\ location.  Then, for each \Planck\ source we selected the brightest H-ATLAS source, which is not always the closest one.  The two sets of flux densities (compared in Fig.~\ref{fig:HatlasFluxes}) are, statistically, remarkably consistent.  It is worth noting that the \Herschel\ GAMA15 field follows quite closely the {\tt BeeP} assumptions, in that the background is homogeneous and slowly-varying, and the foregrounds are well separated.  All {\tt BeeP} sources have \textsf{SRCSIG} values above 8.0 and low values of \textsf{RELTH}, except for one pair that is very close and mutually induces non-Gaussianity.  {\tt BeeP} errors are plotted as found in the catalogue (see Sect.~\ref{ssec:FluxDensityAccuracy}).

\subsection{Background estimates}
\label{sssec:BackgroundEstimates}

As described in Sect.~\ref{subsec:BackgroundProperties}, the goal of the {\tt BeeP} background analysis is not to provide an alternative characterization of \Planck's submillimetre diffuse background thermal properties, but rather to understand the impact of the background-foreground thermal contrast on the {\tt BeeP} selection function.  Nevertheless, it is interesting to check the validity of the {\tt BeeP} background parameters.  For this purpose, we have used the dust temperature and spectral-index maps from \cite{planck2016-XLVIII}, which have been extracted using the {\tt GNILC} algorithm. We applied to these maps the procedure described in Sect.~\ref{subsec:BackgroundProperties} used to compute the pairs $\{I_T,\sigma_{I_T}\}$ and $\{I_\beta,\sigma_{I_\beta}\}$, for direct comparison to the equivalent {\tt BeeP} estimates.

\begin{figure*}[htbp!]
	\begin{center}
		\leavevmode
		\includegraphics[width=0.48\textwidth]{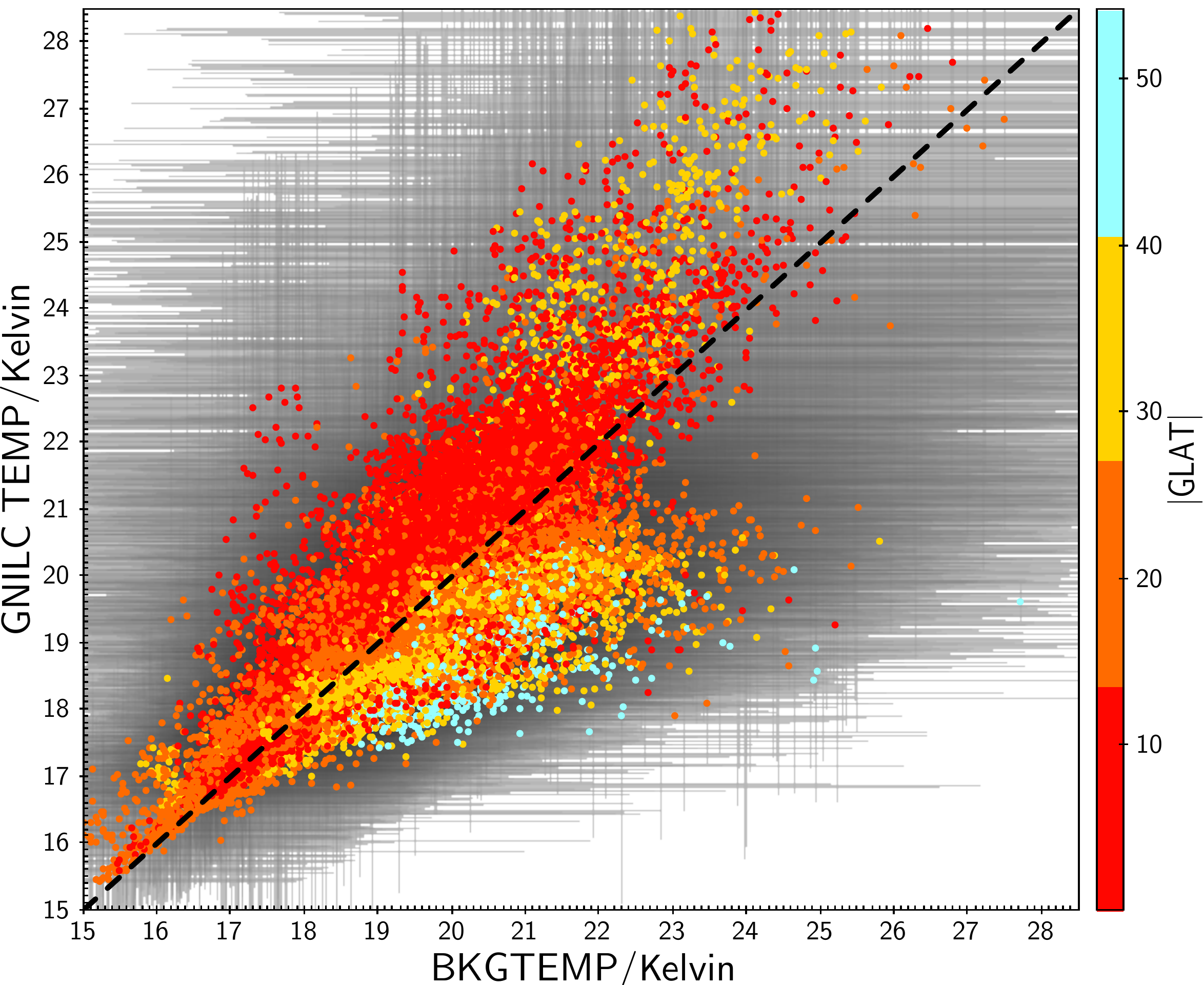}
		\includegraphics[width=0.48\textwidth]{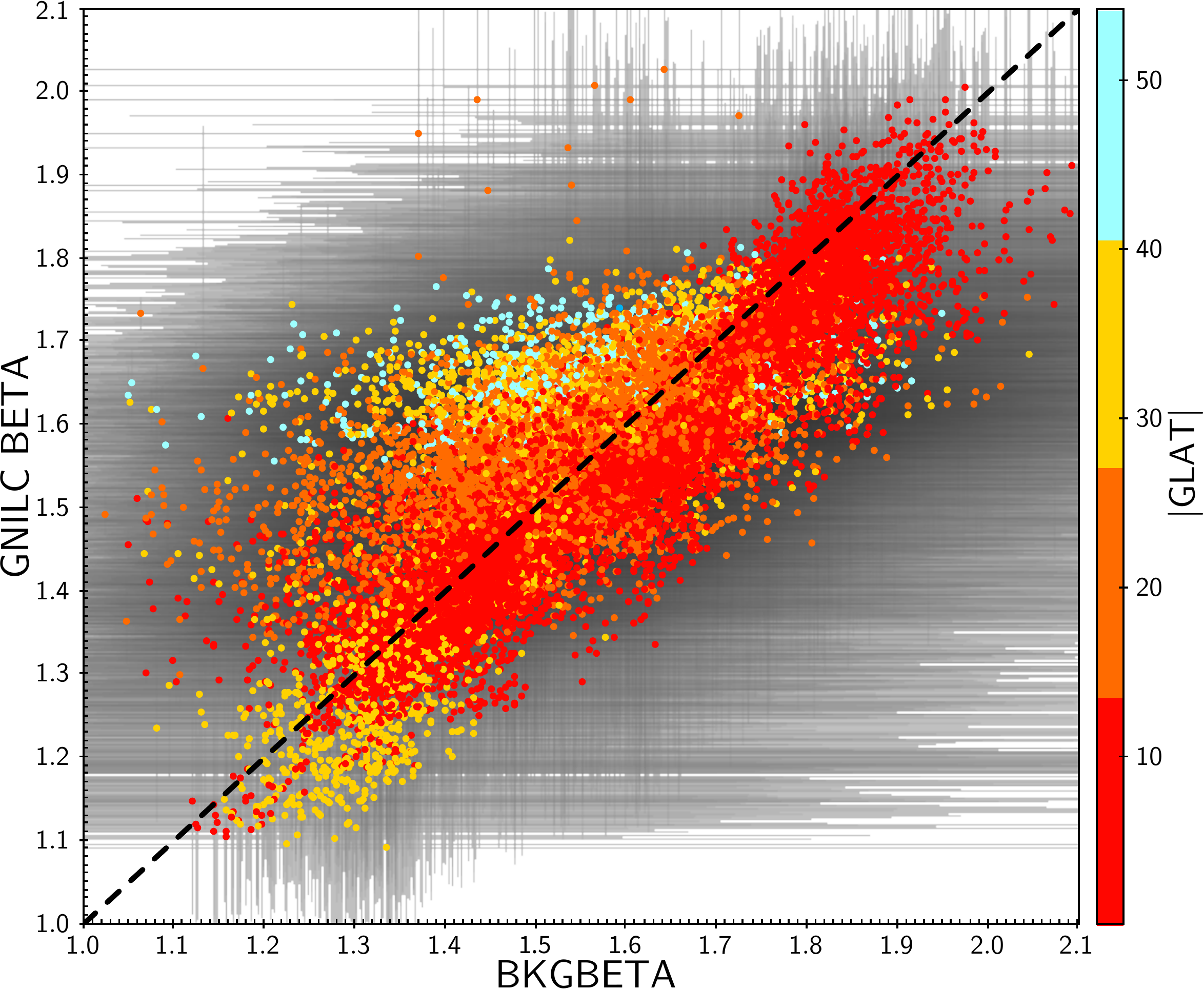}\\
		\includegraphics[width=0.48\textwidth]{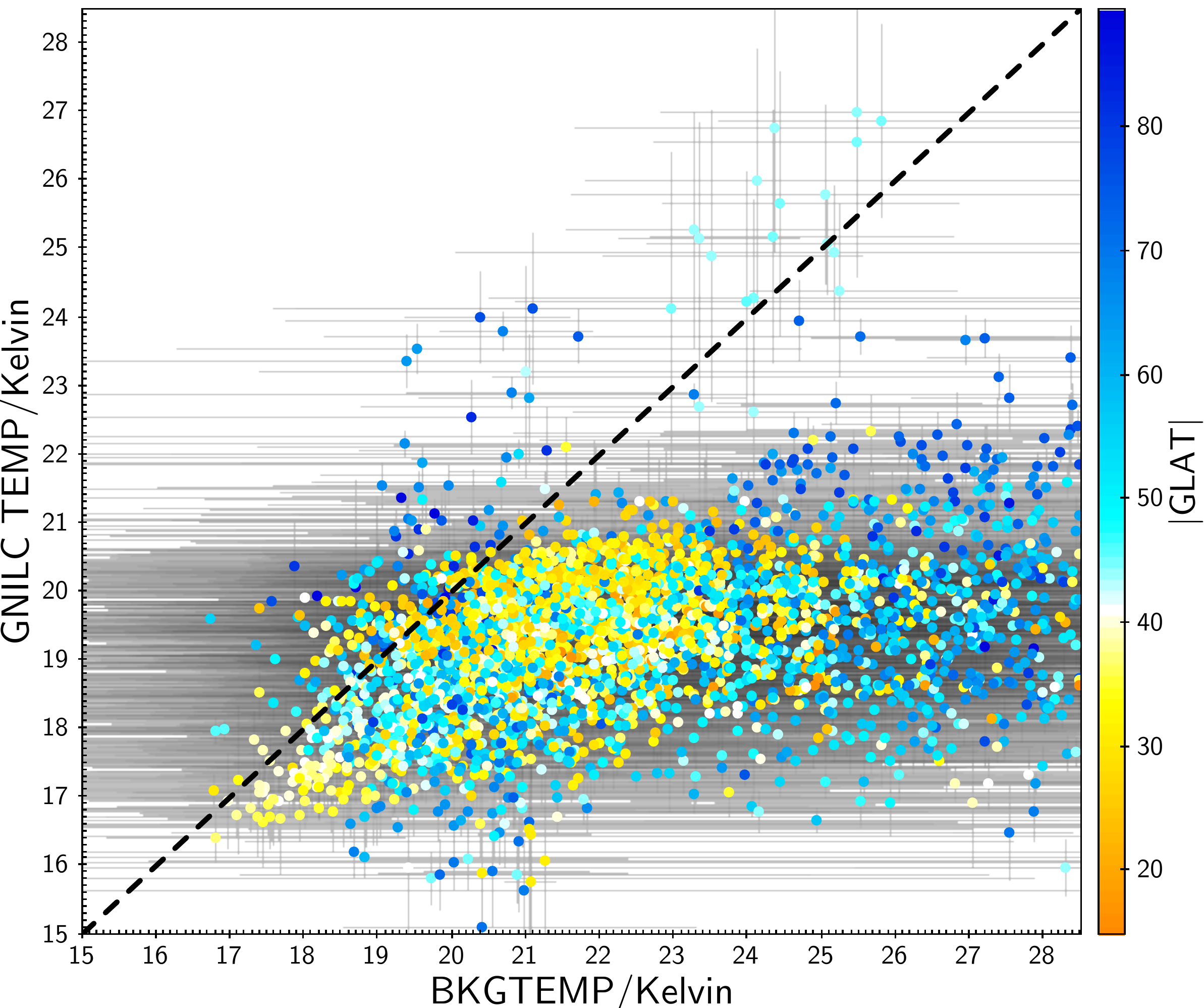}
		\includegraphics[width=0.48\textwidth]{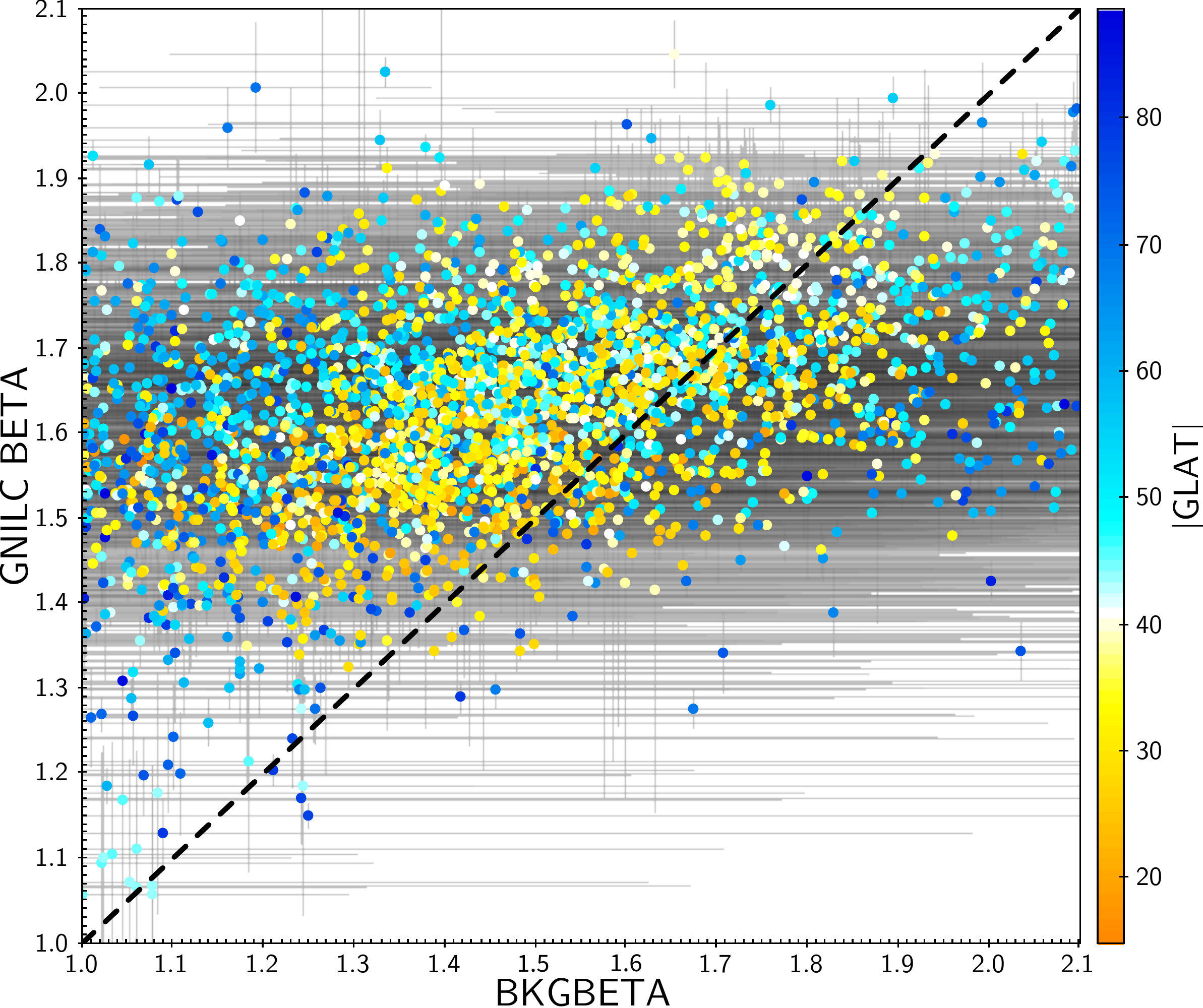}\\
		\includegraphics[width=0.36\textwidth]{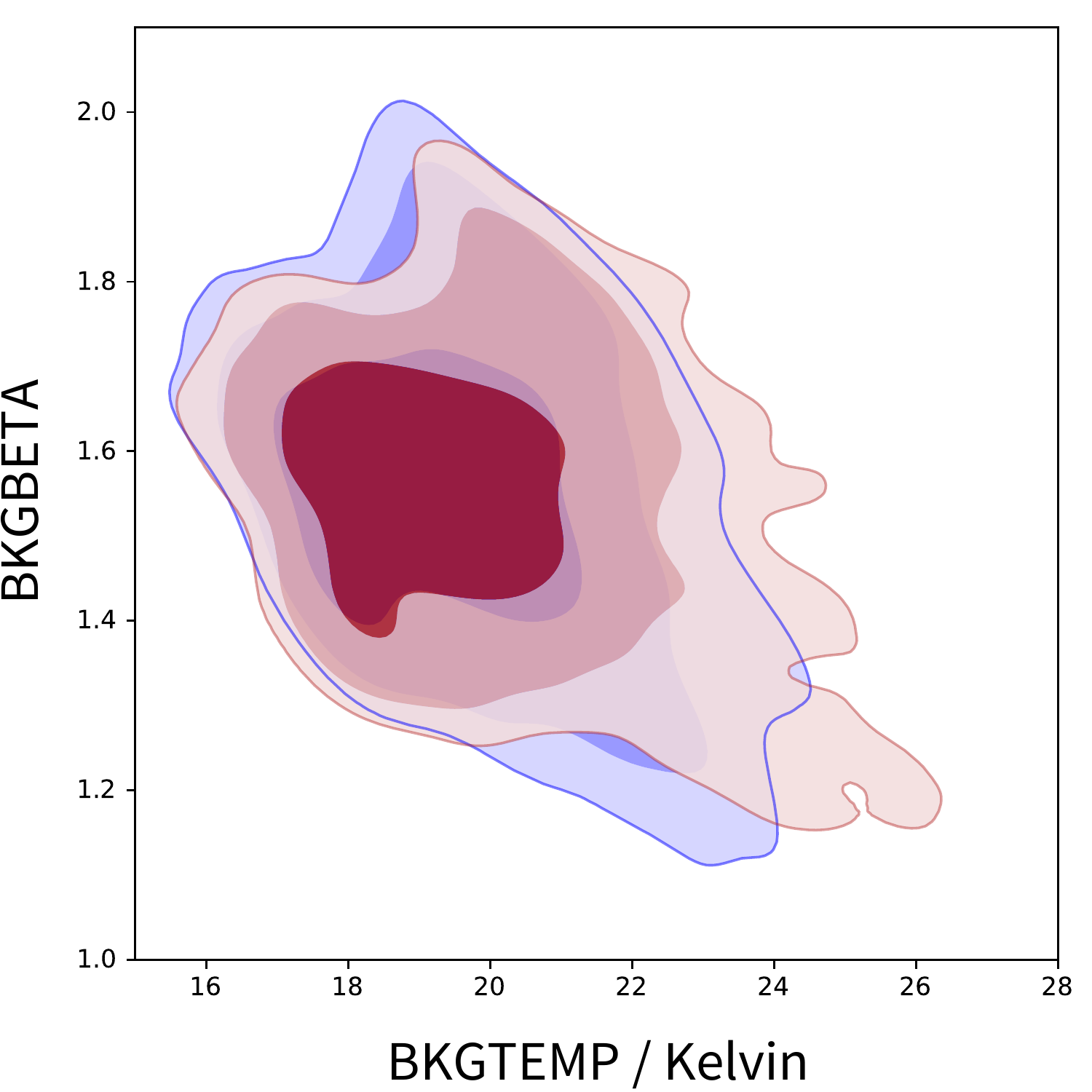}		
		\caption{Comparison of background $T$ (left) and $\beta$ (right) estimates from {\tt BeeP} and {\tt GNILC}.  The colour of the points indicates the object's Galactic latitude, with grey lines being $1\,\sigma$ error bars and the dashed black lines showing equality.  The top row shows sources inside the PCCS2+2E Galactic mask, i.e., regions with strong dust emission.  The middle row shows sources in the PCCS2 set, i.e., high-Galactic-latitude, dust-poor regions.  The bottom row shows the same sources as the top row (i.e., inside the PCCS2+2E Galactic mask), but this time their $T$--$\beta$ distribution. {\tt BeeP} distribution contours are shown in blue ([$68,95,99$]\%) and {\tt GNILC} in red.
		}
		\label{fig:gnilccomparison}
	\end{center}
\end{figure*}

Figure~\ref{fig:gnilccomparison} shows a comparison of the MBB background parameters estimated by {\tt BeeP} with those based on the {\tt GNILC} dust component.  The top row shows sources inside the PCCS2E Galactic mask (where dust emission is dominant).  For these sources there is reasonably good agreement between both estimates.  The bottom row shows the {\tt GNILC} and {\tt BeeP} background estimates in the $T - \beta$ plane for the same set of sources. The parameters ({\tt BeeP} in blue and {\tt GNILC} in red) are in good agreement; however, as the Galactic latitude increases, we start to see some disagreement (light blue points in the top row).  Indeed, in the low-background PCCS2 region (middle row), the {\tt BeeP} and {\tt GNILC} estimates agree less well.  In particular, the higher {\tt BeeP} temperatures are significantly higher than the {\tt GNILC} estimates.  From section~2.3 of \citet{planck2013-p06b} we know that in order to correctly fit the dust emission, we should first remove any other emission (CMB or CIB) and set map zero levels correctly.  However, doing this would have biased the analysis of compact sources (e.g., the CMB is not determined at the location of the PCCS2+2E sources), and therefore the maps used as input to {\tt BeeP} were not adjusted.  This is the cause for the discrepancy with {\tt GNILC} that we observe at high Galactic latitudes, where the relative weight of the CMB component, or even residual CIB anisotropies, is much higher.  For this reason, the {\tt BeeP} MBB parameter estimates of the background at high Galactic latitudes should not be taken to be good measures of the physical properties of dust emission.  Their main purpose is to complement the characterization of the compact objects by adding a physical description of their embedding surroundings.  Nevertheless, in regions of strong dust emission, the {\tt BeeP} MBB parameter estimates are fairly good representations of $T$ and $\beta$ of dust in those regions.

\smallskip

\section{Summary and conclusions}
\label{sec:Conclusions}
 
{\tt BeeP} is a Bayesian algorithm that uses an assumed SED profile to combine observations of a source and its background at multiple frequencies, with the objective of evaluating the reliability of the source detection and estimating its physical properties.  To implement {\tt BeeP}, we developed a fast likelihood code, based on a simplified version of the data model, which overcomes the difficulties posed by the high data volume.

By applying the {\tt BeeP} algorithm to the \Planck\ 2015 maps at 353, 545, and 857\,GHz, and the IRIS map at 3000\,GHz, and assuming a dusty (MBB) SED, we constructed an extension to the \Planck\ 857-GHz PCCS2 and PCCS2E single-channel catalogues, which provides new information on the reliability and physical properties of the sources in the catalogues.  Since multiple frequencies are used, improved detection strength (\textsf{NPSNR}) is achieved.  Our data model permits the construction of a statistic to measure the local non-Gaussianity of the background (\textsf{RELTH}), which was used to correct \textsf{NPSNR}.  The new significance statistic \textsf{SRCSIG} resulting from this process helps to separate foreground compact objects from the background, even in regions of strong and complex backgrounds such as cirrus and filaments.  {\tt BeeP} also determines a new effective source position that incorporates information at all frequencies used,  which is not the same as what is provided in PCCS2+2E.

In addition to its determination of source reliability, {\tt BeeP} provides a characterization of the thermal properties of each source and its background.  As part of the {\tt BeeP} output, we provide a figure for each source that displays the SED curves associated with the posterior parameter samples.  We also provide the joint posterior distributions of the source MBB model parameters in ``triangle'' plots, which give, in each non-diagonal position, the marginalized bi-dimensional posterior PDF of the parameter samples defining the row and the column.  The diagonal locations contain posterior marginalized distributions.  The visualization of the posterior distributions enable a more complete understanding of the uncertainties associated with the source-parameter estimates.  For instance, as expected, there is a strong correlation between the source MBB spectral index and its temperature. 

For the sake of completeness, and to allow for a better comparison with PCCS2+2E data, we also extracted source parameters using a data model where the flux density of the source in each channel is a free independent parameter, the ``Free'' model.  We also provide the MBB characterization of the background surrounding each source.

We tested the limits of our simplified data model and likelihood implementation, using an extensive battery of simulations, ranging from the fiducial, which closely follows the assumptions of the simplified data model, to the most realistic \Planck\ simulations associated with the \Planck\ 2015 data set, namely the full focal plane simulations (FFP8).  To enhance confidence in the results, we also resorted to injection simulations, where a representative set of compact objects was injected into the real maps, extracted, and then cross-matched with the input catalogue.  Analysis of the simulations allowed us to identify some data subsets where the optimality of the algorithm could not be guaranteed.  We established a criterion of quality (\textsf{EST\_QUALITY}), which can be used to filter out these anomalous sources.

The simulations were used to evaluate the effect of beam variations and ellipticity (which are not considered in our basic model) on parameter estimation.  In particular, we find an 11\,\%\ bias in the estimated flux densities, which we corrected in the output catalogue. 
Simulations also allowed us to determine that the uncertainties estimated by {\tt BeeP} for flux densities and positions are unrealistically small for sources with very high values of NPSNR.  We suggest procedures to correct these uncertainties for the small fraction of sources affected, but we did not apply them to the output catalogue -- they should only be used if the rigorous statistical characterization of samples including those sources is required.

Based on our analysis, we define a reliable and accurate subset of PCCS2+2E ({\tt BeeP/base}) containing 26\,083 sources (54.1\,\% of PCCS2+2E), of which 21\,997 are in PCCS2E (50.8\,\%). The estimated contamination level of this subset is between 5\,\%\ and 10\,\%. This, on its own, significantly improves the original PCCS2E, which contains no validated indicator of source reliability.  Further imposing a criterion of contamination below 1\,\%, {\tt BeeP} still ranks 5077 compact objects in the PCCS2E as ``good.''
Although the {\tt BeeP/base} catalogue should be adequate for most purposes, we provide the relevant information needed by a user of our augmented version of PCCS2+2E to select a different subsample fitting specific scientific requirements (suggestions for selection criteria can be found in Sect.~\ref{subsec:BeyondBase}).

{\tt BeeP}'s selection function overlaps with that of the \texttt{CoCoCoDeT} extraction method used to generate the Planck Catalogue of Galactic Cold Clumps (GCC).  The number of common objects between {\tt BeeP/base} and the best quality detections in GCC (\textsf{FLUX\_QUALITY} = 1) contains 3757 sources.  We find good consistency in the thermal source parameters recovered by the two methods, considering the uncertainties in the estimation.  The {\tt BeeP} selection function is broader than that of GCC, even for the same range of temperatures, since the {\tt BeeP} likelihood is not limited by the temperature contrast between a cold source and a warm background.  The {\tt BeeP} catalogue is, therefore, complementary to the GCC.  For the GCC-selected sample, the {\tt BeeP} parameter \textsf{NPSNR} (strength of the detection) is well-correlated with the source-to-background contrast. 

The {\tt BeeP} reference flux-density estimates (at 857\,GHz) were also cross-checked against the PCCS2+2E estimates at 857\,GHz and the \Herschel\ GAMMA15-field catalogue at 350\microns.  The match with the \Herschel\ estimates is reasonably good when we include all sources within the \Planck\ beam.  The consistency with the PCCS2+2E flux-density estimates is also good, with only small biases, of known origin, but with some dispersion that is almost entirely the result of the large uncertainty in the \textsf{PCCS2+2E} aperture-photometry estimates.  The {\tt BeeP} flux-density uncertainty is significantly smaller (by a factor of 2) than that of the \textsf{PCCS2+2E} aperture-photometry estimates.

We also compared the {\tt BeeP} estimates of background parameters against those of the {\tt GNILC} temperature and spectral index dust maps.  In those regions where dust is the dominant component, for instance within the PCCS2+2E Galactic masks, the agreement with the MBB thermal parameters is good.  However, at high Galactic latitudes, where dust is no longer dominant and the CIB is strong, and, especially for the 353-GHz channel, the correlation is not as good, and the {\tt BeeP} parameter estimates are less reliable.

In conclusion, we provide a new data set that characterizes the reliability and thermal properties of all sources in the PCCS2+2E.  We expect this to greatly improve the utility of these catalogues. The results of this analysis will be made publicly available via the Planck Legacy Archive.


\begin{acknowledgements}
The Planck Collaboration acknowledges the support of: ESA; CNES and CNRS/INSU-IN2P3-INP (France); ASI, CNR, and INAF (Italy); NASA and DoE (USA); STFC and UKSA (UK); CSIC, MINECO, JA, and RES (Spain); Tekes, AoF and CSC (Finland); DLR and MPG (Germany); CSA (Canada); DTU Space (Denmark); SER/SSO (Switzerland); RCN (Norway); SFI (Ireland); FCT/MCTES (Portugal); and ERC and PRACE (EU). A description of the Planck Collaboration and a list of its members, indicating which technical or scientific activities they have been involved in, can be found at 
\href{url}{http://www.cosmos.esa.int/web/planck/planck-collaboration}.
We acknowledge support from the ESTEC Faculty Research Project Programme.
\end{acknowledgements}

\bibliographystyle{aat}
\bibliography{references,Planck_bib} 

\appendix
\section{\textsf{BeeP}}
\label{sec:beep}

In this Appendix we provide a more detailed description of our algorithm, which we have named Bayesian Estimation and Extraction Package, and which we refer to as \textsf{BeeP}. 

\subsection{Characterization method}
\label{subsec:characterizationMethod}
The Bayesian system of inference is an extension of deductive logic ( \{ 0=false, 1= true\} ) to a broader class of ``degrees-of-belief'' that consistently maps them into the real interval $[0,1]$ \citep[][chapters~1 and 2]{Jaynes}.
It associates those degrees-of-belief with conditional probabilities.
So, if we represent the quantities we are interest in, like source position, flux, SED etc., by parameter vector $\vec{\Theta}$, the relevant question we can ask is: what is the joint probability distribution of our parameter vector $\vec{\Theta}$, given our data $\vec{d}$ and model assumptions $H$:
\begin{equation}
\Pr(\vec{\Theta} | \vec{d}, H). \label{eq:BI_posteriordefinition}
\end{equation}
It is possible to relate the quantity we are interested in with others that can be computed with the help of Bayes theorem:
\begin{equation}
\Pr(\vec{\Theta} | \vec{d}, H) =
\frac{\Pr(\vec{d}|\,\vec{\Theta},H)\Pr(\vec{\Theta}|H)}
{\Pr(\vec{d}|H)}, \label{eq:BI_Params}
\end{equation}
where $\Pr(\vec{\Theta} | \vec{d}, H)$ is the ``posterior probability'' distribution of $\vec{\Theta}$,
$\Pr(\vec{d}|\,\mathbf{\Theta},H) \equiv \mathcal{L}(\vec{\Theta})$ is the likelihood,
$\Pr(\vec{\Theta}|H) \equiv \pi(\vec{\Theta})$ is the probability distribution of the variables of interest before considering the data, or the ``prior''
and $\Pr(\vec{d}|H)$ is the Bayesian ``evidence,'' which, in this case, does not depend on any variable. 
Therefore, the evidence will only act as a normalizing constant and will be ignored.
So, our main inference equation will read,
\begin{equation}
\Pr(\vec{\Theta} | \vec{d}, H) \propto \mathcal{L}(\vec{\Theta}) \, \pi(\vec{\Theta}).
\label{eq:BI_ParamsShort}
\end{equation}

Once we have defined the likelihood and the prior functions, the parameter manifold $\vec{\Theta}$ is sampled using a Markov-chain Monte Carlo (MCMC) algorithm \citep[][]{MCMCBook}.
Choosing the right MCMC algorithm is still very much a matter of trial and error, since an optimal choice depends very much on the parameters manifold topology.
In our case we expect a very heterogeneous manifold.
Variables like temperature ($T$) and spectral index ($\beta$) are highly correlated and generate deep curved likelihood valleys, particularly for high signal-to-noise (S/N) ratio sources.
The source flux density ($S$) and extension/radius ($r$) variables are expected to be correlated as well; however, the correlation is mostly linear and not very narrow.
Additionally, the position vector variables $(X , Y)$ are completely uncorrelated with all others. 
After reviewing several candidate algorithms, we chose \texttt{MCMC Hammer},\footnote{\url{http://dan.iel.fm/emcee/current/}}.
which is currently popular in astrophysics and well adapted to sample from a likelihood manifold like ours.
However, we did not use the available \texttt{python} code version because it did not show the required performance.
A completely new sampler code was written in \texttt{C++}, based on the same algorithm \citep{MCMC_Hammer}.
When running this code we chose to set all prior distributions to be uniform within a defined range.\footnote{
Given that we are only estimating the parameter posterior distributions, there is no need to define a precise and well-motivated range for the uniform distributions.
We just need to make sure they are wide enough to not truncate a significant fraction of the likelihood volume for the physically possible parameter ranges
(eee Sect.~\ref{subsubsec:Priors}).
}

\subsubsection{Likelihood}
\label{subsubsec:Likelihood}
In Sect.~\ref{sec:SingleSrcModel}, we have described how we build a model for \textit{each} of the sources independently, $\vec{s}_j(\vec{\vec{\Theta}_j})$, which we combine with a model for the background, $\vec{b}_j(x)$, and the noise, $\vec{n}_j(x)$, in the neighbourhood of the source.
Following the same principles, but extending to the full data set, the data model would now read:
\begin{equation}
\label{eq:dataModelApp}
\vec{d}(\vec{x}) = \sum^{N_{{\mathrm{s}}}}_{j=1}
\vec{s}_j(\vec{x};\vec{\Theta}_j) + \vec{b}(\vec{x}) +
\vec{n}(\vec{x}).
\end{equation}
where for convenience we concatenated the individual background and noise quantities into the full sky 
$\vec{b}(\vec{x})$ and $\vec{n}(\vec{x})$
quantities.
Given the assumptions described in Sect.~\ref{sec:MethSourceModel} the likelihood representing \textit{all} compact objects in the map is
\begin{equation}
\label{eq:LikelihoodComplete}
\mathcal{L}(\vec{\Theta})=
\frac{\exp\left\{-\frac{1}{2}\left[\vec{d} - \widehat{\vec{b}} - \vec{s}(\vec{\vec{\Theta}}) \right]^{\rm t} \tens{N}^{-1} \left[\vec{d}- \widehat{\vec{b}} - \vec{s}(\vec{\vec{\Theta}}) \right] \right\}}
{\left(2\pi\right)^{N_{\rm pix}/2} \left|\tens{N}\right|^{1/2}},
\end{equation}
where $\vec{d}$ is the data (pixels), $\widehat{\vec{b}}$ is the generalized background ($\vec{b} + \vec{n}$) and $\tens{N}$ is the generalized background covariance matrix. For compactness, we merged all individual source parameters ($\vec{\Theta}_j$) into $\vec{\Theta}$.
$\tens{N}$ is a huge matrix $N_{\rm pix} \times N_{\rm pix}$, where $N_{\rm pix} \sim 50\,000\,000$.
Any brute force attempt to evaluate the likelihood will undoubtedly be frustrated by the sheer magnitude of the problem.
This is where we take advantage of the homogeneity condition.
Since $\tens{N}$ is the covariance matrix of a homogeneous Gaussian random field, by definition it is ``circulant.'' Therefore, when represented in Fourier space it becomes diagonal.
Performing this transformation, the full-sky source signal (Eq.~\ref{eq:dataModelApp}) in Fourier space reads,
\begin{equation}
\widetilde{\vec{s}}(\vec{\eta};\vec{\Theta}) =
\widetilde{\vec{B}}(\vec{\eta})\sum_{j=1}^{N_{\mathrm{s}}} A_j \vec{f}(\vec{\phi}_j)
\,\widetilde{\tau}\,(-\vec{\eta};\vec{a}_j)\textrm{e}^{\textrm{i} 2\pi{\bm \eta}\cdot\vec{X}_j},
\label{eq:sft}
\end{equation}
where the vector $\widetilde{\vec{B}}(\vec{\eta})$ contains the
Fourier transform of the beam at each frequency,\footnote{The beam transfer function is also convolved with the pixel window function at each frequency. In this particular case the pixel window function does not change across maps.}
$\vec{f}(\vec{\phi}_j)$ contains the Fourier transform of the emission coefficients at each frequency, and $\widetilde{\tau}(\vec{\eta};\vec{a})$ is the Fourier transform of
the template for an unconvolved object at the origin, characterized by
the shape parameters $\vec{a}$.

We now consider the likelihood of the ``no-source'' model $\mathcal{L}_0$, i.e., when $A$, the source amplitude is equal to $0$.
$\mathcal{L}_0$ is a constant, since it does not contain any parameter. 
By taking the logarithm of the $\mathcal{L}(\vec{\Theta})/\mathcal{L}_0$ ratio, we reach a likelihood expression that reads 
\[
\hspace*{-0cm}\ln\left[\mathcal{L}(\vec{\Theta}){L}_0\right] =
\]
\vspace*{-0.2cm}
\[
\sum^{N_{\mathrm{s}}}_j \left\{A_j\mathcal{F}^{-1}
\left[\mathcal{P}_j(\vec{\eta})
\widetilde{\tau}(-\vec{\eta};\vec{a}_j)\right]_{\vec{X}_j} - \tfrac{1}{2}A_j^2 \sum_{\vec{\eta}}
\mathcal{Q}_{jj}(\vec{\eta})
|\widetilde{\tau}(\vec{\eta};\vec{a}_j)|^2\right\}
\]
\vspace*{-0.4cm}
\begin{equation}
- \sum^{N_{\mathrm{s}}}_{i > j}\left\{A_iA_j
\mathcal{F}^{-1}
\left[\mathcal{Q}_{ij}(\vec{\eta})
\widetilde{\tau}(\vec{\eta};\vec{a}_i)
\widetilde{\tau}(-\vec{\eta};\vec{a}_j)\right]_{\vec{X}_i
	- \vec{X}_j} \right\},
\label{eq:LikeFilterComplete}
\end{equation}
where $\mathcal{F}^{-1}[\ldots]_{\vec{x}}$ denotes the inverse Fourier
transform of the quantity in brackets, evaluated at the point $\vec{x}$.
We have also defined the following quantities: 
\begin{itemize}
	\item \textit{Point source response} (or beam shape, i.e., how the data responds to the presence of a point source),\\
	$(\vec{\psi}_{i})_\nu = \widetilde{B}_\nu(\vec{\eta})
	(\vec{f}_{i})_\nu$, with $\nu$ labelling frequency channels;
	\item \textit{Information source} (on the point sources),\\
	$\mathcal{P}_j({\bm \eta}) \equiv \widetilde{\vec{d}}^{\sf T}(\vec{\eta})
	\tens{N}^{-1}(\vec{\eta}) \vec{\psi}(\vec{\eta})$;
	\item \textit{Point-source flux precision matrix}\\
	$\mathcal{Q}_{ij}({\bm \eta}) \equiv
	\widetilde{\vec{\psi}}_i^{\sf T}(\vec{\eta})
	\tens{N}^{-1}(\vec{\eta}) \vec{\psi}_j(\vec{\eta})$,
\end{itemize}
where $\tens{N}$ is the covariance matrix represented in Fourier space.

Let us take a closer look at the second term of Eq.~\eqref{eq:LikeFilterComplete}, the ``cross-term,'' where multiple sources interact.
To help clarify the physical meaning of this expression let us write it for the ``pure'' point source, $\tau(\vec{x},\vec{a}) = \delta(\vec{x})$:
\begin{equation}
\sum_{i>j} A_i A_j
\mathcal{F}^{-1}[\mathcal{Q}_{ij}(\vec{\eta})]_{\vec{X}_i-\vec{X}_j}.
\label{eq:WellSeparetedObjectsPS}
\end{equation}
For simple, uncorrelated backgrounds, $\mathcal{Q}_{ij}(\vec{\eta})$ contains just linear combinations of the instrument beams at each frequency channel. The condition for this expression to become small, when compared with the rest of the likelihood, is the common assumption in astronomy that beam blending effects are negligible. If the sources are well separated such that $| \vec{X}_i-\vec{X}_j |$ is large enough for the multi-channel equivalent beam to have died out, then the contribution of Eq.~\eqref{eq:WellSeparetedObjectsPS} to the likelihood may be safely dropped.\footnote{When the background is uncorrelated, this condition is immediately fulfilled if each pixel contains signal coming from one and only one source. However, this might not be sufficient when there are strong correlations in the background as in the case for \Planck\ data.}
So, assuming
\begin{equation}
\sum^{N_{\mathrm{s}}}_{i > j}\left\{A_iA_j
\mathcal{F}^{-1}
\left[\mathcal{Q}_{ij}(\vec{\eta})
\,\widetilde{\tau}(\vec{\eta};\vec{a}_i)
\,\widetilde{\tau}(-\vec{\eta};\vec{a}_j)\right]_{\vec{X}_i
	- \vec{X}_j} \right\} \approx 0,
\label{eq:WellSeparetedObjects}
\end{equation}
is nothing more than a generalization of the common assumption that objects are well separated so that we can ignore object blending effects caused by the beam.\footnote{Given \Planck's sensitivity, the surface density of sources is such that this condition holds well, except for in the Galactic plane.}
We are left with the likelihood expression, which we sample from
\[
\hspace*{-0cm}\ln\left[\mathcal{L}(\vec{\Theta})/\mathcal{L}_0\right] =
\].
\vspace*{-0.4cm}
\begin{equation}
\sum^{N_{\mathrm{s}}}_j \left\{A_j\mathcal{F}^{-1}
\left[\mathcal{P}_j(\vec{\eta})
\widetilde{\tau}(-\vec{\eta};\vec{a}_j)\right]_{\vec{X}_j} - \tfrac{1}{2}A_j^2 \sum_{\vec{\eta}}
\mathcal{Q}_{jj}(\vec{\eta})
|\widetilde{\tau}(\vec{\eta};\vec{a}_j)|^2\right\}.
\label{eq:LikelihoodFinal}
\end{equation}
Equation~\eqref{eq:LikelihoodFinal} is extremely convenient from a computational point of view.
The likelihood ratio, when neglecting the blending effects, becomes the sum of the individual contributions from each source.
This allows one to use a very convenient ``one source at a time'' approach.

If $\widehat{\vec{\Theta}}_j$ is the set of parameter values that maximizes the likelihood ratio (Eq.~\ref{eq:LikelihoodFinal}) for source $j$ then
\begin{equation}
\label{eq:NPSNR_def}
\ln\left(\frac{\mathcal{L}(\widehat{\vec{\Theta}}_j)}{\mathcal{L}_0}\right) =
\tfrac{1}{2} \sum_{{\boldsymbol \eta}} \mathcal{\widehat{Q}}_{jj}(\vec{\eta})
|\widetilde{\tau}(\vec{\eta};\widehat{\vec{a}}_j)|^2 \widehat{A}_j^2
=
\tfrac{1}{2}\widehat{\textsf{NPSNR}}_j^{~2},
\end{equation}
where we have defined the quantity ${\cal R}$ as the Neyman-Pearson S/N ratio, (corresponding to $\textsf{NPSNR}$ in the catalogue).
This variable is a function of the likelihood ratio, hence ``Neyman-Pearson,'' but since
\begin{equation}
	 \sum_{{\boldsymbol \eta}} \mathcal{\widehat{Q}}_{jj}(\vec{\eta})
	|\widetilde{\tau}(\vec{\eta};\widehat{\vec{a}}_j)|^2 = \frac{1}{\sigma^2},
\end{equation}
where $\sigma^2$ is the variance of the likelihood-ratio background field, it is also a signal-to-noise ratio, or the detection significance level (i.e., ``how many sigma'' this detection is).
As we noted in Sect.~\ref{ssec:FluxDensityAccuracy}, if all our assumptions hold and all source parameters were known except the amplitude ($A$), then $\textsf{NPSNR}$ would indeed be the inverse of the fractional error on amplitude $A / \Delta A$.

\subsubsection{Source-detection significance evaluation: dealing with the deviations from the data model}
\label{ssec:DetSignif}
Although much of this has already been described in 
Sect.~\ref{sec:SingleSrcModel}, we repeat a brief discussion of the evaluation of source significance here, in order to preserve the continuity of Appendix~\ref{sec:beep}.

There are two main data features that break the assumptions in our data model:
\begin{itemize}
	\item background non-Gaussianity;
	\item localized structures.
\end{itemize}
It is well known that diffuse emission from dust, the main background component, is highly non-Gaussian.
One may argue that because we are combining data from several channels, that increases the data volume and because of the central limit theorem the statistics should converge to Gaussian.
Unfortunately, this is only true close to the mode of the distribution.
But detection is all about the positive tail of the background distribution (see e.g., Fig.~\ref{fig:Histogram}), and in this case the non-Gaussianity only decreases very slowly when more data are added \citep[][chapter~2]{RiskAnalys}.
However, an even larger problem comes from localized structures such as cirrus.\footnote{Cirrus is not the only type of localized feature. Extended sources that were identified as compact objects in the PCCS2, but where the actual positions were off the centre, also appear like localized artefacts.}
The likelihood ($\mathcal{L}(\widehat{\vec{\Theta}})$) of a cirrus cloud being confused for a source is small, given that cirrus is rather poorly described as a compact source.
On the other hand, the likelihood of cirrus being a homogeneous Gaussian random field $\mathcal{L}_0$ is also very small, since by definition these structures do not behave as a homogeneous random field.
So, by looking at Eq.~\eqref{eq:NPSNR_def}, one can see that the source significance indicator
\begin{equation*}
\label{eq:NPSNR_prop}
\textsf{NPSNR} \propto \sqrt{\ln\left(\mathcal{L}(\widehat{\vec{\Theta}})/\mathcal{L}_0\right)},
\end{equation*}
might indeed create a strong positive tail event when a cirrus structure is present, even in the absence of a genuine source,
and this might be taken (erroneously) to be an object of interest.
It can be shown that, if all our assumptions hold, under the ``null'' hypothesis of our model (i.e., ``only background is present'') the following field is a white-noise unitary ($\sigma = 1$) Gaussian random field in pixel space ($\vec{X}$):
\begin{equation}
\label{eq:FilteredField}
\frac{\mathcal{F}^{-1} \left[\widehat{\mathcal{P}}(\vec{\eta})
	\widetilde{\tau}(-\vec{\eta};\widehat{\vec{a}})\right]_
	{\vec{X}}}
{\sqrt{\sum_{\vec{\eta}} \widehat{\mathcal{Q}}(\vec{\eta})
	|\widetilde{\tau}(\vec{\eta};\widehat{\vec{a}})|^2}}.
\end{equation}
Then if we added a point source to the centre of this perfect background we would introduce significant outliers in the positive tail of the distribution.
Let us now assume that the positive outlier pixels created by the source are no more than a small fraction of the total number of pixels ($\alpha$).
Then using the quantile definition one would expect that
\begin{equation}
\label{eq:QuantileDef}
\int_{-\infty}^{\textsf{RELTH}} \,\, \frac{\exp\left[-\frac{1}{2}\left(\frac{x}{\sigma}\right)^2\right]}{\sqrt{2 \pi} \sigma} \textrm{d}x = 1-\alpha,
\end{equation}
where { $\textsf{RELTH}$ (reliability threshold) is the $1-\alpha$ distribution quantile}.
$\textsf{RELTH}$ can be read from the actual field histogram and then Eq.~\eqref{eq:QuantileDef} solved for $\sigma$.
If the remaining $ 1-\alpha$ pixels follow a unitary Gaussian distribution then $\sigma = 1$.
However because of the enlarged distribution tails induced by the localized features and the background non-Gaussianity, $\sigma$ is expected to be larger.
Using simulations, we have verified that the number of outlier pixels created by the source is less than 5\,\%\ of the total, so we use $\alpha = 5\,\%$.

Solving Eq.~\eqref{eq:QuantileDef}, $\sigma$ is equal to
\begin{equation}
\label{eq:RelthSigma}
\sigma = k \,\, \textsf{RELTH},
\end{equation}
where $k$ is a pure numerical constant given by 
\begin{equation}
\label{eq:RelthKdef}
k = \frac{1}{\sqrt{2} \, \textrm{erfc}^{-1}(2\alpha)},
\end{equation}
where $\textrm{erfc}^{-1}$ is the inverse complementary error function.
We finally define the ``source significance'' estimator as
\begin{equation}
\label{eq:RelthDef}
\textsf{SRCSIG} = \frac{1}{k} \,\, \frac{\textsf{NPSNR}}{\textsf{RELTH}} ,
\end{equation}
where $k$ is a pure numerical constant given by Eq.~\eqref{eq:RelthKdef}. This value is the same for all sources.
If the histogram of the field given in Eq.~\eqref{eq:FilteredField} is Gaussian, then $\sqrt{2} \, \textrm{erfc}^{-1}(2\alpha) = \textsf{RELTH}$ by definition and $\textsf{SRCSIG} = \textsf{NPSNR}$.
If our assumptions hold then, as predicted, $\textsf{NPSNR}$ is the detection significance.
However when there is non-Gaussianity in the background, either from diffuse components or localized features, then $\textsf{RELTH}$ increases and a penalty is applied to the Gaussian criterion.
This criterion becomes relaxed for high galactic latitudes away from cirrus where the homogeneity and Gaussian assumptions hold well, while in the neighbourhood of the Galactic plane or inside cirrus structures it becomes mores stringent to avoid false positives induced by the non-Gaussianity of the background.\footnote{See Sects.~\ref{ssec:AlgoImplSrcDetec} and \ref{psssec:ArtificialSimulations} for the practicalities of applying Eq.~\eqref{eq:RelthDef}.}

\subsubsection{Priors}
\label{subsubsec:Priors}
We have tried to choose ``non-informative'' priors, constructing them such that the ``maximum a
posteriori'' (MAP) estimator of any quantity  depends exclusively
on the current data set.
One way of expressing this condition is that, when
changing the data, the likelihood shape remains unchanged and only its
location in the parameter space changes \citep[][chapter~1]{BoxTiao}.
Source position and amplitude are ``location'' parameters, at least within small ranges around the likelihood maxima. So all associated priors will be taken as uniform.
The same cannot be said about the source extension parameter \textsf{EXT} (see definition in Section~\ref{ssec:SourceSize}), which is a ``hybrid'' parameter that shifts and scales the likelihood \citep{PwSII}.
To improve the accuracy of the estimates, we use a ``trick'' (see Sect.~\ref{ssec:AlgoImplSrcDetec}) that makes the objects always appear as if they were slightly extended.
The prior on \textsf{EXT} should behave as $\pi(\textsf{EXT}^{-2}$), and this function varies slowly for values of \textsf{EXT} away from $0$. Since we target compact sources (i.e. close to beam-sized), we are able to select a narrow range (between 0.46 and 2.6 pixels), which allows us to replace the functional prior with a uniform one. This trick simplifies the problem without biasing the estimate of the value that maximizes the likelihood ($\widehat{\textsf{EXT}}$).

Regarding the source brightness parameters (flux and spectral index), \cite{Commander} claim that using uniform priors instead of the Jeffreys non-informative priors creates a strong bias on the spectral index estimate.
However, we have carried out an extensive battery of simulations to test this claim, and failed to find such a bias.
Therefore, for simplicity we have kept the uniform prior distribution.
It is important to keep the range of priors large enough to properly explore and characterize uncertainties.
The ranges we selected bracket widely physically motivated values $\{\,\beta \in [0,7]; T \in [3, 150]\,{\rm K}\,\}$.\footnote{We allow $\beta$ to go down to $0$, to 
accommodate for flat spectra synchrotron sources.} We note, however, that the resulting range of values (see Fig.~\ref{fig:DistribTempBeta}) is consistent with physically reasonable values and that the error bars do extend to much wider ranges.

\subsubsection{Covariance matrix estimation: cross-correlation factor}
\label{subsubsec:CovarianceMatrixEstimation}
The background cross-power spectrum matrix $\tens{N}$ is a critical part of the likelihood and our data model assumes we know its true value.\footnote{$\tens{N}$ is a set of $4\times4$ matrices, one for each pixel.}
However, as it is not known a priori, an estimate must be computed.
There are at least two completely different ways of tackling this problem.
One way is by using theoretical models for each of the background components (for diffuse dust emission models see \citealt{planck2016-XLVIII} and \citealt{SchaeferII} for their application to the estimation of background cross-covariance).
However powerful, this technique assumes full-sky statistical isotropy.
A quick look at \Planck\ maps immediately shows that these conditions are severely broken and hence the models are a sub-optimal approximation to real data.

A different approach (and the one we take) is to split the sky into small fields, where the isotropy conditions apply fairly well, and estimate the cross-power spectrum directly from the data.
This method is not without its problems. For instance the data is one single realisation of the random process and not the ensemble average.
In order to improve the estimation quality of the background covariance, we have developed a method based on the work of \citet[][see chapter~9]{RiskAnalys} for time series.
Expressing the problem in Fourier space allows us to treat each pixel \textit{of the same channel}, or Fourier spatial mode, independently.
However, each Fourier mode in channel $k$ ($(\eta_i,\zeta_j)_k$) is correlated with the same Fourier mode in channel $l$ ($(\eta_i,\zeta_j)_l$).
So, assuming that each spatial Fourier mode $(\eta_i,\zeta_j)$ is one datum, and that we have $N$ channels ($N_{\rm ch}$), the covariance estimation quality factor for one single Fourier mode is given by
\begin{equation}
\label{eq:QualityFactor}
Q \equiv N_{\rm rel} / N_{\rm ch},
\end{equation}
where $N_{\rm rel}$ is the number of realisations of that particular Fourier mode.
However, since for each patch we have one single realisation of the background $N_{\rm rel} = 1 \Rightarrow Q = 1/4$, an extremely low value.
So, a simple estimate will be nothing but noise, as intuition would have told us.
Assuming the process is ergodic and the field is homogeneous, it is possible to replace the ensemble average by a spatial average.
So, we enlarge the patch to 16 times its initial area around the targeted source, (see the ``field'' definition in Sect.~\ref{secsec:flatpatches} and Fig.~\ref{fig:Patches} for more details) and we average each background Fourier mode $(\eta_i,\zeta_j)_k$ over the same mode in different sub-regions (see ``patch'').
Since we now have 16 realisations of each individual Fourier mode, we have boosted $Q$ to approximately 4.\footnote{It would have been exactly 4 if the ``field'' were perfectly homogeneous.} This is already a reasonable estimation quality factor. 
However, the covariance matrix only enters the likelihood via its inverse. Even a small error in the estimation might render the inversion unstable.
We have therefore decomposed the covariance matrix into two components:
\begin{equation}
\label{eq:CoVarDecomp}
\begin{bmatrix}
\sigma_{353}^2 & 0 & 0 & 0\\
0  & \sigma_{545}^2 & 0 & 0\\
0  & 0 & \sigma_{857}^2 & 0\\
0  & 0 & 0 & \sigma_{\rm IRIS}^2
\end{bmatrix}
+ \phi \,
\begin{bmatrix}
0 & kl_{353-545} & kl_{353-857} & kl_{\rm 353-IRIS}\\
.  & 0 & kl_{545-857} & kl_{\rm 545-IRIS}\\
.  & . & 0 & kl_{\rm 857-IRIS}\\
.  & . & . & 0
\end{bmatrix},
\end{equation}
where the matrix on the left is the ``independent'' component, the matrix on the right is the ``systemic'' component and $\phi$ is the ``cross-correlation factor'' ($\phi \in [0,1]$).
The independent component would be the covariance matrix if we neglected all cross-correlations between the same spatial Fourier mode in different channels.
The systemic component is obtained as:
\begin{equation}
\label{eq:klmodes}
\sum^N_{i=1} \gamma_j \vec{V}_j \vec{V}^{\sf T}_j,
\end{equation}
where $\gamma_j$ and $\vec{V}_j$ are the eigenvalues and eigenvectors of the Karhunen–Lo\`{e}ve decomposition of the Fourier mode $(\eta_i,\zeta_j)$ covariance matrix.
Firstly, we start by sorting the KL eigenmodes in decreasing order of the respective eigenvalue.
Then we include up to $N$ ($\in [1,4] $) eigenmodes in the sum of Eq.~\eqref{eq:klmodes}.
$N$, the cut-off, is given by the theory of random covariance matrices \citep[][chapter~9]{RiskAnalys}.
After forming the systemic matrix using Eq.~\eqref{eq:klmodes}, we set the diagonal terms to $0$.

Let us now inspect the two extreme cases of the cross-correlation factor:
\begin{itemize}
	\item $\phi = 0$,\\
	the covariance becomes reduced to the independent only, which is equivalent to ignoring all cross-channel correlations;
	\item $\phi = 1$,\\
	we are including the cross-correlation between channels to its full extent only neglecting the modes that are severely contaminated by noise.
\end{itemize}
The behaviour of the background covariance-matrix estimator may be fine tuned using $\phi$.
A low value of $\phi$ improves the quality factor of the estimation at the cost of ignoring a portion of the signal, namely the inter-channel cross-correlation.
A high value indicates inclusion of complete data information at the cost of lower estimation quality.
As we shall see in Sect.~\ref{psssec:InjectSimulations}, the ``cross-correlation factor'' ($\phi$) has proved to be important in the extraction of an unbiased $\{\beta, T\}$ set.

\subsubsection{Incomplete modelling of the data: systematics}
\label{subsubsec:Systematics}
The determination of the covariance matrix of the cross-power spectra is an approximation, and any potential mis-estimation is not being propagated into the source parameter errors.
However, we know that the dynamic range of source flux density and source-detection S/N is enormous, ranging from close to zero to the thousands.
For most of the catalogue sources, the effect of the covariance-matrix estimation error is masked by the intrinsic uncertainty on the source parameters.
However, for the most significant sources the covariance-matrix estimation error is likely to be the dominant effect, and for these we are missing a critical component of the uncertainty in the source parameters.

The rigorous and complete way of modelling the problem would be to include the uncertainty of the covariance-matrix coefficients as sampling variables in our problem and consider the joint likelihood.
The inverse covariance-matrix coefficients\footnote{It is the inverse covariance that is part of our likelihood.} are distributed according to a Wishart distribution \citep[][chapter~8]{BoxTiao}.
In principle we could add this contribution to the source parameters and sample from the joint likelihood.
However, we are dealing with seven or eight source parameters, and adding the inverse covariance likelihood would increase that number to more than $10\,000$.
Sampling from tens of thousands of parameters would slow down the code to the point where it would no longer be possible to tackle a catalogue with more than $40\,000$ sources.
Therefore we do not implement such a scheme. As a consequence, as we go up in source significance, some of the estimated parameter uncertainties will keep artificially decreasing, whereas in reality they should saturate at some minimum level. This effect particularly concerns the estimates of source location and flux density. In Appendix~\ref{sssec:Simulations} we describe a wide range of simulations on which we have tested the limits of our approximation on several parameters, and suggest ways to correct this shortcoming.

\subsection{Algorithm implementation}
\label{sec:AlgoImpl}
As described earlier, the maps that we use as inputs are \Planck\ 2015 data at 353, 545, and 857\,GHz, plus IRIS data at 3000\,GHz.
Here we describe some details of how we treat these data.

\subsubsection{Masks and map sets}
\begin{figure}
	\begin{center}
		\leavevmode
		\includegraphics[width=0.49\textwidth]{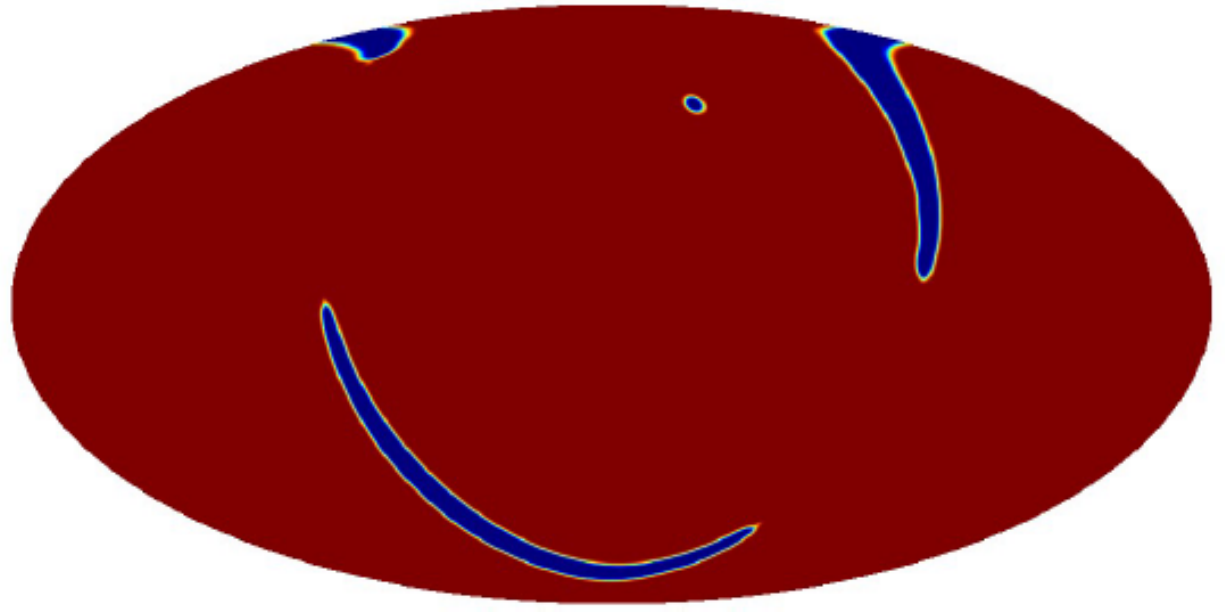}\\
		\includegraphics[width=0.49\textwidth]{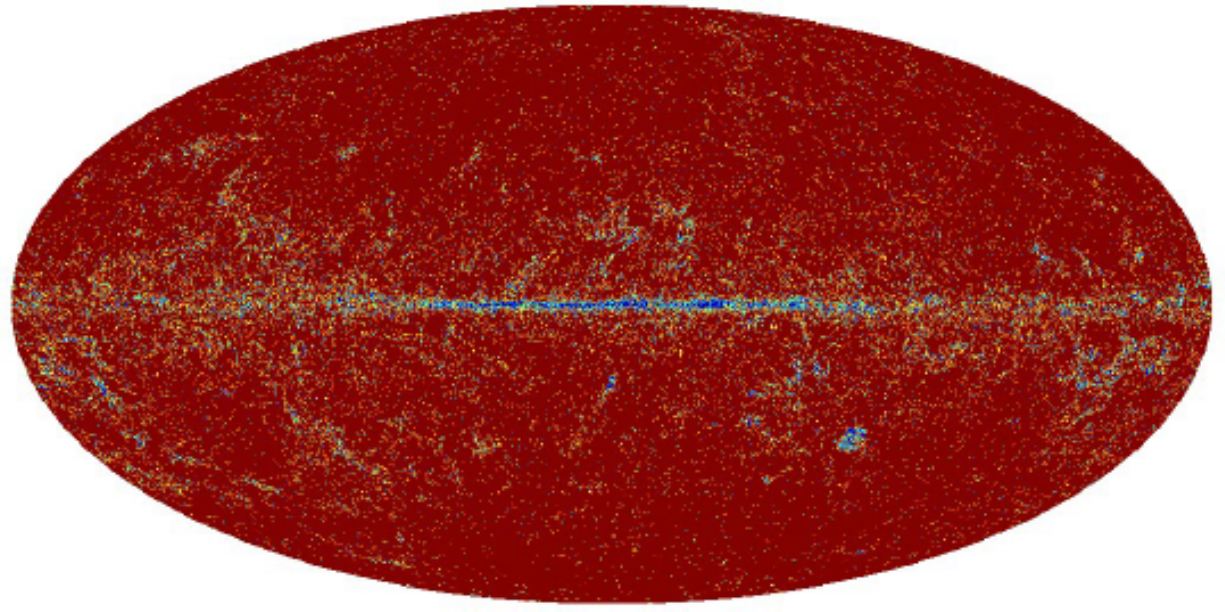}
		\caption{Masks used in our analysis.  The upper panel shows the IRIS mask. The blue regions were not observed by IRAS or contain compact sets of ``bad'' pixels (3.3\,\% of the sky).
		The lower panel is the background mask. The masked regions were later inpainted by diffusing the hole boundary pixels into the interior.}
		\label{fig:Masks}
	\end{center}
\end{figure}

\begin{figure}
	\begin{center}
		\leavevmode
		\includegraphics[width=0.49\textwidth]{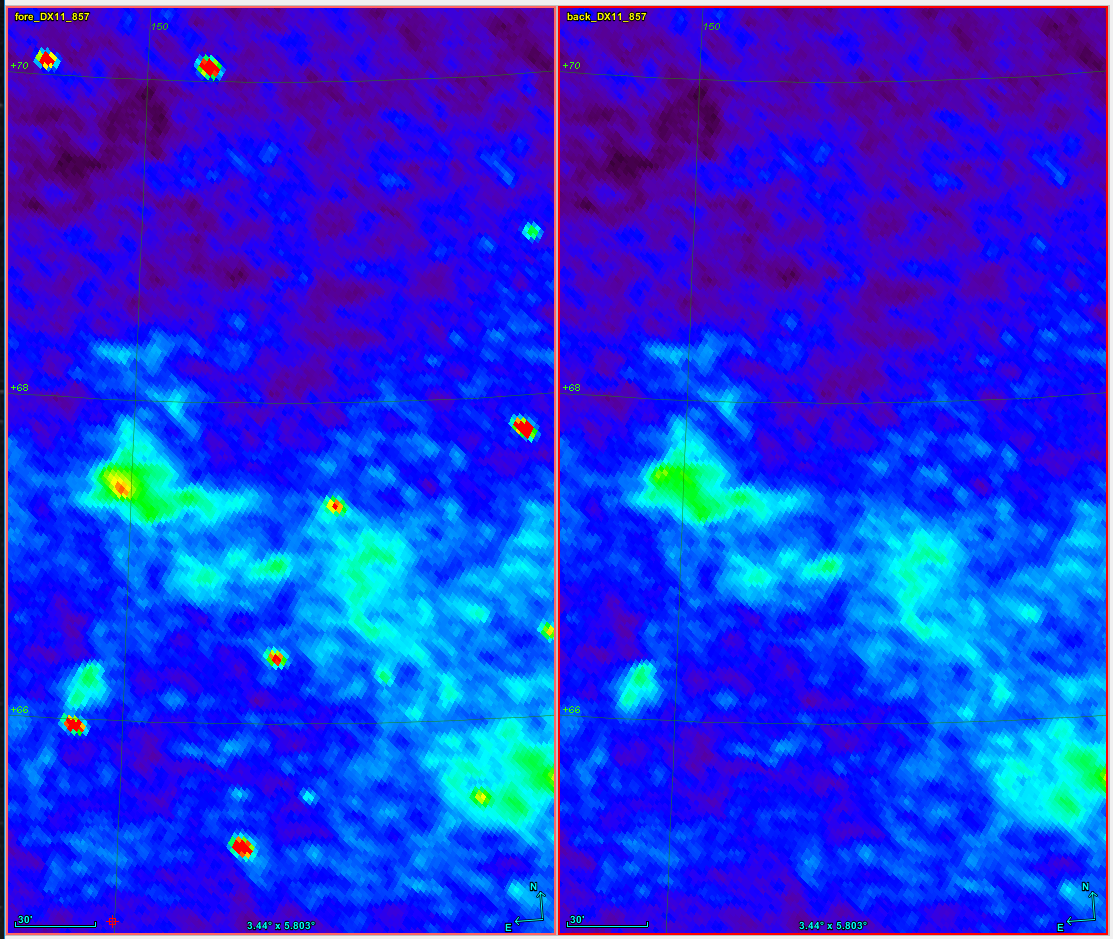}
		\includegraphics[width=0.49\textwidth]{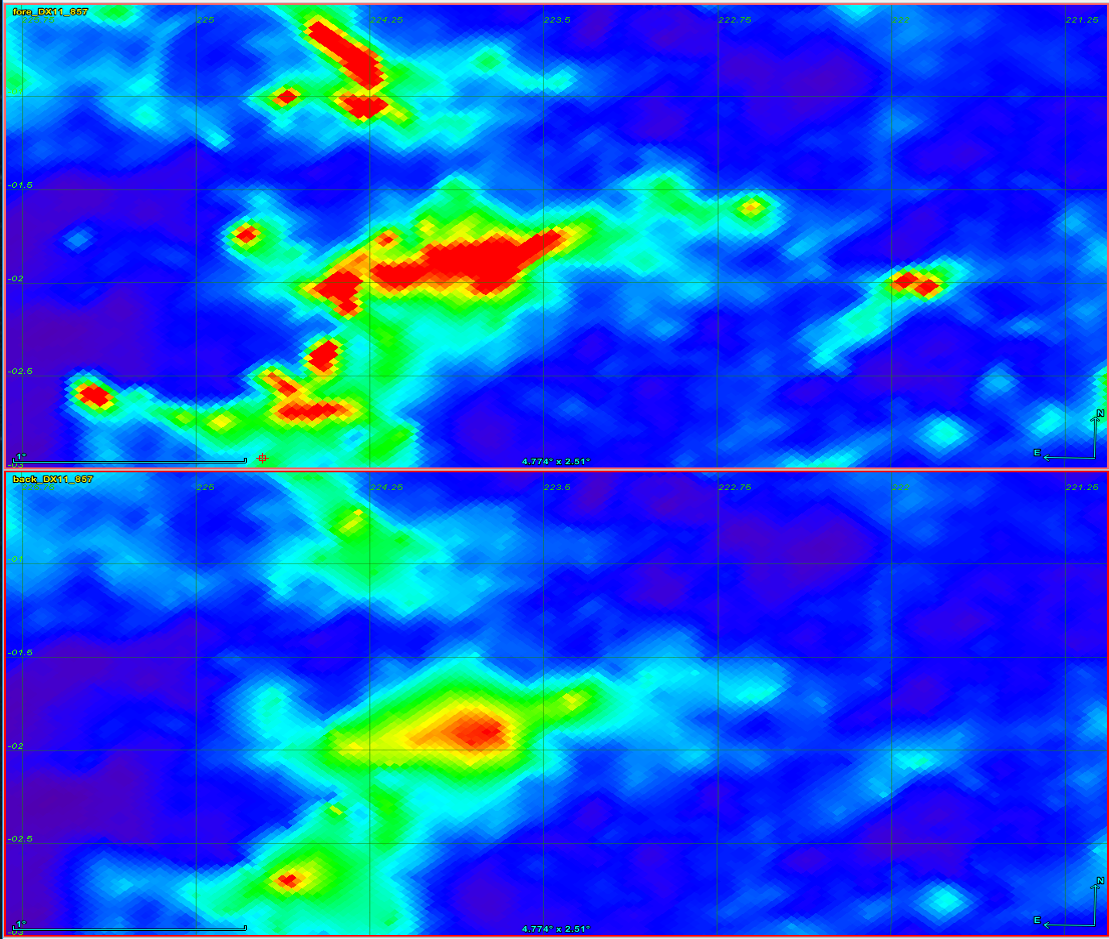}
		\caption{Masking and inpainting effects.  Each of the top panels shows a small ($3\pdeg4\times5\pdeg8$) high Galactic latitude patch cut from \Planck's 857-GHz map. The brightness-colour mapping is the same for both panels.
			The left panel is from the foreground map and the right from the background.
			For this low spatial density field, the ``mask+inpaint'' method recovers the background brightness map very accurately.
			The two lower panels, with foreground above and background below, show a very bright low Galactic latitude region ($4\pdeg8\times2\pdeg5$). In this region of high spatial density of sources the ``mask+inpaint'' process is much less accurate and some degradation of the background can be seen.}
		\label{fig:ForeBackmaps}
	\end{center}
\end{figure}

\textsf{BeeP} creates two types of masks, which are applied to the input maps to generate two types of map sets:

\begin{itemize}
	\item \textbf{IRIS}\\
The ``IRIS'' mask (see Fig.~\ref{fig:Masks}) flags the regions on the IRIS map where there is incomplete data either because those regions were not observed or they contain compact sets of ``ill-conditioned'' pixels.
The total area of this mask is about 3.3\,\% of the full sky.
There are 650 PCCS2+2E sources (1.4\,\%) that are located within the IRIS mask.
It is very difficult to constrain the emission temperatures using \Planck\ data only; therefore objects positioned inside the IRIS mask are flagged in the catalogue as being of lower quality (see Table~\ref{table:QualityPenalties}).
The IRIS mask is applied to the input maps to provide a set of foreground maps (see Fig.~\ref{fig:ForeBackmaps}).
	All likelihood elements (except the background covariance) will be estimated using this data set.
	All injection, and non-injection, simulations only employ the foreground maps data set.\\
	\item \textbf{Background}\\
The sole purpose of the ``background'' mask (see Fig.~\ref{fig:Masks}) is to help in the removal of compact objects in order to create a set of ``background'' only maps.
To construct this mask, firstly we merge all sources contained in the 353--857\,GHz PCCS2+2E catalogues.
We assume that this set of catalogues provides an almost complete sample at the sensitivities we are aiming for.\footnote{The small excess of objects that are not part of these catalogues will then add to the background fluctuation levels and to its non-Gaussianity. This would only make our acceptance statistic even more conservative.}
Then for every source in the merged catalogue we mask all the pixels inside a circle of $7'$ radius.
The $7'$ radius was chosen to provide a good balance between an effective source brightness removal and, especially at low Galactic latitudes where the density of sources is very high, to preserve the statistical properties of the background (see Fig.~\ref{fig:ForeBackmaps}).
The total background area masked are is 4.3\,\%.
The background mask is applied to the input maps to provide a set of background maps (see Fig.~\ref{fig:ForeBackmaps}).
The masked regions are then inpainted.\footnote{\textsf{BeeP} reports the percentage of pixels that were changed by the inpainting routine in a field labelled \textsf{INPIX}.}
The main purpose of this set of maps is the evaluation of the background covariance matrix (cross-power spectrum). 
\end{itemize}

\subsubsection{Projection into flat fields}
\label{secsec:flatpatches}
\begin{figure}
	\begin{center}
		\leavevmode
		\includegraphics[width=0.49\textwidth]{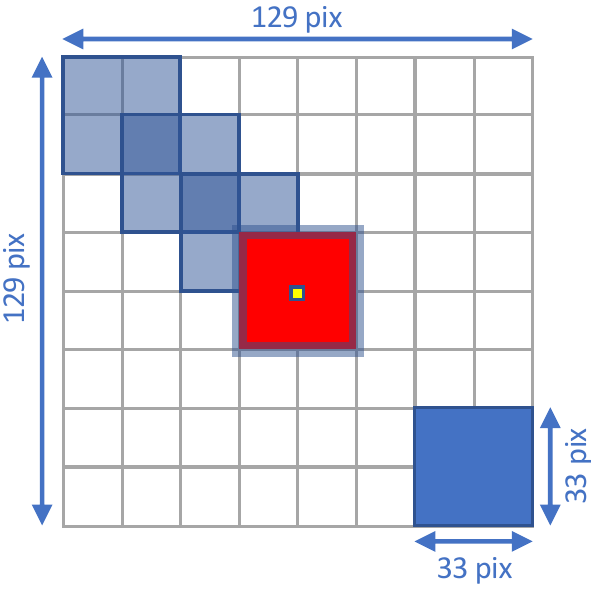}
		\caption{Schematic (not drawn to scale) showing parts of a flat ``field'' ($129 \times 129$ pixels). The covariance matrix is computed at each of the large squares, or ``patches'' ($33 \times 33$ pixels), from the ``background'' map, and then averaged over them. There are $49$ overlapping patches ($7 \times 7$) in each field. These are laid out as shown in the figure.	
		The full likelihood is only evaluated at the interior of the central patch (in red with the PCCS2+2E position at its centre, in yellow). The \textsf{RELTH} statistic is then estimated using the pixels of the red region, leaving a border of four pixels.
		The field/patch $Y$ and $X$ directions, at the centre of the field, match the Galactic coordinate lines of constant latitude and longitude, respectively.
	 	Each individual pixel (not drawn) is $\approx 1\parcm72 \times 1\parcm72$.}
		\label{fig:Patches}
	\end{center}
\end{figure}
For each source in the PCCS2+2E, a flat area of size $3\pdeg69 \times 3\pdeg69$, ($129 \times 129$) pixels, centred on the source position and obtained using a gnomonic projection, is cut from each \Planck\ (353--857\,GHz) and IRIS map. We repeat the procedure for the background map set and the IRIS mask (see Fig.~\ref{fig:Patches}). We call these projected square maps ``fields.''
Each individual pixel in each field is uniformly over-sampled by a factor of 25 to minimize resampling artefacts that could result from the overlap between the map and field grids.\footnote{This is equivalent to a field/patch pixel grid with a resolution 25 times greater than that of \Planck\ HFI, which was later downgraded back to \Planck's original map resolution.} Each sample is computed by bi-linearly interpolating the map pixels.
The combination of the oversampling and the interpolation operations also smooths the map brightness. That effect is accounted for by adding a pixel-window correction to the effective beams.

The field is then divided into $49$ $(7 \times 7)$ overlapping ``patches'' of $33 \times 33$ pixels, as shown in Fig.~\ref{fig:Patches}.\footnote{This procedure is a simple 2-d extension of the quite common equivalent method in time series. If a time series is ergodic and stationary, the ensemble average may be replaced by a time/space average.}
The cross-power spectrum is computed in each of the patches and then averaged over all patches in a given field (see Sect.~\ref{subsubsec:CovarianceMatrixEstimation}).
The IRIS mask is used to down-weight individual cross-spectrum Fourier modes according to the number of pixels removed by the mask.
Since there is overlap between patches, the quality ratio improvement (Eq.~\ref{eq:QualityFactor}) is limited to about a fctor of 16.
Finally, the likelihood/posterior is computed only using the central patch (Fig.~\ref{fig:Patches} in red), which is centred on the PCCS2+2E target object's original position (in yellow).

\subsubsection{Running the likelihood}
\label{ssec:AlgoImplSrcDetec}
For our likelihood runs we have set the ``cross-correlation factor'' (see Sect.~\ref{subsubsec:CovarianceMatrixEstimation}) $\phi$ to $10\,\%$.
This value was selected using simulations (see Sect.~\ref{psssec:InjectSimulations}) to minimize the bias between the object's injected and recovered parameters.
The distribution of the thermal parameters ($T, \beta$) is particularly sensitive to the value of $\phi$.
A high $\phi$ value (${>}\,50\,\%$) generates significant positive bias in $\beta$ and negative bias in $T$, while a low value (${<}\,10\,\%$) has the opposite effect.
The value we have selected (10\,\%) leads to the lowest global bias in the main recovered source parameters, $\beta, T$, and flux density (see Appendix~\ref{subsubsec:CovarianceMatrixEstimation}).

We further assume that the \Planck\ and IRIS background maps are uncorrelated, because the introduction of IRIS results in instabilities in the estimation of the covariance matrix, particularly in regions with a very bright background or with visible artefacts in IRIS.

Having fixed the cross-correlation parameters, we can proceed to run \textsf{BeeP}'s likelihood. We first try to find the posterior maximum inside a square of at most $7 \times 7$ pixels centred on the original PCCS2+2E position.\footnote{As described in Sect.~\ref{sec:MethSourceModel}, we allow the optimal source location to vary from the original PCCS2+2E location by up to 3 pixels.  However, some source positions may end up at a distance of slightly more than 3 pixels---this happens because when the maximum reaches the $3$ pixel boundary, we allow the sampler to explore the region around the boundary.}
It is often the case that the maximum of the posterior does not match the central patch pixel or that we cannot even find a maximum (see e.g., Fig.~\ref{fig:IntroMovement}).
If there is more than one likelihood maximum inside the search region, we always prefer the one closest to the original PCCS2+2E coordinates.
It is useful to note that if a posterior maximum is not found, there is no guarantee that the derived parameter estimates follow the statistical properties predicted in Sect.~\ref{sec:CatalogueCharacteristics}.
Whether a maximum is found or not is reported in the catalogue field \textsf{MAXFOUND}.

The source extension parameter (\textsf{EXT}) 
poses further difficulties to an unbiased recovery of the object parameters (in this case its size).
In the current implementation of \textsf{BeeP}, we have fixed the beam size at each frequency to an average value for the entire catalogue. 
According to this data model, the narrowest feature in the maps must at least have the width of the beam at that channel.
However, in the real maps narrower compact objects may be present.\footnote{This might happen because in some regions of the sky the real beam size is narrower than the average value, or because background or noise fluctuations may cause a beam-sized object in the map to be artificially narrowed.}
These cases create regions of the likelihood manifold with a high concentration of probability (they contain the likelihood peak) that cannot be explored because our source model does not consider ``negative'' radii. As a result, strong deviations in the recovered parameters for these sources can be expected.
To tackle this problem we take advantage of degeneracy between the source and the beam size: a pixel brightness pattern can be the result of a narrow source and a large beam or of the reverse situation.
Our solution consists of implementing simulated beams that are narrower than the average of the real beam, i.e., their FWHM is selected such that a source with an estimated size of $\textsf{EXT} = 0.975$ pixel ($\approx 1\parcm72$) will result in an object on the map that has the same extension as the average (real) beam (see Table~\ref{table:BeamSize}).
This trick is actually quite important for recovering a flux-density-unbiased sample: as we can see in Fig.~\ref{fig:CornerPlot}, there is a positive correlation between the source extension $\textsf{EXT}$ and the flux density $\textsf{SREF}$, which implies that a bias in the estimate of $\textsf{EXT}$ will propagate into $\textsf{SREF}$. 
Narrowing the beam artificially removes most of this bias.
However, in those regions where a feature in the map is narrower than the beam size, the ``source size'' recovered by \textsf{BeeP} (\textsf{EXT} $\lesssim 1\parcm72$) is poorly determined, since it is degenerate with the beam width. In the \textsf{BeeP} catalogue, we report both \textsf{EXT} and a more realistic source size under the field label \textsf{R}.\footnote{For all cases where \textsf{EXT}$\,{<}\,1\parcm72$, we set \textsf{R}$\,{=}\,0.0$. }

\begin{table}[htbp!]
\begingroup
\newdimen\tblskip \tblskip=5pt
\caption{Relation between the instrument beam FWHM and those used in \textsf{BeeP}'s likelihood.}
\label{table:BeamSize}
\nointerlineskip
\vskip -3mm
\footnotesize
\setbox\tablebox=\vbox{
   \newdimen\digitwidth
   \setbox0=\hbox{\rm 0}
   \digitwidth=\wd0
   \catcode`*=\active
   \def*{\kern\digitwidth}
   \newdimen\signwidth
   \setbox0=\hbox{+}
   \signwidth=\wd0
   \catcode`!=\active
   \def!{\kern\signwidth}
\halign{\tabskip 0pt\hbox to 1.0in{#\leaderfil}\hfil\tabskip 0.5em&
         \hfil#\hfil\tabskip 1em&
         \hfil#\hfil\tabskip 0pt\cr
\noalign{\doubleline}
\omit\hfil Channel\hfil&Instrument FWHM& Likelihood FWHM\cr
\noalign{\vskip 5pt\hrule\vskip 5pt}
3000\,GHz & $4\parcm3*$& $1\parcm72$\cr
*857\,GHz & $4\parcm64$& $2\parcm44$\cr
*545\,GHz & $4\parcm83$& $2\parcm79$\cr
*353\,GHz & $4\parcm94$& $2\parcm98$\cr
\noalign{\vskip 5pt\hrule\vskip 4pt}}}
\endPlancktable                    
\endgroup
\end{table}

Once \textsf{BeeP} has found the parameters that maximize the likelihood, the field described in Eq.~\eqref{eq:FilteredField} is generated and the \textsf{RELTH} quantile is evaluated over the red area in Fig.~\ref{fig:Patches}.
We leave a $4$ pixel-wide border to avoid fake likelihood maxima resulting from edge effects,\footnote{``Edge'' effects are the result, in the Fourier transform, of the discontinuities at the borders of the patch in which we are computing the likelihood. The Fourier transform requires that the data must be periodic.} which would artificially increase the background non-Gaussianity.

As described in Sect.~\ref{ssec:ThermalProperties}, we implement two different ``SED models,'' and each of these requires a separate run of \textsf{BeeP}.
For the secondary method (which does not impose an SED correlating frequencies), we assume that the backgrounds are independent and set the ``cross-correlation factor'' to $0$. We use \textsf{BeeP} to estimate the flux density and its uncertainty in each channel independently.
Then we fit an MBB curve to the individual channel flux densities using a Gaussian likelihood with variances estimated in the previous step.

\subsubsection{Sampling the posterior: multi-modality}
\label{ssec:AlgoImplPostMultiModal}
\begin{figure*}
	\begin{center}
		\leavevmode
		\includegraphics[width=0.7\textwidth]{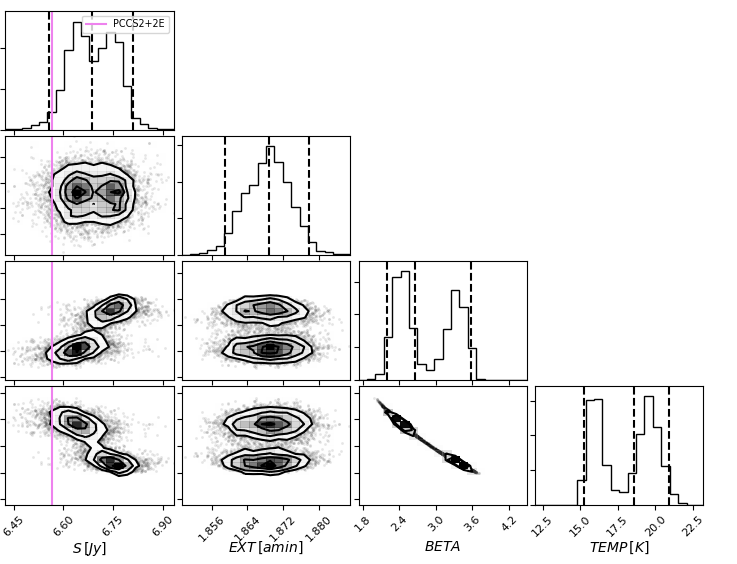}
		\caption{Multi-parameter fits for a particular source.  This fragment of a ``corner'' plot (see Fig.~\ref{fig:CornerPlot}) shows the multi-modal character of some of the posterior distributions. 
		}
		\label{fig:LikeMultiModal}
	\end{center}
\end{figure*}

Unfortunately, owing to the very complex backgrounds, especially close to localized features like the cirrus, some posterior distributions are multi-modal (see Fig.~\ref{fig:LikeMultiModal})
Although there are specialized samplers that can handle multi-modal distributions like \citep[that described in][]{HobsonMCMC2007}, they are far too slow for this problem.
It is recommended by the authors of \texttt{MCMC Hammer} that the efficiency of the sampler can be increased by starting it in the neighbourhood of the posterior mode.
We have opted for running the likelihood maximizer multiple ($10$) times, with initial points scattered across the prior volume,
and then split the initial samples (``walkers'') across the different maxima.
Using this strategy we have almost completely removed the chance of the sampler systematically missing significant parts of the likelihood manifold.\footnote{In the first runs of the algorithm, we were finding, in the $T$--$\beta$ plane, an unexpected high Galactic latitude source population around 20\,K. 
This ``anomalous'' population simply vanished when we adapted the code to account for multi-modal likelihoods. At the same time we also saw an increase in the parameters uncertainty, which supports the fact that we are now exploring a wider likelihood manifold.}
However there are still a few cases where this solution is not effective.
Some of the outliers (see Eq.~\ref{eq:QualityCriterion}), in particular those with a tiny extension (smaller than the beam) or vanishingly small error bars, are the result of the \textsf{MCMC} chains being attracted to strong and very narrow maxima. After being caught inside these narrow local maxima, the chains are {\it not\/} able to explore the entirety of likelihood manifold, and cannot properly account for the correct parameter uncertainty, or find the global maximum.
Another problem is the ``chain correlation length'' and dependence on the sampler ``initial conditions.''
The first problem can be solved by periodically throwing away samples.
However, one of the reason why we have chosen \texttt{MCMC Hammer} is its small correlation length.
We always monitor the correlation length of the chain and when it is higher than the required level we reset the sampler and restart all chains again.
The samples acceptance rate always remains very close to the optimal range of 20--50\,\% \citep{MCMC_Hammer}, except for very rare occasions when it could be as low as $6\,\%$, but never higher than $62\,\%$.\footnote{Only 151 \textsf{BeeP/base} catalogue sources ($0.6\,\%$) have acceptance rates below $20\,\%$.}
The sample acceptance rate is reported in the \textsf{ACCEPT} field of the output catalogue. 
Perhaps surprisingly, after the first runs of the sampler (with very simple examples), we realised that the quality of the generated samples was very dependent on the sampler initial state.
To overcome this difficulty we massively increased the ``burn in'' phase and the problem was solved.
We are currently using 5000 ``walkers,'' 98 ``burn in'' iterations and we only keep the two final ones, generating 10\,000 posterior samples.

\subsubsection{Colour correction}
\label{ssec:AlgoImplColorCorr}
One important advantage of using a multi-channel estimation algorithm is that the effect of the detector finite band-passes (or ``colour-correction''), can be included in the estimation chain.
Although the colour correction is a relatively small adjustment,\footnote{Colour-correction coefficients are of the order a few percent ($\la 10\,\%$). For extreme values of $T$ (${\ga}\,30$K) and $\beta$ (${\ga}\,3.0$) they can reach values in excess of ${\ga}\,20\,\%$, but only for the 353- and 545-GHz channels.} it can introduce a bias in the MBB $T$--$\beta$ estimates if not properly accounted for.
We created a 2-d colour-correction matrix with one axis assigned to ``$\beta$'' and the other ``$T$'' based on the code described in \cite{planck2013-p03d} for each \Planck\ channel at 353--857\,GHz, as well as IRIS. 
The actual correction coefficient is then obtained using bilinear interpolation.

\section{Simulation-based tests}
\label{sssec:Simulations}
In this appendix we describe the different simulations used in this paper.

\subsection{Synthetic background}
\label{psssec:ArtificialSimulations}
This type of simulation tries to recreate a data set that follows our data assumptions as closely as possible.
It is meant to verify the algorithm and code correctness under ideal circumstances.
The outcome serves as a yardstick to assess the robustness of the code/algorithm as we move to more realistic cases where some of these assumptions need to be relaxed.

The diffuse background in these simulations is intended to be as close to a homogeneous Gaussian random process as possible, but with realistic \Planck\ levels and characteristics.
The simulations were generated from one \Planck\ 2015 CMB simulation and four different noise realizations, {taken from the Planck Legacy Archive}. 
The CMB$+$noise maps were scaled in amplitude to match the median level found in each of the four real maps (\Planck\ 2015 345, 545, 857\,GHz, and IRIS 3000\,GHz). This process ensured that the maps have signal amplitudes similar to those found in the real maps, but their statistical properties are Gaussian.
Then we cut the spherical maps into many small patches and we injected a source directly into the centre of each patch. 
All injected sources were simulated to be equally shaped, following a bi-dimensional symmetrical Gaussian profile with constant and very small radius, and then convolved with the PSF at each frequency, which was assumed to be constant and equal to \Planck's average effective beam.
The sources were rendered in very high resolution and projected directly into the patch pixels.
The source SEDs were derived from an MBB law with three free parameters: $T$, temperature; $\beta$, spectral index; and $S_{857}$, flux density at 857\,GHz.
The values for the source SED parameters were then randomly drawn from a preliminary catalogue that had been extracted with \textsf{BeeP} from the real maps.
That precursor catalogue was cross-matched with the PCCS2+2E and the GCC.
The parameter estimates showed a high degree of consistency with both catalogues and were thereafter assumed as representative of the actual sky distribution.
The goal of this type of simulation is to closely replicate the assumptions of our data model, and therefore constitute our ``fiducial case.''

\begin{figure}[htbp!]
	\begin{center}
		\leavevmode
		\includegraphics[width=0.49\textwidth]{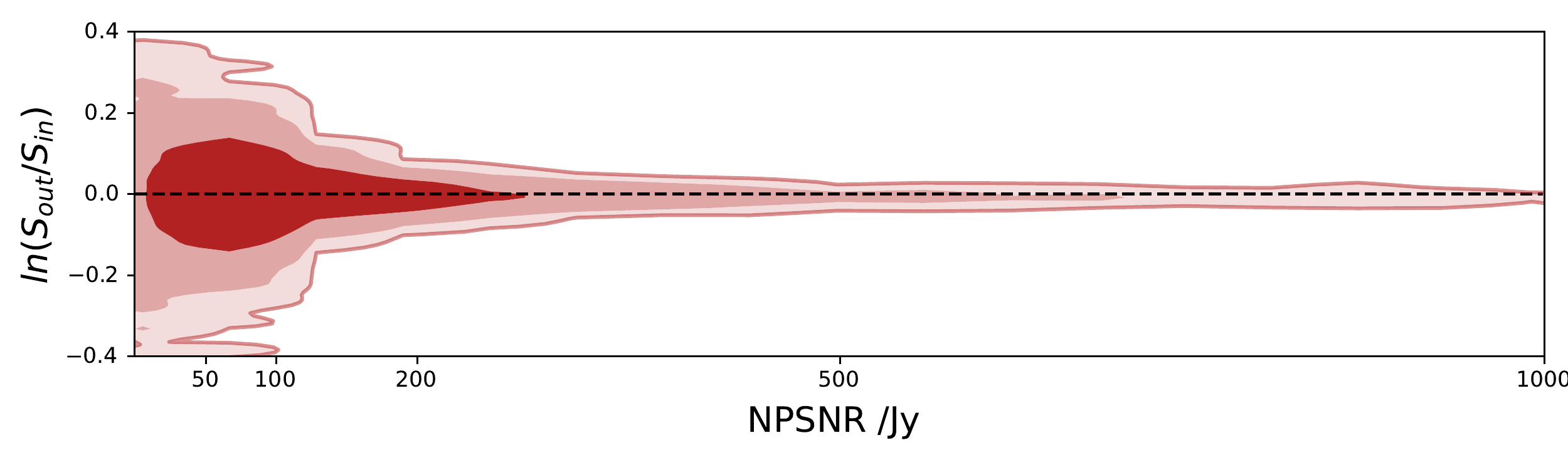}\\
		\includegraphics[width=0.49\textwidth]{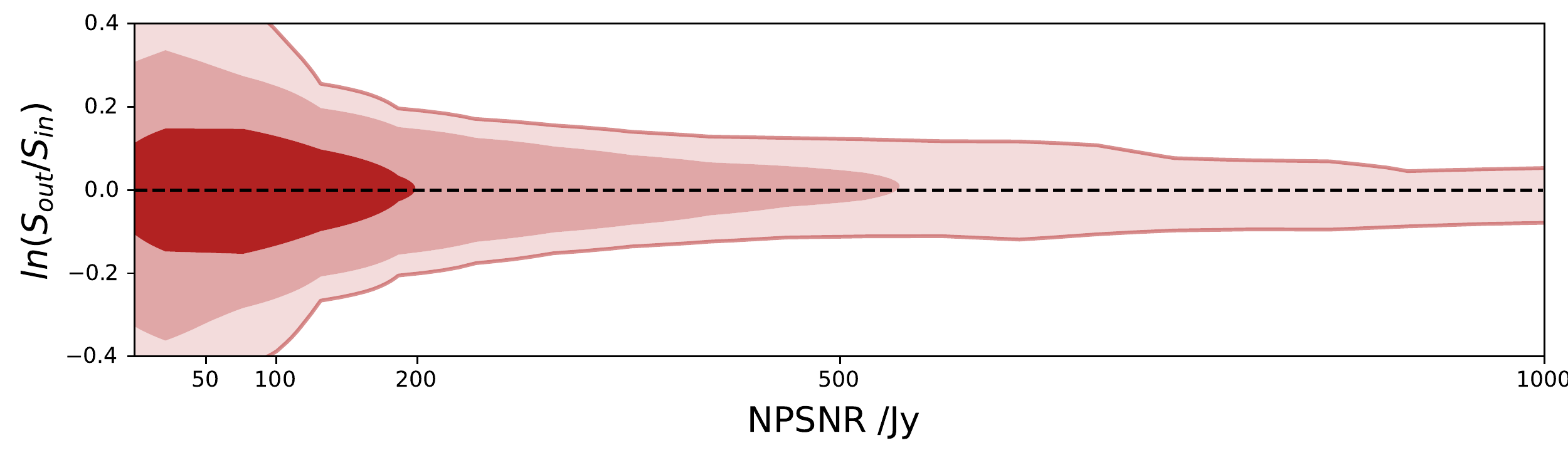}\\
		\includegraphics[width=0.49\textwidth]{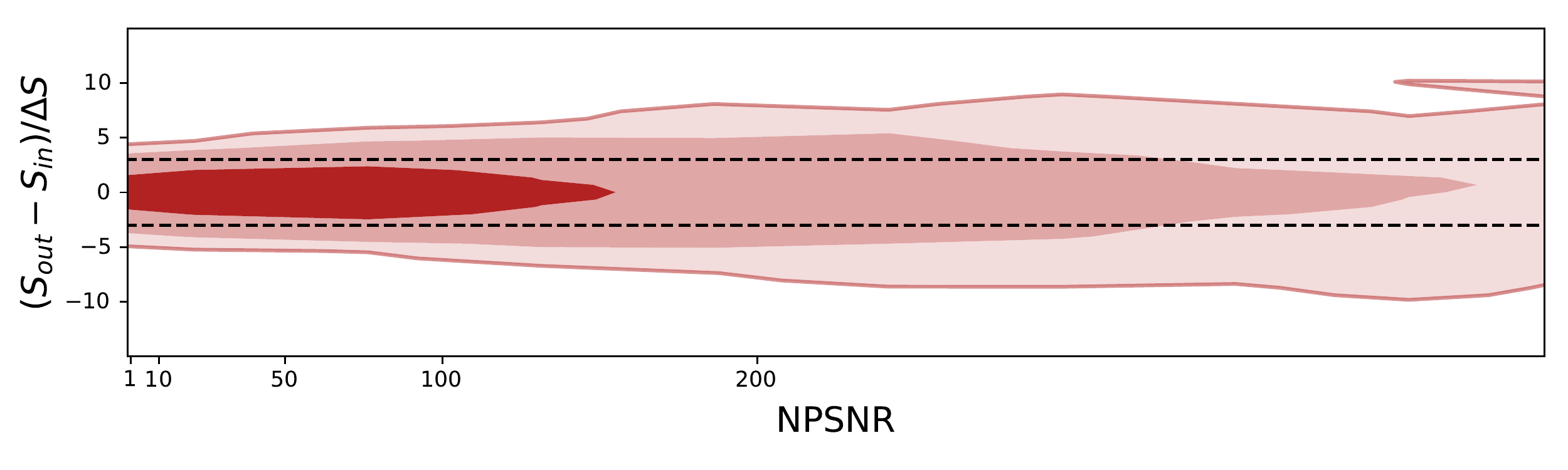}
		\caption{Comparison of input and output flux densities.
			The upper and middle panels show $\ln(S_{\rm out}/S_{\rm in})$  versus \textsf{NPSNR} distribution contours ([$68,95,99$]\%).
			In the upper panel the field and patch are $513\times513$ and $129\times129$ pixels, respectively, and in the middle $129\times129$ and $33\times33$.
			For both cases a source was directly injected in the central pixel of the patch, but always with a small random shift from the pixel centre.
			The lower panel shows the ``normalized error'' $=$ ($S_{\rm out} - S_{\rm in}$) / $\Delta S$ distribution contours for the small patches.
			The horizontal lines in the bottom panel, are the $\pm3\,\sigma$ boundaries.
		}
		\label{fig:PatchDirectLLSS}
	\end{center}
\end{figure}

In Fig.~\ref{fig:PatchDirectLLSS} we make a comparison between the reference flux density of the injected sources with those we retrieved using \textsf{BeeP}.
Owing to the huge dynamic range of the values (five orders of magnitude), computing the distribution of the fractional difference gives a better understanding of their consistency than just the difference.
As explained in Sect.~6.2.4, instead of plotting the fractional difference formula directly ($({\rm out}-{\rm in})/{\rm in}$), we replace it with $\ln({\rm out}/{\rm in})$.
Figure~\ref{fig:PatchDirectLLSS} depicts two flux retrieval cases, with two different patch sizes, to gauge its impact on the recovery precision (see Fig.~\ref{subsubsec:Systematics}).
The top panel shows the comparison when the ``fields'' cut from the homogeneous background sphere were $513\times513$ pixels and the ``patches'' (core region where the likelihood is evaluated) $129\times129$ pixels.
The middle panel of Fig.~\ref{fig:PatchDirectLLSS} shows exactly the same thing, but the dimensions of the fields were this time $129\times129$ and $33\times33$ pixels.
The retrieved flux distributions are similar and both show that after a certain \textsf{NPSNR} threshold the precision of the estimates saturates.
However, the top panel of Fig.~\ref{fig:PatchDirectLLSS} (larger field and patch) shows much less dispersion, especially as we move towards higher \textsf{NPSNR} values, and it reaches saturation much later.
This should not have come as a surprise. As was mentioned in Sect.~\ref{subsubsec:Systematics}, in the high S/N regime, the uncertainty in the parameters recovery is limited by the estimation accuracy of the covariance matrix.
The larger the data set, the more precise the estimation is, the less dispersion the estimates show, the later the onset of the flux accuracy saturation.
However, when tackling the ``real world,'' the smaller the patches the better the background homogeneity assumption actually holds.
From our simulation exercises, fields of $129\times129$ and patches of $33\times33$ pixels seem to provide the best balance between background homogeneity and enough data (i.e., pixels) to guarantee that the error in the statistics we collect do not dominate (for the majority of cases). 
However, when assuming that the covariance matrix had no estimation error, we failed to propagate into the likelihood that extra source of uncertainty arising from the field/patch statistics.
That will necessarily lead to an underestimation of the error bars.
For the low \textsf{NPSNR} regime this is not a problem because the covariance-matrix estimation error is still sub-dominant;
however, at the high end where it completely dominates, the error bars are underestimated (Fig.~\ref{fig:PatchDirectLLSS} lower panel).
\begin{figure}[htbp!]
	\begin{center}
		\leavevmode
		\includegraphics[width=0.49\textwidth]{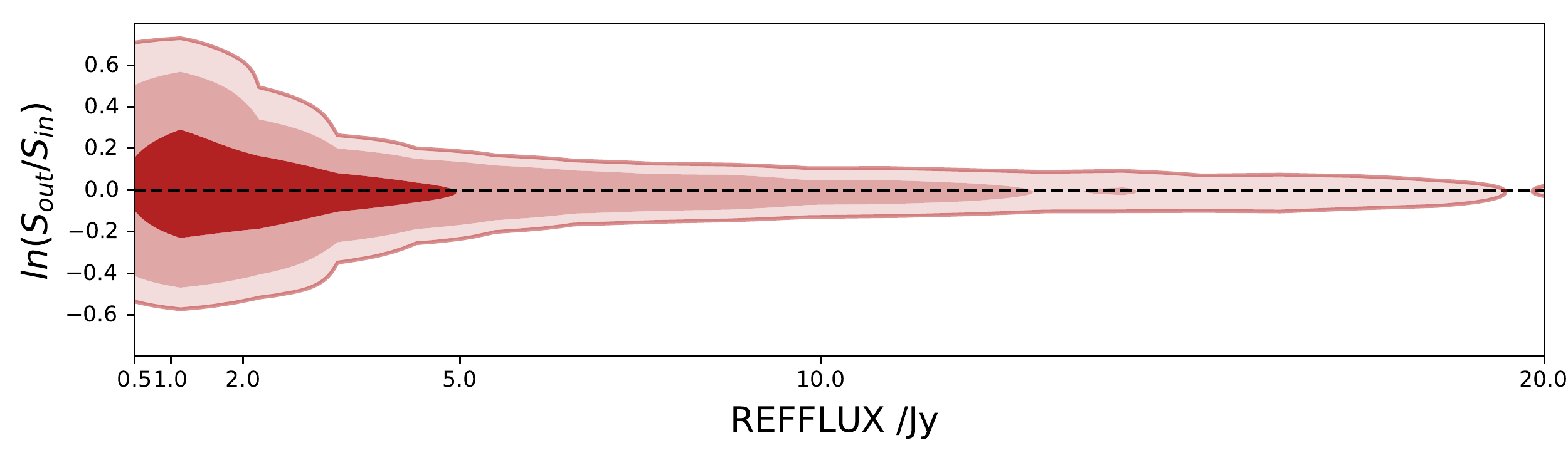}
		\caption{Recovery error versus source brightness. This specifically shows the $\ln(S_{\rm out}/S_{\rm in})$ versus $S_{\rm in}$ distribution contours ([$68,95,99$]\%).
			The sources were injected into the CMB+noise only \healpix\ maps following the same process as when injecting into the real data. The same source was injected multiple times to assess the impact of the different background conditions. 
		}
		\label{fig:cmbNoiseFlux}
	\end{center}
\end{figure}
Since in practical terms it is impossible to propagate the covariance-matrix uncertainty into the likelihood (see Sect.~\ref{subsubsec:Systematics}), we have chosen to keep the likelihood as it is, but later correct the error bars for the unaccounted uncertainty.

Another source of systematic errors could stem from the projection of the compact objects onto the flat fields.
To study the effect of a potential projection distortion, we also injected the sources directly into \Planck's CMB + noise simulated maps at \Planck's HFI native resolution (\healpix\ $N_{\rm side}=2048$).
A minimum distance ($12'$) between injected sources was imposed in order to avoid beam blending effects.
Figure~\ref{fig:cmbNoiseFlux} shows once again the quantity $\ln(S_{\rm out}/S_{\rm in})$ but this time versus $S_{\rm in}$.
We injected the same source multiple times to learn how the different background conditions would affect the extraction.
The background is homogeneous, but its dominant component (the scaled CMB) has a typical correlation length of around $1^{\circ}$, which is much larger than the typical extension of a source, around $10'$.
This implies that some of the injected sources sit on top of (scaled CMB) crests and others on valleys.
\subsection{Injecting simulated sources into real maps}
\label{psssec:InjectSimulations}
{ 
The sky distribution of the PCCS2+2E, our input catalogue, is extremely inhomogeneous.
The PCCS2 contains less than 10\,\% of the sources and covers about half of the sky.
The PCCS2E ($\ga 90\,\%$) sources are located almost entirely within regions of strong, complex background emission.
\cite{planck2014-a35} indicates that a realistic rendering of the cirrus, including localized embedded features, is absolutely crucial for validating any catalogue at the frequencies considered here.
Unfortunately, simulating the submillimetre sky is a formidable task \citep{PlanckSkyModel}, and realistic simulations of such a complex background are not yet available.
Since \textsf{BeeP}'s model of the background is a statistical one, it is critical that the statistical properties of the simulated background match those of the real data.
To achieve this match, we used the actual \Planck\ 2015 maps and injected fake sources directly into them.
This approach is similar to the one previously employed in the production of \Planck's Early Release Compact Source Catalogue \citep{planck2011-1.10}.
The physical parameters of the mock sources correspond to those of the original sources, as extracted from a preliminary run of \textsf{BeeP} on the 2015 maps.

Given the complexity of the Galactic background, the mock sources should ideally be injected exactly on top of the PCCS2+2E catalogue positions.
However, this is only possible if the real source is first removed in such a way that the background where it is to be embedded is left undisturbed; otherwise residuals of the removed source could systematically bias the extraction results.
To try to accomplish this, we mask the pixels around each real source and then inpaint them by diffusing the background into the masked region, starting from its boundaries; the inpainting method is described in \cite{InPaint} and preserves the statistical properties of the field surrounding the inpainted area.
The radius of the inpainting mask is $7'$, and we impose the condition that the minimum distance between any two injected sources should never be smaller than $12'$.
As a consequence, some source positions in the PCCS2+2E do not have any source injected, a situation that happens more frequently at low Galactic latitudes.

We recognize the possible bias of injecting sources into a modified sky. Therefore, to validate the inpainting procedure we also generate a second set of simulations in which each mock source is injected not at the original PCCS2+2E location but in its near neighbourhood.
For this set of injections, we place a mock source within an annulus around the original position, within a radius of $12'$ and outside a radius of $20'$;\footnote{For this case we implement a wider range of distances between the original and the injected source than in the ``no-sources'' simulation, to avoid the risk of systematically creating pairs of identical sources, but retaining a similar background as that of the original source.}
we also ensure that the injected source does not blend with any other source
previously injected or in the PCCS2+2E. This mechanism guarantees that no source is ever injected in an inpainted area. Because of this restriction, in regions of a very high source density, such as the Galactic plane, there may be some PCCS2+2E source locations that are not associated with any injected source.
We note that for these simulations we must also inpaint the original source location, otherwise it would \textit{systematically} increase the non-Gaussianity of the background patch under analysis.
Sources of equivalent flux densities would \textit{always} appear in pairs, making the original PCCS2+2E source systematically increase the background non-Gaussianity of the injected source background.
The annulus, however, lies well outside the inpainted region
and guarantees (given the equivalent beam width) that the condition of Eq.~\eqref{eq:WellSeparetedObjects} always applies and that any background disturbance, such as another source or an inpainted hole, does not perturb the parameter estimation.
As may be seen below, both types of simulations produce statistically similar sets of results (see Table~\ref{table:TempBetaStats}).

For the sake of completeness, we also add a third set of simulations in which the injected mock sources are uniformly distributed on the sky; as in the previous simulations we make sure that sources do not overlap
with any other source in the PCCS2+2E.
For this case, we draw the mock source parameters at random from the PCCS2 sub-catalogue rather than from the full PCCS2+2E (otherwise, we would create an unrealistically bright high-Galactic-latitude population that would systematically increase the catalogue source significance.

The three types of simulations (just described), were employed to calibrate the ``cross-correlation factor'' ($\phi$) (see Sect.~\ref{subsubsec:CovarianceMatrixEstimation}).
We applied \textsf{BeeP} using a set of $\phi$ values, \{0.0, 0.05, 0.10, 0.15, 0.20, 0.30, 0.50\}.  We then looked for bias by comparing recovered to injected source properties.
The simulations showed that the value of $\phi$ used has a strong impact on the recovery of the MBB parameters ($T, \beta$), and must be chosen with care.
What we learned from this exercise\footnote{When $\phi \geq 0.20$ all three sets displayed negative bias in $T$ and a positive in $\beta$.
For $\phi = 0$ all three cases showed exactly the opposite trend.
The value of $\phi$ that minimized the bias for the ``uniform'' injection policy was 0.15.
For the ``in place'' and ``neighbourhood'' policies the best $\phi$ was 0.05.
However, if we selected only the PCCS2 subset for these two cases, the optimal value of $\phi$ would return to 0.15.
}
 is that faint background regions prefer higher ($\phi\approx 0.15$) values and complex, bright regions lower values ($\phi \approx 0.05$).
So we selected $\phi=0.10$ as a balance between the two cases.
In any event, the bias in $T$ and $\beta$ that was observed when $\phi$ was not optimal for a particular subset was always small ($\la 5\,\%$).

With $\phi$ set, we now examine the accuracy of the recovery by \textsf{BeeP} of the physical source parameters in the simulations.
For this purpose, we define now the ``normalized symmetric error'' ($\Delta$) of a parameter as 
\begin{equation}
\Delta = \frac{4 (\theta_{\rm out}-\theta_{\rm in})}{\epsilon_{{\rm h}2\sigma}-\epsilon_{{\rm l}2\sigma}},
\label{eq:NormSymError}
\end{equation}
where $\theta_{\rm in}$ is the injected parameter value, $\theta_{\rm out}$ is \textsf{BeeP}'s best estimate and $\epsilon_{{\rm h}2\sigma}$, $\epsilon_{{\rm l}2\sigma}$ are the $97.5\,\%$ and $2.5\,\%$ values of the distribution, respectively.
In the case of an ideal Gaussian distribution of errors, the variable $\Delta$ should follow a unitary normal distribution.
However, because of the strong inhomogeneity and non-Gaussianity of the background, especially at low Galactic latitudes, it is expected that sub-optimal parameters will be found for certain objects, i.e.,
that in some parameter ranges significant outliers will arise in the distribution of errors.
As one of our goals, perhaps the most important, is to have a well defined set of statistical traits for the catalogue estimates, these extreme outliers need to be identified and removed to avoid biasing or distorting the characterization.

The set of simulations previously described was used to identify such cases,
which appear in the results as sources with either unphysical parameters or vanishingly small parameter uncertainties.
Both manifestations are signs of poor or insufficient posterior sampling, raising the possibility that entire ranges of feasible parameter values are not being sampled by the likelihood exploration tool (for instance when the likelihood is multi-modal).
When this happens, one cannot be sure that the optimal estimate set was obtained and the error bars are certainly underestimated.\footnote{The catalogue also contains a field \textsf{ACCEPT} with the sampler acceptance rate. Very low (${<}\,10\,\%$) or very high (${>}\,80\,\%$) values are signs of a sub-optimal likelihood exploration. For further details see Sect.~\ref{ssec:AlgoImplPostMultiModal}.}
We are now able to define exclusion regions in parameter space. We have found that the vast majority of sources whose estimated parameters do \textit{not} meet the following criterion should be flagged:
\begin{align}
&\textsf{EXT} > 1.46 \wedge \textsf{TEMP} < 60 \wedge \textsf{BETA} < 5
 \nonumber\\
&\quad \wedge (\textsf{TH2SB} - \textsf{TL2SB}) >0.8 \nonumber\\
&\quad \wedge (\textsf{BETAH2SB}-\textsf{BETAL2SB})> 0.25.
\label{eq:QualityCriterionApp}
\end{align}
Such sources are severe outliers in one or more parameters,\footnote{$\Delta_{\beta}$ and $\Delta_{T}$ at least $>7$.} and therefore should be discarded. This is the ``outliers criterion.''\footnote{In the \textsf{BeeP/base} catalogue, only 999 of the sources that pass the ``reliability criterion'' (Eq.~\ref{eq:QualityCriterionApp}; about 2\,\% of the entire PCCS2+2E) are rejected by the ``outliers rejection'' condition.}
In Fig.~\ref{fig:SimulsBetaTemp} we show the distribution of $\Delta$ (Eq.~\ref{eq:NormSymError}) for the MBB parameters \textsf{TEMP} and \textsf{BETA}, as well as for \textsf{SREF}.
In this figure we show the ``same-location'' simulations on the left and the ``neighbourhood'' simulations on the right.
We did not include in this assessment a small fraction of simulated objects that behaved anomalously for other (understood) reasons.\footnote{We omit in particular those whose recovered positions moved by more than $0\parcm8$ (about half a pixel), since the estimates at these new locations cannot be directly compared to those of the injected objects; in fact this is only a tiny fraction of the simulated catalogues (50 sources in the ``same-location'' simulations).
In addition, we have noted that some source locations, especially around extended objects, do not coincide with the actual centre of the object; the removal and inpainting process are not effective in these cases (374 sources).
Finally, close to the Galactic plane where the PCCS2E is hardly complete, in some instances \textsf{BeeP} prefers the location of a nearby object that was not in the PCCS2E. 
Very rarely, this also occurs when the injected source is extremely faint.}

\begin{figure*}[htbp!]
	\begin{center}
		\leavevmode
		\includegraphics[width=0.49\textwidth]{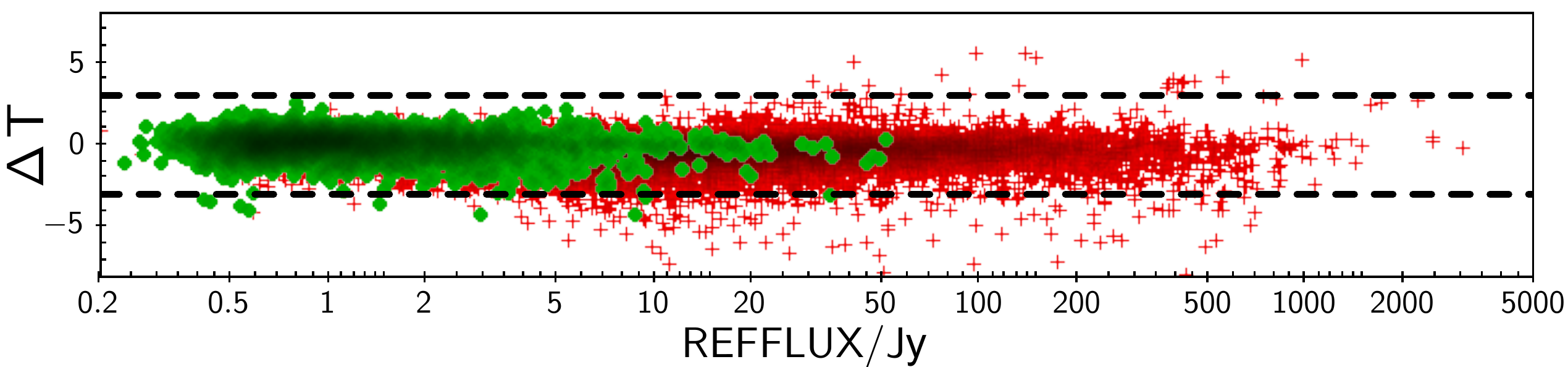}
		\includegraphics[width=0.49\textwidth]{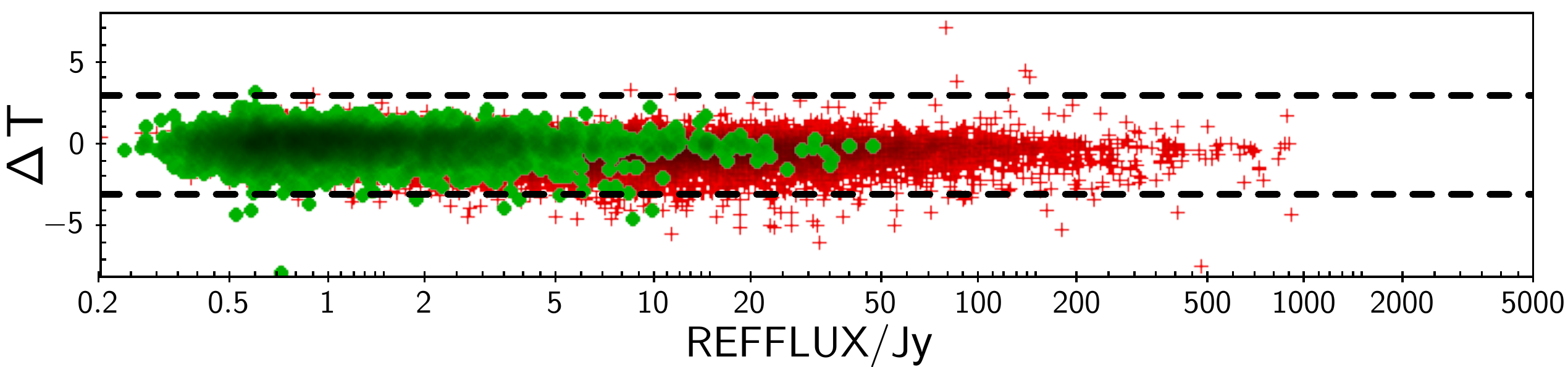}\\
		\includegraphics[width=0.49\textwidth]{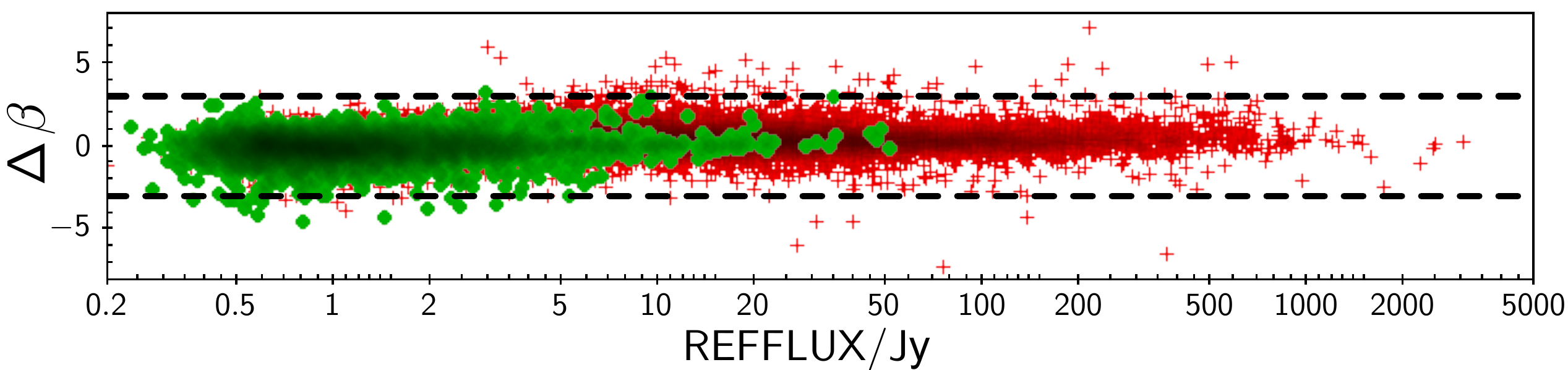}
		\includegraphics[width=0.49\textwidth]{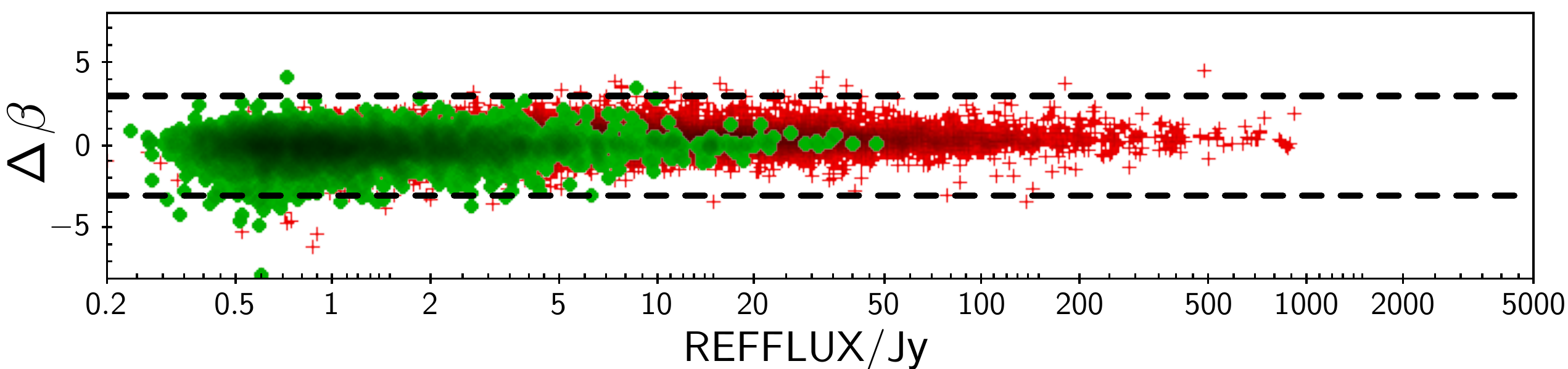}
		\includegraphics[width=0.49\textwidth]{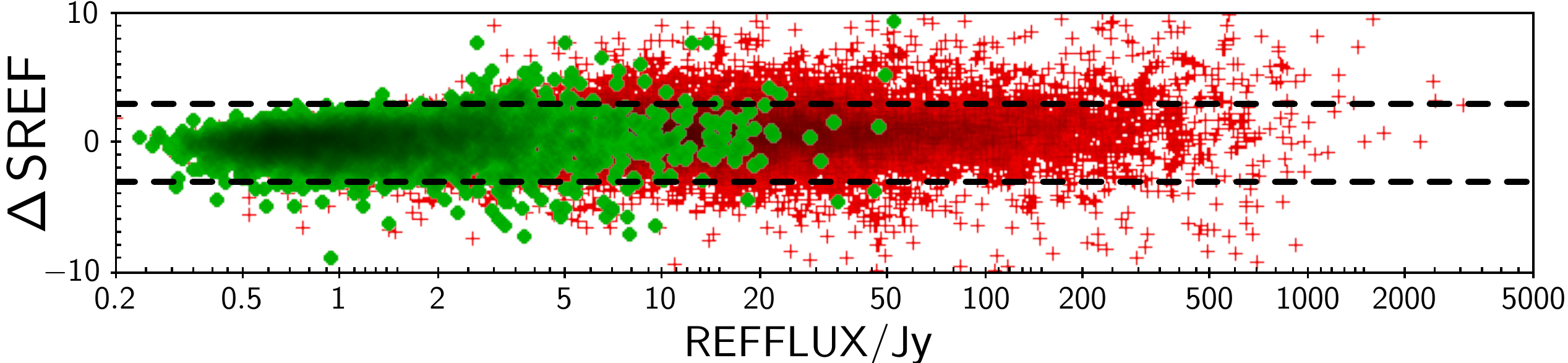}
		\includegraphics[width=0.49\textwidth]{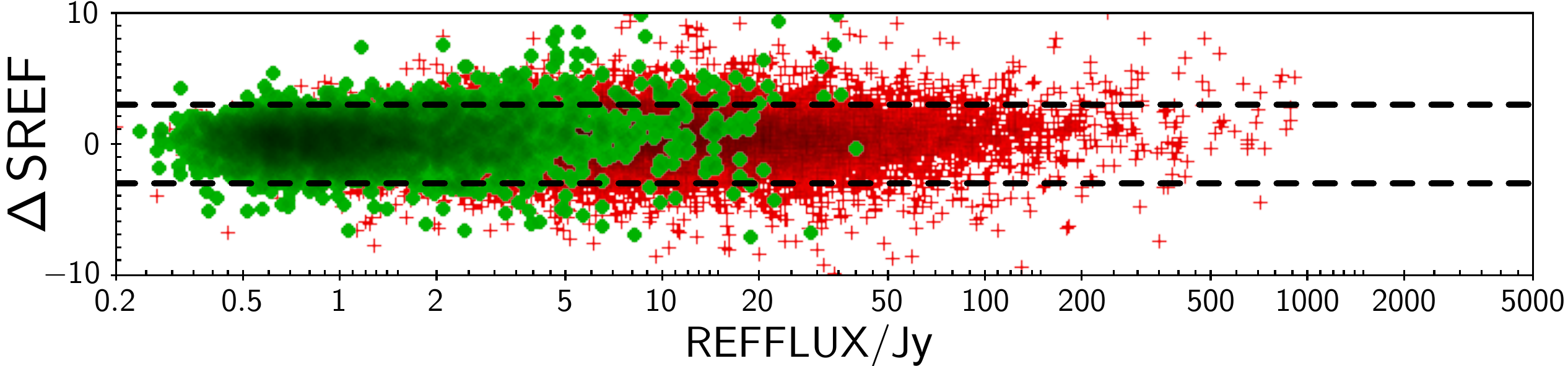}
		\caption{
Errors in simulated source properties.
The $y$-axis shows the normalized symmetric error as defined in Eq.~\eqref{eq:NormSymError} ($\Delta$) and
the $x$-axis the injected source flux density (\textsf{REFFLUX}).
The data set used here is the intersection of the reliability and outliers criteria.
The red crosses show all sources, while the green circles are just the PCCS2.
The horizontal dashed lines are at $y = \{-3 , +3\}$.
The panels on the left were drawn when the simulated sources were injected in the same positions as the PCCS2+2E and those on the right for sources injected in their neighbourhood.
Those sources whose recovered position moved by more than $0\parcm8$ (about half a pixel) from the injection location were removed from the set.
The total number filtered out was about $0.02\,\%$ (50 sources) for the case of injection at the same location; for injection in the neighbourhood, the total number filtered out rose to 1.63\,\% (374 sources).
The close similarity between the two sets of plots shows that the effect of inpainting is small. Table~\ref{table:TempBetaStats} contains a statistical analysis summary of these results.
		}
		\label{fig:SimulsBetaTemp}
	\end{center}
\end{figure*}

\begin{table}[htbp!]
\begingroup
\newdimen\tblskip \tblskip=5pt
\caption{Statistics of the MBB $\{T, \beta\}$ and \textsf{SREF} recovered parameters. The data sets were filtered with the intersection of the ``reliability'' and ``outliers rejection'' criteria.}
\label{table:TempBetaStats}
\nointerlineskip
\vskip -3mm
\footnotesize
\setbox\tablebox=\vbox{
   \newdimen\digitwidth
   \setbox0=\hbox{\rm 0}
   \digitwidth=\wd0
   \catcode`*=\active
   \def*{\kern\digitwidth}
   \newdimen\signwidth
   \setbox0=\hbox{+}
   \signwidth=\wd0
   \catcode`!=\active
   \def!{\kern\signwidth}
\halign{\tabskip 0pt#\hfil\tabskip 1.5em&
         \hfil#\hfil\tabskip 1em&
         \hfil#\hfil\tabskip 1em&
         \hfil#\hfil\tabskip 1em&
         \hfil#\hfil\tabskip 1em&
         \hfil#\hfil\tabskip 0pt\cr
\noalign{\doubleline}
\omit\hfil Data set\hfil& Parameter& $\left\langle\Delta\right\rangle$& Median& $\sigma_{\Delta}$& $\text{SMAD}^{\rm a}$\cr
 & & &$\ln(\theta_{\rm out}/\theta_{\rm in})$& &\cr
\noalign{\vskip 5pt\hrule\vskip 5pt}
 &\multispan5\hfil Injection in neighbourhood\hfil\cr
\noalign{\vskip -5pt}
 &\multispan5\hrulefill\cr
 & \textsf{TEMP}&       $-0.01$& $!1.5$\,\%& 0.74& 0.64\cr
PCCS2&\textsf{BETA}&    $-0.19$& $-6.4$\,\%& 0.83& 0.68\cr
 &\textsf{SREF}&        $!0.30$& $!4.0$\,\%& 1.63& 1.36\cr
\noalign{\vskip -5pt}
 &\multispan5\hrulefill\cr
 & \textsf{TEMP}&       $-0.33$& $-3.2$\,\%& 0.77& 0.68\cr
PCSS2+2E&\textsf{BETA}& $!0.13$& $!3.7$\,\%& 0.74& 0.62\cr
 &\textsf{SREF}&        $!0.35$& $!6.3$\,\%& 1.72& 1.38\cr
\noalign{\vskip 5pt\hrule\vskip 5pt}
 &\multispan5\hfil Injection at same location\hfil\cr
\noalign{\vskip -5pt}
 &\multispan5\hrulefill\cr
 & \textsf{TEMP}&       $-0.03$& $!0.8$\,\%& 0.69& 0.59\cr
PCCS2&\textsf{BETA}&    $-0.15$& $-5.4$\,\%& 0.75& 0.61\cr
 &\textsf{SREF}&        $-0.04$& $!0.2$\,\%& 1.35& 1.00\cr
\noalign{\vskip -5pt}
 &\multispan5\hrulefill\cr
 & \textsf{TEMP}&       $-0.38$& $-3.3$\,\%& 0.80& 0.60\cr
PCCS2+2E&\textsf{BETA}& $!0.16$& $!3.2$\,\%& 0.70& 0.55\cr
 &\textsf{SREF}&        $!0.40$& $!6.5$\,\%& 1.66& 1.12\cr
\noalign{\vskip 5pt\hrule\vskip 4pt}}}
\endPlancktable                    
\tablenote {{\rm a}} Scaled median absolute deviation.\par
\endgroup
\end{table}

Table~\ref{table:TempBetaStats} shows a statistical summary of the offsets in parameters shown in Fig.~\ref{fig:SimulsBetaTemp}.
To reduce sensitivity to the presence of outliers, we have replaced the usual ``average'' and ``standard deviation'' with the more robust ``median'' and ``scaled median absolute deviation'' (\text{SMAD}).

For both types of simulation, Table~\ref{table:TempBetaStats} shows that \textsf{BeeP} recovers $T$ and $\beta$ in a largely unbiased manner.
In addition, Fig. ~\ref{fig:SimulsBetaTemp} shows that the uncertainties in $T$ and $\beta$ are only slightly overestimated. 
They also show a significant correlation (see Fig.~\ref{fig:SimulsBetaTempCorr}).
This is not unexpected: inspecting the posteriors for individual sources (see e.g., Fig.~\ref{fig:CornerPlot}) we see a strong banana-shaped degeneracy between these two parameters.
The flux-density recovery statistics depict a slightly different situation.
We note that in these simulations the injected sources are circularly symmetric and beam shaped, in accord with the data model of \textsf{BeeP}.
Even in this benign situation, inspection of Fig.~\ref{fig:SimulsBetaTemp} and Table~\ref{table:TempBetaStats} indicates that the dispersion of the flux-density estimates is larger than expected.
We must conclude that \textsf{BeeP} underestimates the uncertainty of the recovered values of \textsf{SREF} .
On the other hand, for the ``same location'' simulations, those that should best reproduce the real extraction conditions, in the PCCS2 region, the SMAD statistic shows values equal to or below 1, even for \textsf{SREF}.
When extending to the full PCCS2+2E only a small excess appears.
Based on this we could conclude that, unless a rigorous statistical characterization of the estimates is necessary, the uncertainty values as given in the catalogue, are fit for the purpose.

\begin{figure}[htbp!]
	\begin{center}
		\leavevmode
		\includegraphics[width=0.24\textwidth]{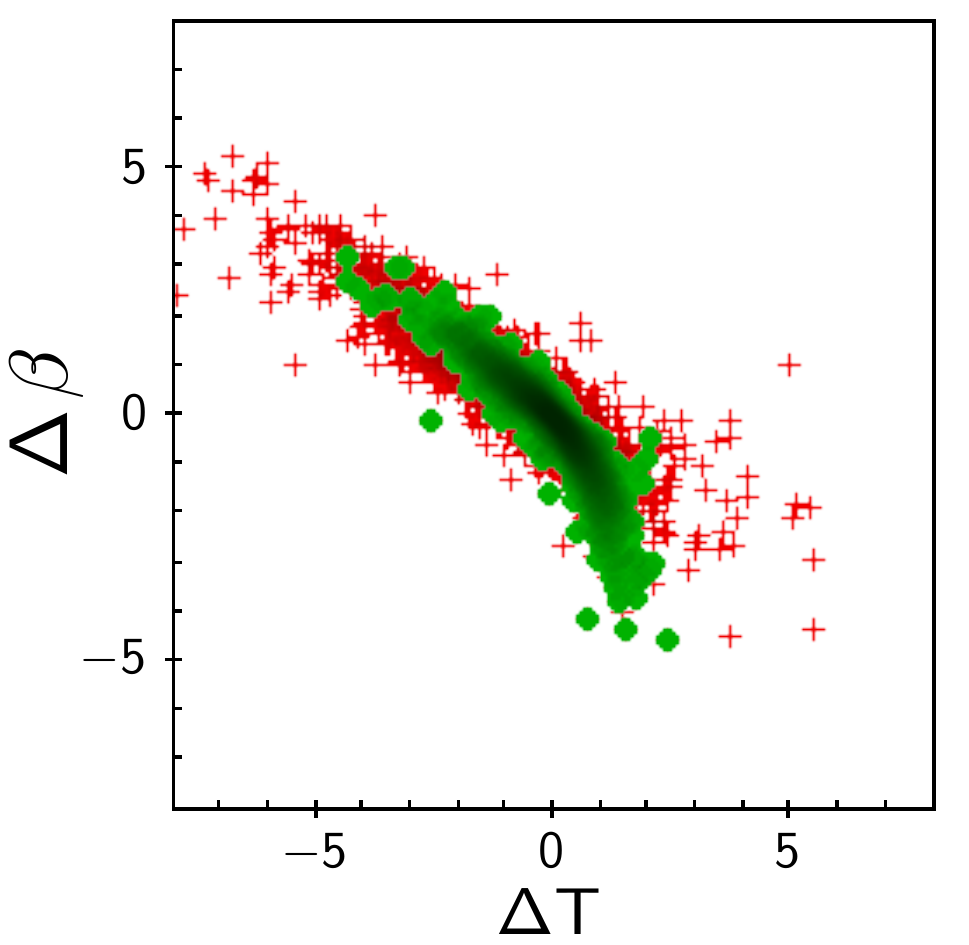}
		\includegraphics[width=0.24\textwidth]{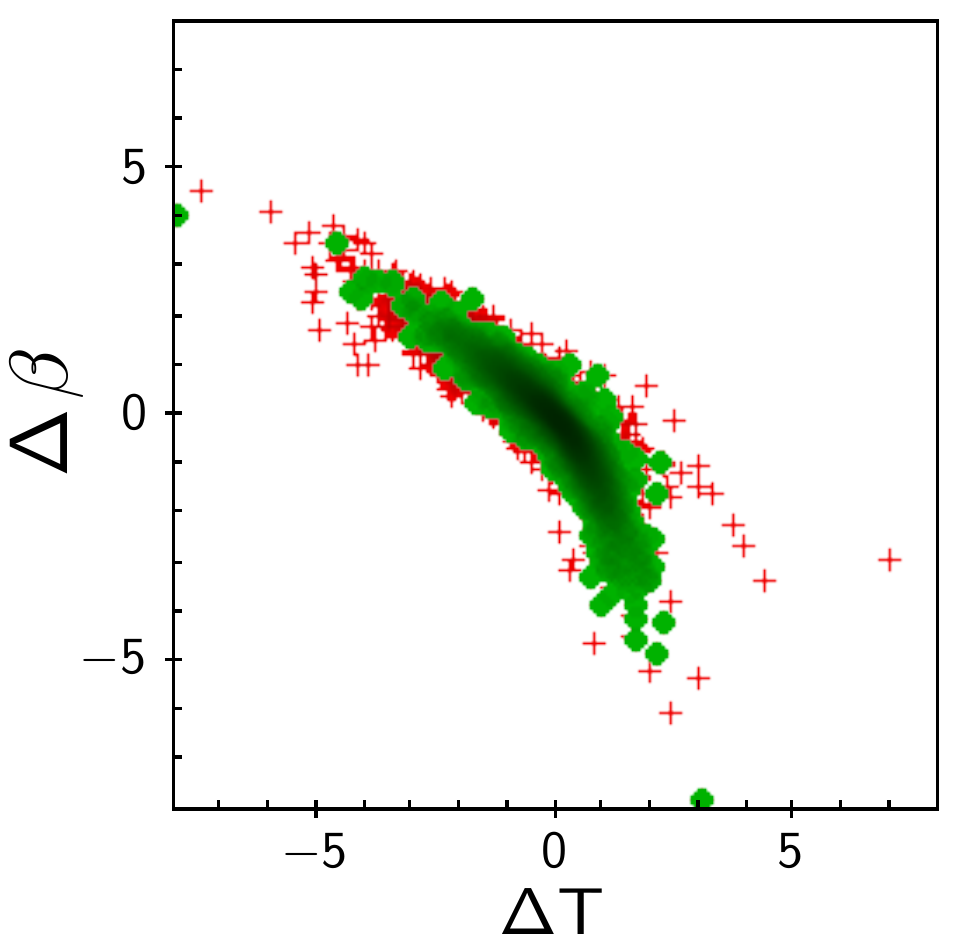}
		\caption{
			Correlation between $\Delta \beta$ and $\Delta T$.
			This uses the same data set and symbols as in Fig.~\ref{fig:SimulsBetaTemp}.
			The left panel shows the correlation for the ``same position'' injection criterion and the right panel for the ``neighbourhood'' criterion.
		}
		\label{fig:SimulsBetaTempCorr}
	\end{center}
\end{figure}

At the same time, we know that these simulations are not realistic enough to provide a proper assessment of the retrieval of the flux density. 
Therefore, we postpone the discussion of flux-density recovery bias to the next sections (see \ref{subsec:IntroducingSourceExtension} and \ref{psssec:FFP8Simulations}).
\begin{figure}[htbp!]
	\begin{center}
		\leavevmode
		\includegraphics[width=0.235\textwidth]{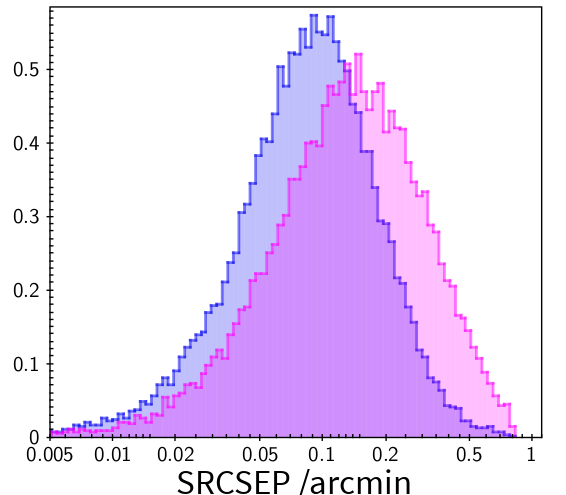}
		\includegraphics[width=0.235\textwidth]{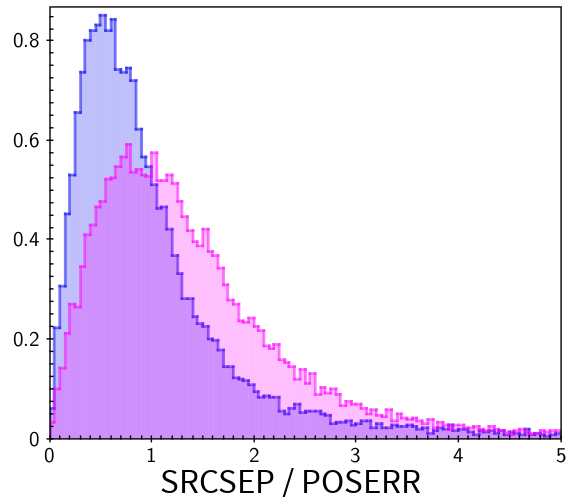}
		\caption{
			Histograms of the separation between the injected and estimated source positions (\textsf{SRCSEP}).
			The left panel shows the absolute deviation and the right one the deviation normalized by the position error bar (\textsf{POSERR}).
			The results of the ``same place'' simulation are in blue and the ``neighbourhood'' ones in pink.
		}
		\label{fig:SrcsepHist}
	\end{center}
\end{figure}

These simulations also allow us to examine the quality of recovery of the source locations. Figure~\ref{fig:SrcsepHist} shows histograms of the separation between the injected and estimated source position (\textsf{SRCSEP}).
When injecting at the same position as the PCCS2+2E (in blue) we find a small bias (around 0\parcm1--0\parcm2). 
If now we normalise \textsf{SRCSEP} with \textsf{POSERR} (right panel), we find that the ``same place'' simulation overerestimates the position error (since we expect the normalized distribution to peak at 1).
On the other hand, the histogram of the normalized separation for the ``neighbourhood'' shows the expected statistical behaviour.\footnote{The normalized position deviation should follow a unitary Rayleigh distribution.}
We believe that this might be the result of 
the inpainting of the original source.
It shows the usefulness of the ``neighbourhood'' simulation, and 
since its results are closer to expectation than the ``same place'' simulation, we use it in what follows.

	\begin{figure*}[htbp!]
		\begin{center}
			\leavevmode
			\includegraphics[width=0.49\textwidth]{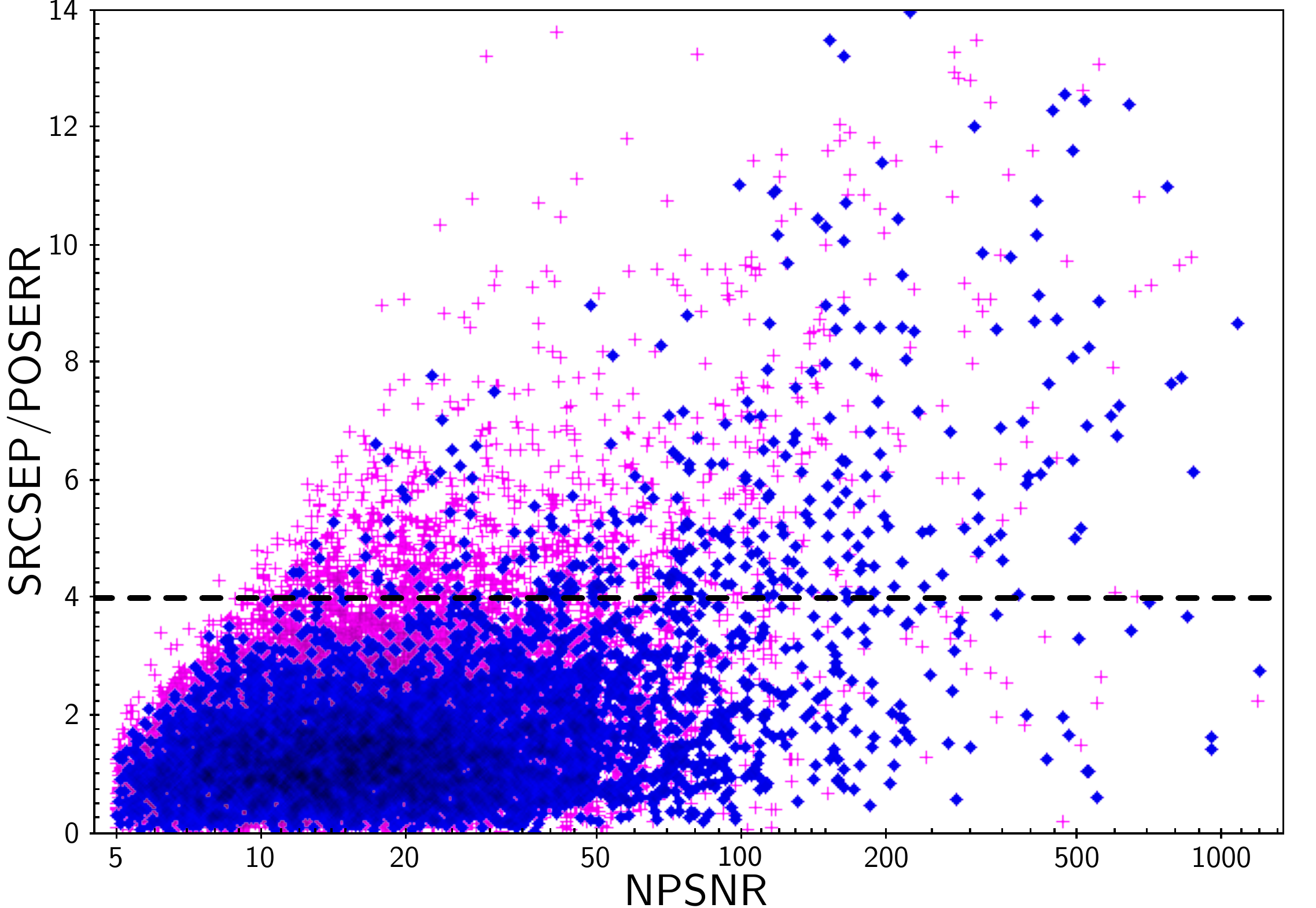}
			\includegraphics[width=0.49\textwidth]{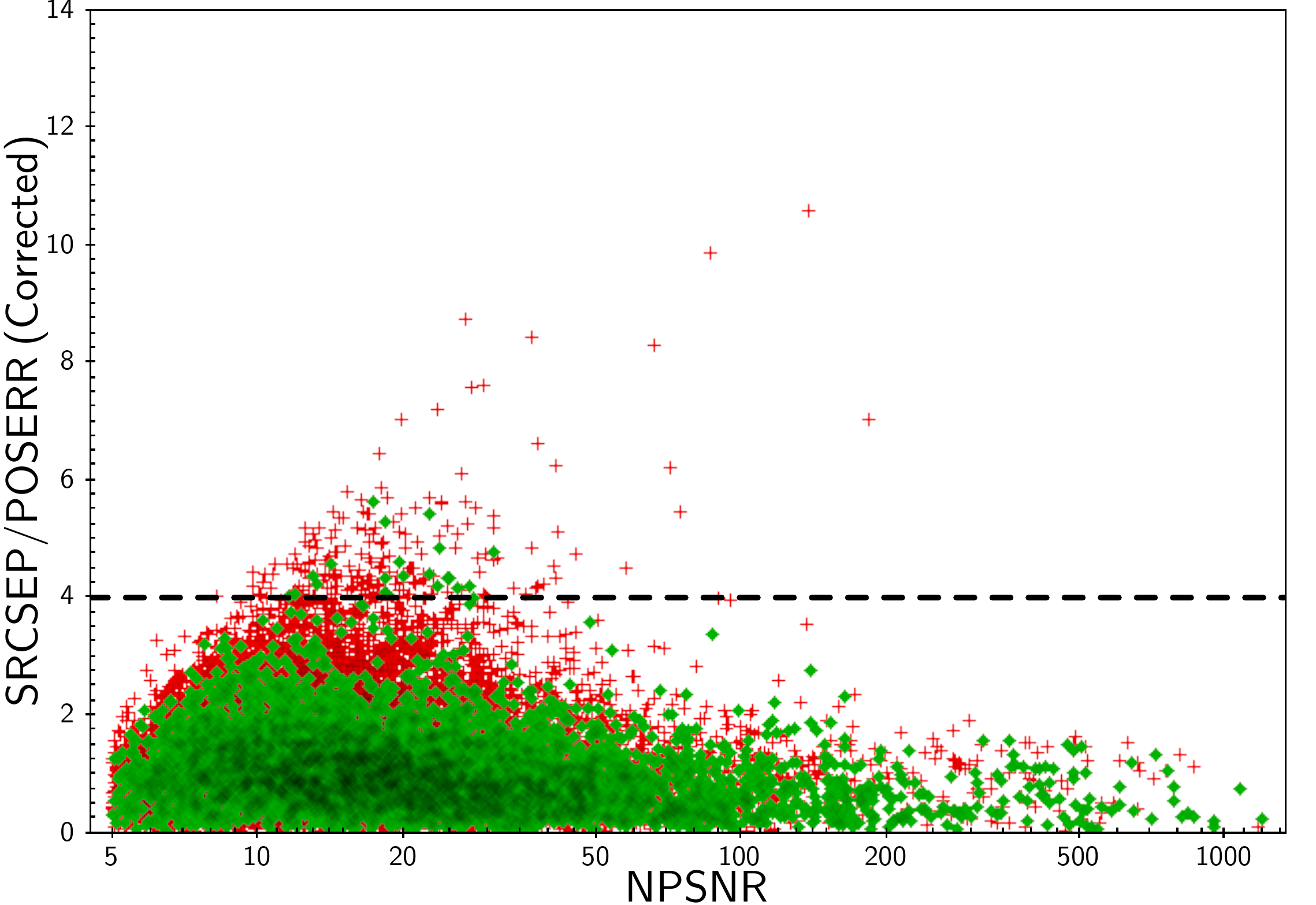}\\
			\includegraphics[width=0.49\textwidth]{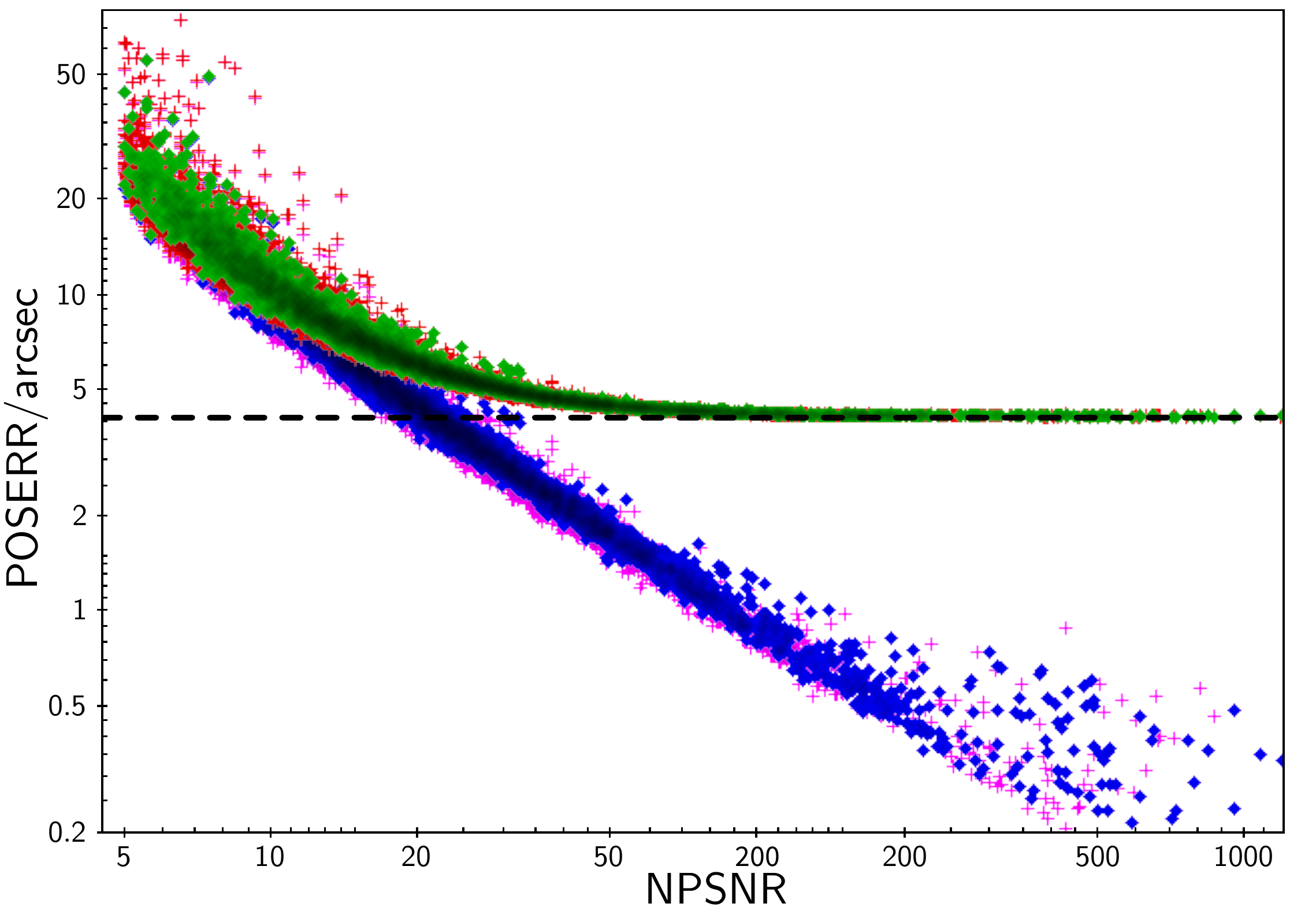}
			\includegraphics[width=0.49\textwidth]{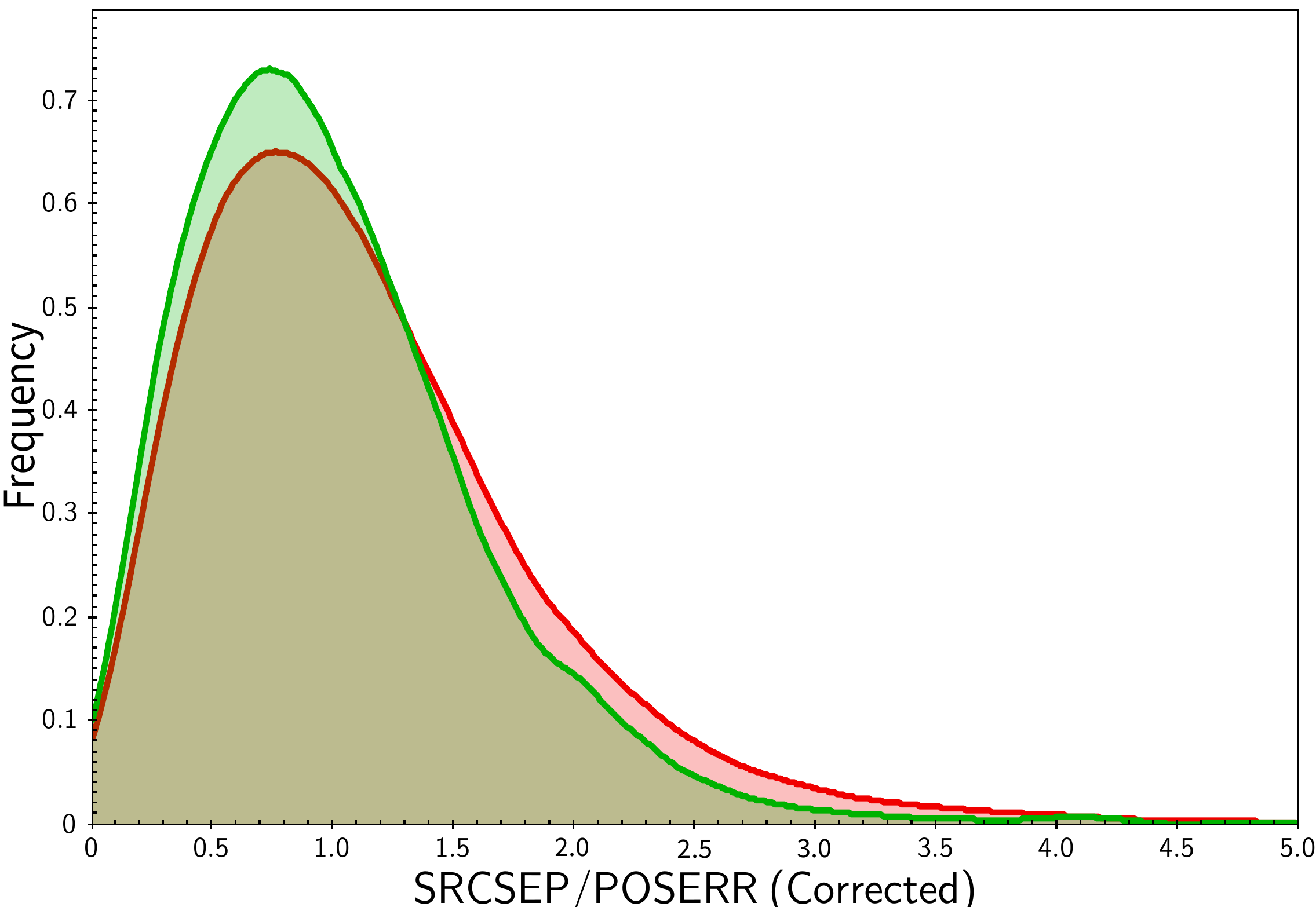}
			\caption{
				Position-recovery accuracy for a simulation with injection in the ``neighbourhood'' of PCCS2+2E source locations.
				The upper left panel shows the distance between the injection and recovered position, normalized by the estimated uncertainty (\textsf{POSERR}), as a function of \textsf{NPSNR}. The blue points represent the PCCS2 and the magenta the full PCCS2+2E.
				The upper right panel shows the same but after applying to \textsf{POSERR} the correction suggested in Eq.~\eqref{eq:PositionCorrectionPCCs2} with $\sigma_0 = 4\parcs11$ (here green is PCCS2 and red is PCCS2+2E).
				In the lower left panel we show \textsf{POSERR} for both sets, corrected and uncorrected, as a function of \textsf{NPSNR}. 
				The horizontal dashed line is the positional precision saturation constant $\sigma_{0}$.
				The lower right panel shows a histogram of the separation between the injected and estimated source position, normalized by the corrected position error bar.
				The distribution is consistent with a unitary Rayleigh distributions with a minor excess in the tail.
			}
			\label{fig:PosAccuracySNR}
		\end{center}
	\end{figure*}

In Fig.~\ref{fig:PosAccuracySNR} (upper left panel), the blue-purple set shows the dependence of the distance between the injection and recovered position normalized by the estimated uncertainty (\textsf{POSERR}), 
as a function of \textsf{NPSNR}.
In this figure it would appear that there are a number of cases where the location is severely mis-estimated (those above the dashed line); when we restrict the catalogue to the PCCS2 set, we see that these cases correspond preferentially to high \textsf{NPSNR} values.
In the lower left panel we can see a very strong correlation between the catalogue \textsf{POSERR} (blue and purple) and \textsf{NPSNR}\footnote{The PCCS2 set (blue points) \textsf{POSERR} dependence on \textsf{NPSNR} is well modelled by
$\textsf{POSERR} \approx \alpha \, \textsf{NPSNR}^{-1.01}$, with $r = -0.98$,
where $\alpha$ is an arbitrary proportionality constant.}.
We note that the estimated positional uncertainty for sources with \textsf{NPSNR}${>}\,20$ is very small (${<}\,1.5$\,\% of the beam size).
It seems clear that the anomalous cases in the upper-left figure are mainly due to a serious underestimation of the positional uncertainty for sources with high \textsf{NPSNR}.  The most likely reason for this is outlined in Sect.~\ref{subsubsec:Systematics}.
In fact, a similar situation was found when producing the PCCS2, and it was handled by adding a term \citep[see equation~7 in][]{planck2014-a35} that forced the positional uncertainty to remain above a threshold. We follow suit by adding a term $\sigma_0$ to our estimate of the full positional uncertainty:
\begin{equation}
	\label{eq:PositionCorrectionPCCs2}
	\sigma^2_{\rm c} = \textsf{POSERR}^{2} + \sigma_{0}^2,
\end{equation}
where $\sigma_{\rm c}$ is the corrected position error bar and $\sigma_{0}$ the saturation constant.
For high \textsf{NPSNR}, $\textsf{POSERR} \rightarrow 0$ and $\sigma_{\rm c} \approx \sigma_{0}$.
In order to determine $\sigma_{0}$, we created a likelihood based on the Rayleigh distribution and we sampled from $\sigma_{0}$, to find
\begin{equation}
	\label{eq:PositionCorrectionValue}
	\sigma_{0} = 4\parcs11 \pm 0\parcs03.
\end{equation}

In Fig.~\ref{fig:PosAccuracySNR} (upper right and lower left panels), we show the corrected position error $\sigma_{\rm c}$ as a function of \textsf{NPSNR} (PCCS2+2E in red and PCCS2 in green).
The horizontal dashed line is $\sigma_0$.
We can see that above $\textsf{NPSNR} \ga 20$ the positional uncertainty stops reducing and instead saturates at $\sigma_0$.
In the lower right panel we show the histograms of the normalized position distribution but now using $\sigma_{\rm c}$.
The PCCS2 distribution (green) is now a good match to a Rayleigh distribution.\footnote{If $\sigma_{\rm c}$ is an accurate description of the actual position errors, then the normalized position distribution should follow a Rayleigh distribution with a scale parameter equal to $\sigma_{\rm c}$.}
The PCCS2+2E (red) is also a good match, though it has a small excess in the tail.
This is the same excess seen in the vertical direction in the upper right panel.
For this simulation, the median position-corrected error bar is $7\parcs8$ or 7.6\,\% of the pixel size.
	
We stress that the vast majority of sources with \textsf{NPNSR}$\,{>}\,20$ have a well-determined positional uncertainty without any correction---i.e., those well below the dashed line in Fig.~\ref{fig:PosAccuracySNR} (upper left panel); applying the correction penalizes those sources unnecessarily. 
	For this reason, the correction on \textsf{POSERR} described here is {\it not\/} applied to the output of \textsf{BeeP}, and should be used only for statistical characterization of samples of sources that contain high \textsf{NPNSR} sources. }

\subsection{Flux density uncertainty correction due to source extension}

\label{subsec:IntroducingSourceExtension}
In every simulation we have described so far, the injected sources have always been beam shaped. 
However, as can be seen in Fig.~\ref{fig:realR_histogram}, PCCS2+2E contains many sources that are at least slightly extended. 

In order to address the effects of source extension, we create a new set of simulations, following exactly the same procedure as described in Sect.~\ref{psssec:InjectSimulations}, i.e., including three types of simulation, each following one of our injection policies.
The only difference is that this time we inject extended rather than point sources. The source size parameter is sampled from a preliminary run of \textsf{BeeP} on the real data.
\begin{figure*}[htbp!]
	\begin{center}
		\leavevmode \hskip 0.2cm
		\includegraphics[width=0.305\textwidth,height=0.14\textwidth]{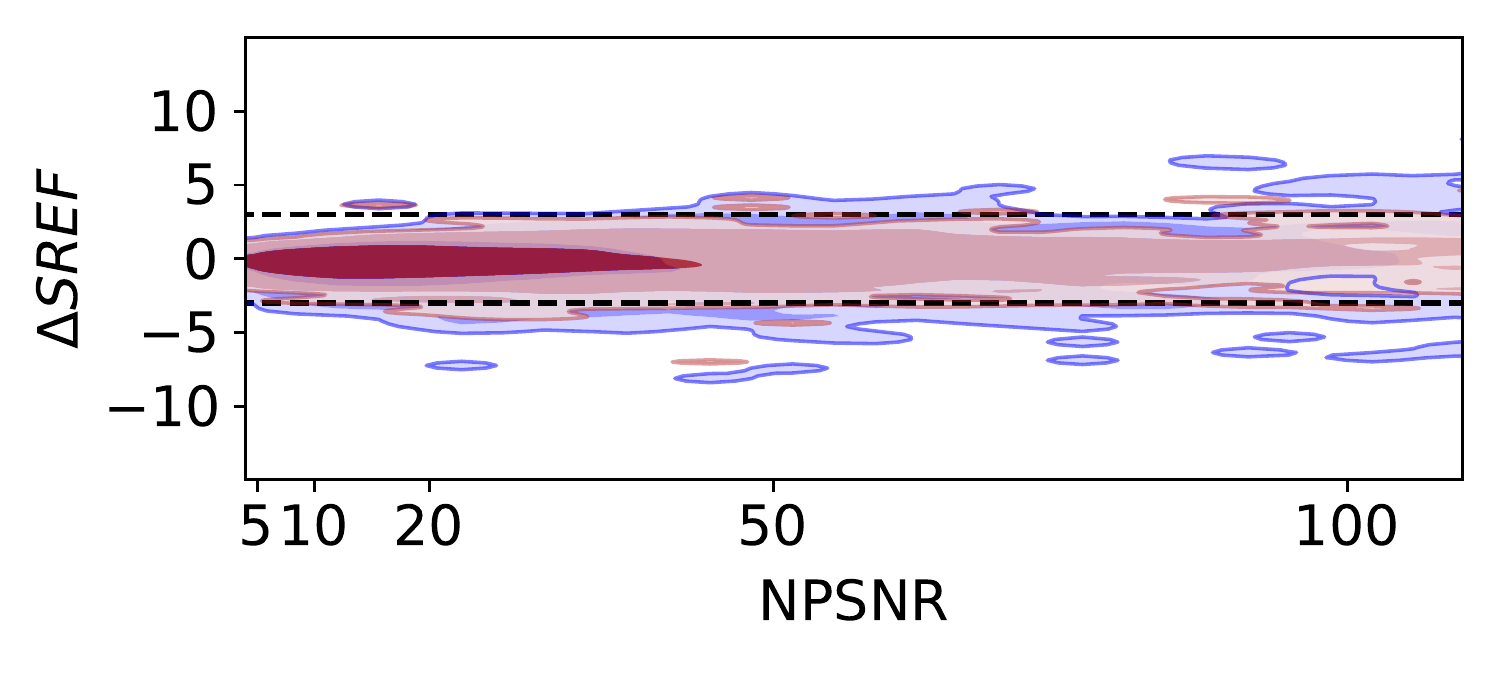} \hskip 0.5cm
		\includegraphics[width=0.305\textwidth,height=0.14\textwidth]{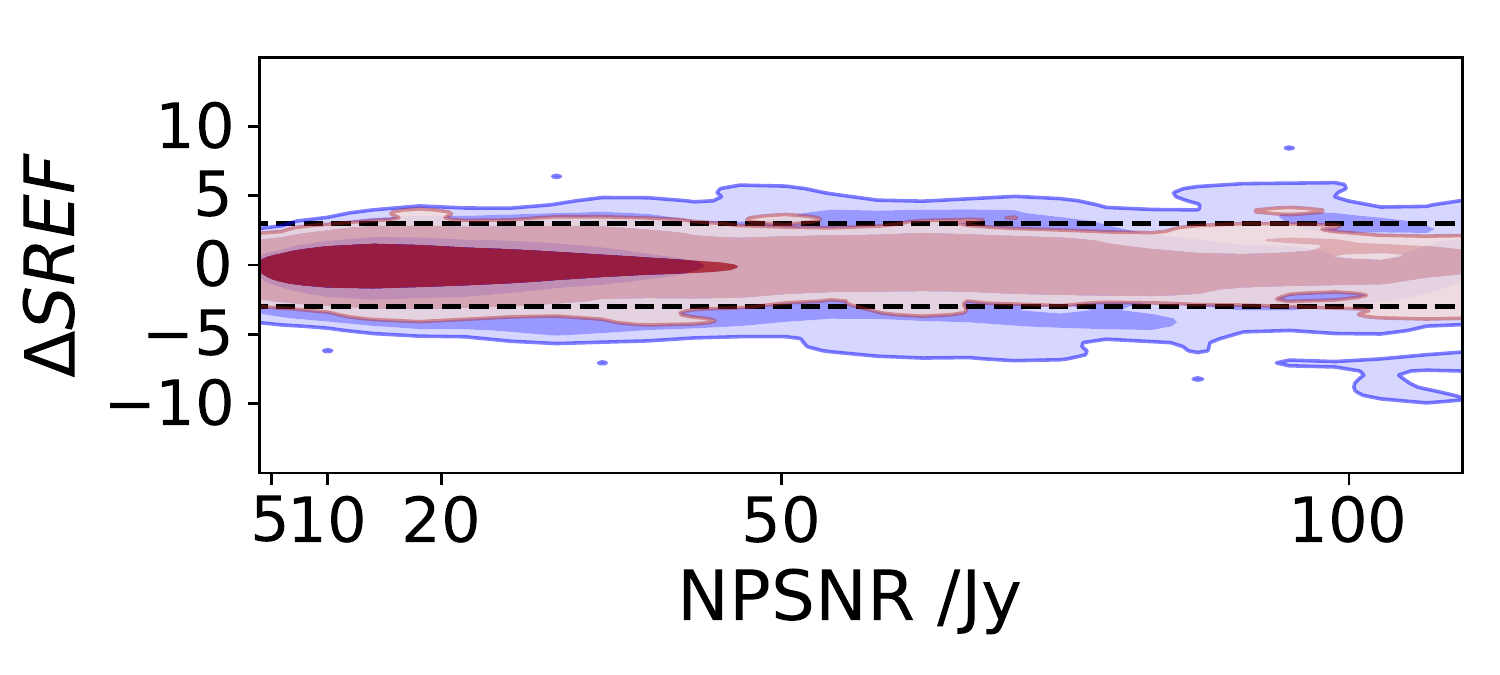} \hskip 0.4cm
		\includegraphics[width=0.305\textwidth,height=0.14\textwidth]{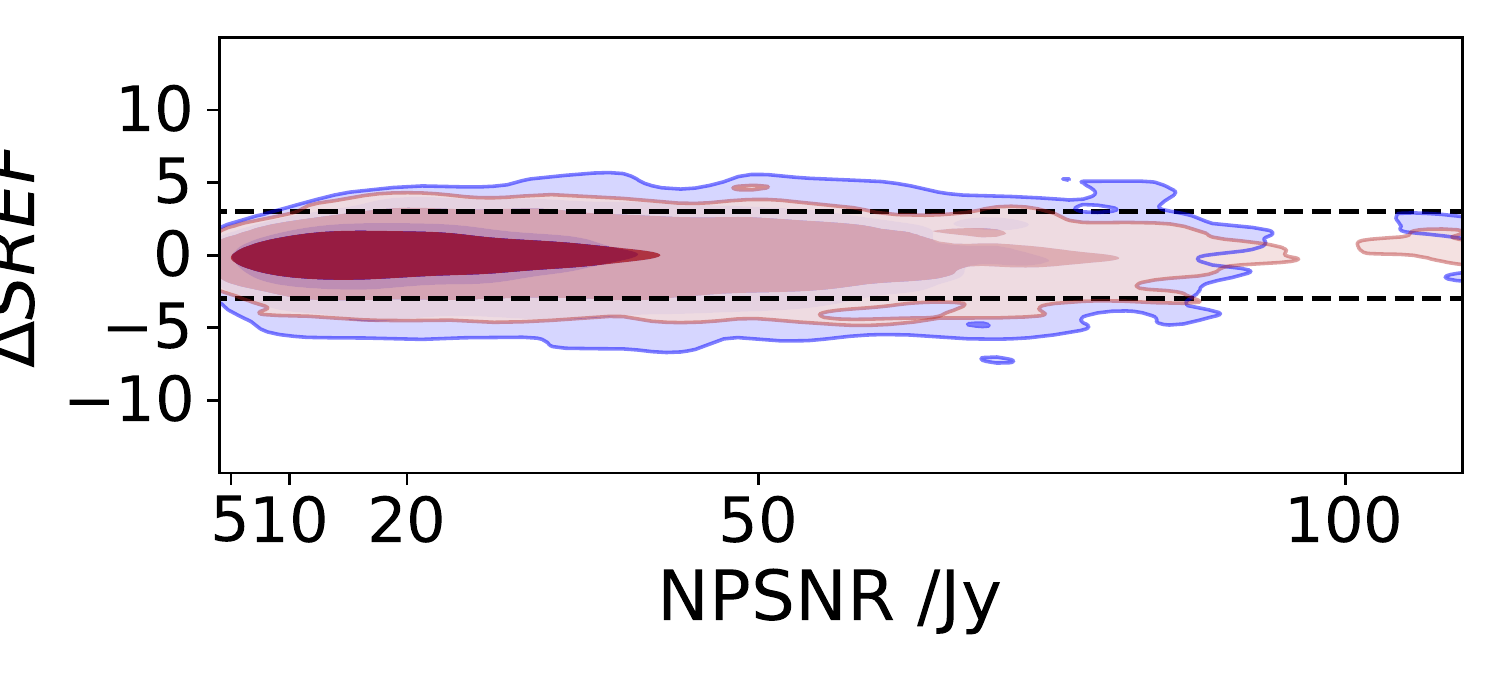}\\
		\includegraphics[width=0.33\textwidth]{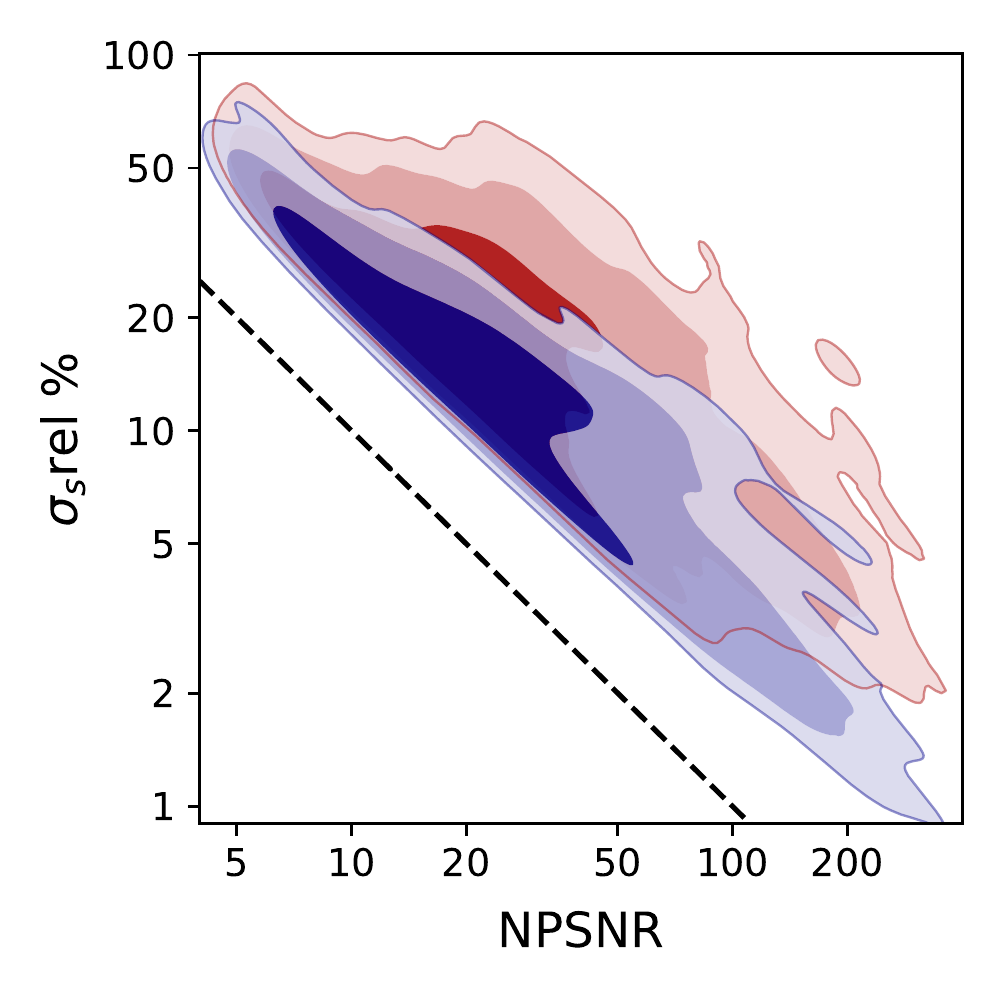}
		\includegraphics[width=0.33\textwidth]{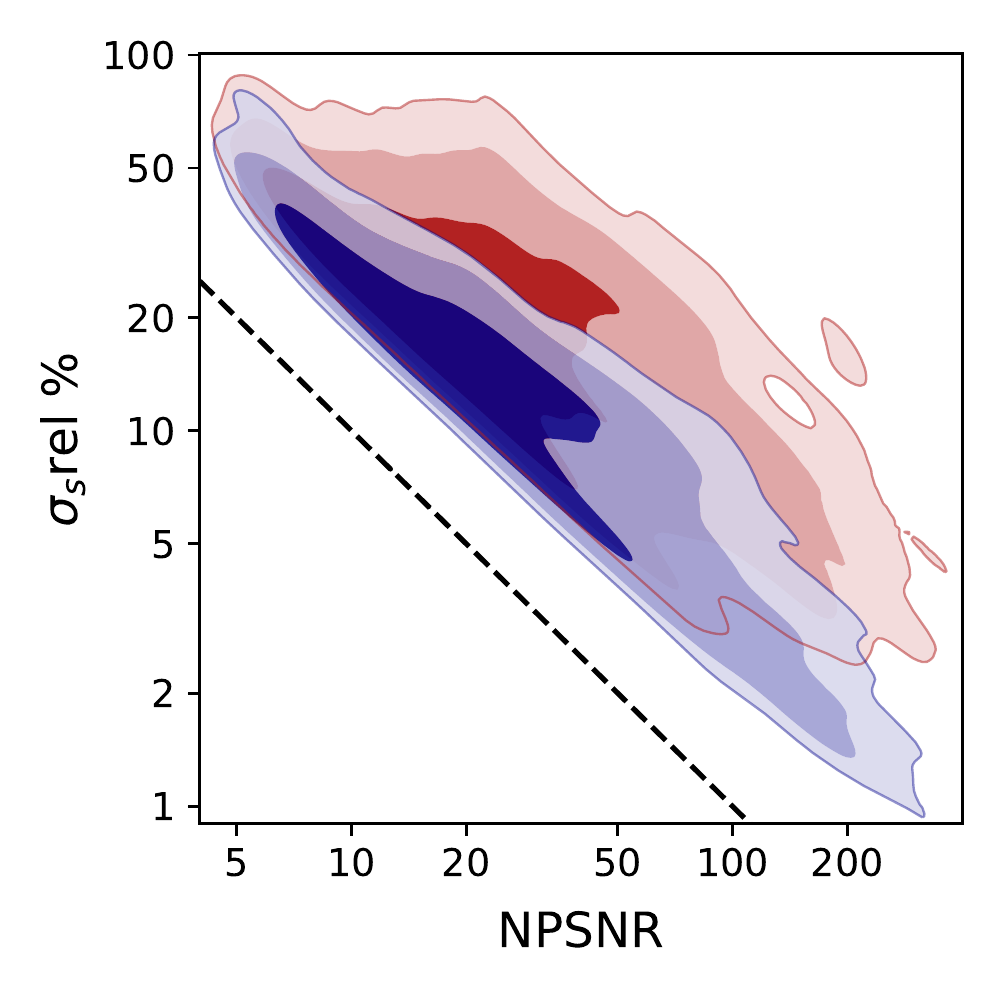}
		\includegraphics[width=0.33\textwidth]{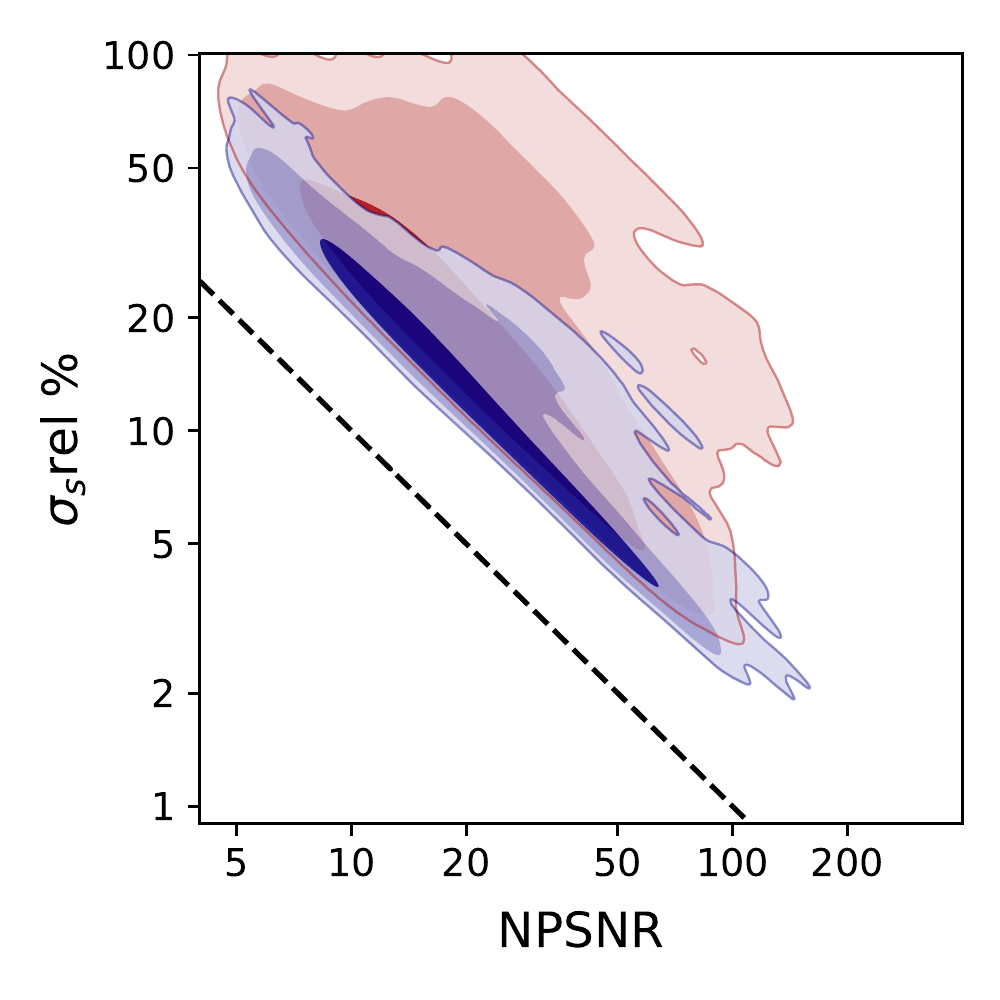}
		\caption{Comparison of injected and recovered flux densities.  The upper row shows the distribution  of the normalized difference (Eq.~\ref{eq:NormSymError}, blue contours, [$68,95,99$]\%) versus \textsf{NPSNR} for three different data sets (from left to right): PCCS2 ``same location''; PCCS2 ``neighbourhood''; and PCCS2E $\wedge \, |b| > 10^\circ$ ``neighbourhood.'' The red contours show the same as the blue ones, but this time for corrected values (Eq.~\ref{eq:CorrectionVar}). The horizontal dashed lines are $\pm3\,\sigma$.
			The lower row shows identical flux-density relative error ($\Delta \textsf{SREF} /\textsf{SREF}$) distribution contours. The black dashed lines here are $\textsf{NPSNR}^{-1}$, the lower limit of the flux-density relative error, which is only achievable if the only unknown parameter in the model is the flux density.
		}
		\label{fig:ConsistencySystematics}
	\end{center}
\end{figure*}

Figure~\ref{fig:ConsistencySystematics} (top row) shows, in blue contours, the normalized difference between the injected and recovered flux densities (Eq.~\ref{eq:NormSymError}) as a function of \textsf{NPSNR}.
In the high \textsf{NPSNR} regime all three types show an excess of deviations ($\ga 3\,\sigma$), reflecting the fact that their distribution is not fully Gaussian.
This is not unexpected, since the same behaviour has been seen for the ``fiducial'' case simulations (see Sect.~\ref{psssec:ArtificialSimulations} and Fig.~\ref{fig:PatchDirectLLSS}), and the same mechanism should be at work here.\footnote{The propagation of the covariance-matrix uncertainty into the likelihood results in an underestimation of the flux-density uncertainty.}
These deviations do not affect significantly the 1$\,\sigma$ levels of the uncertainty distribution, but only its tails. This non-Gaussianity should be taken into account only if it is desired to make a statistical analysis of the flux-density uncertainties of large populations of sources.
To account for this effect, it is possible to add a correction, which we describe in the rest of this subsection. 
But we recommend that  users interested in the 1$\,\sigma$ uncertainty of individual sources do {\it not\/} apply this correction.

Figure~\ref{fig:ConsistencySystematics} suggests that a correction proportional to $\ln(\textsf{NPSNR})$ would be adequate.
This correction is to be added in quadrature to the error bars extracted by \textsf{BeeP}.
For this purpose we have defined a new variable 
\begin{equation}
\label{eq:CorrectionVar}
\sigma_{\rm s} \equiv \sqrt{\sigma_{\rm cat}^2 + (c * \ln(\textsf{NPSNR}))^2},
\end{equation}
where $\sigma_{\rm cat} \equiv (S_{{\rm h}2\sigma} - S_{{\rm l}2\sigma}) / 4$ and $c$ is the flux-density correction constant.
To compute the optimal value of $c$ we follow a similar procedure as that for the positional accuracy.
Let us define a new variable $\xi$ as
\begin{equation}
\label{eq:SCorrLikeVariable}
\xi \equiv \frac{(S_{\rm out}-S_{\rm in}) - b}{\sigma_{\rm s} (c)},
\end{equation}
where $\sigma_{\rm s}$ is the ``corrected'' flux-density uncertainty and $b$ is a ``bias,'' which is added to help symmetrize $\xi$.
If the \textsf{BeeP} uncertainty ($\sigma_{\rm cat}$) were a truthful representation of the uncertainty, in a Gaussian sense, then $\xi$ would follow a normal distribution with $b = 0, c = 0$.
We characterize a Gaussian likelihood for $\xi$ with two parameters $\{b,c\}$.
We sample from $\{b,c\}$ to construct a posterior distribution and then we find the median of both parameters to correct the catalogue. 
We expect that in regions with strong complex backgrounds, the sub-optimality of \textsf{BeeP}'s likelihood will manifest itself more strongly and require larger corrections.
We therefore compute corrections for each of three sky regions:
\begin{itemize}
	\item PCCS2;
	\item PCCS2E $\wedge \, |b| > 10^\circ$;
	\item $|b| \leq 10^\circ$.
\end{itemize}
In Fig.~\ref{fig:ConsistencySystematics} (lower row) we show the flux-density relative accuracy as a function of \textsf{NPNSR}, before (in blue contours) and after applying the correction (in red contours).
The corrected error bars show (as expected) a larger dispersion with \textsf{NPSNR}. 

\begin{table}[htbp!]
\begingroup
\newdimen\tblskip \tblskip=5pt
\caption{Flux density error bar calibration constant ``$c$'' in mJy, for the three types of simulation (see Eqs.~\ref{eq:CorrectionVar} and \ref{eq:SCorrLikeVariable}).}
\label{table:SCorrection}
\nointerlineskip
\vskip -3mm
\footnotesize
\setbox\tablebox=\vbox{
   \newdimen\digitwidth
   \setbox0=\hbox{\rm 0}
   \digitwidth=\wd0
   \catcode`*=\active
   \def*{\kern\digitwidth}
   \newdimen\signwidth
   \setbox0=\hbox{+}
   \signwidth=\wd0
   \catcode`!=\active
   \def!{\kern\signwidth}
\halign{\tabskip 0pt\hbox to 1.0in{#\leaderfil}\hfil\tabskip 0.5em&
         \hfil#\hfil\tabskip 2.5em&
         \hfil#\hfil\tabskip 0.5em&
         \hfil#\hfil\tabskip 1em&
         \hfil#\hfil\tabskip 0.5em&
         \hfil#\hfil\tabskip 0pt\cr
\noalign{\doubleline}
\omit& & \multispan2\hfil Before$^{\rm b}$\hfil& \multispan2\hfil After$^{\rm c}$\hfil\cr
\omit\hfil Data set$^{\rm a}$\hfil& $c$ [mJy]& $\sigma_\xi^{\rm d}$& SMAD$_\xi$$^{\rm e}$& $\sigma_\xi$& SMAD$_\xi$$^{\rm e}$\cr
\noalign{\vskip 2pt\hrule\vskip 5pt}
\omit&\multispan5\hfil Injection at same location\hfil\cr
\noalign{\vskip -5pt}
\omit&\multispan5\hrulefill\cr
P2E$\,{>}\,|10^\circ|\,^{\rm f}$&$103\pm*2$& 1.38& 0.95& 1.19& 0.81\cr
PCCS2& $*48\pm*1$& 1.51& 0.91& 0.91& 0.66\cr
${<}\,|10^\circ|$& $621\pm15$& 1.54& 0.99& 1.28& 0.77\cr
\noalign{\vskip 5pt\hrule\vskip 5pt}
\omit&\multispan5\hfil Injection in neighbourhood\hfil\cr
\noalign{\vskip -5pt}
\omit&\multispan5\hrulefill\cr
P2E$\,{>}\,|10^\circ|$& $168\pm*2$& 1.59& 1.38& 1.17& 0.93\cr
PCCS2& $*57\pm*1$& 1.70& 1.31& 1.00& 0.88\cr
${<}\,|10^\circ|$& $820\pm15$& 1.66& 1.40& 1.17& 0.91\cr
\noalign{\vskip 5pt\hrule\vskip 5pt}
\omit&\multispan5\hfil Uniform distribution\hfil\cr
\noalign{\vskip -5pt}
\omit&\multispan5\hrulefill\cr
P2E$\,{>}\,|10^\circ|$& $*85\pm*2$& 1.37& 1.14& 1.06& 0.91\cr
PCCS2& $*57\pm*1$& 1.38& 1.21& 1.07& 0.93\cr
${<}\,|10^\circ|\,^{\rm g}$& $259\pm15$& 1.60& 1.17& 1.02& 0.86\cr
\noalign{\vskip 5pt\hrule\vskip 4pt}}}
\endPlancktable                    
\tablenote {{\rm a}} ``P2E$ > 10\deg$'' means the PCCS2E data set with $|b| > 10\deg$; ``PCCS2'' means the PCCS2 data set; and ``$<|10\deg$'' means the PCCS2+2E set with $|b| < 10\deg$.\par
\tablenote {{\rm b}} Before applying the correction (as in the catalogue).\par
\tablenote {{\rm c}} After applying the correction.\par
\tablenote {{\rm d}} Standard deviation of $\xi$ (Eq.~\ref{eq:SCorrLikeVariable}).\par
\tablenote {{\rm e}} Scaled median absolute deviation of $\xi$.\par
\tablenote {{\rm f}} The median \textsf{NPSNR} for all subsets is approximately 20, except for ``${<}\,|10^\circ|$'', which is approximately 9 and ``PCCS2E$\,{>}\,|10^\circ|$'', which is approximately 13.\par
\tablenote {{\rm g}} There are only 1066 sources in this subset.\par
\endgroup
\end{table}

Table~\ref{table:SCorrection} contains a summary of statistics of the variable $\xi$ (Eq.~\ref{eq:SCorrLikeVariable}) for the three type of simulations and the three sky regions before and after applying the correction.
If the error bars were correctly describing the flux recovery uncertainty, in a Gaussian sense, then $\sigma_\xi \approx 1$.
Considering that the data statistics are very non-Gaussian, with broad tails, and $\sigma_\xi$ is very sensitive to outliers, we also included the more robust {scaled median absolute deviation (SMAD$_\xi$).
``Before'' applying the correction, all sets, for all simulations, show a clear excess in $\sigma_\xi$.
This is also seen, as expected, in the SMAD for $\xi$, for the cases ``neighbourhood'' and ``uniform,'' but not for ``same location.''
By examining Fig.~\ref{fig:ConsistencySystematics} (left column, in green), one can indeed see that the distribution of estimates for ``same location'' is tighter than for the other types.
We also find for ``same location'' a small bias towards low values, which is often present in this type of simulations (e.g., Fig.~\ref{fig:SimulsBetaTemp}).
We believe this bias could be an effect induced by the inpainting procedure.

On the other hand, ``After'' correcting the flux-density error, all three types of simulations show reasonable values for both statistics, although
in the case of ``same location'' simulations (which is perhaps the most realistic), the improvement is not as good as in the other types.
However, as previously mentioned, the SMAD statistic (the more robust measurement of dispersion for non-Gaussian distributions) applied without any correction (\{b,c\}$= 0$) to the ``same location'' simulation already showed very good values (see Table~\ref{table:SCorrection}).
For most purposes it should therefore not be necessary to apply any correction to the \textsf{BeeP} estimates of flux-density uncertainties. Corrections should be applied only if a strictly Gaussian characterization of the uncertainties is needed, particularly true for high-Galactic-latitude objects. 

\subsection{Planck FFP8 simulations}
\label{psssec:FFP8Simulations}

Possibly the crudest part of the data model implemented by \textsf{BeeP} is that it assumes that the beam shapes are perfectly circularly symmetric and homogeneous across the sky.
In addition, in our injection simulations, the mock sources always have circularly symmetric Gaussian shapes, and in the most sophisticated simulations we also vary their radius. 
However, for the \Planck\ 857-GHz channel, the average beam ellipticity ($\varepsilon \approx 1.39$) is sufficiently high and variable across the sky (dispersion about 10\,\%), to induce systematic flux-density deviations as a result of the model and actual beam-shape mismatch.
Given the huge flux-density dynamic range of the PCCS2+2E, we expect that, especially at the bright end, these effects will have a significant influence on the estimation of flux densities. These systematic effects cannot be directly taken into account by \textsf{BeeP}.

\begin{figure}[htbp!]
	\begin{center}
		\leavevmode
		\includegraphics[width=0.40\textwidth]{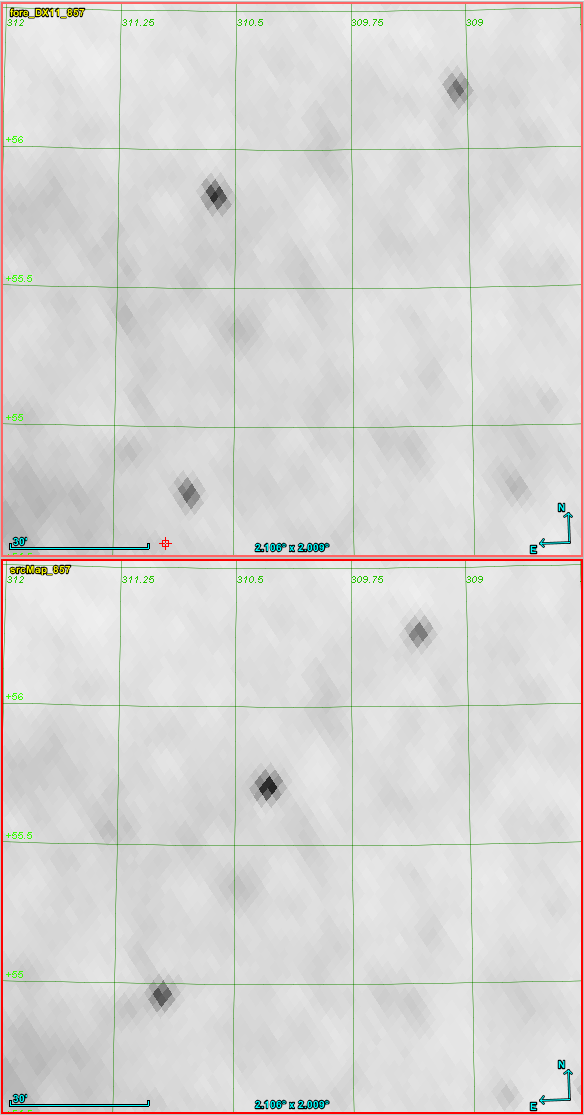}
		\caption{
			Both pictures show a small field ($2\pdeg1\times 2\pdeg0$) centred at $l=309\pdeg68, b= +55\pdeg41$.
			The upper panel is \Planck's 857-GHz map and the lower a \textsf{BeeP} injection simulation where mock sources were added to the same \Planck\ map around the neighbourhood of real sources that were masked and inpainted.
			The simulated sources were rendered using the average effective Gaussian 857-GHz \Planck\ beam.
			The brightness scale is the same on both plots.
			One can see in the lower panel that the injection simulations fail to capture the ellipticity of compact objects in the \Planck\ 857-GHz map (upper panel).
		}
		\label{fig:SrcElliptical}
	\end{center}
\end{figure}

In principle \textsf{BeeP}'s likelihood can easily accommodate more realistic beam shapes, including their spatial variation,\footnote{For example, as described in the FEBeCoP effective beam approach \citep{FeBeCop}.} but the computational cost would be prohibitive.
However, \Planck\ has produced a set of simulations that include a very accurate model of the beam shapes and their variation across the sky,
the ``FFP8'' simulations \citep{planck2014-a14}.
We note two important drawbacks of FFP8: the absence of a 3000-GHz map; and the fact that all simulated compact objects in the maps are exactly beam shaped (they are drawn from a simulated set of zero-extension sources). 
These issues affect the constraining capability of \textsf{BeeP}.
With only \Planck's three high-frequency channels available, \textsf{BeeP} can no longer effectively constrain $T$.
The extra uncertainty propagates to $\beta$ (they are highly correlated) and to a smaller extent to the flux density $S$ (see Fig.~\ref{fig:CornerPlot}).
In spite of these drawbacks, we can use the FFP8 simulations to effectively calibrate the effect of beam shapes on the recovery of source parameters.

\begin{figure}[htbp!]
	\begin{center}
		\leavevmode
		\includegraphics[width=0.49\textwidth, height=4cm]{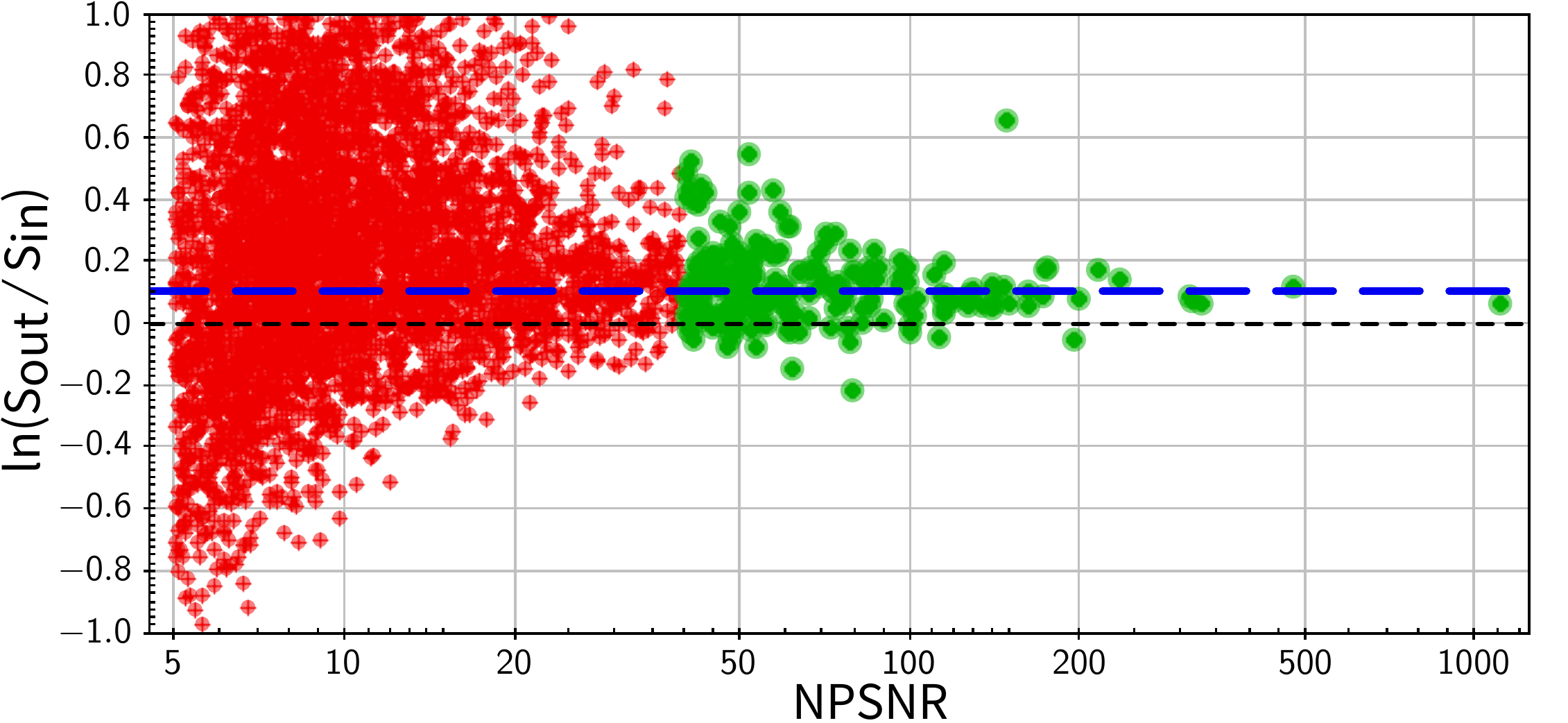}
		\caption{
			Comparison of FFP8's full catalogue of bright source flux densities with \textsf{BeeP}'s \textsf{SREF}.
			Shown in red are the source flux estimates with \textsf{NPSNR}$\,{<}\,40$ and in green those with NPSNR$\,{>}\,40$.
			The horizontal long-dashed blue line is the $\ln(S_{\rm out}/S_{\rm in})$ median (0.104) of the green subset.
			The strong outlier is the result of a mock source blended with a real one.
		}
		\label{fig:FFP8_flux}
	\end{center}
\end{figure}

We are particularly concerned about any systematic bias in the flux-density recovery.
As explained in Sect.~\ref{ssec:AlgoImplSrcDetec}, we already found it necessary to force \textsf{BeeP}'s likelihood to model beam shaped sources with a source extension $\textsf{EXT} \approx 1\parcm72$ (about 1~pixel).  As a result, the flux-density estimation bias was much reduced but not completely eliminated.
Figure~\ref{fig:FFP8_flux} shows the comparison of $\ln(S_{\rm out}/S_{\rm in})$ based on the analysis of the FFP8 maps by \textsf{BeeP}.
The figure shows a bias in this quantity and a large dispersion, particularly at the low \textsf{NPSNR} regime.
The median of $\ln(S_{\rm out}/S_{\rm in})$ is $0.104$,\footnote{To avoid any possible distortion resulting from Eddington-type bias, we restrict the data set to high \textsf{NPSNR} (${>}\,40$) sources only (see green points in Fig.~\ref{fig:FFP8_flux}).} which implies an approximately $+11.0\,\%$ bias in the recovered reference flux density,
both for the flux-density estimates based on the MBB model and those based on the ``Free'' model.

The FFP8 simulations allow us to assess the impact that a realistic beam has on the positional accuracy, as was done for the injection simulations using all four channels and circularly symmetric sources (see Sect.~\ref{psssec:InjectSimulations}).
\begin{figure*}[htbp!]
	\begin{center}
		\leavevmode
		\includegraphics[width=0.49\textwidth]{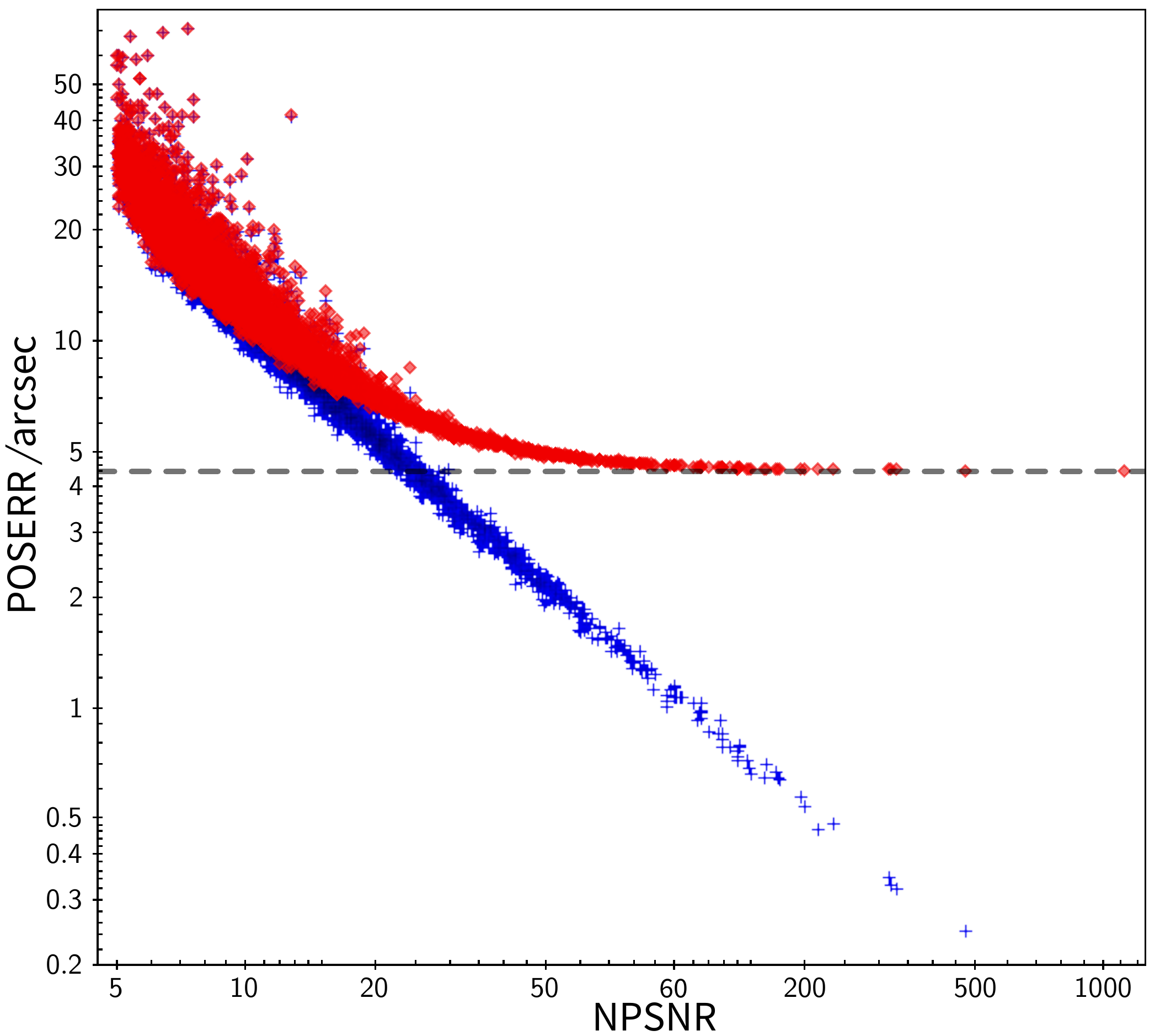}
		\includegraphics[width=0.49\textwidth]{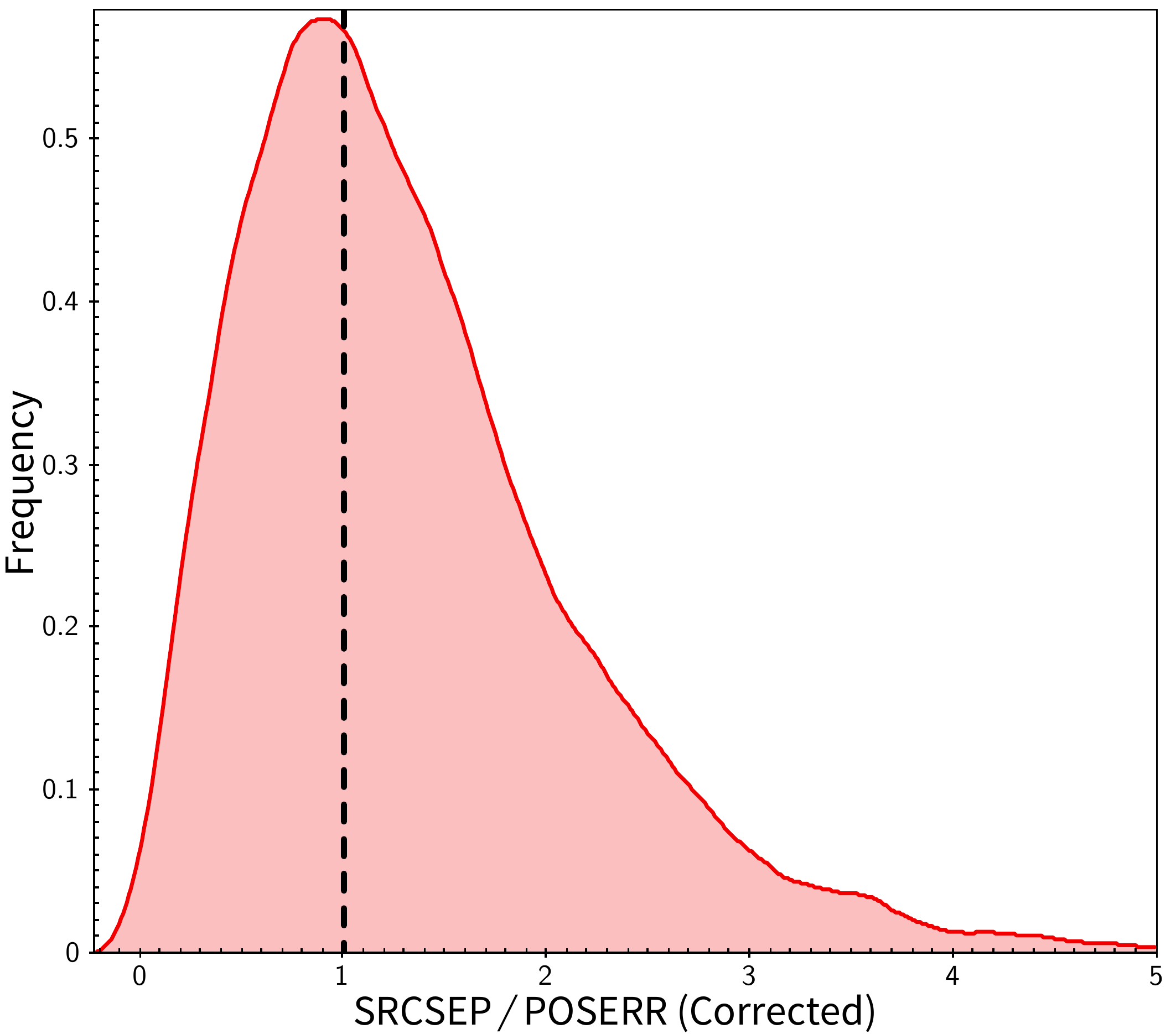}
		\caption{
			FFP8 positional accuracy results.
			{\it Left}: raw (as in the catalogue) and corrected \textsf{POSERR} versus \textsf{NPSNR} (FFP8 bright sources catalogue).
			The correction procedure was the same as in Sect.~\ref{psssec:InjectSimulations}.
			The horizontal line is the saturation constant $\sigma_0$.
			{\it Right}: histogram of the deviations between the injected and recovered positions, normalized by the corrected positional uncertainty.
		}
		\label{fig:FFP8PosAccuracy}
	\end{center}
\end{figure*}
Figure~\ref{fig:FFP8PosAccuracy} (left panel) shows that the positional accuracy of the recovery exhibits similar traits as in the case of the injection simulations.
We can therefore apply the same procedure as in Sect.~\ref{psssec:InjectSimulations} to correct \textsf{POSERR},
and we find a very similar threshold level:
\begin{equation}
\label{eq:PositionCorrectionFFP8}
\sigma_{0} = {4\parcs43}^{+0\parcs17}_{-0\parcs18}.
\end{equation}
The position deviations normalized by the corrected position error bar now follow a unitary Rayleigh distribution (Fig.~\ref{fig:FFP8PosAccuracy}, right panel). 
The median \textsf{POSERR} for the full FFP8 catalogue is around $13\parcs2$ ($13\,\%$ of a pixel), which is larger than that found in the injection simulations ($7\parcs8$, Sect.~\ref{psssec:InjectSimulations}).
This difference must be partly due to the more realistic beam simulations included in FFP8. However, other factors are likely to play a role as well, e.g.,
the FFP8 catalogue has a smaller median \textsf{NPSNR} (9.65) when compared with the injection simulations (15.1); and the absence of the 3000-GHz channel may have an impact as well. 

As in the analysis of Sect.~\ref{psssec:InjectSimulations} , we do not apply this correction to the \textsf{BeeP} output. The correction should be applied only when well-behaved statistical characterization of a sample containing high-\textsf{NPNSR} sources is required.  In this case, we recommend to use the slightly more conservative correction value shown in Eq.~\eqref{eq:PositionCorrectionFFP8}.

%
%
\section{The no-source simulations: frequentist versus Bayesian approaches}
\label{sec:ContaminationBayes}
In Appendix~\ref{sec:beep} 
we quite closely followed a frequentist framework:
\begin{itemize}
	\item we define the ``null hypothesis'' to be that no source is present (only background);
	\item we define a data-based statistic SRCSIG and its cumulative distribution {\it assuming\/} that the null hypothesis is true;
	\item we reject the null hypothesis for extreme values of SRCSIG, using a single-tailed test.
\end{itemize}
Rejecting the null hypothesis means that the data do not support, at a certain level (``tail probability''), the background-only hypothesis. Therefore one chooses the alternative hypothesis, namely a source is present. 
Figure~\ref{fig:Reliability} shows the level at which we reject the null hypothesis as a function of \textsf{SRCSIG}.
We call this ``contamination'' because at that threshold of SRCSIG,
we still expect the null hypothesis (no real source) to be true a certain fraction of the time.  That fraction is given as a percentage of the total using the $y$-axis of Fig.~\ref{fig:Reliability}.	
This is a commonly employed method of measuring the contamination of a catalogue.

When alternatively using the Bayesian framework, instead of dealing with each source individually, we prefer to address the broader question of ``catalogue contamination.''
Catalogue contamination can be defined as the percentage of false sources in the sub-catalogue defined by a given threshold of the selection statistic:
\begin{equation}
\label{secApp:ContamDef}
\Pr({\rm F} | {\rm C} ; \theta),
\end{equation}
where ${\rm F}$ means a ``false'' source, ${\rm C}$ means ``it is part of the catalogue,'' and $\theta$ is the threshold.
Using conditional probability rules and Bayes theorem, the definition of Eq.~\eqref{secApp:ContamDef} can be expressed as
\begin{equation}
\Pr({\rm F} | {\rm C}) \equiv \Gamma = \left(1 + \frac{\Pr({\rm T}) \Pr({\rm C} | {\rm T} )}{\Pr({\rm F}) \Pr({\rm C} | {\rm F})}\right),
\end{equation}
where ${\rm T}$ means a \textit{true} source and $\Pr({\rm T})$ and $\Pr({\rm F})$ are the prior probabilities (before running \textsf{BeeP}) of a source being real or spurious. $\Pr({\rm C} | {\rm F} )$ is what we compute from the ``no-source'' simulation and $\Pr({\rm C} | {\rm T} )$ is the completeness. We have dropped the threshold from the expression to make it more readable, although all the factors are dependent on it.
The quantity $\Gamma$ is a proper probability $\in [0,1]$.
Let us define
\begin{equation}
\label{eq:ContAlphaDef}
\alpha \equiv \frac{\Pr({\rm T}) \Pr({\rm C} | {\rm T} )}{\Pr({\rm F})}.
\end{equation}
Then $\Gamma$ reads,
\begin{equation}
\label{eq:ContGammaAlpha}
\Gamma = \frac{ \Pr({\rm C} | {\rm F} )}{ \Pr({\rm C} | {\rm F} ) + \alpha}
\end{equation}
If $\alpha \gtrsim 1$ then $\Gamma \la \Pr({\rm C} | {\rm F} )$ and the no-source simulations give a good estimate of an upper bound on the expected catalogue contamination.
It is interesting to note that the catalogue completeness is also present when computing the catalogue contamination using a Bayesian approach.

A useful catalogue must always have the following properties:
\begin{itemize}
	\item completeness $\equiv \Pr({\rm C} | {\rm T} ) \approx 1$;
	\item contamination $\equiv \Pr({\rm C} | {\rm F} ) \approx 0$.
\end{itemize}
If $\Pr({\rm T})/\Pr({\rm F}) \approx 1$, i.e., no prior bias, then $\Gamma \approx \Pr({\rm C} | {\rm F} )$, as in a frequentist result.
In the extreme case of a \textsf{SRCSIG} threshold of zero, then $\Pr({\rm C} | {\rm T} ) = \Pr({\rm C} | {\rm F} ) = 1$ (i.e., we accept everything) and $\Gamma = 1/2$, the value one would expect if $\Pr({\rm T})/\Pr({\rm F}) \approx 1$.

In \cite{planck2014-a35} (bottom right panel of figure~7) there are no reliability values provided for the PCCS2E. However, assuming that the PCCS2E reliability is as low as 70\,\% at 1\,Jy and $\Pr({\rm C} | {\rm T} ) \approx 0.4$, then $\alpha \approx 0.93$ and $\Gamma$ (Eq.~\ref{eq:ContGammaAlpha}) is $\approx \Pr({\rm C} | {\rm F} )$, just like our prediction for \textsf{BeeP}'s catalogue contamination.
When completeness is very low ($\Pr({\rm C} | {\rm T} ) \approx 0$) or $\Pr({\rm T}) \ll \Pr({\rm F})$, then false objects are dominant and $\alpha \approx 0$ implies $\Gamma \approx 1$.
In this case, even with a good rejection of false detections, the catalogue contamination can reach very high values.\footnote{For an extreme (but realistic) example, see p.\,1132 of \citet{MikesBook}.}
We cannot completely rule out this scenario at very low Galactic latitudes ($|b| < 1^{\circ}$) close to the Galactic centre. In this region, the properties of PCCS2+2E are not well defined and false detections could dominate.

\section{Source examples}
\label{sssec:Examples}

In this appendix we show a few representative examples of SEDs resulting from the analysis of \textsf{BeeP}.  Specifically we show:
\begin{itemize}
\item three archetypal nearby galaxies, Arp\,220, M\,100, and NGC\,895 \citep{HerschelPlanckDusty};
\item one source (J091828.6+514223) from the \Planck\ list of high-redshift candidates \citep{planck2015-XXXIX}, also detected in the Herschel Lensing Survey overview \citep{egami10}, which is a strongly lensed galaxy at $z=5.2$ \citep{Combes2012};
\item one source from the GEMS catalogue \citep[\Planck\ dusty Gravitationally-enhanced submillimetre sources,][]{canaveras15}, PLCK\,G138.6+62.0;
\item two sources with non-thermal SEDs that cannot be fit to an MBB spectrum,
namely M1 (the Crab Nebula, a supernova remnant) and 3C\,273 (a blazar);
\item the brightest source in our ATLAS comparison field (HATLAS J144011.1-001719); 
\item one of the coldest Galactic clumps extracted from the PGCC \citep{planck2014-a37}, IRDC MSXDC G033.69$-$00.01;
\item Orion\,A IRC\,2, an archetypal infrared source in the Orion A molecular cloud.
\end{itemize}

\begin{figure*}[htbp!]
	\begin{center}
		\leavevmode
		\includegraphics[width=0.80\textwidth]{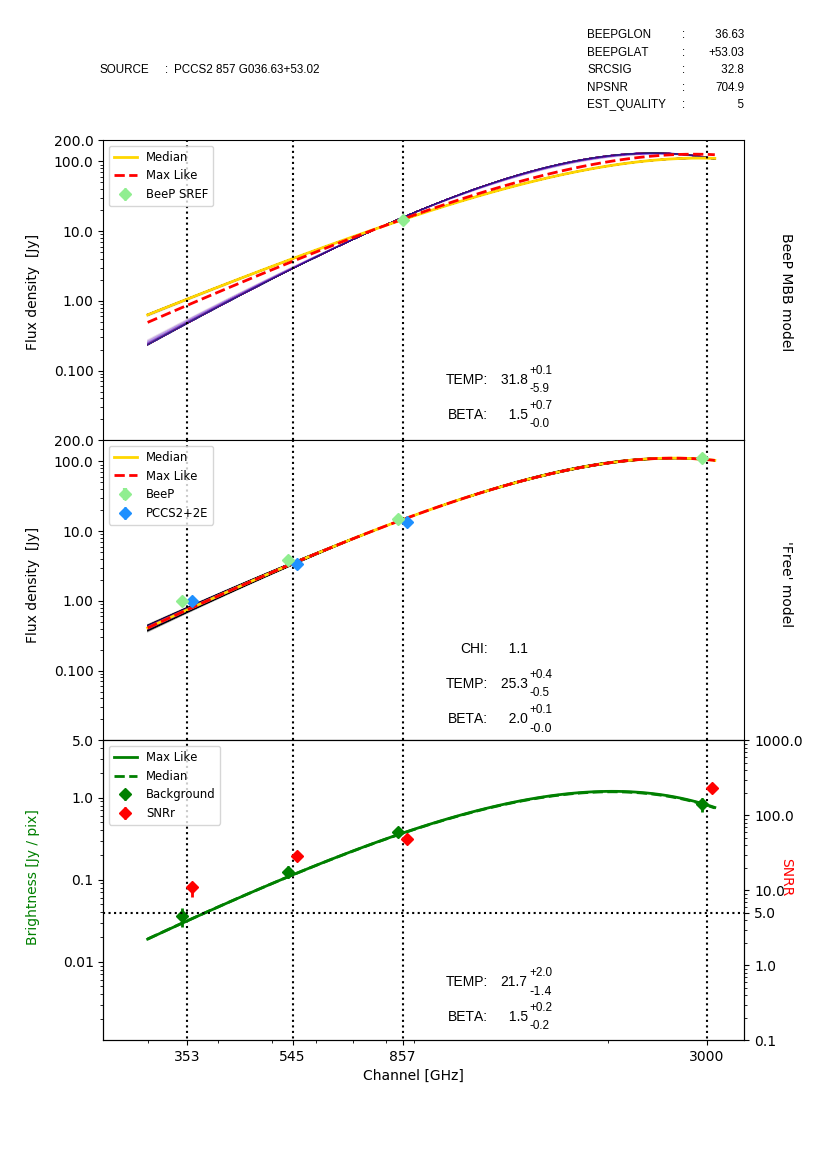}
		\caption{Results of {\tt BeeP} analysis for Arp\,220. See the caption of Fig.~\ref{fig:SEDPlot} for a full description of the contents of this figure. This case is a very clean example of a well-determined model for source and background.} 	
		\label{fig:FirstExample}
	\end{center}
\end{figure*}

\begin{figure*}[htbp!]
        \begin{center}
                \leavevmode
                \includegraphics[width=0.80\textwidth]{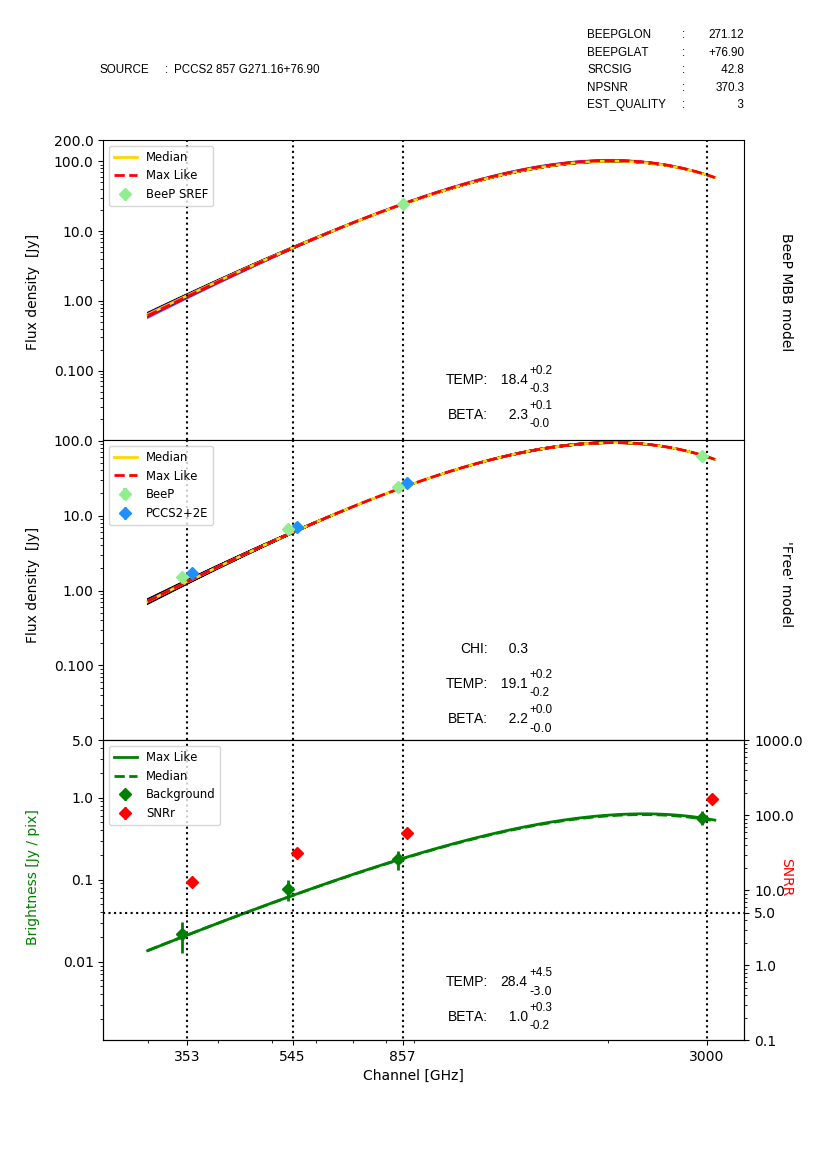}
                \caption{Results of {\tt BeeP} analysis for M\,100. See the caption of Fig.~\ref{fig:SEDPlot} for a full explanation. This case is interesting because {\tt BeeP} has reduced \textsf{EST\_QUALITY} due to the extremely low uncertainties in both temperature and spectral index (Sect.~\ref{ssec:EstimateValuesQualityCriterion}), in spite of the fact that the SEDs fit the data very well. However the $\chi^2$ value of the Free-model fit (middle panel) is not far from the expected unity-per-degree of freedom level, and so this is one of those exceptional cases where the very low uncertainties reflect a very good fit, rather than the fact that the sampler has not been able to explore the parameter space.}
                \label{fig:SecondExample}
        \end{center}
\end{figure*}

\begin{figure*}[htbp!]
        \begin{center}
                \leavevmode
                \includegraphics[width=0.80\textwidth]{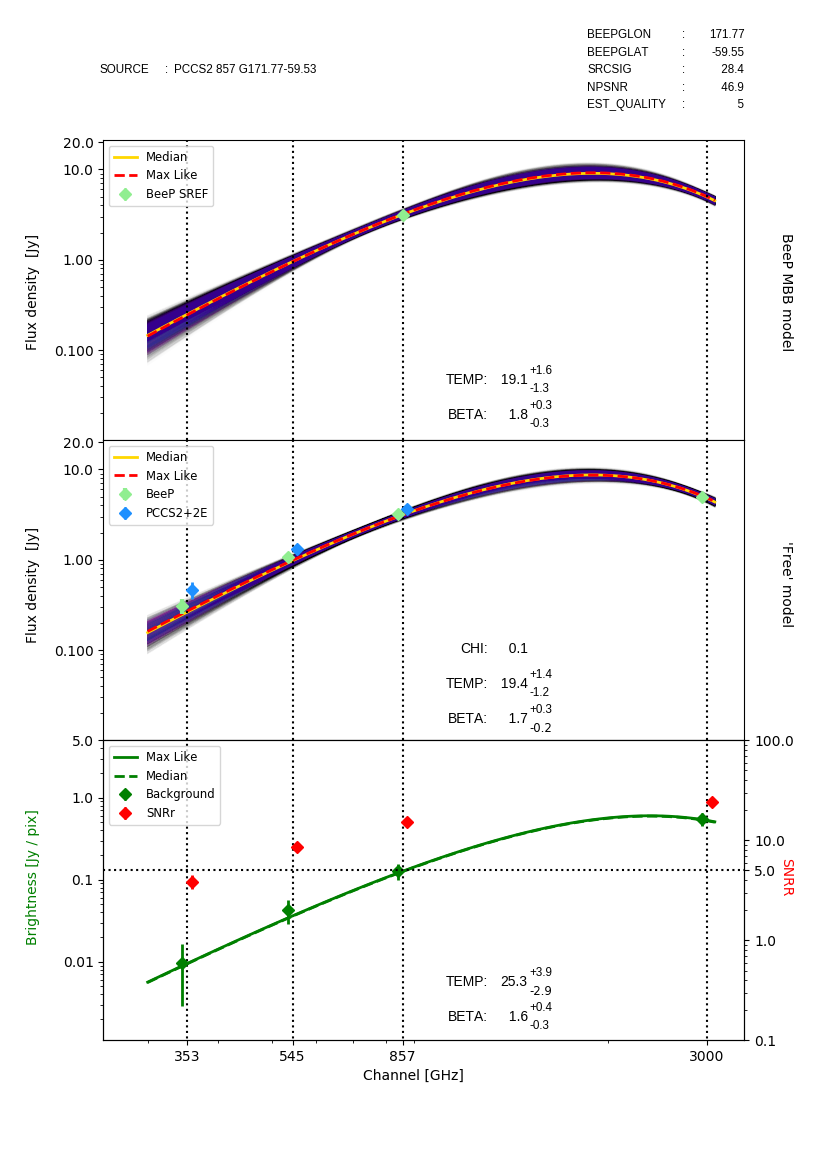}
                \caption{Results of {\tt BeeP} analysis for NGC\,895. See the caption of Fig.~\ref{fig:SEDPlot} for a full explanation. This case is also a very clean example of a well-determined model for source and background.}
                \label{fig:ThirdExample}
        \end{center}
\end{figure*}

\begin{figure*}[htbp!]
        \begin{center}
                \leavevmode
                \includegraphics[width=0.80\textwidth]{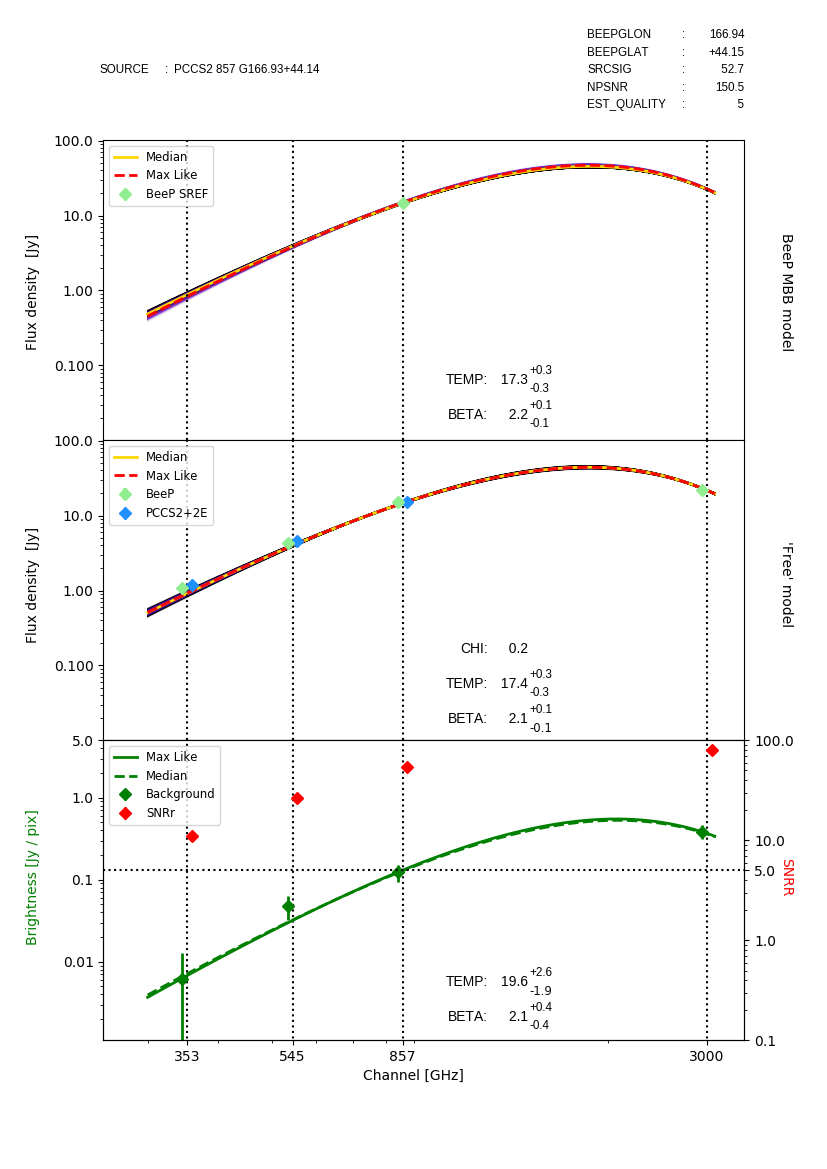}
                \caption{Results of {\tt BeeP} analysis for J091828.6+514223. See the caption of Fig.~\ref{fig:SEDPlot} for a full description.
                This is a strongly lensed galaxy at $z=5.2$ and appears as a relatively cold dusty source.}
                \label{fig:FourthExample}
        \end{center}
\end{figure*}

\begin{figure*}[htbp!]
        \begin{center}
                \leavevmode
                \includegraphics[width=0.80\textwidth]{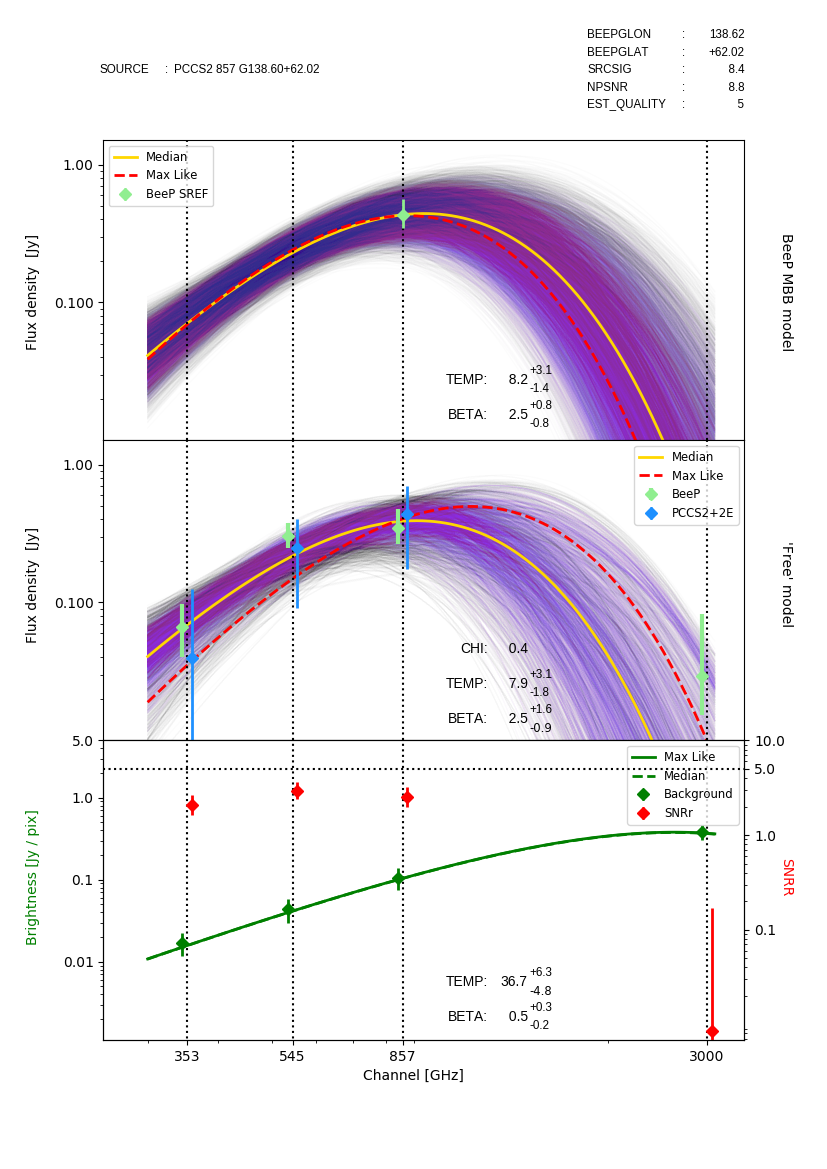}
                \caption{Results of {\tt BeeP} analysis for GEMS PLCK\,G138.6+62.0. See the caption of Fig.~\ref{fig:SEDPlot} for a full description. This is a source with fairly low S/N ratio with respect to the background, but {\tt BeeP} is able to find a good model for it. The Free model flux densities recovered by {\tt BeeP} have much lower uncertainties than those found in the PCCS2+2E catalogue.}
                \label{fig:FifthExample}
        \end{center}
\end{figure*}

\begin{figure*}[htbp!]
        \begin{center}
                \leavevmode
                \includegraphics[width=0.80\textwidth]{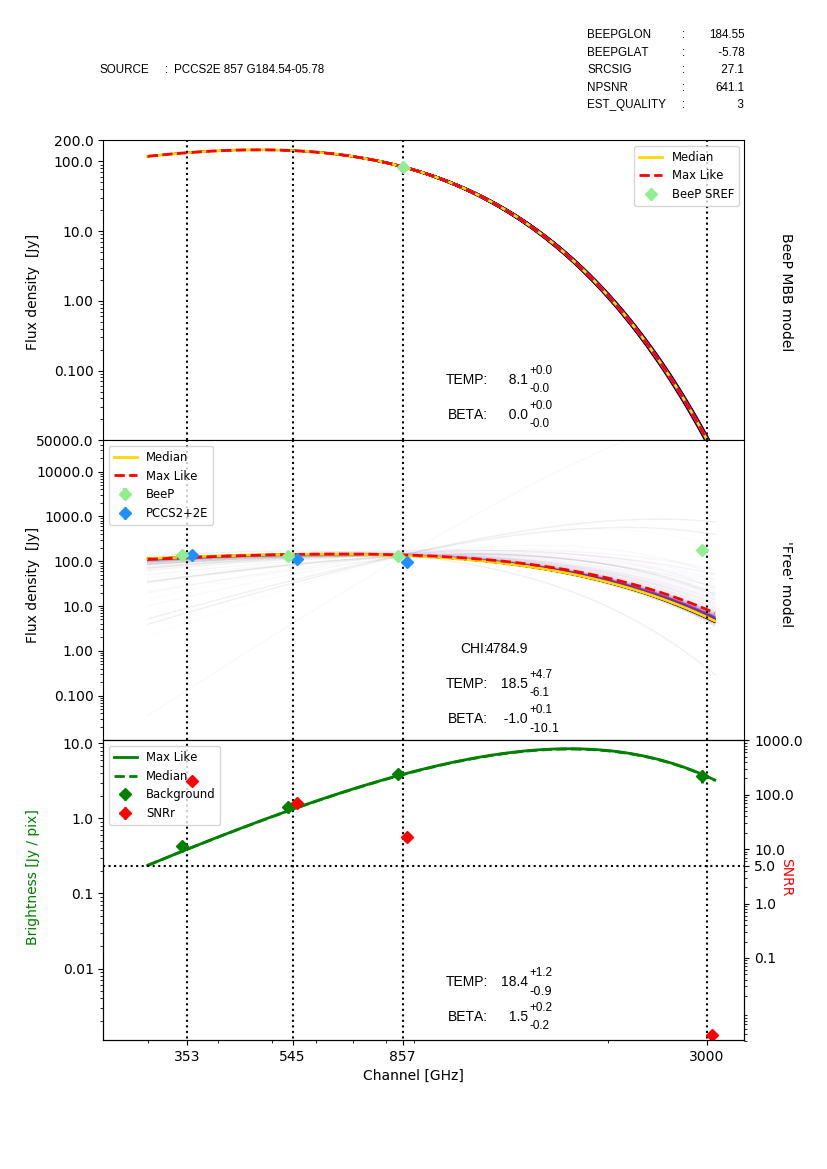}
                \caption{Results of {\tt BeeP} analysis for M\,1 (Crab Nebula). See the caption of Fig.~\ref{fig:SEDPlot} for a full description. This is a non-thermal source and {\tt BeeP}  has reduced \textsf{EST\_QUALITY} accordingly; the full likelihood is not able to find a reasonable value for the SED parameters, in particular the spectral index. The Free model fit does find parameters, since it is less constrained, but the high $\chi^2$ value indicates a very poor fit.}
                \label{fig:SixthExample}
        \end{center}
\end{figure*}

\begin{figure*}[htbp!]
        \begin{center}
                \leavevmode
                \includegraphics[width=0.80\textwidth]{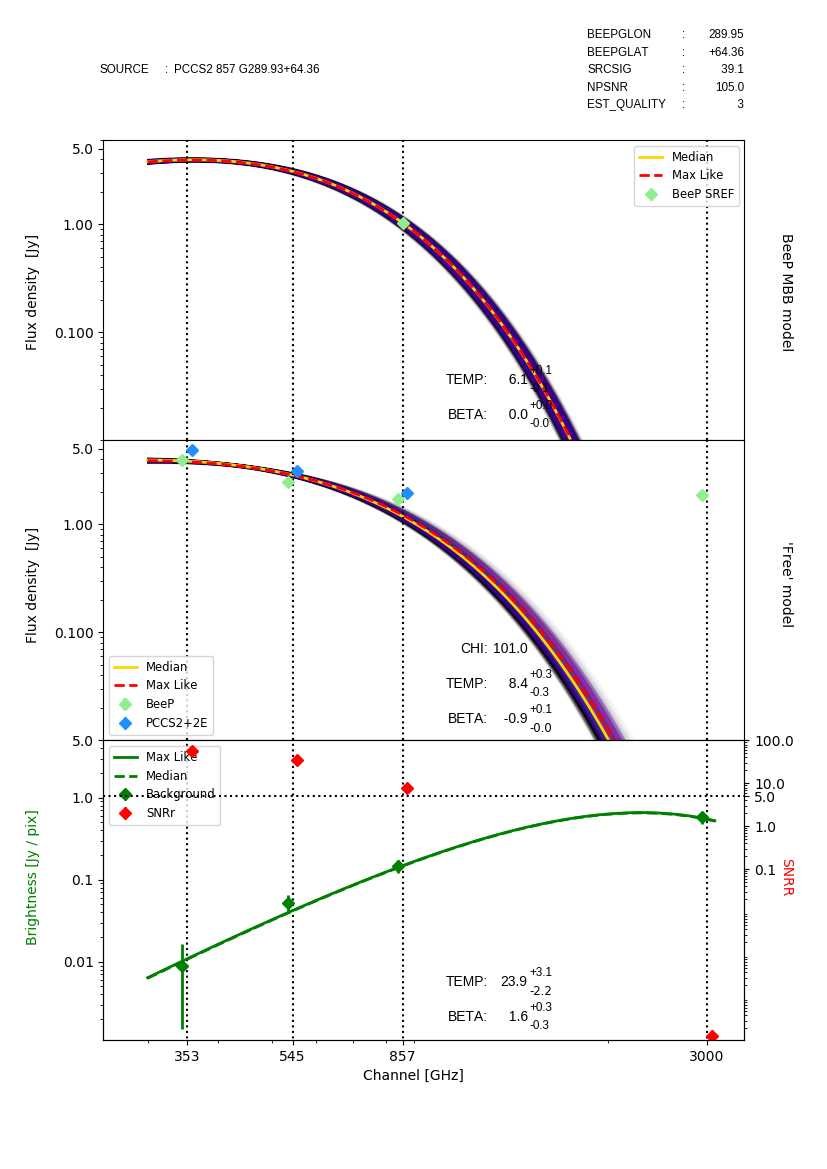}
                \caption{Results of {\tt BeeP} analysis for 3C\,273.  See the caption of Fig.~\ref{fig:SEDPlot} for a full description. As in the previous figure, this is a non-thermal source, and {\tt BeeP} also obtains very poor results (though not as extreme as in the previous case). The flux density of the source in the IRIS map is highly anomalous, but also has very low S/N ratio with respect to the well-determined background.}
                \label{fig:SeventhExample}
        \end{center}
\end{figure*}

\begin{figure*}[htbp!]
        \begin{center}
                \leavevmode
                \includegraphics[width=0.80\textwidth]{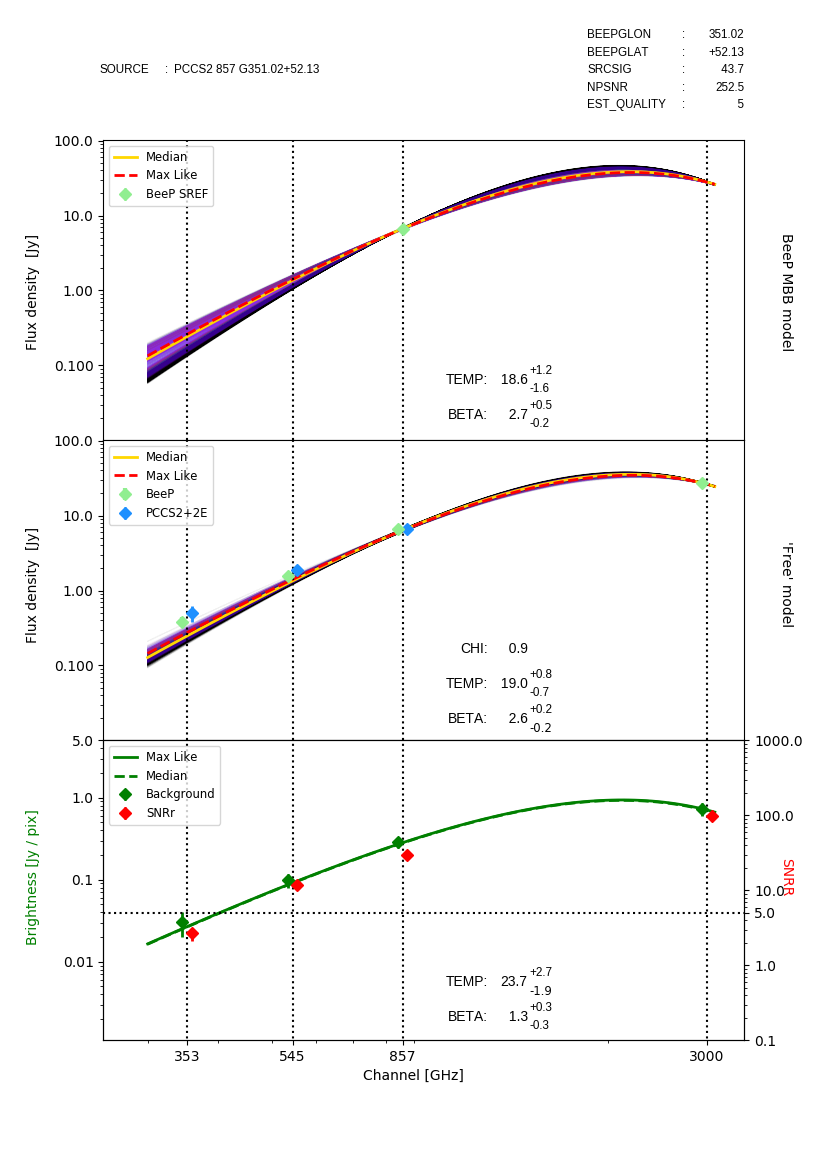}
                \caption{Results of {\tt BeeP} analysis for the brightest source in our ATLAS comparison field (HATLAS J144011.1-001719).  See the caption of Fig.~\ref{fig:SEDPlot} for a full description. Overall this is a clean case of a cold dusty source on a fairly warm background, and {\tt BeeP} obtains good results.}
                \label{fig:SeventhExample}
        \end{center}
\end{figure*}

\begin{figure*}[htbp!]
        \begin{center}
                \leavevmode
                \includegraphics[width=0.80\textwidth]{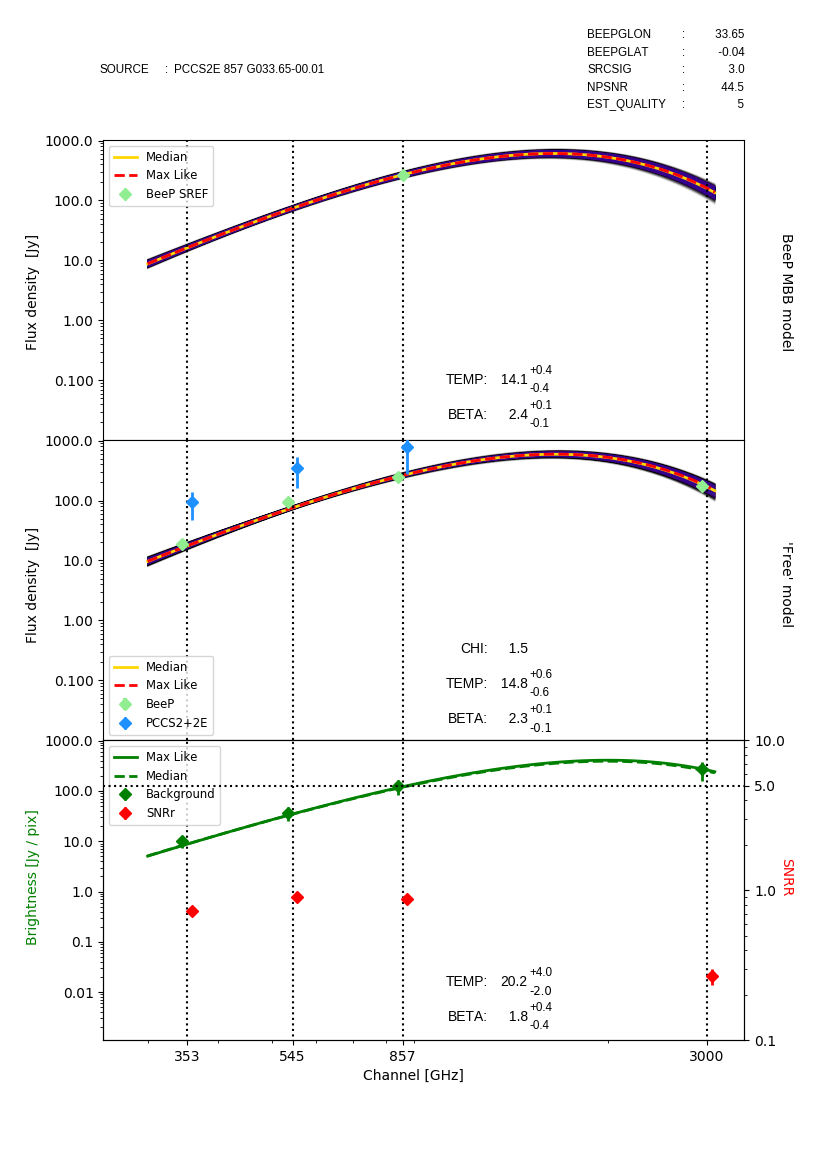}
                \caption{Results of {\tt BeeP} analysis for cold clump IRDC\,MSXDC G033.69$-$00.01.   See the caption of Fig.~\ref{fig:SEDPlot} for a full description.  We note the low temperature of this source (about 14\,K). {\tt BeeP}  obtains a good fit for this cold source on a warm background, but we see that the recovered flux densities (middle panel) are well below those obtained by PCCS2+2E. Examination of the source maps shows that it is surrounded by bright complex structure, which has confused the aperture photometry used by PCCS2+2E; indeed other flux-density algorithms (e.g., DETFLUX in PCCS2) obtain values closer to those of {\tt BeeP}.}
                \label{fig:SeventhExample}
        \end{center}
\end{figure*}

\begin{figure*}[htbp!]
        \begin{center}
                \leavevmode
                \includegraphics[width=0.80\textwidth]{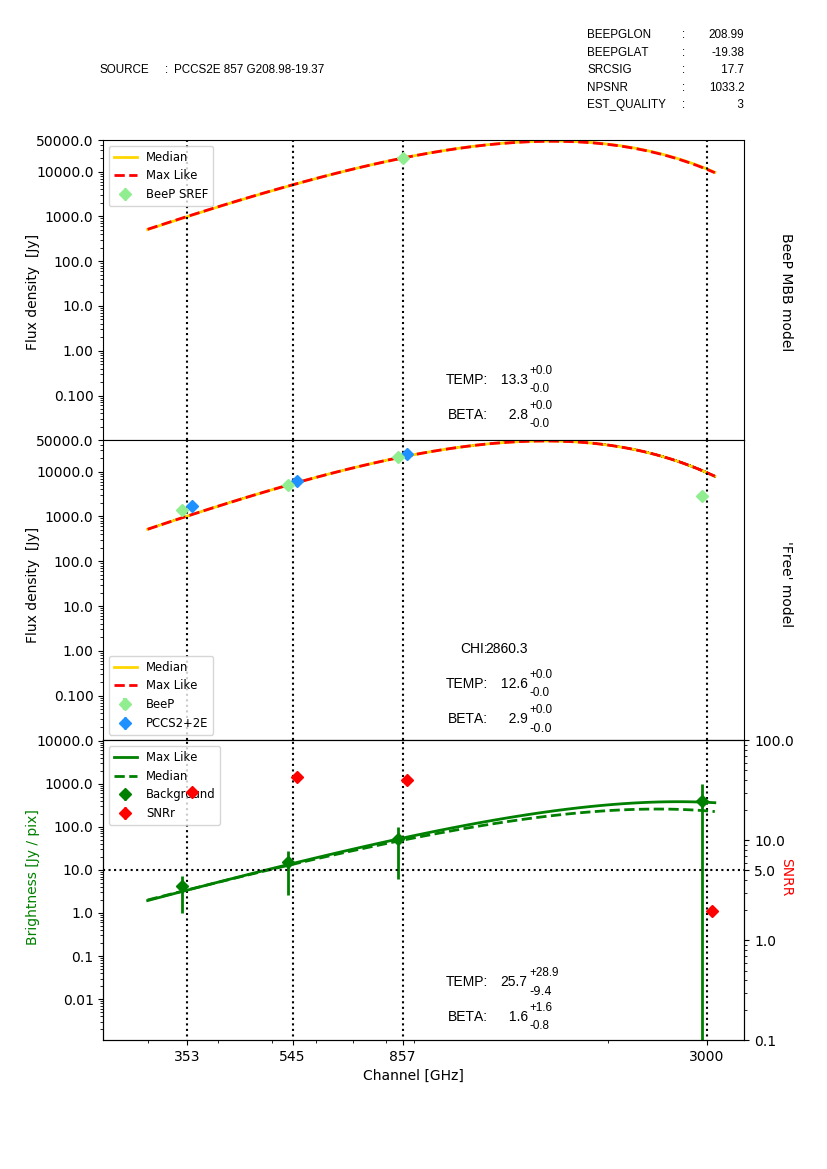}
                \caption{Results of {\tt BeeP} analysis for Orion IRC\,2.  See the caption of Fig.~\ref{fig:SEDPlot} for a full description. This case is similar to that of Fig.~\ref{fig:SecondExample}, where {\tt BeeP}  has reduced \textsf{EST\_QUALITY} due to the very low parameter uncertainties. However, in this case the $\chi^2$ of the Free model fit is very high, and this is clearly due to the fact that very tight constraints coming from the \Planck\ data do not allow a satisfactorily fit to the low flux density in the IRIS map.}
                \label{fig:SeventhExample}
        \end{center}
\end{figure*}

\end{document}